\documentclass[twocolumn, showpacs, aps, pre, superscriptaddress, amsmath, amssymb, nofootinbib,longbibliography,10pt]{revtex4-2}

\usepackage[T1]{fontenc}
\usepackage{lmodern}
\usepackage{graphicx}
\usepackage{upgreek}
\usepackage[mathscr]{eucal}
\usepackage{amsmath}
\usepackage{amssymb}
\usepackage{bm}

\usepackage[dvipsnames]{xcolor}
\usepackage{hyperref}
\hypersetup{colorlinks=true,citecolor = blue,urlcolor = black}
\usepackage[many]{tcolorbox}
\usepackage{cleveref}

\graphicspath{ {Figures/} }

\begin{document}

\title{Form and function in biological filaments: A physicist's review}

\author{Jan Cammann}
\thanks{equal contribution}
\affiliation{Interdisciplinary Centre for Mathematical Modelling and Department of Mathematical Sciences, Loughborough University, Loughborough, Leicestershire LE11 3TU, United Kingdom}
\author{Hannah Laeverenz-Schlogelhofer}
\thanks{equal contribution}
\affiliation{Living Systems Institute and Department of Mathematics and Statistics, University of Exeter, Stocker Road, Exeter EX4 4QD, UK}
\author{Kirsty Y. Wan}
\email[]{K.Y.Wan2@exeter.ac.uk}
\affiliation{Living Systems Institute and Department of Mathematics and Statistics, University of Exeter, Stocker Road, Exeter EX4 4QD, UK}
\author{Marco G. Mazza}
\email[]{M.G.Mazza@lboro.ac.uk}
\affiliation{Interdisciplinary Centre for Mathematical Modelling and Department of Mathematical Sciences, Loughborough University, Loughborough, Leicestershire LE11 3TU, United Kingdom}

\date{\today}

\begin{abstract}
Nature uses elongated shapes and filaments to build stable structures, generate motion, and allow complex geometric interactions. 
In this Review, we examine the role of biological filaments across different length scales. From the molecular scale, where cytoskeletal filaments provides a robust but dynamic cellular scaffolding, over the scale of cellular appendages like cilia and flagella, to the scale of filamentous microorganisms like cyanobacteria, among the most successful genera on Earth, and even to the scale of elongated animals like worms and snakes, whose motility modes inspire robotic analogues. We highlight the general mechanisms that couple form and function. We discuss physical principles and models, such as classical elasticity and the non-reciprocity of active matter, that can be used to trace unifying themes linking these systems across about nine orders of magnitude in length.

\end{abstract}

\maketitle

\section{Introduction}

The profound connection between form and function in biology has been quantified at least since D'Arcy Wentworth Thompson's influential work \cite{d2010growth}. Because geometrical features influence and direct biological activity through physical forces, it seems worthwhile to explore the link between geometry and biological function. In this Review, we examine a simple geometrical shape --the filament-- which is pervasive in the architecture of living organisms, and discuss the function of filamentous shapes and the physical advantages they offer.  

Filamentous shapes permeate every length scale in the biological world. Natural evolution has found that the geometrical properties of long aspect-ratio shapes can be used in a multiplicity of useful ways. 
The most important filamentous shape is the DNA double helix molecule \cite{watson1953molecular}, that carries a self-description of the organism and provides a biomolecular realization of a Turing-von Neumann tape \cite{al2023turing}.
The biopolymers forming the cytoskeleton provide structural integrity to the cell and form an intracellular transportation system. 
Cytoskeletal networks generate cytokinesis \cite{eggert2006animal}, migration \cite{paluch2006dynamic,lammermann2008rapid}, and apical contraction \cite{martin2009pulsed}, and even generate cytoplasmic streaming in the \textit{Drosophila} oocyte \cite{dutta2024self}.

At length scales of a few microns, cilia and flagella can generate fluid flow and active motion due to their hydrodynamical interactions \cite{lighthill1976flagellar,gilpin2020multiscale,lauga2009hydrodynamics} and can also synchronize to generate metachronal waves \cite{vilfan2006hydrodynamic,gilpin2020multiscale}. 
Microorganisms such as filamentous fungi form complex networks of hyphae following available water and resources, and can move cytoplasm along the network \cite{klein2004filamentous}. 
Filamentous fungi are also the group of fungi most often used in biotechnological processes and include perhaps the most important fungus in the history of humanity: \textit{Penicillium rubens}, which was used by Alexander Fleming to extract penicillin \cite{fleming1929antibacterial,fleming1941penicillin,houbraken2011fleming}.
The slime mold \textit{Physarum polycephalum} performs cytoplasmic flow across its network akin to a generalised form of peristalsis, which is optimised when the wavelength of the peristaltic wave is of the order of the size of the network \cite{alim2013random}; this filamentous network can even encode memory about nutrient location in its morphology \cite{kramar2021encoding}. 
In the prokaryotic domain, cyanobacteria were among the first microorganisms to straddle the transition from unicellular to multicellular organisms \cite{grosberg2007evolution,claessen2014bacterial}, and include numerous filamentous species.

\begin{figure*}
    \centering
    \includegraphics[width=0.99\linewidth]{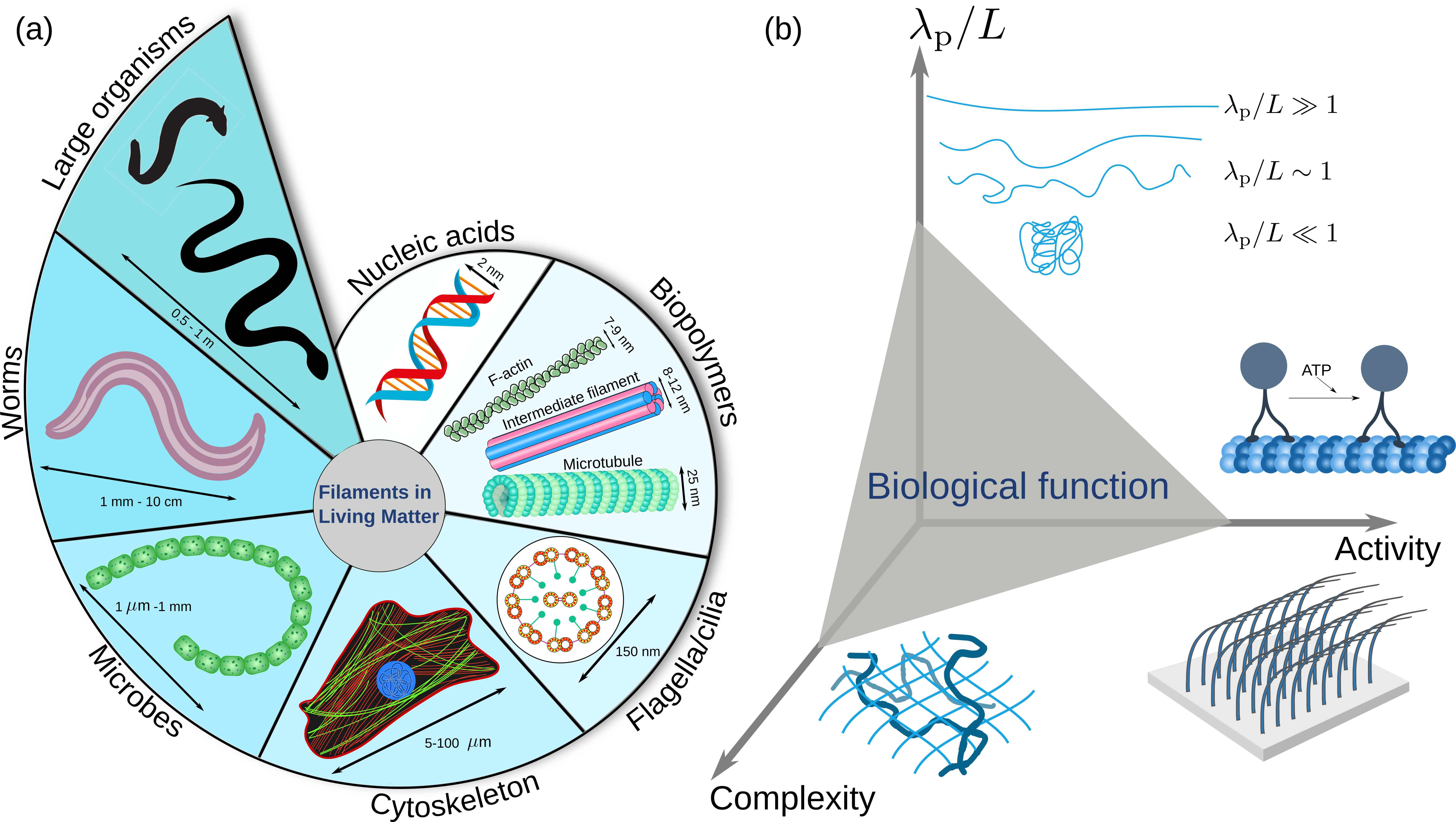}
    \caption{Filamentous shapes in biology. (a) Growing complexity of filaments across orders of magnitude in length, from nucleic acids that store information, to biopolymers that provide structural integrity to cells and form networks spanning the entire cytoskeleton. 
    Flagella and cilia self-assemble from polymers to generate motion. 
    Bacterial cells such as cyanobacteria self-organize into filamentous shapes. Animals, from worms and nematodes to snakes and eels utilize elongated shapes to achieve locomotion and collective behaviour. (b) Conceptual classification of biological function emerging from the three main physical properties of biological filaments: (\textit{i}) the ratio of persistence length to contour length, which determines whether a filament is flexible, semi-flexible, or rigid; (\textit{ii}) structural complexity, such as twisting or cross-linking into networks, assembly into cilia carpets, organisation of muscle fibres or even an entire organism; (\textit{iii}) activity, that is, the nonequilibrium conversion of chemical energy (ATP) into persistent motion. An archetypical example is the motion of kinesin on microtubules.}
    \label{fig:filaments_diagram}
\end{figure*}

At much larger length scales, organisms such as worms, snakes, and eels have elongated bodies that grant them ecological advantages. Worms, for example, can form entangled `blobs' that exhibit shear-thinning rheology \cite{deblais2020rheology} and reversible self-assembly \cite{patil2023ultrafast}.
The physical properties of single filaments can be combined to create emergent complexity and functionality, such as in networks of cross-linking cytoskeletal filaments \cite{gardel2004elastic} or in slime molds that can form more optimal foraging networks than Steiner's minimal tree \cite{nakagaki2004smart}.  

The systems described above span about nine orders of magnitude in size: from nucleic acids in the nanometre range in diameter, to snakes and eels in the metre range.    Figure \ref{fig:filaments_diagram}(a) shows a schematic representation of different biological filaments across these different length scales, and also illustrates the growth in complexity of biofilaments. Despite this apparent diversity, in this work, we posit that it is possible to identify and discuss general physical features of elongated shapes (from macromolecules to whole organisms) that depend on the geometrical form of filaments and, in turn, influence their biological function. 

In this context, we ask the following questions. Can we find similarities in the function of the systems depicted in Fig.~\ref{fig:filaments_diagram}(a)? For example, how do cytoskeletal filaments generate structural integrity in the cell? How do groups of cilia generate fluid flow, or how do large elongated organisms generate undulatory locomotion? Understanding how physical systems interact via simple rules to generate interesting collective behaviour can and will have novel applications. 
The condensed-matter motto of `\textit{More is Different}' proposed by Philip Anderson \cite{anderson1972more} will be a guiding principle in the present work. 
For example, a self-organizing robot exhibits behaviour intermediate between solid and liquid with adaptive capabilities such that it can flow over obstacles \cite{saintyves2024self}. This combination of flexibility and structural integrity is reminiscent of a cell's cytoskeleton, or even of snake locomotion. 
Each of these filamentous systems can serve as inspiration for new biosynthetic materials \cite{fladung2024lies}.
Examples include strain-stiffening metamaterials using PDMS slats that mimic cellular behaviour \cite{taale2023minimalistic}, or the use of carbon nanotubes to study the effect of crowding and filament stiffness on their microscopic dynamics \cite{fakhri2010brownian,fakhri2009diameter}.

Can we find some general motifs to unify systems as disparate as those highlighted above? In Fig.~\ref{fig:filaments_diagram}(b), we propose a conceptual diagram containing three main principles that govern the biological function of filaments based on their physical properties. 
We identify three main axes: (\textit{i}) \textit{activity}: every living organism uses adenosine triphosphate (ATP) as the currency of energy. Active matter is  defined as any system that converts energy into persistent or systematic motion \cite{ramaswamy2010mechanics,marchetti2013hydrodynamics}. Thus, active matter systems are intrinsically out of equilibrium, and have access to a variety of steady states not allowed in equilibrium systems.

(\textit{ii}) The \textit{ratio of persistence length to  contour length} is a fundamental physical parameter that determines the type of biological activity of a filament: from the rigid microtubules that provide the scaffolding of the cell to more flexible filaments like actin (or arguably the body of snakes and eels) which are more prone to reconfigurations.  
Physically, the rigidity and length of different biofilaments determine the class of models used to describe their mechanical and dynamical properties. If we consider the angle $\theta$ between the local tangent and a reference axis, thermal and biological fluctuations will bend a filament randomly, such that $\langle (\Delta \theta)^2 \rangle = \Delta s / \lambda_\mathrm{p}$ where $s$ is the arc length along the filament, and $\lambda_\mathrm{p}$ is called the filament's persistence length, which  is determined by the relative strength of the random fluctuations and the filament's bending rigidity \cite{doi1988theory} (see Sec.~\ref{sec:cytoskeleton}). On the one hand, if the observation length scale (or total length) is much smaller than $\lambda_\mathrm{p}$ a filament can be approximated as a smooth line enabling an elastodynamic approach.
In this regime, we can mathematically describe filaments as a curve in three-dimensional (3D) space, whose local orientation changes smoothly along their length. Box \ref{box:frenet} describes the Frenet--Serret equations used in differential geometry to characterize a curve.

On the other hand, at small length scales, such as inside the cell, the polymeric nature of biofilaments is consequential and this character is amplified by the nonequilibrium nature of the fluctuations. For example,  variations in polymer stiffness and connectivity of actomyosin networks 
by actin filament bundling and cross-linking proteins can systematically tune the mechanical response of actin networks from extensile to contractile
\cite{stam2017filament,banerjee2020actin}.

\begin{tcolorbox}[boxrule=0.2mm, colbacktitle=cyan!5!white, coltitle=black, every float=\centering, colback=cyan!5!white, title=\textsf{Box I |  Geometry of filaments}, label={box:frenet}]
\begin{minipage}[t]{1.0\linewidth}
    \vspace*{0pt}  
{\small \textsf{A filament can be described as a portion of a three-dimensional curve, whose local orientation changes smoothly along its length. The curve needs to be sufficiently smooth, that is, it is at least thrice continuously differentiable. The Frenet frame or trihedron is a local, right-handed, orthogonal coordinate system with basis $(\bm{t},\bm{n},\bm{b})$. 
We describe the curve with the position vector $\bm{r}(s)\equiv (x(s), y(s), z(s))$, where $s=\int_0^s \Vert  \frac{d\bm{r}(s')}{ds'} \Vert ds'$ is the arc length of the curve. The tangent unit vector to the curve is defined as $\bm{t}\equiv \frac{d \bm{r}}{ds}$. The normal unit vector $\bm{n}\equiv \frac{d\bm{t}}{ds}/|\frac{d\bm{t}}{ds}|$, and always satisfies $\bm{n}\cdot \bm{t}=0$. Furthermore, $\kappa\equiv |\frac{d\bm{t}}{ds}|$ is the local curvature of the curve $\bm{r}(s)$. Finally, the binormal vector $\bm{b}\equiv \bm{t} \times \bm{n}$ together with $\bm{t}$ and $\bm{n}$ form an orthonormal basis for space. These three vectors obey the Frenet--Serret equations \cite{millman1977elements,novikovfomaneko}
\[  \frac{d\bm{t}}{ds}=  \kappa \bm{n}\,, \quad
    \frac{d\bm{n}}{ds}=  -\kappa \bm{t} + \tau \bm{b}  \,, \quad
    \frac{d\bm{b}}{ds}=  -\tau \bm{n}\,, 
\]
from where $\tau$ is the torsion. Thus, only two parameters are required to describe a curve in 3D space. The Frenet trihedron can be useful in the description of the helical motion of free-swimming organisms \cite{crenshaw1993orientation,friedrich2009steering,bente2020high}, sperm cells \cite{jikeli2015sperm,friedrich2007chemotaxis}, chiral active particles \cite{sevilla2016diffusion}, and for gliding motility \cite{lettermann2024geometrical}.}}
\end{minipage}
\begin{minipage}[t]{\linewidth}
    \vspace*{0pt}
    \centering
    \includegraphics[width=0.6\linewidth]{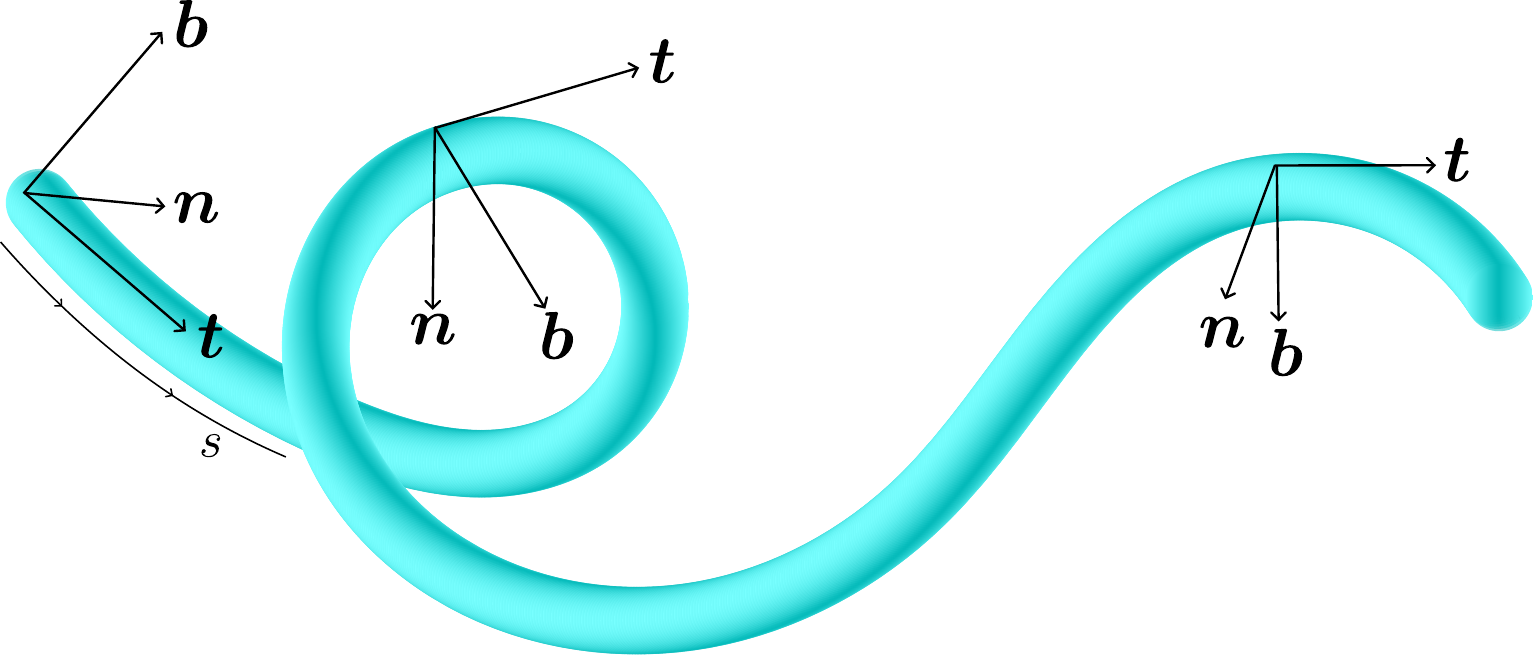}
\end{minipage}
\end{tcolorbox}

(\textit{iii}) The \textit{complexity} of the structural organisation. Filaments can link and bundle together in different ways and achieve novel mechanical functionalities.
Single globular actin proteins (G-actins) assemble in filaments (F-actin), which together with cross-linking proteins and a myriad of other molecules form  cytoskeletal networks. 
The fascin bundling protein can form filopodia -- tightly packed, parallel filaments of actin that exert protrusive forces in migrating cells \cite{mattila2008filopodia}, whereas the disordered actomysoin cortex can sustain tensive stresses. At larger scales, arrays of cilia can exhibit metachronal waves ( travelling waves of synchronized cilia phase \cite{solovev2022synchronization,poon2025dynamics} which generate self-propulsion in microorganisms or mucociliary clearance in ciliated epithelia \cite{wanner1996mucociliary}.
Multiple cells, in turn, can assemble into simple filaments, such as the \textit{Oscillatoria lutea} cyanobacteria, or form even more complex filamentous organisms, such as nematodes, eels and snakes. 
We discuss [Fig.~\ref{fig:filaments_diagram}(b)] in more detail below.

This Review is organised as follows.
Section~\ref{sec:cytoskeleton} discusses the role of biofilaments inside the cell. Section
~\ref{sec:cilia} reviews the role of filamentous cellular appendages, such as the locomotor function of cilia and flagella. In Sec.~\ref{sec:microbes} we discuss the biological properties of filamentous bacteria. Section~\ref{sec:snakes_on_a_plane} deals with macroscopic elongated organisms, such as snakes and eels. Finally, in Sec.~\ref{sec:roadmap} we outline a roadmap of future work and outstanding questions at this exciting frontier shared by biology and physics, to establish a common framework describing the physics of biological filaments (Fig.~\ref{fig:filaments_diagram}).

\section{Biopolymers \& cytoskeletal filaments} \label{sec:cytoskeleton}

The cytoskeleton is a hierarchically built, 3D polymer-based scaffold that spans the entire interior of a cell and defines cell morphology \cite{huber2013emergent}. The basic building blocks are few key proteins that assemble together into a broad range of filaments \cite{fletcher2010cell}. These filaments together with a host of proteins that bind to the filaments' sides or ends form a network that provides mechanical stability to the cell \cite{pegoraro2017mechanical}. 
Using only a few building blocks allows great flexibility and economy of materials. For example, cytoskeletal polymers achieve the same rigidity (elastic modulus) of gel-like substances with two orders of magnitude less material \cite{kollmannsberger2011linear,pegoraro2017mechanical}. 

At the molecular level, cytoskeletal filaments are linear polymers: monomers that are bound together as an unbranched chain. Physically, they can be distinguished by their stiffness. Thermal fluctuations induce random deformations of the filaments; thus, the thermal energy $k_\mathrm{B}T$, where $k_\mathrm{B}$ is Boltzmann's constant, and $T$ the temperature, is a natural scale to classify the ease of bending. The ratio
of bending stiffness $\beta_\mathrm{f}$ (see  Box~\ref{box:elastodyn}) to thermal energy $\lambda_\mathrm{p}=\frac{\beta_\mathrm{f}}{k_\mathrm{B}T}$, called  \textit{persistence length}, has dimensions of a length, and represents the typical length scale over which fluctuations in the filament's tangent decay. The persistence length can in fact be measured from the equilibrium fluctuations of the filament. Under thermal fluctuations, the tangent $\bm{t}$ to different parts of the filament performs a random walk, and obeys the diffusion equation. The correlation between two different points will then exhibit an exponential decay $\langle \bm{t}(s+\Delta s) \cdot\bm{t}(s)  \rangle =e^{-\Delta s/\lambda_\mathrm{p}}$.

By comparing $\lambda_\mathrm{p}$ with the filament length $L$ we can classify filaments as follows. 
If $\lambda_\mathrm{p} \ll L$ the filament is considered \textit{flexible} (the filament will exhibit highly coiled configurations); if $\lambda_\mathrm{p} \approx L$ the filament is termed \textit{semiflexible}, coiling and knotting are not present but thermal fluctuations can induce bending \cite{kroy1996force}; if  $\lambda_\mathrm{p} \gg L$ the filament is \textit{stiff}, and it appears as a stiff rod.

Biological filaments are subject to different types of deformations: extensional and compressional forces and moments that cause the filaments to bend and twist. Any deformation mode costs energy, which is continuously used in the cytoskeleton to maintain the cell in a nonequilibrium state and allows dynamic control and reconfiguration \cite{banerjee2020actin}. 
The theory of elasticity can help identify the associated bending energy $E_b$ \cite{landau2012theory}, see Box~\ref{box:elastodyn}. 

\begin{tcolorbox}[boxrule=0.2mm, colbacktitle=cyan!5!white, coltitle=black, every float=\centering, colback=cyan!5!white, title=\textsf{Box II | Elastodynamics},label={box:elastodyn}] 
{\small \textsf{Let us consider a small section of a macroscopic thin rod.  We assume that bending is in the $zx$ plane, where the $z$-axis is parallel to the unperturbed rod (deviations of a slightly bent curve from a plane, its torsion, are higher order than its curvature \cite{landau2012theory}). Because external forces are zero, it can be shown that a thin volume element of the rod deforms only by extension (or compression), although this amount varies from point to point along the rod. The relative extension of an element of length $dz$ is $x/R_c$, where $R_c$ is the local radius of curvature. The strain tensor component $u_{zz}=x/R_c$, and the stress tensor component $\sigma_{zz}=E\,u_{zz}$ \cite{landau2012theory}, where $E$ is Young's modulus. The bending energy per unit length $L$ is then $E_\mathrm{b}/L=\frac{1}{2}\sigma_{zz}u_{zz}=\frac{1}{2}E x^2/R_c^2$. Integrating now over the cross-sectional area yields $E_\mathrm{b}=\frac{1}{2}\beta_\mathrm{f} \kappa^2 L$, where $\beta_\mathrm{f}\equiv EI$ is the flexural rigidity or bending stiffness of the filament, $I\equiv\int x^2 dA$ the second area moment or the area moment of inertia. 
For a solid cylinder of radius $R$, $\beta_\mathrm{f}=\frac{1}{4}\pi R^4 E$, and for a hollow cylinder with inner radius $R_i$, $\beta_\mathrm{f}=\frac{1}{4}\pi (R^4-R_i^4) E$. In general, however, the radius of curvature of the bending need not be constant but might vary along the contour of the filament. 
For dynamical processes, the Euler--Lagrange equations for an energy function that includes kinetic energy, the elastic energy $E_b$ and an external load $q$ yield the Euler--Bernoulli dynamic beam equation \cite{civalek2011bending,kuvcera2017vibrations}
\begin{equation}\label{eq:euler_bernoulli}
     \rho_f A \frac{\partial^2 w}{\partial t^2}+ \frac{\partial^2}{\partial z^2}\left(\beta_\mathrm{f} \frac{\partial^2 w}{\partial z^2}\right) = q(z)
\end{equation}
where $w(z)$ is the deflection of the filament from the equilibrium along its length $z$, $\rho_f$ the mass density, $A$ the cross section, and $q(z)$ the load. These equations can be used to characterize buckling, e.g., in microtubules observed in living cells \cite{mehrbod2011significance,pallavicini2017characterization}.
}}
\end{tcolorbox}

Generally, one expects the bending energy to be proportional to the square of the deformation, i.e., of the curvature; if the symmetry of the system does not distinguish positive and negative curvature, linear terms in the curvature cannot be present. To lowest order, the energy of deformation is then quadratic 
\begin{equation}
    E_b=\frac{1}{2}\beta_\mathrm{f}\int_0^L \left( \frac{\partial^2 \bm{r}}{\partial s^2} \right)^2 ds\,,
\end{equation}
which is known as the \textit{worm-like chain} model or  Kratky--Porod model \cite{kratky1949rontgenuntersuchung}.

We can modify the Euler--Bernoulli Eq.~\eqref{eq:euler_bernoulli} (see Box~\ref{box:elastodyn}) and apply it to microscopic filaments that experience viscous drag and whose inertia is negligible \cite{wiggins1998flexive}.  Considering model A dissipative dynamics \cite{hohenberg1977theory,chaikin1995principles}
\begin{equation}\label{eq:modA}
    \gamma \frac{\partial w}{\partial t}= -\frac{\delta \mathcal{H}_b}{\delta w}+\xi(z,t)\,,
\end{equation}
where $\mathcal{H}_b=\frac{1}{2}\beta_\mathrm{f}\int_0^L \left( \frac{\partial^2 w}{\partial z^2} \right)^2 dz$ is the Hamiltonian associated to the bending modes, $\xi(z,t)$ is the thermal noise, which we take as independent in any point along the filament, with zero mean and white noise correlator $\langle \xi(z,t)\xi(z',t')  \rangle=2k_\mathrm{B}T\gamma\delta(z-z')\delta(t-t')$, and $\delta(\cdot)$ is the Dirac delta distribution. The hydrodynamic drag coefficient $\gamma$, famously first calculated by Lamb \cite{lamb1932hydrodynamics,jayaweera1965behaviour} in the case of a cylinder of length $L$ and radius $a$, for $a/L\ll1$, reads 
$\gamma={4\pi \mu}/\left[{\frac{1}{2}-\gamma_\mathrm{E}-\ln\left(\frac{1}{8}\mathrm{Re}\right)}\right]$, where $\mu$ is the fluid's viscosity, $\gamma_\mathrm{E}$ is the Euler--Mascheroni constant and $\mathrm{Re}$ is the Reynolds number (see Sec.~\ref{sec:cilia}). 
At low Reynolds numbers ($\mathrm{Re}\ll 1$), the parallel and transverse drag coefficients are $\gamma_\parallel=(2\pi \mu L)/\ln(L/a)$, $\gamma_\perp=2\gamma_\parallel$ \cite{doi1988theory}. 
Explicitly, Eq.~\eqref{eq:modA} yields the Langevin equation \cite{broedersz2014modeling}
\begin{equation}
    \gamma \frac{\partial w}{\partial t}= -\beta_\mathrm{f} \frac{\partial^4w}{\partial z^4}+\xi(z,t)\,.
\end{equation}
If the filament is under tension of magnitude $f$, the above equation is modified by the inclusion of a second derivative \cite{hallatschek2005propagation,hallatschek2007tension,winkler2020physics}
\begin{equation}
    \gamma \frac{\partial w}{\partial t}= -\beta_\mathrm{f} \frac{\partial^4w}{\partial z^4} +f \frac{\partial^2 w}{\partial z^2}+\xi(z,t)\,.
\end{equation}

\begin{table*}
    \centering
    \begin{tabular}{lcccc}
    \hline
    Biofilament  & Shear modulus $G$ $\quad$ & Persistence length $\lambda_\mathrm{p}$ $\quad$ & Young's modulus $E$ & Bending stiffness $\beta_\mathrm{f}$\\
    \hline
   units & Pa  & $\mu$m  & kPa & N m$^2$   \\  \noalign{\smallskip}
  
        Actin        & 283 \cite{janmey1991viscoelastic} ($\mathsection$) & $(10-20)$ \cite{boal2012mechanics} & $(1-3)\times 10^6$ \cite{mofrad2009rheology} &  $7.3 \times 10^{-26}$   \cite{gittes1993flexural}\\
        Fibrin       & 104 \cite{janmey1991viscoelastic} ($\mathsection$) & $(0.1-1)$ \cite{boal2012mechanics} &  $(1-15)\times 10^3$ \cite{collet2005elasticity} \\
        Vimentin     & 32 \cite{janmey1991viscoelastic} ($\mathsection$) & 0.4 \cite{schopferer2009desmin} & & $(4-12)\times 10^{-27}$ \cite{walter2011elastic}\\
        Keratin      &  & $(0.3-0.4)$ \cite{lichtenstern2012complex} & $6\times 10^3 - 10^8$ \cite{fudge2003mechanical,mckittrick2012structure}  \\
        Microtubules  & 34 \cite{janmey1991viscoelastic} ($\mathsection$) & $(4-6)\times 10^3$ \cite{boal2012mechanics}  & $(1-50)  \times 10^6 $ \cite{gittes1993flexural,sharma2020length} &  $2.2 \times 10^{-23}$ \cite{gittes1993flexural}\\
        Cellulose     &                          &                  &   $138 \times 10^6$ \cite{nishino1995elastic} \\
        Bacterial flagella   &  & $1-40$ \cite{trachtenberg1992rigidity}  & $(0.1-380)\times 10^6$ \cite{shen2022bending} & $(0.1-100) \times 10^{-23}$ \cite{shen2022bending}\\
        Bacterial flag. hook & & $(10-130)\times10^{-3}$ \cite{sen2004elasticity,nord2022dynamic} & $(0.01-1)\times10^6$ \cite{nord2022dynamic} & $(0.001-300)\times10^{-26}$ (*) \cite{sen2004elasticity,nord2022dynamic}\\
        Motile cilia & & $(0.7-15)\times10^5${(\textdagger)} & $(0.01-120)\times 10^6$(\textdaggerdbl) & $(0.3-6)\times 10^{-21}$ \cite{okuno1979direct,ishijima1994flexural}\\
        Fil. cyanobacteria & & 
        $(0.4-11.2)\times10^3$ \cite{faluweki2022structural, boal2010shape}
        & $(16-46)\times 10^3$ \cite{faluweki2022structural} 
        &$(1-20)\time10^{-17}$ \cite{faluweki2022structural, kurjahn2024quantifying}\\
        \textit{C. elegans} (adult) & & & $(80-140)$(**)\cite{backholm2013viscoelastic} & $\sim10^{-13}$  \cite{backholm2013viscoelastic}
        \\
        Eels and snakes   & & & $100-1000$ \cite{guo2008limbless,zhang2021friction} &$10^{-2} - 10^{-3}$ \cite{long1998muscles}
    \end{tabular}
    \caption{Physical parameters characterizing examples of biological   filaments, from cytoskeletal structural components and flagella, to filamentous organisms, up to the scale of snakes. ($\mathsection$) For shear-modulus measurements, strains were imposed by rheological measurements ranged from 1 to $5\%$ and frequencies from 1 to 10 rad/s \cite{janmey1991viscoelastic}. (*) Large range due to torsional load dependent stiffening \cite{nord2022dynamic}. (\textdagger) Estimated from $\lambda_\mathrm{p} = \beta_\mathrm{f} / (k_B T)$. ({\textdaggerdbl}) Estimated from $\beta_\mathrm{f}$. See text for a discussion about this large range. 
    (**) Approximating \textit{C. elegans} as uniform rods. Missing values have not yet been measured (or we are not aware of them), or are not easily defined (e.g.\ shear modulus of a snake). We also note that persistence length $\lambda_\mathrm{p}$ and bending stiffness $\beta_\mathrm{f}$ are not always directly related through thermal fluctuations on account of the nonequilibrium nature of the active processes driving biofilaments.}
    \label{tab:rheology}
\end{table*}

In addition to elasticity, a fundamental aspect of biological filaments is the fact that they are active, that is, they consume ATP to produce motion. Specifically, the cytoskeleton is highly dynamic: its constituents can be assembled and disassembled as the need arises due to permanent ATP consumption and energy dissipation. 
The overdamped dynamics of a particle can be described by a Langevin equation
\begin{equation}
    \dot{\bm{x}}=\bm{F}(\bm{x})+\bm{\xi}(t)\,,
\end{equation}
where $\bm{F}$ represents the deterministic drift driving the system, and $\bm{\xi}$ is a vector whose components are zero-mean white Gaussian noise with variance $\langle\xi_i(t)\xi_j(t')\rangle=2D_{ij} \delta(t-t')$, and $D_{ij}$ is the diffusion tensor. 
The drift may contain a conservative term in addition to a nonconservative one representing the self-propulsion $\bm{F}(\bm{x})=(\bm{v}-\nabla U)$. By coarse-graining, one can obtain the Fokker--Planck equation for the probability density $P(\bm{x},t)$ written in conservative form
\begin{equation}
\frac{\partial P(\bm{x},t)}{\partial t}+ \nabla \cdot \bm{J}=0\,,
\end{equation}
where $\bm{J}=\bm{F}P-\nabla \cdot (\bm{D}P)$ is the probability flux. If the process has reached steady state (i.e., after some time $\frac{\partial P_\mathrm{ss}}{\partial t}=0$), by Helmholtz--Hodge decomposition, the drift field can be expressed as the sum of reversible and irreversible parts $\bm{F}=\bm{F}_\mathrm{rev}+\bm{F}_\mathrm{irr}$, where $\bm{F}_\mathrm{rev}=\bm{D}\cdot \nabla \ln P_\mathrm{ss}+\nabla\cdot\bm{D}$ and $\bm{F}_\mathrm{irr}=\bm{J}_\mathrm{ss}/P_\mathrm{ss}$. For constant noise, the reversible part can be written as the gradient of a potential (thus is curl-free), while $\nabla\cdot \bm{F}_\mathrm{irr}=0$ so it is the curl of some vector \cite{zia2007probability,da2023entropy}. The nonequilibrium probability flux, and in particular the rotational drift $\bm{F}_\mathrm{irr}$ is a characteristic and strong measure of how far active matter systems are from equilibrium \cite{battle2016broken,herminghaus2017phase,cammann2021emergent}.

\begin{figure}[t]
    \centering
    \includegraphics[width=0.9\linewidth]{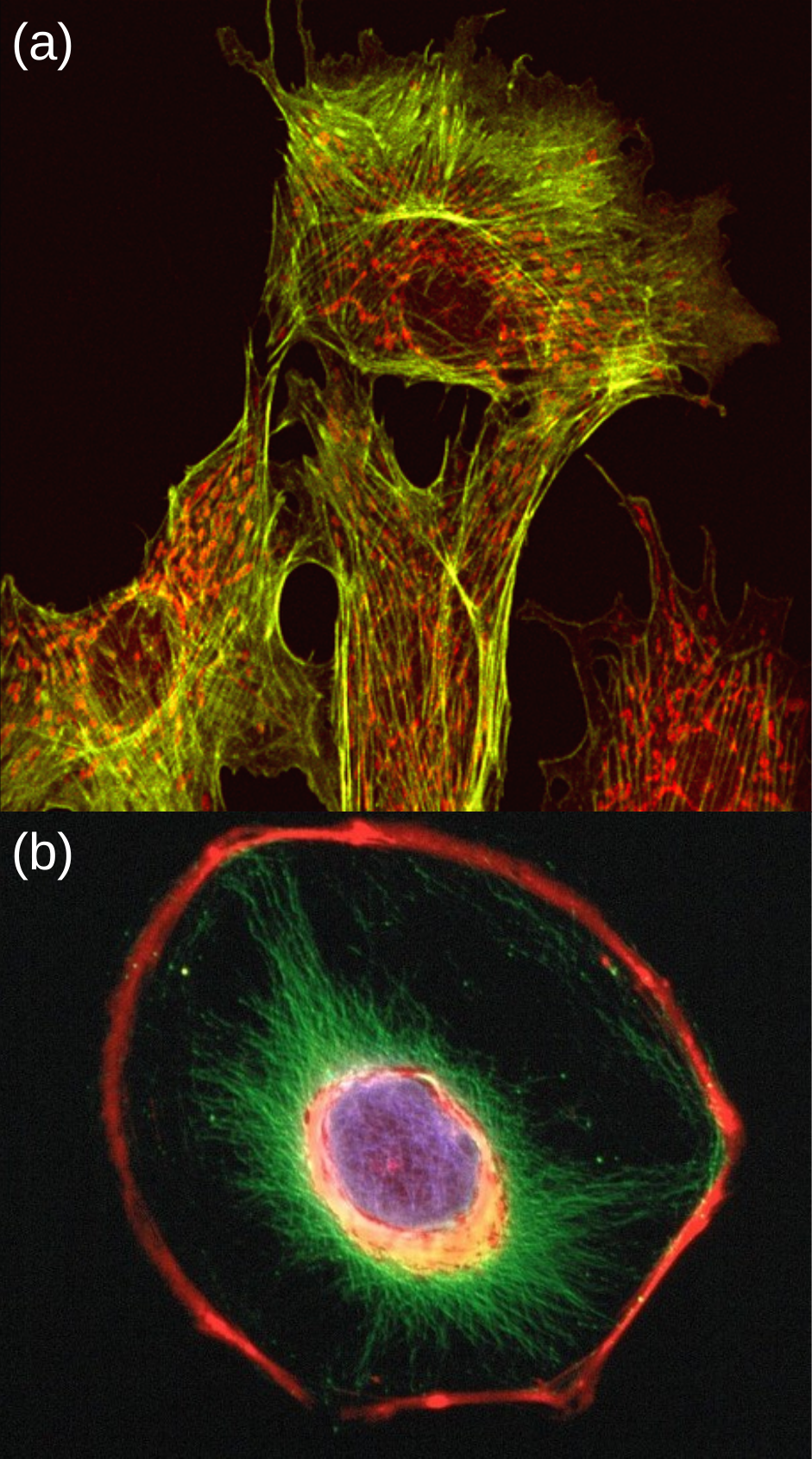}
    \caption{Cytoskeletal filaments. (a) Endothelial cells from cow lung with two components of the cytoskeleton: microtubules (green) and actin filaments (red) \cite{CIL219}.
    (b) Confocal microscope image of a mouse 3T3 fibroblast cell. The nucleus has been stained blue, and two components of the cytoskeleton, actin microfilaments and intermediate filaments, are stained red and green respectively \cite{CIL39004}.}
    \label{fig:cytoskel}
\end{figure}

Cytoskeletal filaments are grouped into three main classes: microtubules, actin filaments, and intermediate filaments \cite{fletcher2010cell}, see Fig.~\ref{fig:cytoskel}. 
The main differences among these three types of filaments are their mechanical stiffness, polarity, and the molecular motors associated with them \cite{fletcher2010cell}. 
We now discuss these three classes of biofilaments in turn.

$\bullet$ \textit{Microtubules} are hollow and polar tubular structures, which grow by the addition of tubulin dimers ($8$ nm in length) bound to GTP (guanosine-5'-triphosphate) molecules \cite{brouhard2018microtubule,zhou2023structural}. 
Microtubules are the stiffest cytoskeletal filaments, see Table~\ref{tab:rheology}. They are formed by $\alpha$- and $\beta$-tubulin protein dimers.  The protein dimers  polymerize into linear chains, called protofilaments, which arranged as a hollow cylinder constitute a microtubule. Most eukaryotic cells contain microtubules with 13 protofilaments, characterised by  an internal diameter of about $15$ nm and an external diameter of about $25$ nm \cite{boal2012mechanics,dustin2012microtubules}. 
Microtubules perform crucial functions in the life of a cell. During cell division, microtubules nucleate from the centrioles that organize the mitotic spindle \cite{azimzadeh2010building}. Microtubules are also the building blocks of eukaryotic cilia and flagella \cite{grimstone1966observations,amos1974arrangement}, see Sec.~\ref{sec:cilia}.

The considerable stiffness of microtubules is reflected in a persistent length $\lambda_\mathrm{p}$ of about 5 mm, which can span the entire length of an animal cell \cite{fletcher2010cell}, and a Young's modulus ranging from one to 50 GPa, depending on their length, or on the formation of network structures \cite{gittes1993flexural,sharma2020length}, see Table~\ref{tab:rheology}. Classic experiments employed thermal fluctuations to measure the flexural rigidity of microtubules $\beta_\mathrm{f}=2.2\times10^{-23}$ N m$^2$ \cite{gittes1993flexural}. The contour length of microtubules is typically about $10~\mu$m (but in axons they can even reach $100~\mu$m) \cite{pampaloni2006thermal,bray2000cell}. Thus, the ratio $\lambda_\mathrm{p}/L\gg 1$, indicating a strong structural role. In fact, microtubules are involved in key (re)structural processes of the cell; for example, they provide the tracks for kinesin and dynein-driven transport of organelles and vesicles within the cell \cite{welte2004bidirectional,barlan2017microtubule}, and are central components of cell mechanics during tissue morphogenesis \cite{singh2018polarized,roper2020microtubules}.

Microtubules are highly dynamic systems, where the cell-regulated nonequilibrium processes of polymerisation and depolymerisation \cite{dogterom1993physical,fygenson1994phase} of a microtubule coexist (known as `\textit{dynamic instability}' \cite{mitchison1984dynamic}) and extend to a large fraction of the microtubules in a cell. 
The tau protein (short for `tubulin associated unit') maintain the stability of microtubules in the axons of neurons \cite{weingarten1975protein}, however, when  hyperphosphorylation occurs, the microtubules disintegrate and the tau proteins form aggregates (neurofibrillary tangles, which are tangles of tau filaments) that are associated with neurodegenerative diseases \cite{palmqvist2020discriminative,chen2024mathematical}.

A single microtubule never reaches a steady state length but persists in prolonged states of polymerisation and depolymerisation that interconvert infrequently \cite{desai1997microtubule}. 
The cell uses the free energy of GTP hydrolysis to drive the nonequilibrium polymerisation; this free energy is $\approx 7.5$ kcal mol$^{-1}$ under standard conditions, or $\approx 12.5$ kcal mol$^{-1}$ \textit{in vivo} \cite{nelson2008lehninger,desai1997microtubule}.

Recent measurements have shown that as microtubules grow or shrink, they undergo conformational changes. The GTP-tubulin dimers are initially curved at an angle of $\approx 12^\circ$; after binding they straighten but exert a stress on the rest of the filament structure; finally, after GTP hydrolysis the $\alpha$-tubulin shortens by $3$ \r{A} and releases the stress, which
can be harnessed to perform work \cite{coue1991microtubule,koshland1988polewards,brouhard2018microtubule}. These conformational changes in the dimers generate long-range interactions along the microtubules that modulate their growth \cite{brouhard2018microtubule}. 

Perhaps the most important example of microtubule self-organisation is the mitotic spindle, which partitions the set of chromosomes from a cell into two copies for the daughter cells \cite{wittmann2001spindle,gadde2004mechanisms}. The polarity of microtubules is essential in this process, as each half of the spindle contains uniformly oriented microtubules, with their minus-ends at the pole (towards the centrosome) and their plus-ends extending away. 
The initial growth of microtubules around centrosomes is based on their dynamic instability, with random phases of growth and shrinkage probing the cytoplasm for kinetochores (the so-called `search-and-capture' model \cite{kirschner1986beyond}). 
That many classes of kinesins play a role in the spindle formation is evidence that the mitotic spindle is a nonequilibrium self-organised structure \cite{gadde2004mechanisms}. 

Microtubules and actin filaments strongly interact. And this interaction is responsible for regulating cell shape and polarity \cite{dogterom2019actin}. Microtubules are anchored to actin filaments, which maintain the position of the microtubule aster \cite{xie2022contribution}. Additionally, very recent work reports evidence that microtubules may serve as sensors and signalling platforms that guide the organization of the cell around the centrosome \cite{schaeffer2025Microtubule}.

$\bullet$ \textit{Actin filaments} are composed of (globular) G-actin monomers which readily assemble into long strings called F-actin (``F'' for filamentous) of about 8 nm in width \cite{boal2012mechanics}. Actin filaments are about 300 times less stiff than microtubules, with $\beta_\mathrm{f}=7.3\times10^{-26}$ N m$^2$ \cite{gittes1993flexural},   a persistence length  $\lambda_\mathrm{p}\approx 18~\mu$m \cite{gittes1993flexural}, and with a Young's modulus $E\simeq (1-3)$ GPa, which is comparable with Young's modulus for bone ($9$ GPa) \cite{mofrad2009rheology}. Interestingly, actin is less resistant to bending than microtubules only because it is thinner, and not because it is more compliant than tubulin \cite{gittes1993flexural}. It is also worth comparing the Young's moduli of actin and tubulin. The measurements of $\beta_\mathrm{f}$ for a microtubule implies a Young's modulus $E\sim 1.2$~GPa, while for actin $E\sim 2.6$~GPa \cite{gittes1993flexural}. This is comparable with $E\sim 2-8$~GPa of silk (\textit{Bombyx mori}) \cite{lepore2016effect,greco2022artificial,wainwright1982mechanical}, or with $E\sim 2.5$~GPa of keratin (from birds' feathers) \cite{bonser1995young}. 
Interestingly, the values of these Young's moduli are of the same order of magnitude. 
The persistence length $\lambda_\mathrm{p}$ of actin is of the order of $10-20~\mu$m and its contour length is about the same \cite{burlacu1992distribution,gittes1993flexural,karimi2021hydrodynamic}. Thus, the ratio $\lambda_\mathrm{p}/L\approx 1$, making actin a semiflexible polymer. Because of these physical properties, actin exhibits many diverse functions in the cell: from cell movement to tension generation \cite{banerjee2020actin}.

Both actin filaments and microtubules are polar filaments, that is, they have a positive and a negative end due to an asymmetry at the molecular level. Hence, they can function as train tracks along which molecular motors move. Dynein and kinesin are the motor proteins associated to microtubules; by means of ATP hydrolysis they can transport vesicles, mRNA, and other proteins. Myosins are the motor proteins associated to actin. They are a superfamily \cite{hartman2012myosin}: at least 35 classes of motor proteins associated to actin filaments are known \cite{odronitz2007drawing} and used for vesicle transport; they are responsible for muscle contraction in myocytes \cite{sweeney2010structural}, and are macromolecular filaments. Some members of this superfamily, the non-muscle myosin II (NM II) are even involved in cell adhesion and migration \cite{vicente2009non}.

Actin is present in high concentrations in the actomyosin cortex of animal cells, a layer attached to the inner surface of the cell membrane, and its function is to generate tension and regulate force and cell shape \cite{etienne2015cells,saha2016determining,kumar2021actomyosin}. Actin is the main constituent of lamellipodia, two-dimensional structures responsible for movement of e.g.\ epithelial cells \cite{cardamone2011cytoskeletal}, and is involved in cell adhesion \cite{Schwarz2013,garcia2015physics}.  The versatility of actin is but an example of the fact that the molecular organisation of biofilaments is extremely flexible; as a further example, amyloid fibrils can change bending rigidity by four orders of magnitude due to varying backbone interactions in the cross-$\beta$ structure \cite{knowles2007role}.

Actin plays a major role during embryonic development, as the cells have to undergo a specific sequence of physical and biological changes \cite{heisenberg2013forces}. 
Physically, cell cortical forces, cell adhesion and motility together produce self-organisation both at the tissue level and within single cells, e.g., reorientation of the polarity plane \cite{aigouy2010cell}. Another physical factor involved in tissue morphogenesis is actin-myosin flows (or cortical flow), that is, a cell-scale internal flow of actin components of the cortex \cite{bray1988cortical}. Various aspects are at play here: the character of the actin-myosin flow dynamics (pulsatile versus continuous) and direction (centripetal or anisotropic)  influence force generation; additionally, to transmit forces to other cells, this flow needs to be coupled to adhesion complexes at the cell membrane \cite{heisenberg2013forces}. Although much has been understood about force generation in cells and tissue, there are still many open questions about how the exquisitely complex self-organised process is controlled by the cell.

$\bullet$ \textit{Intermediate filaments} (IFs) are the most flexible of the three classes of cytoskeletal filaments, but they are also a very diverse class of filaments \cite{koster2015intermediate}. Estimates of their persistence length range from  $\lambda_\mathrm{p}=415\pm45$ nm  \cite{schopferer2009desmin} to $\lambda_\mathrm{p}=2.1\pm 0.1~\mu$m \cite{Noeding2012} (both studies for human  vimentin). The typical contour lengths for IFs are much larger than $\lambda_\mathrm{p}$. Thus, $\lambda_\mathrm{p}/L \ll 1$, which makes IFs flexible polymers. They are the least studied of the cytoskeletal filaments. Their main function is to  provide mechanical strength to cells, but they appear to be involved in cell shape and motility and the positioning of organelles \cite{lowery2015intermediate}.

There are six classes of IFs  \cite{parry1992intermediate}, which include, e.g., keratin, desmin,  vimentin, and nuclear lamin. Structurally, IFs exhibit a hierarchical structure \cite{chernyatina2015intermediate}. The alpha-helical domains form a coiled-coil structural motif, composed of two parallel in-register chains. Two chains aggregate in parallel to generate an IF molecule, and then two such molecules assemble in an antiparallel fashion to form a four-chain structural unit \cite{parry1992intermediate}.  IF proteins are highly charged, thus representing versatile polyampholytes with multiple functions due to their strong self-assembling ability \cite{herrmann2016intermediate}. 
IFs are built as  complex networks of ionic interactions formed by the acidic rod and the basic head domain, and, unlike microtubules and actin filaments, IFs are non-polar, and lack enzymatic activity \cite{herrmann2016intermediate}. IFs form networks of filaments with important mechanical properties: upon being mechanically stressed, the network becomes stiffer \cite{janmey1991viscoelastic,wen2011polymer,herrmann2016intermediate}.

\textit{Cytoskeletal networks and viscoelastic properties}. 
The viscoelastic properties of cytoskeletal filaments are closely linked to their biologic function \cite{mofrad2009rheology,pritchard2014mechanics}, because their extreme aspect ratio  and rod-like structure dominate the rheological behaviour of the cell, and changes in their structure may cause gel-sol transitions observed when cells are activated or begin to move \cite{janmey1991viscoelastic}. Measurements of the storage modulus $G'$ for F-actin and fibrin  suggest that the molecular basis of the viscoelasticity of these two types of networks is similar and may be approximated by the steric hindrance of diffusion of long, interpenetrated filamentous polymers \cite{janmey1991viscoelastic}. 
While the viscoelastic measurements of these biofilaments can be explained with isotropic models, there is ample evidence that microtubules and F-actin undergo a transition from isotropic to liquid crystalline solutions under the
influence of shear stresses \cite{kerst1990liquid}, other macromolecules \cite{suzuki1989osmoelastic}, or even
spontaneously \cite{hitt1990microtubule,Viamontes2006,hess2017non}, because of the strong thermodynamic drive for long rod-like filaments to form liquid-crystalline arrays \cite{onsager1949effects,flory1956phase}.  

Cytoskeletal filaments play a fundamental role in the mechanical properties and the biological function of the cell. This functional importance is due to not just the physical properties of isolated filaments, but, perhaps more crucially, to their structural organisation.  Most cytoskeletal filaments form networks, and by controlling their architecture and connectivity cells achieve remarkable material properties \cite{wagner2006cytoskeletal,gardel2008mechanical}.

Actin filaments are assembled by actin-binding proteins into branched, cross-linked networks; the list of such proteins is vast but includes, e.g., $\alpha$-actinin, $\beta$-spectrin, filamin, and fimbrin. These complex, dynamically-reassembling, dendritic networks generate force at the cell periphery that creates membrane protrusions (lamellipodia), that in turn produce the cell's directed motility, essential in processes such as embryonic development, wound healing, and the immune response \cite{pollard2003cellular}.

Generally, the elasticity of semi-flexible polymers can have two separate origins: 1.\ \textit{entropic elasticity}, which is due to the reduction in the number of available microstates when a filament is stretched vis-a-vis in an equilibrium state subject to thermal fluctuations; 2.\ \textit{enthalpic elasticity}, which is caused by the change in the equilibrium distance of the molecular bonds in a filament, such as when a filament is deformed (enthalpic elasticity does not depend on thermal fluctuations). The specific details of the network will dictate the relative weight of these two types of elasticity.

The worm-like chain model can be used to derive an important result in the rheology of cytoskeletal filaments \cite{marko1995stretching}. Considering a filament under strong tension $f$ aligned with the $z$-axis, there will only be small fluctuations of the tangent vector $\bm{t}\equiv (t_x,t_y,t_z)$ around $\hat{\bm{z}}$. We can then write $t_z= 1-\bm{t}_\perp^2/2+\mathcal{O}(\bm{t}_\perp^4)$. The bending energy is then
\begin{equation}
    \frac{E_b}{k_\mathrm{B}T}=\frac{1}{2} \int_0^L \left[\lambda_\mathrm{p}\left(\frac{\partial \bm{t}_\perp}{\partial s}\right)^2+f\bm{t}_\perp^2\right] ds -fL\,,
\end{equation}
where $f$ acts as a Lagrangian multiplier to fix the end-to-end extension $z_\mathrm{e}$. We can rewrite the last expression in Fourier modes $\tilde{\bm{t}}_\perp(q)=\int e^{iqs} \bm{t}_\perp(s)\,ds$ as
\begin{equation}\label{eq:bend_fouriermode}
    \frac{E_b}{k_\mathrm{B}T}=\frac{1}{2} \int  \left [\lambda_\mathrm{p}q^2 +f \right] \tilde{\bm{t}}_\perp^2 \frac{dq}{2\pi} -fL\,. 
\end{equation}
From Eq.~\eqref{eq:bend_fouriermode} and from equipartition, one can derive the extension $z_\mathrm{e}$ \cite{marko1995stretching}. The relation between $z_\mathrm{e}$ and the tension $f$ can be approximated very well by the simple formula \cite{bustamante1994entropic}
\begin{equation}\label{eq:force_interpolation}
    \frac{f\lambda_\mathrm{p}}{k_\mathrm{B}T}=\frac{z_\mathrm{e}}{L}+\frac{1}{4(1-z_\mathrm{e}/L)^2}-\frac{1}{4}\,.
\end{equation}
We note that differentiating $f$ in Eq.~\eqref{eq:force_interpolation} with respect to $z_\mathrm{e}$ gives $f'(z_\mathrm{e})\sim f^{3/2}$. While this result was derived for single semiflexible polymers \cite{marko1995stretching,fixman1973polymer}, in a biopolymer network the differential modulus $K'\equiv \partial \sigma/\partial \gamma_\mathrm{s}$ measures the nonlinear change of the elastic modulus with the applied shear stress $\sigma$, where $\gamma_\mathrm{s}$ is the strain. In fact, the prediction $K'\sim \sigma^{3/2}$ is observed in networks of cross-linked and bundled actin filaments \cite{gardel2004elastic} and in IF networks \cite{lin2010origins}. This phenomenon is termed entropic stiffening \cite{burla2019mechanical}, because the elastic modulus grows with shear stress.

The networks assembled with semi-flexible cytoskeletal filaments have unusual viscoelastic properties \cite{pegoraro2017mechanical,storm2005nonlinear,wagner2006cytoskeletal,gardel2008mechanical,chaubet2020dynamic,lorenz2022multiscale}. 
Upon increasing stress, networks of actin filaments show stress-stiffening behaviour up to a critical stress of $\sigma_c\sim 270$ Pa, after which the networks enters a stress-softening regime that is reversible, that is, if the stress is reduced no hysteresis is observed \cite{chaudhuri2007reversible}. 
A microscopic mechanism that explains such behaviour is based on the initial resistance to stretching due to entropic elasticity in the biofilament network, leading to the stress-stiffening regime. Above $\sigma_c$, some filaments under compressive stresses will buckle, which decreases the elasticity of the network and leads to the stress-softening regime \cite{chaudhuri2007reversible}. This is at variance with the behaviour of flexible polymers networks, which exhibit stress softening only due to failure and is not reversible \cite{mofrad2009rheology}. 
Furthermore, cytoskeletal filaments show negative normal stresses, that is, when sheared they tend to pull inward \cite{janmey2007negative}. 
A negative normal stress is unusual (when a typical material is sheared, it expands in the direction normal to the shear, i.e., it experiences a positive normal stress), but it can be rationalised with a simple model of nonlinear elasticity \cite{storm2005nonlinear,janmey2007negative}. Such filament exhibits strain-stiffening behaviour: it resists extension more strongly as the extension grows. In a random, isotropic network of such filaments under shear, some filaments will be compressed while others will be stretched. The stretched filaments will exert a stronger tensile force (negative stress) than the compressed filaments (positive stress); hence, the net stress in the normal direction will be negative \cite{mackintosh1995elasticity,storm2005nonlinear,janmey2007negative}.
This is to be contrasted with flexible-polymer networks, which exhibit positive normal stress. 
The nonlinear viscoelastic features of cytoskeletal filaments contribute to generate the structural integrity and strength of the cell.

Biofilaments exist also outside of the cell. In animal cells, collagen is the most important filament found in the extracellular matrix \cite{sun2021mechanics}. There are many types of collagen molecules, but e.g.\ in humans more than $90\%$ are type I collagen, found in skin and tendons. Type I collagen forms fibres in a hierarchical structures. The amino acids chains form together a triple helix, called tropocollagen, with a diameter of $\approx 1.5$ nm. Multiple tropocollagens bundle together into protofibrils (diameter $\approx 6-25$ nm). Multiple protofibrils further bundle together and form collagen fibril (up to $1 \mu$m in length, $\approx 100$ nm in diameter). The fibrils can then form thicker fibres, which self-organize into a disordered network, the collagen matrix \cite{sun2021mechanics}. The collagen matrix exhibits nonlinear elasticity, with an elastic modulus depending on concentration, pH, and ionic strength, and ranging from few Pa up to hundreds of MPa \cite{sheu2001characterization,dutov2016measurement}. 

In plant cells, the most abundant and important filament is cellulose. This is a linear polysaccharide chain of glucose monomers. The numerous hydrogen bonds within and between chains gives them considerable stiffness and generates robust fibrils, which are used in the cell walls of plant cells. Depending on its microstructure, the Young's modulus of cellulose ranges from $\approx 25$ GPa \cite{eichhorn2001young} (for microcrystalline cellulose) to $114$ GPa (for bacterial cellulose) \cite{hsieh2008estimation} or the classic measurement $138$ GPa \cite{nishino1995elastic}.

A review of cell biopolymers cannot be complete without mentioning nucleic acids: DNA and RNA.  The physics of the DNA helix, or generally of nucleic acids, is surprisingly complex \cite{strzelecka1988multiple}, and is inspiring a number of applications in nanotechnology \cite{ramsay1998dna,seeman2010nanomaterials,sacca2012dna,seeman2017dna,praetorius2017biotechnological,engelen2021advancing}; but, unfortunately, we do not have the space to address it here. There are, however, numerous excellent reviews on DNA's and RNA's structure and biochemical interactions \cite{kornyshev2007structure,wong2010electrostatics,teif2011condensed, travers2015dna,michieletto2019role} and also on application of topology to DNA \cite{chevizovich2020review,michieletto2024kinetoplast}

The cytoskeleton is a formidable subject to study theoretically or  computationally, because of the sheer number of individual units involved and their complex physical and biochemical interactions. However, considerable progress has been made recently \cite{gao2015multiscale,rincon2017kinesin,bun2018disassembly}, and software packages are now available that simulate cytoskeletal structures of various shapes and their dynamics \cite{nedelec2007collective,popov2016medyan,freedman2017versatile,fiorenza2021modeling,yan2022toward}. Often cytoskeletal filaments are modelled as rigid rods, however, a versatile model marries ideas from polymer physics and active matter to simulate active filaments 
\cite{isele2015self, winkler2017active, winkler2020physics, philipps2022tangentially, sinaasappel2024locomotion, zhu2024non, mokhtari2019dynamics,sinaasappel2024locomotion, theeyancheri2023active}. Such model polymers are commonly represented as a semi-flexible chain of $N$ colloids connected by springs. In an overdamped regime, the equations of motion for the monomer positions $\bm{r}_\alpha$ read
\begin{equation}\label{eq:active_polymer_drdt}
    \gamma \frac{d\bm{r}_\alpha}{dt} = -\nabla_\alpha U + \bm{F}_{\mathrm{a},\alpha} + \bm{\xi}_\alpha(t)\,,
\end{equation}
where $\gamma$ is the monomers' drag coefficient, $U = U_\text{bond}+U_\text{angle}+U_{\alpha,\beta}$ the configurational potential energy, with contributions from the bonds between neighbouring monomers
\begin{equation}\label{eq:U_bond}
    U_\text{bond} = \frac{k}{2}\sum_{\alpha=1}^{N-1}\left(|\bm{r}_\alpha-\bm{r}_{\alpha+1}|-r_0\right)^2 \,\text{,}
\end{equation}
with spring constant $k$ and equilibrium bond length $r_0$, from a bending energy
\begin{equation}\label{eq:U_angle}
    U_\text{angle}=\frac{\kappa}{4}\sum_{\alpha=1}^{N-2}  \left(2\bm{r}_{\alpha+1}-\bm{r}_\alpha-\bm{r}_{\alpha+2}\right)^2 \,\text{,}
\end{equation} 
with bending rigidity $\kappa$
and from pairwise steric interactions between monomers $U_{\alpha,\beta}$ commonly modelled as a Weeks--Chandler--Andersen potential \cite{weeks1971role}. In Eq.~\eqref{eq:active_polymer_drdt}, $\bm{\xi}(t)$ is a white noise term describing thermal fluctuations with $\langle \xi_{\alpha,i}(t)\xi_{\beta,j}(t') \rangle = 2\gamma k_\text{B}T\delta(t-t')\delta_{i,j}\delta_{\alpha,\beta}$. The active force term $\bm{F}_{\mathrm{a}_\alpha}$ drives the system out of equilibrium. A common choice is a tangential driving force $\bm{F}_{\mathrm{a},\alpha}=f(\bm{r}_{\alpha+1}-\bm{r}_\alpha)$, however, depending on the filament's driving mechanism other choice of $\bm{F}_{\mathrm{a},\alpha}$ are possible. This model has been successfully applied to larger systems, such as flagella \cite{chelakkot2014flagellar}, cyanobacteria \cite{kurjahn2024collective}, and worms \cite{nguyen2021emergent}.

A different approach to the theoretical modelling of cytoskeletal filaments is the active nematic framework. \textit{In vitro} assays of microtubules driven by kinesin motors can self-organise into arrays of asters and vortices \cite{ndlec1997self}. Addition of a depleting agent, leads to the formation of bundles in a quasi-2D layer of microtubules and molecular motors, which produce chaotic flows \cite{sanchez2012spontaneous}. This system has been the experimental archetype for the theory of active nematics. The theory considers a coarse-grained order parameter that describes the degree of nematic order, that is, the degree of alignment between filaments with head-tail symmetry. A local director $\bm{n}$, with nematic symmetry $\bm{n} \leftrightarrow - \bm{n}$ reflects the average local filament alignment direction and a second-rank, traceless tensor is the order parameter $\mathbf{Q}=\frac{d}{d-1}S(\bm{n}\otimes \bm{n}-\mathbf{I}/d)$, where $d$ is the dimensionality of space, $q$ the magnitude of nematic order, and $\mathbf{I}$ the identity tensor. The dynamics of the active gel is governed by the nematodynamics of $\mathbf{Q}$ and the incompressible Navier--Stokes equation for the flow field $\bm{u}$
\begin{align}
     &\frac{\partial \mathbf{Q}}{\partial t}+ \bm{u}\cdot\nabla \mathbf{Q}- \mathbf{S}=\Gamma \mathbf{H} \,, \label{eq:active_nematics_q}\\
     & \rho \left( \frac{\partial \bm{u}}{\partial t} + \bm{u}\cdot \nabla \bm{u} \right)=\nabla \cdot \mathbf{\Pi}\,, \quad \nabla \cdot \bm{u}=0 \label{eq:active_nematics_u}\,,
\end{align}
where $\mathbf{S}$ is the corotation term,  $\mathbf{H}$ the molecular field, and  $\mathbf{\Pi}$ is the stress tensor, which includes the hydrostatic pressure, the viscous and elastic contributions, and, crucially, the active stress tensor $\mathbf{\Pi}^\mathrm{act}=-\zeta \mathbf{Q}$ (see, e.g.\ \cite{doostmohammadi2018active} for the full definitions of $\mathbf{S}$, $\mathbf{H}$, and $\mathbf{\Pi}$). The coupling of the flow field with the derivative of the nematic tensor $\mathbf{Q}$ generates the rich phenomenology of active nematics and their topological defects \cite{marenduzzo2007hydrodynamics,giomi2013defect,thampi2013velocity,prost2015active,decamp2015orientational,wu2017transition,duclos2020topological,serra2023defect}.

\begin{tcolorbox}[boxrule=0.2mm, colbacktitle=cyan!5!white, coltitle=black, colback=cyan!5!white, title=\textsf{Box III | Drag anisotropy}, label={box:drag_aniso}]
{\small \textsf{
Drag anisotropy refers to the directional dependence of drag forces acting on an object as it moves through a medium. Due to their elongated form, filaments experience less hydrodynamic drag when pulled along their length than when pulled perpendicular to it. Drag anisotropy can be quantified as the ratio 
\begin{equation} \label{eq:drag_anisotropy}
    \xi = \frac{\zeta_\perp}{\zeta_\parallel},
\end{equation}
where $\zeta_\perp$ and $\zeta_\parallel$ are the friction coefficients for motion perpendicular and parallel to the object's length.
Typically, for filaments $\zeta_\perp\approx 2 \zeta_\parallel$.
The total drag force density (drag per unit length) opposing a filament's motion at low Reynolds number (i.e.\ viscosity-dominated regime) is given by $\textbf{F}_\text{d} = -( \zeta_\parallel \textbf{v}_\parallel +  \zeta_\perp \textbf{v}_\perp )$, where $\textbf{v}_\parallel$ and $\textbf{v}_\perp$ are the projections of the local velocity into directions parallel and perpendicular to the segment \cite{lauga2009hydrodynamics, hancock1953self}.
}}
\end{tcolorbox}

\section{Cellular appendages} \label{sec:cilia}

In a letter sent to the Royal Society of London in 1676, Antonie van Leeuwenhoek reported the discovery of protozoa and the observation of cilia and flagella. He wrote: ``\textit{I also discovered a second sort of animalcules, whose figure was an oval, ... provided with diverse incredibly thin little feet, or little legs [cilia], which were moved very nimbly ..., and wherewith they brought off incredibly quick motions}'' (cf.\ translation of letter 18 \cite{dobell1958antony,haimo1981cilia}).

Today, our language has changed a bit, and we describe cilia and flagella as slender, hair-like appendages that
extend from the main cell body and perform diverse functions, such as enabling cell movement \cite{beeby2020propulsive,bondoc2023methods}, generating flows \cite{eichele2020cilia,huang2015microscale,ling2024flow}, sensing environmental stimuli \cite{marshall2006cilia,ferreira2019cilium} and attachment to surfaces \cite{ryu2010unsteady,jarrell2013surface}. While there are a myriad of different filamentous appendages found in biology, here our focus is on those most widely studied, with the aim to capture the diversity of their structures, their main properties, mechanisms and functionalities.

Across the tree of life, the most well studied cellular appendages are those used for motility, in particular the flagella in prokaryotes and cilia in eukaryotes. 
These filaments whip, wave, or rotate through the action of molecular motors that transform the cell’s chemical energy into mechanical energy \cite{beeby2020propulsive,gilpin2020multiscale}. The resulting non-reciprocal motion enables net propulsion within the cell’s microscale, viscosity-dominated environment \cite{purcell1997life,lauga2009hydrodynamics}.

\textit{A low Reynolds number world.}
At the microscale in which these filaments operate, inertia is insignificant and viscous forces dominate. In fluid dynamics, this is the world of ‘Low Reynolds numbers’. The Reynolds number describes the ratio of inertial to viscous forces and is defined as $\text{Re} = \frac{\rho U L}{\mu}$, where $\rho$ is the fluid density, $\mu$ its viscosity, and $U$ and $L$ are characteristic velocity and length scales, respectively. In the world of planktonic cells and filamentous appendages, typically $\textrm{Re}\sim 10^{-5}-10^{-2}$. 
In some extreme cases, where motions transiently reach ultrafast velocities, the local $\textrm{Re}$ can be much higher, up to $\sim 10^{3}$ \cite{chang2024biophysical}. Cellular scale motions are thus mostly within a low $\textrm{Re}$ regime. In this viscosity-dominated environment, cells overcome and even exploit drag to generate motion \cite{purcell1997life,lauga2009hydrodynamics}. Cells can achieve this due to the anisotropic drag (defined in Box III) of slender appendages, which allows propulsive forces to be generated perpendicular to the local velocity of the filament \cite{lauga2009hydrodynamics}. 
Additionally, drag-based thrust relies on the ability of filaments to perform time-periodic deformations that are also non-reciprocal, that is, the sequences of motion differ under a time-reversal \cite{purcell1997life,lauga2009hydrodynamics}, crucial for creating a non-zero time-averaged propulsive force \cite{purcell1997life,lauga2009hydrodynamics}. This can be achieved via the rotation of helical flagella or the whip-like beating of cilia.

\textit{Multiscale dynamics.} 
The activity of cellular filamentous appendages encompasses dynamics across multiple scales \cite{gilpin2020multiscale}, offering physicists many opportunities to develop frameworks and provide insights into how these dynamics integrate and interact across scales. Nanoscale molecular motors drive shape changes and motions of individual filaments, while these shape changes drive fluid flows \cite{gilpin2020multiscale,beeby2020propulsive}. 
Resistive force theory (RFT) and slender body theory (SBT) are two hydrodynamic models of fluid-structure interactions for slender bodies within a low Reynolds number environment (see reviews \cite{lauga2009hydrodynamics,bondoc2023methods} and references therein). Both are used to estimate the propulsive forces generated by slender appendages based on their waveforms and resulting microswimmer velocities. While RFT is a simpler approximation that neglects long-range hydrodynamic interaction, SBT is more accurate but also more computationally intensive.
The basic idea of SBT for Stokes flow is to replace the disturbance due to a body with a suitably chosen line distribution of Stokeslets \cite{batchelor1970slender}. Because the Stokes equation $\mu\nabla^2 \bm{u}-\nabla p=0$, $\nabla \cdot \bm{u}=0$ is linear, a Green's function exists; for a point-like forcing $\bm{F}\delta(\bm{r})$ at the origin in an unbounded fluid, the fluid velocity at point $\bm{r}$ is 
\begin{equation}\label{eq:stokeslet}
   {u}_i(\bm{r})= \frac{F_j}{8\pi\mu} \left( \frac{\delta_{ij}}{|\bm{r}|} + \frac{r_i r_j}{|\bm{r}|^3}  \right)\,,
\end{equation}
where $\delta_{ij}$ is the Kronecker delta, and Einstein's summation convention is used. The expression in Eq.\eqref{eq:stokeslet} is called a `\textit{Stokeslet}'. In the presence of a body with length $L$ much larger than its diameter $2a$, the key idea of SBT is to approximate the flow with a line density of Stokeslets
\begin{align}
    {u}_{i}^\mathrm{SBT}(\bm{r})= & \int_0^L \frac{{F}_j(s)}{8\pi\mu}  \Bigg[ \frac{\delta_{ij}}{|\bm{r}-\bm{R}(s)|}   \\
    & + \frac{({r}_i-{R}_i(s))({r}_j-{R}_j(s))}{|\bm{r}-\bm{R}(s)|^3}\Bigg]   ds\,,
\end{align}
where $s$ and and $\bm{R}(s)$ are the arc length and position vector along the filament, respectively \cite{batchelor1970slender,keller1976slender,lauga2009hydrodynamics}. SBT has been successufully used to reproduce experimental results  with better precision than RFT \cite{rodenborn2013propulsion}.

Mechanical coupling, both via hydrodynamic interactions and the internal cell cytoskeleton, enables coordination of multiple appendages \cite{wan2024mechanisms,wan2018coordination}, and motivates a range of work that considers cellular appendages as coupled oscillators \cite{uchida2017synchronization,bruot2016realizing}. 
Arrays of cilia can synchronise into metachronal waves \cite{poon2023ciliary,meng2021conditions,wan2024mechanisms,elgeti2013emergence}, while the bundling and unbundling dynamics of groups of bacterial flagella are key for their run-and-tumble motility behaviour \cite{grognot2021morethanpropellors,wadhwa2022bacterial}.
Large numbers of cilia can also group together to form compound ciliary structures, such as the leg-like `cirri' the ciliate \textit{Euplotes} uses for walking \cite{laeverenz2024bioelectric} and the comb plates of ctenophores that comprise tens of thousands of cilia \cite{jokura2022two}.
The individual and collective activity of filamentous appendages drive fluid transport, self-propulsion and organism-environment interactions. 
Groups of microswimmers can display collective behaviours such as bacterial swarming \cite{beer2019statistical,wadhwa2022bacterial} and formation of bioconvection patterns that arise from hydrodynamic instabilities \cite{bees2020advances,schwarzendahl2018maximum,vachier2019dynamics,schwarzendahl2019hydrodynamic,schwarzendahl2022active}.

\begin{figure}[t]
    \centering
    \includegraphics[width=\linewidth]{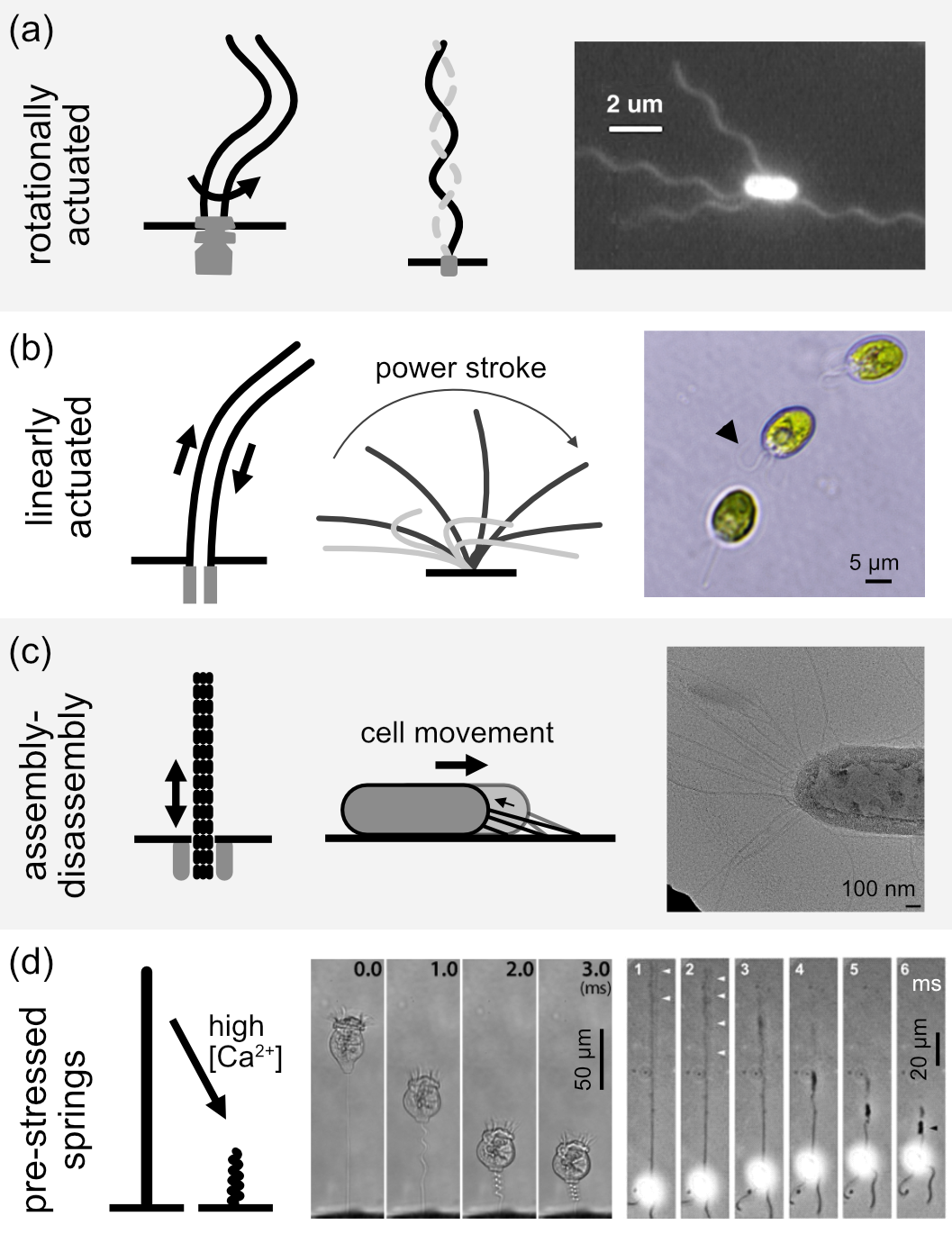}
    \caption{Four principles of activation in filamentous cellular appendages. (a) Rotationally actuated filaments are helical filaments rotated at their base, such as bacterial flagella seen here in an image reproduced from \cite{turner2000real} of a fluorescently labelled \textit{Escherichia coli} cell.
    (b) Linearly actuated bending is the mode of activation found in the cilia of eukaryotes, with typical waveforms consisting of power and recovery strokes. The microalga \textit{Chlamydomonas} uses two cilia for motility, here shown as three overlaid frames from a movie of its swimming behaviour.
    (c) Assembly-disassembly activity involves the dynamic extension and retraction of filaments, such as in pili used to generate twitching motility in bacteria and archaea. An example cryo-electron micrograph of a negatively stained wild-type \textit{Thermus thermophilus} cell shows long flexible pili filaments \cite{neuhaus2020cryo}.
    (d) Pre-stressed springs respond to changes in intracellular calcium resulting in rapid movements, as seen in the time series of the contractile motion of the stalk of \textit{Vorticella} \cite{ryu2010unsteady} and the haptoneme of a haptophyte alga \cite{nomura2019microtubule}.}
    \label{fig:appengages}
\end{figure}

Considering the variety of mechanisms cells employ to actuate their filamentous appendages, these can be broadly classified into four different principles of activity: rotational actuation of passive helical propellors, linear motor actuation of flexible filaments to generate bending, assembly-disassembly dynamics, and pre-stressed springs; see Fig.~\ref{fig:appengages} for the corresponding schematics and examples.

$\bullet$ \textit{Rotationally actuated filaments.}
The rotation of passive filaments by cells involves a rotary complex embedded in the cell membrane at the base of the filament where stator and rotary components interact to generate torque that is transmitted to a helical filament \cite{chwang1971note}. There are two main examples of such filaments: the bacterial flagella and the archaea’s archaella \cite{beeby2020propulsive}. 
Their molecular structures differ. The bacterial flagellum is a hollow cylinder self-assembled from the flagellin protein into a chiral structure and connected to the rotary motor via a short flexible hook at its base \cite{silverman1977bacterial,coombs2002periodic}. The archaea's archaella have a solid multi-stranded helical structure \cite{cvirkaite2023evolution}. However, they are both $\sim$10-20 nm wide and up to $\sim$10 $\mu$m in length. Archaella use an ATPase to generate torque via ATP hydrolysis \cite{streif2008flagellar,albers2015archaellum}, while the flagella’s rotary complex takes a more structurally intricate form and is powered by the proton motive force \cite{berg2003rotary,manson1977protonmotive,wadhwa2022bacterial}. In both cases, the torque generated by the embedded molecular machinery results in a rotation of the connected helical filament that in turn acts as a propeller for cell motility \cite{beeby2020propulsive}.

Although these helical filaments are often considered rigid, flexibility plays a crucial functional role \cite{brumley2015flagella}. Flexibility of the basal hook enables flagella bundling \cite{brown2012flagellar} and, through a buckling instability, allows abrupt directional changes via an off-axis motion of the flagellum known as a `flick' \cite{son2013bacteria}. 
Intuitively, peritrichously-flagellated bacteria should have negligible swimming speeds because of the flagella distribution around the cell body. However, an elastohydrodynamic bending instability is responsible for a conformational change in the hook bending and results in a net propulsive force \cite{riley2018swimming,halte2025bacterial}.
Additionally, flagella can undergo polymorphic transformations ---shape changes driven by cooperative conformational changes in flagellin--- depending on factors like torsional load, pH, temperature and ionic strength \cite{lauga2016bacterial}. These expand the functional versatility of flagella, such as enabling run-and-tumble behaviour and escape from traps \cite{turner2000real,kuhn2017bacteria}.

$\bullet$ \textit{Linearly actuated filaments.}
In contrast, cilia (sometimes referred to as eukaryotic flagella) are typically $\sim$250 nm wide and up to $\sim$100 $\mu$m long, and are actuated by linear stepper motors distributed along their entire length \cite{beeby2020propulsive}. Cilia have a highly conserved structure across eukaryotes known as the axoneme, which comprises 9 microtubule doublets surrounding a central pair of microtubules \cite{nicastro2006molecular}. Ciliary bending and twisting is powered by dynein motors that walk along the 9 outer doublets. Since the microtubules are usually fixed to the cell body at the base of the cilium, this linear forcing generates differential sliding between neighbouring doublets resulting in bending of the axoneme \cite{lindemann2010flagellar,satir2014structural}. The mechanism by which dynein motor activity generates ciliary beating is still under debate, with several biophysical models proposed that largely fall under two categories: (1) mechanical feedback within the axoneme regulates dynein activity \cite{lindemann2010flagellar,riedel2007molecular,cass2023reaction}, (2) continuous dynein activity along the cilium results in dynamic instabilities \cite{woodhams2022generation}. 
Dynein motors are typically distributed along the whole length of the cilium, giving a high degree of flexibility in the types of waveforms that are possible to produce by coordinating the dynein activity in different ways. Compared with the basally actuated rotation of a helical filament, which only has a small set of parameters that can be controlled (e.g. speed and direction of rotation), cilia have the potential to display a much more complex and nuanced diversity of waveforms, with calcium playing a central role in the regulation of ciliary activity \cite{eckert1972bioelectric}. The ability of cilia to change their waveform is critical for enabling ciliated organisms to switch locomotor gaits and perform complex motion \cite{wan2018time,laeverenz2024bioelectric}. 
Although cilia are highly conserved, there are several examples of variations or additions to the standard 9+2 axoneme structure that exemplify how cilia can be adapted to perform specific functions \cite{ginger2008swimming}. For example, the cilium of the parasite \textit{Trypanosoma brucei} has a paraflagellar rod, a lattice-like filament that runs along the length of the axoneme, thought to affect axonemal beating by imposing mechanical constraints and contributing to signaling pathways \cite{langousis2014motility}. Excavates, species of phagotrophic flagellates, are characterised by possessing vaned cilia within a ventral groove whose confined beating is thought to be particularly efficient for generating feeding currents \cite{suzuki2024foraging}.

$\bullet$ \textit{Assembly-disassembly dynamics.}
An entirely different principle of activity is that seen in the type IV pili of bacteria and archaea, in which cycles of extension, surface binding and retraction enable cells to perform functions such as twitching motility, surface sensing and DNA uptake \cite{craig2019typeIV}. This activity is enabled by ATP hydrolysis, which powers the rapid assembly and disassembly of pilin subunits \cite{ellison2022type}.

$\bullet$ \textit{Pre-stressed springs.}
Lastly, pre-stressed springs composed of filaments and tubules power many of the fastest cellular movements \cite{mahadevan2000motility}. There are two main types: a microtubule-based system found in some microalgae (i.e.\ the haptoneme in haptophytes) \cite{nomura2019microtubule} and a myoneme based system found in some ciliates (e.g.\ the spasmoneme of vorticella) \cite{ryu2017vorticella,floyd2023unified}. In both cases, rapid coiling is thought to be driven by an increase in Ca$^{2+}$ concentration, which changes the energetically preferred conformation of the filament from elongated to coiled \cite{kawachi1994ca2mediated,floyd2023unified}. For the haptoneme, some evidence suggests that this is driven by conformational changes in microtubule associated proteins that bind with Ca$^{2+}$ \cite{nomura2019microtubule}, while for the spasmoneme it is hypothesised to be a combined electrostatic and entropic effect \cite{mahadevan2000motility}.

\textit{Elastodynamic properties.} 
A key mechanical parameter for characterising the movements of cellular appendages is the bending stiffness (see Box \ref{box:elastodyn}), which can be estimated from the deformation induced by forces and torques applied using techniques such as optical tweezers \cite{darnton2007force}, microneedles \cite{xu2016flexural}, magnetic beads \cite{hill2010force} or microfluidic flow devices \cite{shen2022bending}. Using such approaches the bending stiffness for the bacterial flagella has been estimated to be in the range $1-1000$ pN $\mu$m$^2$ (see Table S1 in \cite{shen2022bending}), while estimates for eukaryotic motile cilia fall within the 300-6000 pN $\mu$m$^2$ range \cite{okuno1979direct,hill2010force,xu2016flexural,ishijima1994flexural}. The Young's modulus $E$ is generally estimated indirectly from experimentally measured values of the bending stiffness. This requires approximating the second area moment $I$, which is challenging when a filament is not easily defined as a solid or hollow cylinder, such as in the case for cilia. For motile cilia, two values are commonly used: $I=1.3\times10^{-31}$ m$^4$ \cite{holwill1965motion} and $I=3\times10^{-29}$ m$^4$ \cite{baba1972flexural}, the latter assumed more relevant for tighter connections between the microtubules of the axoneme \cite{okuno1979direct}. This gives a large range of estimates $E \sim (0.01-120)$ GPa.

\textit{Appendages as cellular antennae.} 
In addition to their role in motility and fluid transport, appendages can also play important roles in sensory reception and signal transduction that benefit from the geometric and mechanical properties of their filamentous form \cite{marshall2006cilia,hickey2021ciliary,ferreira2019cilium,craig2019typeIV}. 
Both motile and non-motile cilia can provide sensory functions, possessing chemical receptors and mechanically-sensitive ion channels that enable them to act as the cell's antenna \cite{marshall2006cilia,bloodgood2010sensory}.
Primary cilia, typically non-motile and structurally distinct from motile cilia, are specialised sensory organelles critical for coordinating signal transduction pathways during development and homeostasis \cite{mill2023primary}.
Filamentous appendages can extend far beyond the cell surface, reaching more well-mixed and therefore more `representative' regions of the surrounding environment, making them effective as chemosensory probes \cite{marshall2006cilia}. Additionally, their high aspect-ratio and high surface area to volume ratio, improves their capture rate and chemosensory effectiveness \cite{hickey2021ciliary}.
Their elongated shape also means that filamentous appendages can be considered as extended cantilevers, facilitating their sensitivity to mechanical stimuli \cite{ferreira2019cilium}.
Beyond mechanical interactions, recent work suggests a potential role of primary cilia in the regulation of neuronal and glial activity  \cite{tereshko2021ciliary,ott2024ultrastructural}.

\section{Filamentous microbes} \label{sec:microbes}

\begin{figure}
    \centering
    \includegraphics[width=\linewidth]{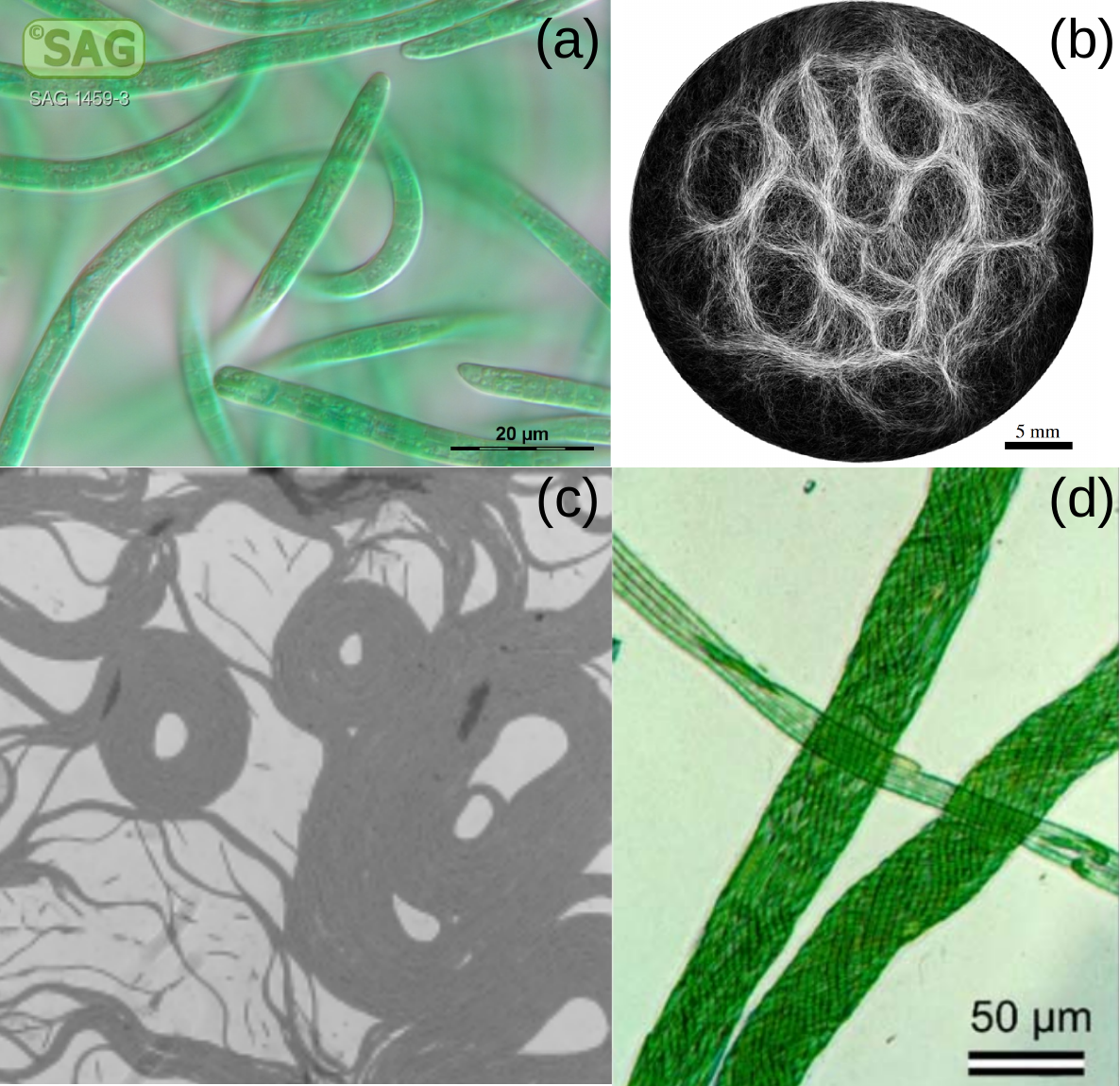}
    \caption{Examples of filamentous cyanobacteria self-organisation. (a) Many cyanobacteria species grow into filaments made out of a chain of identical cells \cite{SAG}. These filaments are capable of organizing into intricate structures like (b) reticulate mats \cite{faluweki2022structural}, (c) spirals \cite{gong2023active}, and (d) even ropes \cite{garcia2009evolution}.}
    \label{fig:cyanos}
\end{figure}

Various microbial organisms have evolved a filamentous body shape. Examples of such organisms that have attracted the interest of physicists include various species of filamentous cyanobacteria \cite{faluweki2023active, faluweki2022structural,kurjahn2024collective, kurjahn2024quantifying, cammann2024topological, ehlers1996cyanobacteria}, filamentous phages \cite{dogic1997smectic, dogic2006ordered, purdy2004isotropic}, and nematodes,  \cite{bilbao2013nematode, gart2011collective, backholm2013viscoelastic, niebur1991theory} that straddle the boundary between the microscopic and macroscopic world. 
As there are too many filamentous microorganisms to list, this section will focus on discussing more generally the implications of a filamentous body shape at the microscale for a few important classes of microorganisms. 

Having a body with a large aspect-ratio like a filament comes with some advantages over microbes shaped as spheres or short rods. 
An elongated shape increases the chance of meeting one another in disordered environments \cite{arguedas2022elongation}.
Thanks to a filamentous shape, microbial organisms can self-propel without appendages. Rather than using flagella or cilia to generate motility, they can bend their whole body to generate thrust thanks to the different drag coefficients that a slender body experiences in the direction tangential and perpendicular to its motion in a viscous fluid \cite{lauga2009hydrodynamics, hancock1953self} (see Box III). 
A filamentous shape has been observed to be a viable strategy to avoid predation by protists by simply being too long to be ingested easily \cite{pernthaler2005predation} (this approach is however futile at a certain size difference between devourer and the devoured, as is evident in the popularity of spaghetti as a human food source).
The increased surface-to-volume ratio of a filamentous shape allows for easier uptake of nutrients \cite{young2006selective}, but also allows stronger attachment to substrates \cite{weiss1995effect} in bacteria, because of a larger contact patch, compared to a spherical one.

A filament's most crucial mechanical property is its bending stiffness. It is known that one can draw conclusions about the forces at play from purely geometrical observations (see Box~\ref{box:elastodyn}). Techniques to measure the bending stiffness $\beta_\mathrm{f}$ and Young's modulus $E$ of gliding filaments with the help of microfluidics \cite{faluweki2022structural} and micropipette techniques \cite{kurjahn2024quantifying} have been developed (see Table~\ref{tab:rheology}).

Filamentous cyanobacteria often exist in a surface-bound but motile state, called gliding \cite{hoiczyk2000gliding,wilde2015motility,tchoufag2019mechanisms}. The underlying mechanism that allows the generation of propulsive forces in this state is still a matter of debate \cite{mcbride2001bacterial, read2007nanoscale, dhahri2013situ, koiller2010acoustic, halfen1970gliding}, as no cyanobacteria species are known to possess flagella \cite{wilde2015motility}. The main mechanisms that are debated are: type IV pili extension and retraction \cite{khayatan2015evidence, wilde2015motility}, fibrils \cite{halfen1970gliding,read2007nanoscale}, ejection of a jet of extracellular slime from junctional pores \cite{hoiczyk1995envelope,hoiczyk1998junctional,wolgemuth2004junctional,dhahri2013situ} (at least for some species), and acoustic streaming \cite{ehlers1996cyanobacteria,koiller2010acoustic,ehlers2011could}. 
By observing the self-buckling of a filament, the forces at play in its propulsion can be inferred via the bending stiffness \cite{kurjahn2024quantifying}. 
Such observations combined with the independence of propulsion speed on filament length imply that all cells within a cyanobacteria filament contribute roughly equally to its propulsion.

Cyanobacteria locomotion exhibits another interesting feature: path-tracking motion, where the body follows the path laid by the forward end of the filament \cite{faluweki2023active, cammann2024topological}. This type of locomotion in worms is also known as  metameric locomotion \cite{du2022model}. Physically, information about a filament's orientation persists over the time it takes a filament to traverse its own length \cite{du2022model}. Although their motility is induced by polar forces, filaments themselves are seemingly nematic, with no head or tail end. In fact, cyanobacteria have been observed to reverse their direction of motion frequently, as a strategy to navigate towards more favourable conditions, and even to form communities. \cite{wilde2015motility, rosko2024cellular, abbaspour2023effects,pfreundt2023controlled}. Given the delocalised nature of the filaments propulsive mechanism, a high degree of intercellular coordination is required to reverse the direction of motion, with the mechanisms of communication between cells not yet understood.
Confined active filaments with low bending rigidity have been shown to curl up into tightly wound spirals (see Fig.~\ref{fig:cyanos}(c)), effectively trapping themselves, greatly impeding their mobility \cite{duman2018collective, prathyusha2018dynamically, isele2015self, lin2014dynamics}. Utilising reversals is a strategy to escape this trapped state \cite{gong2023active} as well as other forms of geometric confinement \cite{kurzthaler2021geometric}.

Having a high aspect-ratio allows microbes to effectively sample changing conditions within space. While organisms with short aspect-ratio bodies commonly have to move around to find favourable conditions, essentially sampling the environment at random to inform their movement through e.g.\ chemo- or phototaxis, elongated filaments can sample gradients along their body and react before they have wholly immersed themselves in an unfavourable environment. Cyanobacteria, for instance, have been observed to react to light shining only onto parts of their body \cite{nultsch1983partial, wilde2017light}, showing scotophobic responses, well before the whole filament loses access to light by reversing its gliding direction \cite{kurjahn2024collective}. 

Having an elongated body may even allow organisms to be in two places with different conditions at once. Cable bacteria are filamentous, multicellular bacteria belonging to the family \textit{Desulfobulbaceae}. They can conduct electrons over centimetre scales \cite{pfeffer2012filamentous,marzocchi2014electric}. Through their filamentous shape they can couple sulfide oxidation in deeper, anoxic marine
sediment with oxygen reduction in surface layers \cite{nielsen2010electric,bjerg2023cable, risgaard2015cable}. Electrical activity is also involved in a microbial feat of engineering: some electroactive bacteria produce nanowires to transfer electrons to extracellular electron acceptors \cite{reguera2005extracellular}. These are, e.g., the \textit{Geobacter sulfurreducens}  or the \textit{Shewanella oneidensis} MR-1 bacteria \cite{el2010electrical,pirbadian2014shewanella,schkolnik2015situ,logan2019electroactive}. 

Active filaments can explore spaces inaccessible to plumper organisms with the same volume. Semiflexible active filaments have been shown to effectively explore complex and crowded environments \cite{mokhtari2019dynamics, zhu2024non, sinaasappel2024locomotion, fazelzadeh2023active, theeyancheri2023active}. The ability to reverse has been shown to further enhance the mobility of active filaments in porous environments \cite{kurzthaler2021geometric}. Many soil dwelling microbes are filamentous in shape, in addition to their ability to explore a porous environment it has been speculated that their shape may enable them to fix their position in space, by wrapping around soil particles \cite{young2006selective}, allowing them to resist being washed away by percolating water. Entangled filamentous microbes have been observed to stabilize their habitats and protect them against erosion \cite{kurtz2001stabilization, pluis1994algal}. Furthermore, nonmotile cells exploring a polymeric fluid can form long cables as they proliferate \cite{Gonzalez2025}.

Their filamentous nature allows filamentous organisms such as cyanobacteria to organise into structures much larger than an individual filament \cite{faluweki2023active, cammann2024topological, shepard2010undirected, tamulonis2014model}. Their filamentous shape allows them to glide over one another without losing contact with the substrate. The dynamics of gliding filaments result in the formation of a network-like reticulate pattern of filament bundles \cite{faluweki2023active, cammann2024topological, tamulonis2014model} (see Fig.~\ref{fig:cyanos}(b)). Under different conditions, filaments have been observed to form aster-like aggregates \cite{pfreundt2023controlled}. 
These relatively simple, mostly two-dimensional structures may serve as a starting point for filaments to organize themselves into even more intricate three-dimensional structures, ranging from complex patterned biomats \cite{shepard2010undirected, cuadrado2018field} to the filaments weaving themselves into ropes \cite{garcia2009evolution} (see Fig.~\ref{fig:cyanos}(d)), bridging large distances. 
The physical mechanisms behind this level of self-organisation and how the mechanical properties of these structures compare to that of a single filaments are yet to be understood, but can shed light on biofilm formation \cite{mazza2016physics,sauer2022biofilm}. 
It is plausible to presume that such complex assemblies could exhibit unusual rheological properties, just like those observed in cytoskeletal networks (see Sec.~\ref{sec:cytoskeleton}).

\section{Undulatory locomotion at the macroscopic scale} \label{sec:snakes_on_a_plane}

\begin{figure}
    \centering
    \includegraphics[width=\linewidth]{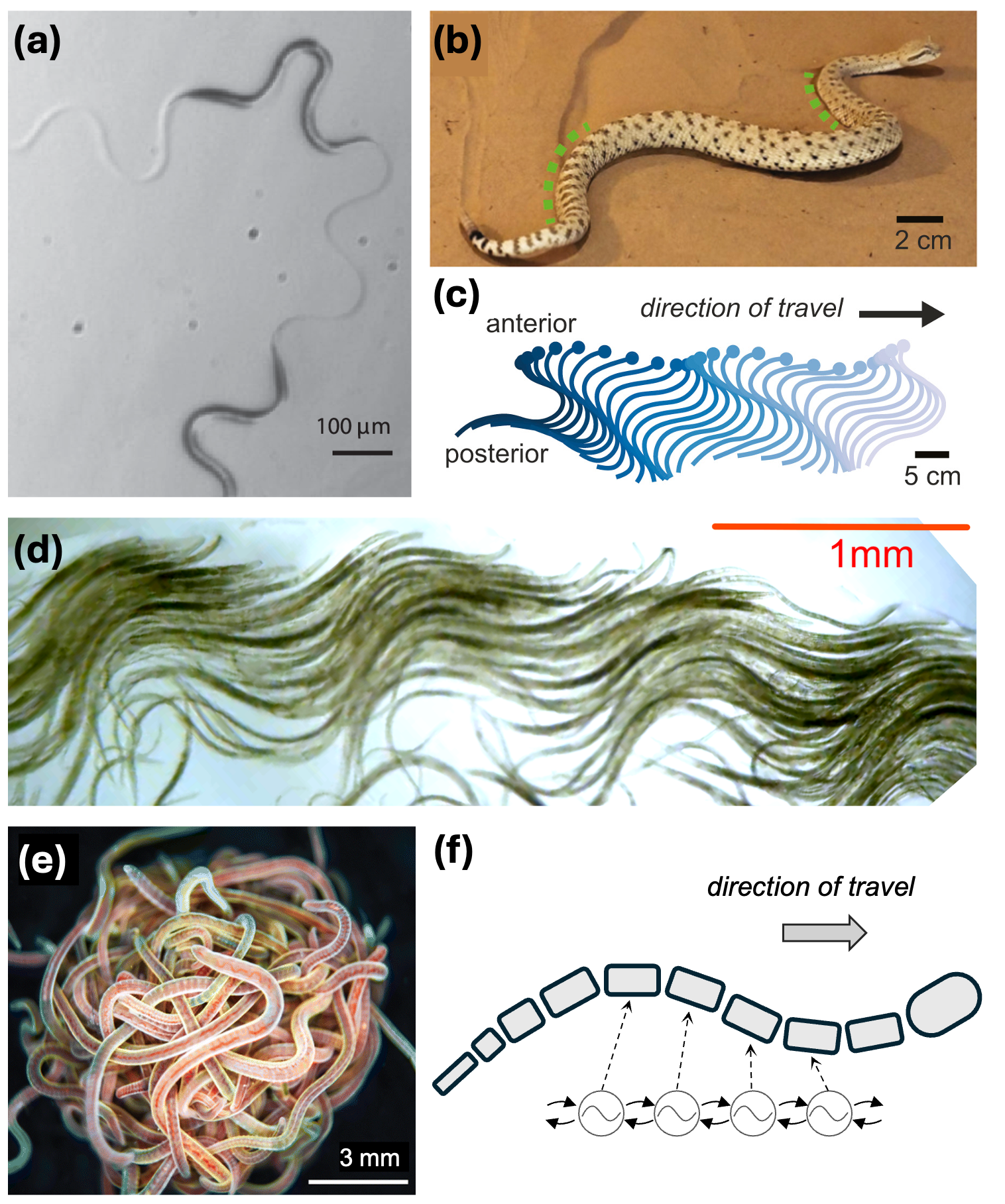}
    \caption{Undulatory locomotion of active filaments across scales in meso and macroscopic organisms. (a) \textit{C. elegans} crawling movement on agar \cite{padmanabhan2012locomotion}. (b) Sidewinding locomotion of a snake, and (c) the timetrace of tracked points along the body of the snake as it moves on sand \cite{rieser2021functional}. (d) Vinegar eels self-organise into metachronal waves (image courtesy of Alice Quillen \cite{peshkov2022synchronized}). (e) Topological entanglement in worm blobs \cite{patil2023ultrafast}. (f) Bio-inspired snake-like robots with central-pattern-generator-like (CPG) control modules (see for example \cite{wu2010cpg}).}
    \label{fig:snakes}
\end{figure}

At the organismal scale, animals with a slender and deformable body shape are also `active filaments', leading to the natural question of whether the analysis and framework presented in the previous sections could still apply. 
Diverse animals locomote by propagating undulatory waves. These body deformations are used to propel the organisms forward, although the actuation mechanisms may be very different across species and spatial scales (Figure \ref{fig:snakes}). 
Well-studied examples include crawling worms \cite{Costa2024,Stephens2008}, swimming eels \cite{Gillis1998}, and gliding snakes \cite{Jayne2020}.
The study of both limbed and limbless movement has paved the way for understanding the origins and principles of neuro-mechanical control in diverse organisms, particularly the interactions between neural patterning and muscular activity \cite{Grillner1991}. 
Undulatory or sinusoidal locomotion strategies evolved across different phyla, often according to a basic neural circuit design involving mutually inhibitory halves with the left and right sides deforming alternately \cite{Katz2016}.
These insights into the neural basis of behaviour have led to the development of novel bio-inspired mechanisms for actuating synthetic systems including robots \cite{Ramdya2023}.

Since propagating wave-like deformations are inherently non-reciprocal, they form the basis of diverse motility strategies through different media. 
For instance an undulating wave propagating head to tail down a sperm flagellum would break time reversibility since the filament will adopt a different sequence of shapes when time is reversed. 
However, non-reciprocity is a necessary but not sufficient condition \cite{lauga2009hydrodynamics} for net propulsion, as the latter also requires additional constraints on the number and placement of flagellar waves. 
Despite the large difference in scale compared to microscopic organisms such as flagellates and cyanobacteria (see Sec.~\ref{sec:cilia} and \ref{sec:microbes}), the undulatory locomotion of macroscopic animals such as nematodes and snakes can still be considered to operate in an overdamped regime where inertia is negligible, and where friction (instead of fluid viscosity) dominates. 
This is particularly true of locomotion in sand or granular media, where a resistive force theory based on drag anisotropy (see Box~III) has been derived empirically, inspired by the movement of slender bodies through viscous fluids \cite{Hatton2011,Zhang2014}.

At the individual organism scale, distinct undulatory behaviours can arise from the same underlying structure and neural circuitry. 
A given body shape can be decomposed into a small number of principal components or oscillation modes, reflecting the intrinsically low-dimensional nature of the posture space, and allowing diverse undulatory shape modes to be reconstructed with only a small number of parameters \cite{Yemini2013}. 
In the example of \textit{C. elegans} crawling, its motion in 2D is stereotyped and well-captured by just six basis modes, or eigenworms \cite{Stephens2008,Costa2024}. 
Frequency and amplitude thus uniquely define a gait, which in turn determines the distance traversed by the organism in one cycle. 
Notions such as geometric phase can help understand and predict the mapping between these gaits and eventual displacement in the lab frame \cite{Rieser2024}.
These dynamics are not totally unconstrained, however, undulatory locomotion across swimming sperm, nematodes, and even fish in overdamped environments may be subject to a universal dispersion relationship \cite{Pierce2024,SnchezRodrguez2023}. 
Thus, active filaments can achieve highly reconfigurable dynamics with minimal neuromechanical coupling.

Collectively, filamentous organisms also engage in novel behaviours that may not be accessible by individuals. 
At sufficiently high densities, these filaments self-assemble to form adaptive active materials with both solid-like and liquid-like properties, that are capable of emergent functions such as directed locomotion or smart navigation. 
Macroscopic `blobs' of the 1 cm-long California blackworm \textit{Lumbriculus variegatus} undergo dynamic collective movements to avoid desiccation or thermal stress without the need for any centralised control \cite{OzkanAydin2021}. 
The entanglement, or `braiding', process enables fast and reversible tangling and untangling of worm collectives. 
Intriguingly, \textit{L. variegatus} blobs can take minutes to assemble but only milliseconds to disentangle. 
A recent study proposed an explanation for this phenomenon by experimentally imaging the 3D trajectories of the worms and mapping the dynamics onto a 3D filament model of Kirchhoff rods \cite{patil2023ultrafast}. 
The unique, activity-dependent rheology of entangled blobs of \textit{Tubifex tubifex} worms was also studied and shown to be shear-thinning, but with a different scaling compared to polymer melts \cite{deblais2020rheology}. 
These properties can even be harnessed to enable activity-dependent separation (chromatography) of worm mixtures \cite{Heeremans2022}. For example, in \textit{T.~tubifex} the diffusivity is larger at low activity (i.e., low temperature) than at higher activity \cite{sinaasappel2024locomotion}. 
In another free-swimming nematode \textit{Turbatrix aceti} (vingar eels), concentrated suspensions of these organisms display large-scale metachronal waves surprisingly similar to those found in ensembles of cilia and flagella \cite{peshkov2022synchronized,quillen2021metachronal}. 
Understanding how these deformable filamentous bodies actively control the topology and rheology of the collective could inspire novel designs of functionally tunable synthetic materials \cite{Deblais2023}.

More generally, undulatory or anguilliform locomotion in elongated bodies allows for direct interactions between an organism and the surrounding medium, which can yield novel principles of biomimetic/bioinspired locomotion reliant on embodied design with few mechanical components. 
Self-propulsion can also be easily adapted for movement through different environments, for example for snakes the same basic locomotion mechanism can be equally effective for locomotion on land as well as in water \cite{taylor1952,Stin2024}.
In more extreme cases, `flying' snakes from the genus \textit{Chrysopelea} use body undulations to generate lift to glide stably over several metres through the air \cite{Yeaton2020}.

From a control theoretic perspective, undulatory locomotion is universal perhaps because it reduces the need for more complex neural control mechanisms and can be achieved simply by coordinating alternating waves of (usually muscle) contraction and relaxation. 
In some cases, oscillations emerge spontaneously as a result of instabilities, for example, dynamics reminiscent of ciliary beating have been observed in macroscopic chains of self-propelled microrobots, with the transition mapped to a supercritical Hopf bifurcation \cite{Zheng2023}. Indeed substrate-attached, microscopic elastic filaments forced at one end also exhibit oscillations in different regimes \cite{de2017spontaneous, bayly2016steady, man2020cilia}.
The propagation of waves thus provides one of the simplest possible appendage or body actuation mechanisms, and naturally breaks time-reversal symmetry (particularly relevant for inertia-free swimming).
This results in highly adaptable and flexible control of movement where both the speed and direction of travel can be readily changed in response to stimuli.

\section{Roadmap \& Key Open Questions}\label{sec:roadmap}

Filamentous structures are ubiquitous in biology from cytoskeletal filaments to propulsive appendages, and to whole organisms. Despite spanning several orders of magnitude in scale and complexity, surprising parallels can be observed in how biological form translates into physiological function. 
In this Review, we have sought to outline the core physical principles that govern these relationships and common frameworks for studying filamentous systems. 
As cutting-edge experimental technologies continue to deliver ever finer resolution of the molecular composition and architecture of the constituent parts of organelles, cells, and organisms, there is increasing need to integrate these insights with functional assays and biophysical models. 
At the same time, we should also be mindful that in many cases the natural context or habitat in which organisms live could also contribute to any observed phenomena \cite{franks2016social}. 
Therefore, there is also a growing need to study and understand the behaviour of living systems `in the wild', not just under controlled laboratory settings.

Critical to many modelling and simulation approaches is detailed knowledge of the emergent material properties of the many types of filaments. 
Yet, such measurements are not always available as they often rely on access to specialised technology, such as high-resolution imaging, atomic force microscopes, micropipette force sensors \cite{backholm2013viscoelastic}, or can only be performed on certain filaments, or in non-living systems or systems that have been rendered passive. 
This apparent disconnect requires more holistic, physics-inspired understanding of the context-dependent material properties of a filament and filamentous network, particularly due to their out-of-equilibrium activity. 
As a concrete example, the measurements of physical parameters such as Young's modulus $E$ or the bending stiffness $\beta_\mathrm{f}$ of cytoskeletal filaments as shown in Table~\ref{tab:rheology} strongly depends on the chemical and physical context. 

Recently, theories of active polymers \cite{winkler2020physics, winkler2017active} and active nematics \cite{doostmohammadi2018active} have identified interesting collective  behaviour (such as bands, asters, and topological defects \cite{liverpool2001viscoelasticity,duman2018collective,bianco2018globulelike,prathyusha2018dynamically,vliegenthart2020filamentous}) that reproduce experiments with purified systems of microtubules and molecular motors \cite{ndlec1997self,sanchez2012spontaneous,kruse2004asters,sumino2012large,huber2018emergence,sciortino2021pattern}. 
While these systems shed light on the nonequilibrium dynamics of cytoskeletal filaments, the challenge that lies ahead is to develop these theories towards more realistic conditions like those inside the cell, which is a 3D crowded environment. 
Thus, these theoretical frameworks should be developed in tandem with experimental innovations. Identifying appropriate dimensionless quantities from filament properties (Table~\ref{tab:rheology}), will enable the development of theories describing common phenomena in the collective dynamics of filaments. Thus, efforts in characterising filaments, completing and expanding the data in Table~\ref{tab:rheology} is required to develop a unified understanding of filament dynamics across scales.

Often, theoretical models developed to describe a specific system at one length scale can be applied across scales with little to no adaptation, as long as the mechanical properties of the filaments of interest are known. For instance, the active polymer model described in Eq.~\eqref{eq:active_polymer_drdt}--\eqref{eq:U_angle} has been successfully used to describe the dynamics of other filaments like flagella \cite{chelakkot2014flagellar}, cyanobacteria \cite{kurjahn2024collective} or even worms \cite{nguyen2021emergent}. The theory of nematics and their topological defects (see Eq.~\eqref{eq:active_nematics_q}-\eqref{eq:active_nematics_u}) has been used to describe microtubules \cite{decamp2015orientational,doostmohammadi2018active}, elongated bacteria \cite{doostmohammadi2016defect, repula2024photosynthetically}, and stem cells \cite{kawaguchi2017topological}. The mathematical description of metameric locomotion used for filamentous bacteria \cite{faluweki2023active, cammann2024topological} has also been suggested as potential description of the motion of annelids and myriapods \cite{du2022model}. Identifying commonalities and differences between filaments at different scales will allow the development of experimentally convenient proxy systems.

Beyond a consideration of possible shared tools, technologies and frameworks to analyse such complex spatiotemporal dynamics, the comparison begs the question of whether and to what extent the analytical or computational tools developed for modelling microscale active matter, can be adapted to study and predict dynamics at the macroscopic scale \cite{Zhang2014,ishimoto2025robust}. 
Modified versions of the classic resistive force theory model for cylinders moving through fluids have been used to develop neuromechanical models of \textit{C. elegans} crawling movement \cite{keaveny2017predicting} with moderate empirical support \cite{rabets2014direct}, but the validity of these assumptions are yet to be fully demonstrated more generally for other organisms, particularly in the regime of $Re\sim1$ where inertial effects are expected to play some role. 
Recent progress has been made on formulating slender body theories (originally developed for movement through Stokes flow) but applicable to larger anguilliform swimmers \cite{iosilevskii2020hydrodynamics}.
As we have already established, wave propagation is a particularly effective and universal means of breaking time-reversal symmetry and achieving self-propulsion across scales, yet the ultrastructure and biological mechanism of actuation can differ greatly across the organisms. 
Thus, it remains an open question how the same range of movement patterns can arise from completely different neuromuscular control architectures or even aneural mechanisms such as coordinated activity of molecular motors (in the case of cilia and flagella).

The study of active or self-propelled filaments in biology have led to advancement of new mathematical and physical theories, notably in the areas of topological braiding and entanglement, nonequilibrium statistical physics, and extreme mechanics such as supercoiling. 
However, one constraint in understanding filamentous systems is the practical challenge of acquiring good resolution data, and the lack of efficient computational tools to track filaments, particularly from 3D image stacks.  
There is also the open question of which metrics are most appropriate to describe the deformation of filaments and their collective organisation; can we identify measures of effective activity that bridge various orders of magnitude in length scale? 
We think that devising universally-applicable, relatively system-independent measures for such systems requires further innovation.

It is also instructive to return to the initial point of this Review and ask why filaments are a recurring motif across living systems?
Mechanical forces are being increasingly recognised as shaping factors in biology \cite{ladoux2017mechanobiology}, and geometry does create and influence mechanical interactions.
We speculate that, because living systems increase entropy production \cite{martyushev2006maximum,davies2013self} while producing more complex structures (Fig.~\ref{fig:filaments_diagram}) during their evolution, a generalised form of entropy might be a useful quantity in the search of system-independent measures. 
Because configurational degrees of freedom (positions, orientations) can be measured, it is possible to define structural complexity (or entropy) and use it to characterize living systems. Consider a function $\psi(\bm{x})$ representing the configuration of a system, for example, the atomic positions in a protein, the orientation of filaments in a network, or the muscle fibers in a metazoan. The complexity of an ensemble  can be quantified by means of a generalised Shannon--Boltzmann entropy \cite{adami2000evolution,adami2002complexity,bonchev2005quantitative}
${H}=-\int D\psi \, P(\psi(\bm{x})) \ln P(\psi(\bm{x}))$, where $\int D\psi$ represents a path-integral sum over all possible configurations. We note that slender rods have five degrees of freedom (compare with the three degrees of freedom for a spherically symmetric object); furthermore, a flexible elongated shape has five degrees of freedom per unit length. Such explosion of degrees of freedom might constitute an entropic incentive for biological systems to evolve and use biological filaments.
Not only do single units admit complex shape deformations, filaments can also easily self-assemble to form higher-order structures, or be reconfigured into networks of increasing complexity. These active physical structures have the capacity to perform complex functions without sophisticated control or centralised computation \cite{Wang2023,Moreau2024}. Future applications, particularly in the realm of synthetic biology or bioengineering, may seek to harness the mechanical intelligence inherent in filamentous structures, to create novel adaptive materials. 

\section{Acknowledgments}

This work was funded by the European Research Council (ERC) under the European Union's Horizon 2020 research and innovation programme grant 853560 EvoMotion, and UK Advanced Research and Invention Agency's Nature Computes Better programme (K.Y.W).
We also acknowledge support by UKRI/
Wellcome grant EP/T022000/1-PoLNET3 (J.C.).
We thank an anonymous referee for helpful  suggestions on Fig.~\ref{fig:filaments_diagram} and Prof Alice Quillen for the use of the image in Fig.~\ref{fig:snakes}(d).


\begin{thebibliography}{414}%
\makeatletter
\providecommand \@ifxundefined [1]{%
 \@ifx{#1\undefined}
}%
\providecommand \@ifnum [1]{%
 \ifnum #1\expandafter \@firstoftwo
 \else \expandafter \@secondoftwo
 \fi
}%
\providecommand \@ifx [1]{%
 \ifx #1\expandafter \@firstoftwo
 \else \expandafter \@secondoftwo
 \fi
}%
\providecommand \natexlab [1]{#1}%
\providecommand \enquote  [1]{``#1''}%
\providecommand \bibnamefont  [1]{#1}%
\providecommand \bibfnamefont [1]{#1}%
\providecommand \citenamefont [1]{#1}%
\providecommand \href@noop [0]{\@secondoftwo}%
\providecommand \href [0]{\begingroup \@sanitize@url \@href}%
\providecommand \@href[1]{\@@startlink{#1}\@@href}%
\providecommand \@@href[1]{\endgroup#1\@@endlink}%
\providecommand \@sanitize@url [0]{\catcode `\\12\catcode `\$12\catcode
  `\&12\catcode `\#12\catcode `\^12\catcode `\_12\catcode `\%12\relax}%
\providecommand \@@startlink[1]{}%
\providecommand \@@endlink[0]{}%
\providecommand \url  [0]{\begingroup\@sanitize@url \@url }%
\providecommand \@url [1]{\endgroup\@href {#1}{\urlprefix }}%
\providecommand \urlprefix  [0]{URL }%
\providecommand \Eprint [0]{\href }%
\providecommand \doibase [0]{https://doi.org/}%
\providecommand \selectlanguage [0]{\@gobble}%
\providecommand \bibinfo  [0]{\@secondoftwo}%
\providecommand \bibfield  [0]{\@secondoftwo}%
\providecommand \translation [1]{[#1]}%
\providecommand \BibitemOpen [0]{}%
\providecommand \bibitemStop [0]{}%
\providecommand \bibitemNoStop [0]{.\EOS\space}%
\providecommand \EOS [0]{\spacefactor3000\relax}%
\providecommand \BibitemShut  [1]{\csname bibitem#1\endcsname}%
\let\auto@bib@innerbib\@empty
\bibitem [{\citenamefont {Thompson}(1917)}]{d2010growth}%
  \BibitemOpen
  \bibfield  {author} {\bibinfo {author} {\bibfnamefont {D.~W.}\ \bibnamefont
  {Thompson}},\ }\href@noop {} {\emph {\bibinfo {title} {On Growth and Form}}}\
  (\bibinfo  {publisher} {Cambridge University Press},\ \bibinfo {year}
  {1917})\BibitemShut {NoStop}%
\bibitem [{\citenamefont {Watson}\ and\ \citenamefont
  {Crick}(1953)}]{watson1953molecular}%
  \BibitemOpen
  \bibfield  {author} {\bibinfo {author} {\bibfnamefont {J.~D.}\ \bibnamefont
  {Watson}}\ and\ \bibinfo {author} {\bibfnamefont {F.~H.~C.}\ \bibnamefont
  {Crick}},\ }\bibfield  {title} {\bibinfo {title} {Molecular structure of
  nucleic acids: a structure for deoxyribose nucleic acid},\ }\href@noop {}
  {\bibfield  {journal} {\bibinfo  {journal} {Nature}\ }\textbf {\bibinfo
  {volume} {171}},\ \bibinfo {pages} {737} (\bibinfo {year}
  {1953})}\BibitemShut {NoStop}%
\bibitem [{\citenamefont {Al-Hashimi}(2023)}]{al2023turing}%
  \BibitemOpen
  \bibfield  {author} {\bibinfo {author} {\bibfnamefont {H.~M.}\ \bibnamefont
  {Al-Hashimi}},\ }\bibfield  {title} {\bibinfo {title} {Turing, von {N}eumann,
  and the computational architecture of biological machines},\ }\href@noop {}
  {\bibfield  {journal} {\bibinfo  {journal} {Proc. Natl. Acad. Sci. USA}\
  }\textbf {\bibinfo {volume} {120}},\ \bibinfo {pages} {e2220022120} (\bibinfo
  {year} {2023})}\BibitemShut {NoStop}%
\bibitem [{\citenamefont {Eggert}\ \emph {et~al.}(2006)\citenamefont {Eggert},
  \citenamefont {Mitchison},\ and\ \citenamefont {Field}}]{eggert2006animal}%
  \BibitemOpen
  \bibfield  {author} {\bibinfo {author} {\bibfnamefont {U.~S.}\ \bibnamefont
  {Eggert}}, \bibinfo {author} {\bibfnamefont {T.~J.}\ \bibnamefont
  {Mitchison}},\ and\ \bibinfo {author} {\bibfnamefont {C.~M.}\ \bibnamefont
  {Field}},\ }\bibfield  {title} {\bibinfo {title} {Animal cytokinesis: from
  parts list to mechanisms},\ }\href@noop {} {\bibfield  {journal} {\bibinfo
  {journal} {Annu. Rev. Biochem.}\ }\textbf {\bibinfo {volume} {75}},\ \bibinfo
  {pages} {543} (\bibinfo {year} {2006})}\BibitemShut {NoStop}%
\bibitem [{\citenamefont {Paluch}\ \emph {et~al.}(2006)\citenamefont {Paluch},
  \citenamefont {Sykes}, \citenamefont {Prost},\ and\ \citenamefont
  {Bornens}}]{paluch2006dynamic}%
  \BibitemOpen
  \bibfield  {author} {\bibinfo {author} {\bibfnamefont {E.}~\bibnamefont
  {Paluch}}, \bibinfo {author} {\bibfnamefont {C.}~\bibnamefont {Sykes}},
  \bibinfo {author} {\bibfnamefont {J.}~\bibnamefont {Prost}},\ and\ \bibinfo
  {author} {\bibfnamefont {M.}~\bibnamefont {Bornens}},\ }\bibfield  {title}
  {\bibinfo {title} {Dynamic modes of the cortical actomyosin gel during cell
  locomotion and division},\ }\href@noop {} {\bibfield  {journal} {\bibinfo
  {journal} {Trends Cell Biol.}\ }\textbf {\bibinfo {volume} {16}},\ \bibinfo
  {pages} {5} (\bibinfo {year} {2006})}\BibitemShut {NoStop}%
\bibitem [{\citenamefont {L{\"a}mmermann}\ \emph {et~al.}(2008)\citenamefont
  {L{\"a}mmermann}, \citenamefont {Bader}, \citenamefont {Monkley},
  \citenamefont {Worbs}, \citenamefont {Wedlich-S{\"o}ldner}, \citenamefont
  {Hirsch}, \citenamefont {Keller}, \citenamefont {F{\"o}rster}, \citenamefont
  {Critchley}, \citenamefont {F{\"a}ssler} \emph
  {et~al.}}]{lammermann2008rapid}%
  \BibitemOpen
  \bibfield  {author} {\bibinfo {author} {\bibfnamefont {T.}~\bibnamefont
  {L{\"a}mmermann}}, \bibinfo {author} {\bibfnamefont {B.~L.}\ \bibnamefont
  {Bader}}, \bibinfo {author} {\bibfnamefont {S.~J.}\ \bibnamefont {Monkley}},
  \bibinfo {author} {\bibfnamefont {T.}~\bibnamefont {Worbs}}, \bibinfo
  {author} {\bibfnamefont {R.}~\bibnamefont {Wedlich-S{\"o}ldner}}, \bibinfo
  {author} {\bibfnamefont {K.}~\bibnamefont {Hirsch}}, \bibinfo {author}
  {\bibfnamefont {M.}~\bibnamefont {Keller}}, \bibinfo {author} {\bibfnamefont
  {R.}~\bibnamefont {F{\"o}rster}}, \bibinfo {author} {\bibfnamefont {D.~R.}\
  \bibnamefont {Critchley}}, \bibinfo {author} {\bibfnamefont {R.}~\bibnamefont
  {F{\"a}ssler}}, \emph {et~al.},\ }\bibfield  {title} {\bibinfo {title} {Rapid
  leukocyte migration by integrin-independent flowing and squeezing},\
  }\href@noop {} {\bibfield  {journal} {\bibinfo  {journal} {Nature}\ }\textbf
  {\bibinfo {volume} {453}},\ \bibinfo {pages} {51} (\bibinfo {year}
  {2008})}\BibitemShut {NoStop}%
\bibitem [{\citenamefont {Martin}\ \emph {et~al.}(2009)\citenamefont {Martin},
  \citenamefont {Kaschube},\ and\ \citenamefont
  {Wieschaus}}]{martin2009pulsed}%
  \BibitemOpen
  \bibfield  {author} {\bibinfo {author} {\bibfnamefont {A.~C.}\ \bibnamefont
  {Martin}}, \bibinfo {author} {\bibfnamefont {M.}~\bibnamefont {Kaschube}},\
  and\ \bibinfo {author} {\bibfnamefont {E.~F.}\ \bibnamefont {Wieschaus}},\
  }\bibfield  {title} {\bibinfo {title} {Pulsed contractions of an
  actin--myosin network drive apical constriction},\ }\href@noop {} {\bibfield
  {journal} {\bibinfo  {journal} {Nature}\ }\textbf {\bibinfo {volume} {457}},\
  \bibinfo {pages} {495} (\bibinfo {year} {2009})}\BibitemShut {NoStop}%
\bibitem [{\citenamefont {Dutta}\ \emph {et~al.}(2024)\citenamefont {Dutta},
  \citenamefont {Farhadifar}, \citenamefont {Lu}, \citenamefont
  {Kabacao{\u{g}}lu}, \citenamefont {Blackwell}, \citenamefont {Stein},
  \citenamefont {Lakonishok}, \citenamefont {Gelfand}, \citenamefont
  {Shvartsman},\ and\ \citenamefont {Shelley}}]{dutta2024self}%
  \BibitemOpen
  \bibfield  {author} {\bibinfo {author} {\bibfnamefont {S.}~\bibnamefont
  {Dutta}}, \bibinfo {author} {\bibfnamefont {R.}~\bibnamefont {Farhadifar}},
  \bibinfo {author} {\bibfnamefont {W.}~\bibnamefont {Lu}}, \bibinfo {author}
  {\bibfnamefont {G.}~\bibnamefont {Kabacao{\u{g}}lu}}, \bibinfo {author}
  {\bibfnamefont {R.}~\bibnamefont {Blackwell}}, \bibinfo {author}
  {\bibfnamefont {D.~B.}\ \bibnamefont {Stein}}, \bibinfo {author}
  {\bibfnamefont {M.}~\bibnamefont {Lakonishok}}, \bibinfo {author}
  {\bibfnamefont {V.~I.}\ \bibnamefont {Gelfand}}, \bibinfo {author}
  {\bibfnamefont {S.~Y.}\ \bibnamefont {Shvartsman}},\ and\ \bibinfo {author}
  {\bibfnamefont {M.~J.}\ \bibnamefont {Shelley}},\ }\bibfield  {title}
  {\bibinfo {title} {Self-organized intracellular twisters},\ }\href@noop {}
  {\bibfield  {journal} {\bibinfo  {journal} {Nat. Phys.}\ }\textbf {\bibinfo
  {volume} {20}},\ \bibinfo {pages} {1} (\bibinfo {year} {2024})}\BibitemShut
  {NoStop}%
\bibitem [{\citenamefont {Lighthill}(1976)}]{lighthill1976flagellar}%
  \BibitemOpen
  \bibfield  {author} {\bibinfo {author} {\bibfnamefont {J.}~\bibnamefont
  {Lighthill}},\ }\bibfield  {title} {\bibinfo {title} {Flagellar
  hydrodynamics},\ }\href@noop {} {\bibfield  {journal} {\bibinfo  {journal}
  {SIAM Rev.}\ }\textbf {\bibinfo {volume} {18}},\ \bibinfo {pages} {161}
  (\bibinfo {year} {1976})}\BibitemShut {NoStop}%
\bibitem [{\citenamefont {Gilpin}\ \emph {et~al.}(2020)\citenamefont {Gilpin},
  \citenamefont {Bull},\ and\ \citenamefont {Prakash}}]{gilpin2020multiscale}%
  \BibitemOpen
  \bibfield  {author} {\bibinfo {author} {\bibfnamefont {W.}~\bibnamefont
  {Gilpin}}, \bibinfo {author} {\bibfnamefont {M.~S.}\ \bibnamefont {Bull}},\
  and\ \bibinfo {author} {\bibfnamefont {M.}~\bibnamefont {Prakash}},\
  }\bibfield  {title} {\bibinfo {title} {The multiscale physics of cilia and
  flagella},\ }\href {https://doi.org/10.1038/s42254-019-0129-0} {\bibfield
  {journal} {\bibinfo  {journal} {Nat. Rev. Phys.}\ }\textbf {\bibinfo {volume}
  {2}},\ \bibinfo {pages} {74} (\bibinfo {year} {2020})}\BibitemShut {NoStop}%
\bibitem [{\citenamefont {Lauga}\ and\ \citenamefont
  {Powers}(2009)}]{lauga2009hydrodynamics}%
  \BibitemOpen
  \bibfield  {author} {\bibinfo {author} {\bibfnamefont {E.}~\bibnamefont
  {Lauga}}\ and\ \bibinfo {author} {\bibfnamefont {T.~R.}\ \bibnamefont
  {Powers}},\ }\bibfield  {title} {\bibinfo {title} {The hydrodynamics of
  swimming microorganisms},\ }\href
  {https://doi.org/10.1088/0034-4885/72/9/096601} {\bibfield  {journal}
  {\bibinfo  {journal} {Rep. Prog. Phys.}\ }\textbf {\bibinfo {volume} {72}},\
  \bibinfo {pages} {096601} (\bibinfo {year} {2009})}\BibitemShut {NoStop}%
\bibitem [{\citenamefont {Vilfan}\ and\ \citenamefont
  {J{\"u}licher}(2006)}]{vilfan2006hydrodynamic}%
  \BibitemOpen
  \bibfield  {author} {\bibinfo {author} {\bibfnamefont {A.}~\bibnamefont
  {Vilfan}}\ and\ \bibinfo {author} {\bibfnamefont {F.}~\bibnamefont
  {J{\"u}licher}},\ }\bibfield  {title} {\bibinfo {title} {Hydrodynamic flow
  patterns and synchronization of beating cilia},\ }\href@noop {} {\bibfield
  {journal} {\bibinfo  {journal} {Phys. Rev. Lett.}\ }\textbf {\bibinfo
  {volume} {96}},\ \bibinfo {pages} {058102} (\bibinfo {year}
  {2006})}\BibitemShut {NoStop}%
\bibitem [{\citenamefont {Klein}\ and\ \citenamefont
  {Paschke}(2004)}]{klein2004filamentous}%
  \BibitemOpen
  \bibfield  {author} {\bibinfo {author} {\bibfnamefont {D.~A.}\ \bibnamefont
  {Klein}}\ and\ \bibinfo {author} {\bibfnamefont {M.~W.}\ \bibnamefont
  {Paschke}},\ }\bibfield  {title} {\bibinfo {title} {Filamentous fungi: the
  indeterminate lifestyle and microbial ecology},\ }\href@noop {} {\bibfield
  {journal} {\bibinfo  {journal} {Microb. Ecol.}\ }\textbf {\bibinfo {volume}
  {47}},\ \bibinfo {pages} {224} (\bibinfo {year} {2004})}\BibitemShut
  {NoStop}%
\bibitem [{\citenamefont {Fleming}(1929)}]{fleming1929antibacterial}%
  \BibitemOpen
  \bibfield  {author} {\bibinfo {author} {\bibfnamefont {A.}~\bibnamefont
  {Fleming}},\ }\bibfield  {title} {\bibinfo {title} {On the antibacterial
  action of cultures of a penicillium, with special reference to their use in
  the isolation of {B}. influenzae},\ }\href@noop {} {\bibfield  {journal}
  {\bibinfo  {journal} {Br. J. Exp. Pathol.}\ }\textbf {\bibinfo {volume}
  {10}},\ \bibinfo {pages} {226} (\bibinfo {year} {1929})}\BibitemShut
  {NoStop}%
\bibitem [{\citenamefont {Fleming}(1941)}]{fleming1941penicillin}%
  \BibitemOpen
  \bibfield  {author} {\bibinfo {author} {\bibfnamefont {A.}~\bibnamefont
  {Fleming}},\ }\bibfield  {title} {\bibinfo {title} {Penicillin},\ }\href@noop
  {} {\bibfield  {journal} {\bibinfo  {journal} {Brit. Med. J.}\ }\textbf
  {\bibinfo {volume} {2}},\ \bibinfo {pages} {386} (\bibinfo {year}
  {1941})}\BibitemShut {NoStop}%
\bibitem [{\citenamefont {Houbraken}\ \emph {et~al.}(2011)\citenamefont
  {Houbraken}, \citenamefont {Frisvad},\ and\ \citenamefont
  {Samson}}]{houbraken2011fleming}%
  \BibitemOpen
  \bibfield  {author} {\bibinfo {author} {\bibfnamefont {J.}~\bibnamefont
  {Houbraken}}, \bibinfo {author} {\bibfnamefont {J.~C.}\ \bibnamefont
  {Frisvad}},\ and\ \bibinfo {author} {\bibfnamefont {R.~A.}\ \bibnamefont
  {Samson}},\ }\bibfield  {title} {\bibinfo {title} {Fleming’s penicillin
  producing strain is not \textit{{P}enicillium chrysogenum} but
  \textit{{P}.~rubens}},\ }\href@noop {} {\bibfield  {journal} {\bibinfo
  {journal} {IMA Fungus}\ }\textbf {\bibinfo {volume} {2}},\ \bibinfo {pages}
  {87} (\bibinfo {year} {2011})}\BibitemShut {NoStop}%
\bibitem [{\citenamefont {Alim}\ \emph {et~al.}(2013)\citenamefont {Alim},
  \citenamefont {Amselem}, \citenamefont {Peaudecerf}, \citenamefont
  {Brenner},\ and\ \citenamefont {Pringle}}]{alim2013random}%
  \BibitemOpen
  \bibfield  {author} {\bibinfo {author} {\bibfnamefont {K.}~\bibnamefont
  {Alim}}, \bibinfo {author} {\bibfnamefont {G.}~\bibnamefont {Amselem}},
  \bibinfo {author} {\bibfnamefont {F.}~\bibnamefont {Peaudecerf}}, \bibinfo
  {author} {\bibfnamefont {M.~P.}\ \bibnamefont {Brenner}},\ and\ \bibinfo
  {author} {\bibfnamefont {A.}~\bibnamefont {Pringle}},\ }\bibfield  {title}
  {\bibinfo {title} {Random network peristalsis in physarum polycephalum
  organizes fluid flows across an individual},\ }\href@noop {} {\bibfield
  {journal} {\bibinfo  {journal} {Proc. Natl. Acad. Sci. USA}\ }\textbf
  {\bibinfo {volume} {110}},\ \bibinfo {pages} {13306} (\bibinfo {year}
  {2013})}\BibitemShut {NoStop}%
\bibitem [{\citenamefont {Kramar}\ and\ \citenamefont
  {Alim}(2021)}]{kramar2021encoding}%
  \BibitemOpen
  \bibfield  {author} {\bibinfo {author} {\bibfnamefont {M.}~\bibnamefont
  {Kramar}}\ and\ \bibinfo {author} {\bibfnamefont {K.}~\bibnamefont {Alim}},\
  }\bibfield  {title} {\bibinfo {title} {Encoding memory in tube diameter
  hierarchy of living flow network},\ }\href@noop {} {\bibfield  {journal}
  {\bibinfo  {journal} {Proc. Natl. Acad. Sci. USA}\ }\textbf {\bibinfo
  {volume} {118}},\ \bibinfo {pages} {e2007815118} (\bibinfo {year}
  {2021})}\BibitemShut {NoStop}%
\bibitem [{\citenamefont {Grosberg}\ and\ \citenamefont
  {Strathmann}(2007)}]{grosberg2007evolution}%
  \BibitemOpen
  \bibfield  {author} {\bibinfo {author} {\bibfnamefont {R.~K.}\ \bibnamefont
  {Grosberg}}\ and\ \bibinfo {author} {\bibfnamefont {R.~R.}\ \bibnamefont
  {Strathmann}},\ }\bibfield  {title} {\bibinfo {title} {The evolution of
  multicellularity: a minor major transition?},\ }\href@noop {} {\bibfield
  {journal} {\bibinfo  {journal} {Annu. Rev. Ecol. Evol. Syst.}\ }\textbf
  {\bibinfo {volume} {38}},\ \bibinfo {pages} {621} (\bibinfo {year}
  {2007})}\BibitemShut {NoStop}%
\bibitem [{\citenamefont {Claessen}\ \emph {et~al.}(2014)\citenamefont
  {Claessen}, \citenamefont {Rozen}, \citenamefont {Kuipers}, \citenamefont
  {S{\o}gaard-Andersen},\ and\ \citenamefont
  {Van~Wezel}}]{claessen2014bacterial}%
  \BibitemOpen
  \bibfield  {author} {\bibinfo {author} {\bibfnamefont {D.}~\bibnamefont
  {Claessen}}, \bibinfo {author} {\bibfnamefont {D.~E.}\ \bibnamefont {Rozen}},
  \bibinfo {author} {\bibfnamefont {O.~P.}\ \bibnamefont {Kuipers}}, \bibinfo
  {author} {\bibfnamefont {L.}~\bibnamefont {S{\o}gaard-Andersen}},\ and\
  \bibinfo {author} {\bibfnamefont {G.~P.}\ \bibnamefont {Van~Wezel}},\
  }\bibfield  {title} {\bibinfo {title} {Bacterial solutions to
  multicellularity: a tale of biofilms, filaments and fruiting bodies},\
  }\href@noop {} {\bibfield  {journal} {\bibinfo  {journal} {Nat. Rev.
  Microbiol.}\ }\textbf {\bibinfo {volume} {12}},\ \bibinfo {pages} {115}
  (\bibinfo {year} {2014})}\BibitemShut {NoStop}%
\bibitem [{\citenamefont {Deblais}\ \emph {et~al.}(2020)\citenamefont
  {Deblais}, \citenamefont {Woutersen},\ and\ \citenamefont
  {Bonn}}]{deblais2020rheology}%
  \BibitemOpen
  \bibfield  {author} {\bibinfo {author} {\bibfnamefont {A.}~\bibnamefont
  {Deblais}}, \bibinfo {author} {\bibfnamefont {S.}~\bibnamefont {Woutersen}},\
  and\ \bibinfo {author} {\bibfnamefont {D.}~\bibnamefont {Bonn}},\ }\bibfield
  {title} {\bibinfo {title} {Rheology of entangled active polymer-like
  \textit{{T}.~Tubifex} worms},\ }\href
  {https://doi.org/10.1103/PhysRevLett.124.188002} {\bibfield  {journal}
  {\bibinfo  {journal} {Phys. Rev. Lett.}\ }\textbf {\bibinfo {volume} {124}},\
  \bibinfo {pages} {188002} (\bibinfo {year} {2020})}\BibitemShut {NoStop}%
\bibitem [{\citenamefont {Patil}\ \emph {et~al.}(2023)\citenamefont {Patil},
  \citenamefont {Tuazon}, \citenamefont {Kaufman}, \citenamefont
  {Chakrabortty}, \citenamefont {Qin}, \citenamefont {Dunkel},\ and\
  \citenamefont {Bhamla}}]{patil2023ultrafast}%
  \BibitemOpen
  \bibfield  {author} {\bibinfo {author} {\bibfnamefont {V.~P.}\ \bibnamefont
  {Patil}}, \bibinfo {author} {\bibfnamefont {H.}~\bibnamefont {Tuazon}},
  \bibinfo {author} {\bibfnamefont {E.}~\bibnamefont {Kaufman}}, \bibinfo
  {author} {\bibfnamefont {T.}~\bibnamefont {Chakrabortty}}, \bibinfo {author}
  {\bibfnamefont {D.}~\bibnamefont {Qin}}, \bibinfo {author} {\bibfnamefont
  {J.}~\bibnamefont {Dunkel}},\ and\ \bibinfo {author} {\bibfnamefont {M.~S.}\
  \bibnamefont {Bhamla}},\ }\bibfield  {title} {\bibinfo {title} {Ultrafast
  reversible self-assembly of living tangled matter},\ }\href@noop {}
  {\bibfield  {journal} {\bibinfo  {journal} {Science}\ }\textbf {\bibinfo
  {volume} {380}},\ \bibinfo {pages} {392} (\bibinfo {year}
  {2023})}\BibitemShut {NoStop}%
\bibitem [{\citenamefont {Gardel}\ \emph {et~al.}(2004)\citenamefont {Gardel},
  \citenamefont {Shin}, \citenamefont {MacKintosh}, \citenamefont {Mahadevan},
  \citenamefont {Matsudaira},\ and\ \citenamefont {Weitz}}]{gardel2004elastic}%
  \BibitemOpen
  \bibfield  {author} {\bibinfo {author} {\bibfnamefont {M.~L.}\ \bibnamefont
  {Gardel}}, \bibinfo {author} {\bibfnamefont {J.~H.}\ \bibnamefont {Shin}},
  \bibinfo {author} {\bibfnamefont {F.}~\bibnamefont {MacKintosh}}, \bibinfo
  {author} {\bibfnamefont {L.}~\bibnamefont {Mahadevan}}, \bibinfo {author}
  {\bibfnamefont {P.}~\bibnamefont {Matsudaira}},\ and\ \bibinfo {author}
  {\bibfnamefont {D.~A.}\ \bibnamefont {Weitz}},\ }\bibfield  {title} {\bibinfo
  {title} {Elastic behavior of cross-linked and bundled actin networks},\
  }\href@noop {} {\bibfield  {journal} {\bibinfo  {journal} {Science}\ }\textbf
  {\bibinfo {volume} {304}},\ \bibinfo {pages} {1301} (\bibinfo {year}
  {2004})}\BibitemShut {NoStop}%
\bibitem [{\citenamefont {Nakagaki}\ \emph {et~al.}(2004)\citenamefont
  {Nakagaki}, \citenamefont {Yamada},\ and\ \citenamefont
  {Hara}}]{nakagaki2004smart}%
  \BibitemOpen
  \bibfield  {author} {\bibinfo {author} {\bibfnamefont {T.}~\bibnamefont
  {Nakagaki}}, \bibinfo {author} {\bibfnamefont {H.}~\bibnamefont {Yamada}},\
  and\ \bibinfo {author} {\bibfnamefont {M.}~\bibnamefont {Hara}},\ }\bibfield
  {title} {\bibinfo {title} {Smart network solutions in an amoeboid organism},\
  }\href@noop {} {\bibfield  {journal} {\bibinfo  {journal} {Biophys. Chem.}\
  }\textbf {\bibinfo {volume} {107}},\ \bibinfo {pages} {1} (\bibinfo {year}
  {2004})}\BibitemShut {NoStop}%
\bibitem [{\citenamefont {Anderson}(1972)}]{anderson1972more}%
  \BibitemOpen
  \bibfield  {author} {\bibinfo {author} {\bibfnamefont {P.~W.}\ \bibnamefont
  {Anderson}},\ }\bibfield  {title} {\bibinfo {title} {More is {D}ifferent:
  {B}roken symmetry and the nature of the hierarchical structure of science.},\
  }\href@noop {} {\bibfield  {journal} {\bibinfo  {journal} {Science}\ }\textbf
  {\bibinfo {volume} {177}},\ \bibinfo {pages} {393} (\bibinfo {year}
  {1972})}\BibitemShut {NoStop}%
\bibitem [{\citenamefont {Saintyves}\ \emph {et~al.}(2024)\citenamefont
  {Saintyves}, \citenamefont {Spenko},\ and\ \citenamefont
  {Jaeger}}]{saintyves2024self}%
  \BibitemOpen
  \bibfield  {author} {\bibinfo {author} {\bibfnamefont {B.}~\bibnamefont
  {Saintyves}}, \bibinfo {author} {\bibfnamefont {M.}~\bibnamefont {Spenko}},\
  and\ \bibinfo {author} {\bibfnamefont {H.~M.}\ \bibnamefont {Jaeger}},\
  }\bibfield  {title} {\bibinfo {title} {A self-organizing robotic aggregate
  using solid and liquid-like collective states},\ }\href@noop {} {\bibfield
  {journal} {\bibinfo  {journal} {Sci. Robot.}\ }\textbf {\bibinfo {volume}
  {9}},\ \bibinfo {pages} {eadh4130} (\bibinfo {year} {2024})}\BibitemShut
  {NoStop}%
\bibitem [{\citenamefont {Fladung}\ \emph {et~al.}(2024)\citenamefont
  {Fladung}, \citenamefont {Berkes}, \citenamefont {Alletzhaeusser},
  \citenamefont {Chen}, \citenamefont {Munding}, \citenamefont {Tanaka},
  \citenamefont {Wegener},\ and\ \citenamefont {Bastmeyer}}]{fladung2024lies}%
  \BibitemOpen
  \bibfield  {author} {\bibinfo {author} {\bibfnamefont {M.}~\bibnamefont
  {Fladung}}, \bibinfo {author} {\bibfnamefont {A.}~\bibnamefont {Berkes}},
  \bibinfo {author} {\bibfnamefont {T.}~\bibnamefont {Alletzhaeusser}},
  \bibinfo {author} {\bibfnamefont {Y.}~\bibnamefont {Chen}}, \bibinfo {author}
  {\bibfnamefont {N.}~\bibnamefont {Munding}}, \bibinfo {author} {\bibfnamefont
  {M.}~\bibnamefont {Tanaka}}, \bibinfo {author} {\bibfnamefont
  {M.}~\bibnamefont {Wegener}},\ and\ \bibinfo {author} {\bibfnamefont
  {M.}~\bibnamefont {Bastmeyer}},\ }\bibfield  {title} {\bibinfo {title} {What
  lies beyond--insights into elastic microscaffolds with metamaterial
  properties for cell studies},\ }\href@noop {} {\bibfield  {journal} {\bibinfo
   {journal} {Curr. Opin. Biomed. Eng.}\ ,\ \bibinfo {pages} {100568}}
  (\bibinfo {year} {2024})}\BibitemShut {NoStop}%
\bibitem [{\citenamefont {Taale}\ \emph {et~al.}(2023)\citenamefont {Taale},
  \citenamefont {Schmidt}, \citenamefont {Taheri}, \citenamefont {Timmermann},\
  and\ \citenamefont {Selhuber-Unkel}}]{taale2023minimalistic}%
  \BibitemOpen
  \bibfield  {author} {\bibinfo {author} {\bibfnamefont {M.}~\bibnamefont
  {Taale}}, \bibinfo {author} {\bibfnamefont {M.}~\bibnamefont {Schmidt}},
  \bibinfo {author} {\bibfnamefont {F.}~\bibnamefont {Taheri}}, \bibinfo
  {author} {\bibfnamefont {M.}~\bibnamefont {Timmermann}},\ and\ \bibinfo
  {author} {\bibfnamefont {C.}~\bibnamefont {Selhuber-Unkel}},\ }\bibfield
  {title} {\bibinfo {title} {A minimalistic, synthetic cell-inspired
  metamaterial for enabling reversible strain-stiffening},\ }\href@noop {}
  {\bibfield  {journal} {\bibinfo  {journal} {Adv. Mater. Technol.}\ }\textbf
  {\bibinfo {volume} {8}},\ \bibinfo {pages} {2201441} (\bibinfo {year}
  {2023})}\BibitemShut {NoStop}%
\bibitem [{\citenamefont {Fakhri}\ \emph {et~al.}(2010)\citenamefont {Fakhri},
  \citenamefont {MacKintosh}, \citenamefont {Lounis}, \citenamefont {Cognet},\
  and\ \citenamefont {Pasquali}}]{fakhri2010brownian}%
  \BibitemOpen
  \bibfield  {author} {\bibinfo {author} {\bibfnamefont {N.}~\bibnamefont
  {Fakhri}}, \bibinfo {author} {\bibfnamefont {F.~C.}\ \bibnamefont
  {MacKintosh}}, \bibinfo {author} {\bibfnamefont {B.}~\bibnamefont {Lounis}},
  \bibinfo {author} {\bibfnamefont {L.}~\bibnamefont {Cognet}},\ and\ \bibinfo
  {author} {\bibfnamefont {M.}~\bibnamefont {Pasquali}},\ }\bibfield  {title}
  {\bibinfo {title} {Brownian motion of stiff filaments in a crowded
  environment},\ }\href@noop {} {\bibfield  {journal} {\bibinfo  {journal}
  {Science}\ }\textbf {\bibinfo {volume} {330}},\ \bibinfo {pages} {1804}
  (\bibinfo {year} {2010})}\BibitemShut {NoStop}%
\bibitem [{\citenamefont {Fakhri}\ \emph {et~al.}(2009)\citenamefont {Fakhri},
  \citenamefont {Tsyboulski}, \citenamefont {Cognet}, \citenamefont {Weisman},\
  and\ \citenamefont {Pasquali}}]{fakhri2009diameter}%
  \BibitemOpen
  \bibfield  {author} {\bibinfo {author} {\bibfnamefont {N.}~\bibnamefont
  {Fakhri}}, \bibinfo {author} {\bibfnamefont {D.~A.}\ \bibnamefont
  {Tsyboulski}}, \bibinfo {author} {\bibfnamefont {L.}~\bibnamefont {Cognet}},
  \bibinfo {author} {\bibfnamefont {R.~B.}\ \bibnamefont {Weisman}},\ and\
  \bibinfo {author} {\bibfnamefont {M.}~\bibnamefont {Pasquali}},\ }\bibfield
  {title} {\bibinfo {title} {Diameter-dependent bending dynamics of
  single-walled carbon nanotubes in liquids},\ }\href@noop {} {\bibfield
  {journal} {\bibinfo  {journal} {Proc. Natl. Acad. Sci. USA}\ }\textbf
  {\bibinfo {volume} {106}},\ \bibinfo {pages} {14219} (\bibinfo {year}
  {2009})}\BibitemShut {NoStop}%
\bibitem [{\citenamefont {Ramaswamy}(2010)}]{ramaswamy2010mechanics}%
  \BibitemOpen
  \bibfield  {author} {\bibinfo {author} {\bibfnamefont {S.}~\bibnamefont
  {Ramaswamy}},\ }\bibfield  {title} {\bibinfo {title} {The mechanics and
  statistics of active matter},\ }\href@noop {} {\bibfield  {journal} {\bibinfo
   {journal} {Annu. Rev. Condens. Matter Phys.}\ }\textbf {\bibinfo {volume}
  {1}},\ \bibinfo {pages} {323} (\bibinfo {year} {2010})}\BibitemShut {NoStop}%
\bibitem [{\citenamefont {Marchetti}\ \emph {et~al.}(2013)\citenamefont
  {Marchetti}, \citenamefont {Joanny}, \citenamefont {Ramaswamy}, \citenamefont
  {Liverpool}, \citenamefont {Prost}, \citenamefont {Rao},\ and\ \citenamefont
  {Simha}}]{marchetti2013hydrodynamics}%
  \BibitemOpen
  \bibfield  {author} {\bibinfo {author} {\bibfnamefont {M.~C.}\ \bibnamefont
  {Marchetti}}, \bibinfo {author} {\bibfnamefont {J.-F.}\ \bibnamefont
  {Joanny}}, \bibinfo {author} {\bibfnamefont {S.}~\bibnamefont {Ramaswamy}},
  \bibinfo {author} {\bibfnamefont {T.~B.}\ \bibnamefont {Liverpool}}, \bibinfo
  {author} {\bibfnamefont {J.}~\bibnamefont {Prost}}, \bibinfo {author}
  {\bibfnamefont {M.}~\bibnamefont {Rao}},\ and\ \bibinfo {author}
  {\bibfnamefont {R.~A.}\ \bibnamefont {Simha}},\ }\bibfield  {title} {\bibinfo
  {title} {Hydrodynamics of soft active matter},\ }\href@noop {} {\bibfield
  {journal} {\bibinfo  {journal} {Rev. Mod. Phys.}\ }\textbf {\bibinfo {volume}
  {85}},\ \bibinfo {pages} {1143} (\bibinfo {year} {2013})}\BibitemShut
  {NoStop}%
\bibitem [{\citenamefont {Doi}\ and\ \citenamefont
  {Edwards}(1988)}]{doi1988theory}%
  \BibitemOpen
  \bibfield  {author} {\bibinfo {author} {\bibfnamefont {M.}~\bibnamefont
  {Doi}}\ and\ \bibinfo {author} {\bibfnamefont {S.~F.}\ \bibnamefont
  {Edwards}},\ }\href@noop {} {\emph {\bibinfo {title} {The theory of polymer
  dynamics}}},\ Vol.~\bibinfo {volume} {73}\ (\bibinfo  {publisher} {Oxford
  University Press},\ \bibinfo {year} {1988})\BibitemShut {NoStop}%
\bibitem [{\citenamefont {Stam}\ \emph {et~al.}(2017)\citenamefont {Stam},
  \citenamefont {Freedman}, \citenamefont {Banerjee}, \citenamefont {Weirich},
  \citenamefont {Dinner},\ and\ \citenamefont {Gardel}}]{stam2017filament}%
  \BibitemOpen
  \bibfield  {author} {\bibinfo {author} {\bibfnamefont {S.}~\bibnamefont
  {Stam}}, \bibinfo {author} {\bibfnamefont {S.~L.}\ \bibnamefont {Freedman}},
  \bibinfo {author} {\bibfnamefont {S.}~\bibnamefont {Banerjee}}, \bibinfo
  {author} {\bibfnamefont {K.~L.}\ \bibnamefont {Weirich}}, \bibinfo {author}
  {\bibfnamefont {A.~R.}\ \bibnamefont {Dinner}},\ and\ \bibinfo {author}
  {\bibfnamefont {M.~L.}\ \bibnamefont {Gardel}},\ }\bibfield  {title}
  {\bibinfo {title} {Filament rigidity and connectivity tune the deformation
  modes of active biopolymer networks},\ }\href@noop {} {\bibfield  {journal}
  {\bibinfo  {journal} {Proc. Natl. Acad. Sci. USA}\ }\textbf {\bibinfo
  {volume} {114}},\ \bibinfo {pages} {E10037} (\bibinfo {year}
  {2017})}\BibitemShut {NoStop}%
\bibitem [{\citenamefont {Banerjee}\ \emph {et~al.}(2020)\citenamefont
  {Banerjee}, \citenamefont {Gardel},\ and\ \citenamefont
  {Schwarz}}]{banerjee2020actin}%
  \BibitemOpen
  \bibfield  {author} {\bibinfo {author} {\bibfnamefont {S.}~\bibnamefont
  {Banerjee}}, \bibinfo {author} {\bibfnamefont {M.~L.}\ \bibnamefont
  {Gardel}},\ and\ \bibinfo {author} {\bibfnamefont {U.~S.}\ \bibnamefont
  {Schwarz}},\ }\bibfield  {title} {\bibinfo {title} {The actin cytoskeleton as
  an active adaptive material},\ }\href@noop {} {\bibfield  {journal} {\bibinfo
   {journal} {Annu. Rev. Condens. Matter Phys.}\ }\textbf {\bibinfo {volume}
  {11}},\ \bibinfo {pages} {421} (\bibinfo {year} {2020})}\BibitemShut
  {NoStop}%
\bibitem [{\citenamefont {Millman}\ and\ \citenamefont
  {Parker}(1977)}]{millman1977elements}%
  \BibitemOpen
  \bibfield  {author} {\bibinfo {author} {\bibfnamefont {R.~S.}\ \bibnamefont
  {Millman}}\ and\ \bibinfo {author} {\bibfnamefont {G.~D.}\ \bibnamefont
  {Parker}},\ }\href@noop {} {\emph {\bibinfo {title} {Elements of differential
  geometry}}}\ (\bibinfo  {publisher} {Prentice-Hall},\ \bibinfo {year}
  {1977})\BibitemShut {NoStop}%
\bibitem [{\citenamefont {Novikov}\ and\ \citenamefont
  {Fomenko}(1990)}]{novikovfomaneko}%
  \BibitemOpen
  \bibfield  {author} {\bibinfo {author} {\bibfnamefont {S.~P.}\ \bibnamefont
  {Novikov}}\ and\ \bibinfo {author} {\bibfnamefont {A.~T.}\ \bibnamefont
  {Fomenko}},\ }\href@noop {} {\emph {\bibinfo {title} {Basic Elements of
  Differential Geometry and Topology}}}\ (\bibinfo  {publisher} {Kluwer
  Academic Publishing},\ \bibinfo {year} {1990})\BibitemShut {NoStop}%
\bibitem [{\citenamefont {Crenshaw}\ and\ \citenamefont
  {Edelstein-Keshet}(1993)}]{crenshaw1993orientation}%
  \BibitemOpen
  \bibfield  {author} {\bibinfo {author} {\bibfnamefont {H.~C.}\ \bibnamefont
  {Crenshaw}}\ and\ \bibinfo {author} {\bibfnamefont {L.}~\bibnamefont
  {Edelstein-Keshet}},\ }\bibfield  {title} {\bibinfo {title} {Orientation by
  helical motion-{II}. {C}hanging the direction of the axis of motion},\
  }\href@noop {} {\bibfield  {journal} {\bibinfo  {journal} {Bull. Math.
  Biol.}\ }\textbf {\bibinfo {volume} {55}},\ \bibinfo {pages} {213} (\bibinfo
  {year} {1993})}\BibitemShut {NoStop}%
\bibitem [{\citenamefont {Friedrich}\ and\ \citenamefont
  {J{\"u}licher}(2009)}]{friedrich2009steering}%
  \BibitemOpen
  \bibfield  {author} {\bibinfo {author} {\bibfnamefont {B.~M.}\ \bibnamefont
  {Friedrich}}\ and\ \bibinfo {author} {\bibfnamefont {F.}~\bibnamefont
  {J{\"u}licher}},\ }\bibfield  {title} {\bibinfo {title} {Steering chiral
  swimmers along noisy helical paths},\ }\href@noop {} {\bibfield  {journal}
  {\bibinfo  {journal} {Phys. Rev. Lett.}\ }\textbf {\bibinfo {volume} {103}},\
  \bibinfo {pages} {068102} (\bibinfo {year} {2009})}\BibitemShut {NoStop}%
\bibitem [{\citenamefont {Bente}\ \emph {et~al.}(2020)\citenamefont {Bente},
  \citenamefont {Mohammadinejad}, \citenamefont {Charsooghi}, \citenamefont
  {Bachmann}, \citenamefont {Codutti}, \citenamefont {Lef{\`e}vre},
  \citenamefont {Klumpp},\ and\ \citenamefont {Faivre}}]{bente2020high}%
  \BibitemOpen
  \bibfield  {author} {\bibinfo {author} {\bibfnamefont {K.}~\bibnamefont
  {Bente}}, \bibinfo {author} {\bibfnamefont {S.}~\bibnamefont
  {Mohammadinejad}}, \bibinfo {author} {\bibfnamefont {M.~A.}\ \bibnamefont
  {Charsooghi}}, \bibinfo {author} {\bibfnamefont {F.}~\bibnamefont
  {Bachmann}}, \bibinfo {author} {\bibfnamefont {A.}~\bibnamefont {Codutti}},
  \bibinfo {author} {\bibfnamefont {C.~T.}\ \bibnamefont {Lef{\`e}vre}},
  \bibinfo {author} {\bibfnamefont {S.}~\bibnamefont {Klumpp}},\ and\ \bibinfo
  {author} {\bibfnamefont {D.}~\bibnamefont {Faivre}},\ }\bibfield  {title}
  {\bibinfo {title} {High-speed motility originates from cooperatively pushing
  and pulling flagella bundles in bilophotrichous bacteria},\ }\href@noop {}
  {\bibfield  {journal} {\bibinfo  {journal} {eLife}\ }\textbf {\bibinfo
  {volume} {9}},\ \bibinfo {pages} {e47551} (\bibinfo {year}
  {2020})}\BibitemShut {NoStop}%
\bibitem [{\citenamefont {Jikeli}\ \emph {et~al.}(2015)\citenamefont {Jikeli},
  \citenamefont {Alvarez}, \citenamefont {Friedrich}, \citenamefont {Wilson},
  \citenamefont {Pascal}, \citenamefont {Colin}, \citenamefont {Pichlo},
  \citenamefont {Rennhack}, \citenamefont {Brenker},\ and\ \citenamefont
  {Kaupp}}]{jikeli2015sperm}%
  \BibitemOpen
  \bibfield  {author} {\bibinfo {author} {\bibfnamefont {J.~F.}\ \bibnamefont
  {Jikeli}}, \bibinfo {author} {\bibfnamefont {L.}~\bibnamefont {Alvarez}},
  \bibinfo {author} {\bibfnamefont {B.~M.}\ \bibnamefont {Friedrich}}, \bibinfo
  {author} {\bibfnamefont {L.~G.}\ \bibnamefont {Wilson}}, \bibinfo {author}
  {\bibfnamefont {R.}~\bibnamefont {Pascal}}, \bibinfo {author} {\bibfnamefont
  {R.}~\bibnamefont {Colin}}, \bibinfo {author} {\bibfnamefont
  {M.}~\bibnamefont {Pichlo}}, \bibinfo {author} {\bibfnamefont
  {A.}~\bibnamefont {Rennhack}}, \bibinfo {author} {\bibfnamefont
  {C.}~\bibnamefont {Brenker}},\ and\ \bibinfo {author} {\bibfnamefont {U.~B.}\
  \bibnamefont {Kaupp}},\ }\bibfield  {title} {\bibinfo {title} {Sperm
  navigation along helical paths in 3d chemoattractant landscapes},\
  }\href@noop {} {\bibfield  {journal} {\bibinfo  {journal} {Nat. Commun.}\
  }\textbf {\bibinfo {volume} {6}},\ \bibinfo {pages} {7985} (\bibinfo {year}
  {2015})}\BibitemShut {NoStop}%
\bibitem [{\citenamefont {Friedrich}\ and\ \citenamefont
  {J{\"u}licher}(2007)}]{friedrich2007chemotaxis}%
  \BibitemOpen
  \bibfield  {author} {\bibinfo {author} {\bibfnamefont {B.~M.}\ \bibnamefont
  {Friedrich}}\ and\ \bibinfo {author} {\bibfnamefont {F.}~\bibnamefont
  {J{\"u}licher}},\ }\bibfield  {title} {\bibinfo {title} {Chemotaxis of sperm
  cells},\ }\href@noop {} {\bibfield  {journal} {\bibinfo  {journal} {Proc.
  Natl. Acad. Sci. USA}\ }\textbf {\bibinfo {volume} {104}},\ \bibinfo {pages}
  {13256} (\bibinfo {year} {2007})}\BibitemShut {NoStop}%
\bibitem [{\citenamefont {Sevilla}(2016)}]{sevilla2016diffusion}%
  \BibitemOpen
  \bibfield  {author} {\bibinfo {author} {\bibfnamefont {F.~J.}\ \bibnamefont
  {Sevilla}},\ }\bibfield  {title} {\bibinfo {title} {Diffusion of active
  chiral particles},\ }\href@noop {} {\bibfield  {journal} {\bibinfo  {journal}
  {Phys. Rev. E}\ }\textbf {\bibinfo {volume} {94}},\ \bibinfo {pages} {062120}
  (\bibinfo {year} {2016})}\BibitemShut {NoStop}%
\bibitem [{\citenamefont {Lettermann}\ \emph {et~al.}(2024)\citenamefont
  {Lettermann}, \citenamefont {Ziebert},\ and\ \citenamefont
  {Schwarz}}]{lettermann2024geometrical}%
  \BibitemOpen
  \bibfield  {author} {\bibinfo {author} {\bibfnamefont {L.}~\bibnamefont
  {Lettermann}}, \bibinfo {author} {\bibfnamefont {F.}~\bibnamefont
  {Ziebert}},\ and\ \bibinfo {author} {\bibfnamefont {U.~S.}\ \bibnamefont
  {Schwarz}},\ }\bibfield  {title} {\bibinfo {title} {A geometrical theory of
  gliding motility based on cell shape and surface flow},\ }\href@noop {}
  {\bibfield  {journal} {\bibinfo  {journal} {Proc. Natl. Acad. Sci. USA}\
  }\textbf {\bibinfo {volume} {121}},\ \bibinfo {pages} {e2410708121} (\bibinfo
  {year} {2024})}\BibitemShut {NoStop}%
\bibitem [{\citenamefont {Mattila}\ and\ \citenamefont
  {Lappalainen}(2008)}]{mattila2008filopodia}%
  \BibitemOpen
  \bibfield  {author} {\bibinfo {author} {\bibfnamefont {P.~K.}\ \bibnamefont
  {Mattila}}\ and\ \bibinfo {author} {\bibfnamefont {P.}~\bibnamefont
  {Lappalainen}},\ }\bibfield  {title} {\bibinfo {title} {Filopodia: molecular
  architecture and cellular functions},\ }\href@noop {} {\bibfield  {journal}
  {\bibinfo  {journal} {Nat. Rev. Mol. Cell Biol.}\ }\textbf {\bibinfo {volume}
  {9}},\ \bibinfo {pages} {446} (\bibinfo {year} {2008})}\BibitemShut {NoStop}%
\bibitem [{\citenamefont {Solovev}\ and\ \citenamefont
  {Friedrich}(2022)}]{solovev2022synchronization}%
  \BibitemOpen
  \bibfield  {author} {\bibinfo {author} {\bibfnamefont {A.}~\bibnamefont
  {Solovev}}\ and\ \bibinfo {author} {\bibfnamefont {B.~M.}\ \bibnamefont
  {Friedrich}},\ }\bibfield  {title} {\bibinfo {title} {Synchronization in
  cilia carpets: multiple metachronal waves are stable, but one wave
  dominates},\ }\href@noop {} {\bibfield  {journal} {\bibinfo  {journal} {New
  J. Phys.}\ }\textbf {\bibinfo {volume} {24}},\ \bibinfo {pages} {013015}
  (\bibinfo {year} {2022})}\BibitemShut {NoStop}%
\bibitem [{\citenamefont {Poon}\ \emph {et~al.}(2025)\citenamefont {Poon},
  \citenamefont {J{\'e}kely},\ and\ \citenamefont {Wan}}]{poon2025dynamics}%
  \BibitemOpen
  \bibfield  {author} {\bibinfo {author} {\bibfnamefont {R.~N.}\ \bibnamefont
  {Poon}}, \bibinfo {author} {\bibfnamefont {G.}~\bibnamefont {J{\'e}kely}},\
  and\ \bibinfo {author} {\bibfnamefont {K.~Y.}\ \bibnamefont {Wan}},\
  }\bibfield  {title} {\bibinfo {title} {Dynamics and emergence of metachronal
  waves in the ciliary band of a metazoan larva},\ }\href@noop {} {\bibfield
  {journal} {\bibinfo  {journal} {bioRxiv}\ ,\ \bibinfo {pages} {2025}}
  (\bibinfo {year} {2025})}\BibitemShut {NoStop}%
\bibitem [{\citenamefont {Wanner}\ \emph {et~al.}(1996)\citenamefont {Wanner},
  \citenamefont {Salath{\'e}},\ and\ \citenamefont
  {O'Riordan}}]{wanner1996mucociliary}%
  \BibitemOpen
  \bibfield  {author} {\bibinfo {author} {\bibfnamefont {A.}~\bibnamefont
  {Wanner}}, \bibinfo {author} {\bibfnamefont {M.}~\bibnamefont
  {Salath{\'e}}},\ and\ \bibinfo {author} {\bibfnamefont {T.~G.}\ \bibnamefont
  {O'Riordan}},\ }\bibfield  {title} {\bibinfo {title} {Mucociliary clearance
  in the airways.},\ }\href@noop {} {\bibfield  {journal} {\bibinfo  {journal}
  {Am. J. Respir. Crit. Care Med.}\ }\textbf {\bibinfo {volume} {154}},\
  \bibinfo {pages} {1868} (\bibinfo {year} {1996})}\BibitemShut {NoStop}%
\bibitem [{\citenamefont {Huber}\ \emph {et~al.}(2013)\citenamefont {Huber},
  \citenamefont {Schnau{\ss}}, \citenamefont {R{\"o}nicke}, \citenamefont
  {Rauch}, \citenamefont {M{\"u}ller}, \citenamefont {F{\"u}tterer},\ and\
  \citenamefont {K{\"a}s}}]{huber2013emergent}%
  \BibitemOpen
  \bibfield  {author} {\bibinfo {author} {\bibfnamefont {F.}~\bibnamefont
  {Huber}}, \bibinfo {author} {\bibfnamefont {J.}~\bibnamefont {Schnau{\ss}}},
  \bibinfo {author} {\bibfnamefont {S.}~\bibnamefont {R{\"o}nicke}}, \bibinfo
  {author} {\bibfnamefont {P.}~\bibnamefont {Rauch}}, \bibinfo {author}
  {\bibfnamefont {K.}~\bibnamefont {M{\"u}ller}}, \bibinfo {author}
  {\bibfnamefont {C.}~\bibnamefont {F{\"u}tterer}},\ and\ \bibinfo {author}
  {\bibfnamefont {J.}~\bibnamefont {K{\"a}s}},\ }\bibfield  {title} {\bibinfo
  {title} {Emergent complexity of the cytoskeleton: from single filaments to
  tissue},\ }\href@noop {} {\bibfield  {journal} {\bibinfo  {journal} {Adv.
  Phys.}\ }\textbf {\bibinfo {volume} {62}},\ \bibinfo {pages} {1} (\bibinfo
  {year} {2013})}\BibitemShut {NoStop}%
\bibitem [{\citenamefont {Fletcher}\ and\ \citenamefont
  {Mullins}(2010)}]{fletcher2010cell}%
  \BibitemOpen
  \bibfield  {author} {\bibinfo {author} {\bibfnamefont {D.~A.}\ \bibnamefont
  {Fletcher}}\ and\ \bibinfo {author} {\bibfnamefont {R.~D.}\ \bibnamefont
  {Mullins}},\ }\bibfield  {title} {\bibinfo {title} {Cell mechanics and the
  cytoskeleton},\ }\href@noop {} {\bibfield  {journal} {\bibinfo  {journal}
  {Nature}\ }\textbf {\bibinfo {volume} {463}},\ \bibinfo {pages} {485}
  (\bibinfo {year} {2010})}\BibitemShut {NoStop}%
\bibitem [{\citenamefont {Pegoraro}\ \emph {et~al.}(2017)\citenamefont
  {Pegoraro}, \citenamefont {Janmey},\ and\ \citenamefont
  {Weitz}}]{pegoraro2017mechanical}%
  \BibitemOpen
  \bibfield  {author} {\bibinfo {author} {\bibfnamefont {A.~F.}\ \bibnamefont
  {Pegoraro}}, \bibinfo {author} {\bibfnamefont {P.}~\bibnamefont {Janmey}},\
  and\ \bibinfo {author} {\bibfnamefont {D.~A.}\ \bibnamefont {Weitz}},\
  }\bibfield  {title} {\bibinfo {title} {Mechanical properties of the
  cytoskeleton and cells},\ }\href@noop {} {\bibfield  {journal} {\bibinfo
  {journal} {Cold Spring Harb. Perspect. Biol.}\ }\textbf {\bibinfo {volume}
  {9}},\ \bibinfo {pages} {a022038} (\bibinfo {year} {2017})}\BibitemShut
  {NoStop}%
\bibitem [{\citenamefont {Kollmannsberger}\ and\ \citenamefont
  {Fabry}(2011)}]{kollmannsberger2011linear}%
  \BibitemOpen
  \bibfield  {author} {\bibinfo {author} {\bibfnamefont {P.}~\bibnamefont
  {Kollmannsberger}}\ and\ \bibinfo {author} {\bibfnamefont {B.}~\bibnamefont
  {Fabry}},\ }\bibfield  {title} {\bibinfo {title} {Linear and nonlinear
  rheology of living cells},\ }\href@noop {} {\bibfield  {journal} {\bibinfo
  {journal} {Annu. Rev. Mater. Res.}\ }\textbf {\bibinfo {volume} {41}},\
  \bibinfo {pages} {75} (\bibinfo {year} {2011})}\BibitemShut {NoStop}%
\bibitem [{\citenamefont {Kroy}\ and\ \citenamefont
  {Frey}(1996)}]{kroy1996force}%
  \BibitemOpen
  \bibfield  {author} {\bibinfo {author} {\bibfnamefont {K.}~\bibnamefont
  {Kroy}}\ and\ \bibinfo {author} {\bibfnamefont {E.}~\bibnamefont {Frey}},\
  }\bibfield  {title} {\bibinfo {title} {Force-extension relation and plateau
  modulus for wormlike chains},\ }\href@noop {} {\bibfield  {journal} {\bibinfo
   {journal} {Phys. Rev. Lett.}\ }\textbf {\bibinfo {volume} {77}},\ \bibinfo
  {pages} {306} (\bibinfo {year} {1996})}\BibitemShut {NoStop}%
\bibitem [{\citenamefont {Landau}\ \emph {et~al.}(2012)\citenamefont {Landau},
  \citenamefont {Pitaevskii}, \citenamefont {Kosevich},\ and\ \citenamefont
  {Lifshitz}}]{landau2012theory}%
  \BibitemOpen
  \bibfield  {author} {\bibinfo {author} {\bibfnamefont {L.~D.}\ \bibnamefont
  {Landau}}, \bibinfo {author} {\bibfnamefont {L.~P.}\ \bibnamefont
  {Pitaevskii}}, \bibinfo {author} {\bibfnamefont {A.~M.}\ \bibnamefont
  {Kosevich}},\ and\ \bibinfo {author} {\bibfnamefont {E.~M.}\ \bibnamefont
  {Lifshitz}},\ }\href@noop {} {\emph {\bibinfo {title} {Theory of elasticity:
  volume 7}}},\ Vol.~\bibinfo {volume} {7}\ (\bibinfo  {publisher} {Elsevier},\
  \bibinfo {year} {2012})\BibitemShut {NoStop}%
\bibitem [{\citenamefont {Civalek}\ and\ \citenamefont
  {Demir}(2011)}]{civalek2011bending}%
  \BibitemOpen
  \bibfield  {author} {\bibinfo {author} {\bibfnamefont {{\"O}.}~\bibnamefont
  {Civalek}}\ and\ \bibinfo {author} {\bibfnamefont {{\c{C}}.}~\bibnamefont
  {Demir}},\ }\bibfield  {title} {\bibinfo {title} {Bending analysis of
  microtubules using nonlocal {E}uler--{B}ernoulli beam theory},\ }\href@noop
  {} {\bibfield  {journal} {\bibinfo  {journal} {Appl. Math. Model.}\ }\textbf
  {\bibinfo {volume} {35}},\ \bibinfo {pages} {2053} (\bibinfo {year}
  {2011})}\BibitemShut {NoStop}%
\bibitem [{\citenamefont {Ku{\v{c}}era}\ \emph {et~al.}(2017)\citenamefont
  {Ku{\v{c}}era}, \citenamefont {Havelka},\ and\ \citenamefont
  {Cifra}}]{kuvcera2017vibrations}%
  \BibitemOpen
  \bibfield  {author} {\bibinfo {author} {\bibfnamefont {O.}~\bibnamefont
  {Ku{\v{c}}era}}, \bibinfo {author} {\bibfnamefont {D.}~\bibnamefont
  {Havelka}},\ and\ \bibinfo {author} {\bibfnamefont {M.}~\bibnamefont
  {Cifra}},\ }\bibfield  {title} {\bibinfo {title} {Vibrations of microtubules:
  Physics that has not met biology yet},\ }\href@noop {} {\bibfield  {journal}
  {\bibinfo  {journal} {Wave Motion}\ }\textbf {\bibinfo {volume} {72}},\
  \bibinfo {pages} {13} (\bibinfo {year} {2017})}\BibitemShut {NoStop}%
\bibitem [{\citenamefont {Mehrbod}\ and\ \citenamefont
  {Mofrad}(2011)}]{mehrbod2011significance}%
  \BibitemOpen
  \bibfield  {author} {\bibinfo {author} {\bibfnamefont {M.}~\bibnamefont
  {Mehrbod}}\ and\ \bibinfo {author} {\bibfnamefont {M.~R.~K.}\ \bibnamefont
  {Mofrad}},\ }\bibfield  {title} {\bibinfo {title} {On the significance of
  microtubule flexural behavior in cytoskeletal mechanics},\ }\href@noop {}
  {\bibfield  {journal} {\bibinfo  {journal} {PLoS One}\ }\textbf {\bibinfo
  {volume} {6}},\ \bibinfo {pages} {e25627} (\bibinfo {year}
  {2011})}\BibitemShut {NoStop}%
\bibitem [{\citenamefont {Pallavicini}\ \emph {et~al.}(2017)\citenamefont
  {Pallavicini}, \citenamefont {Monastra}, \citenamefont {Bardeci},
  \citenamefont {Wetzler}, \citenamefont {Levi},\ and\ \citenamefont
  {Bruno}}]{pallavicini2017characterization}%
  \BibitemOpen
  \bibfield  {author} {\bibinfo {author} {\bibfnamefont {C.}~\bibnamefont
  {Pallavicini}}, \bibinfo {author} {\bibfnamefont {A.}~\bibnamefont
  {Monastra}}, \bibinfo {author} {\bibfnamefont {N.~G.}\ \bibnamefont
  {Bardeci}}, \bibinfo {author} {\bibfnamefont {D.}~\bibnamefont {Wetzler}},
  \bibinfo {author} {\bibfnamefont {V.}~\bibnamefont {Levi}},\ and\ \bibinfo
  {author} {\bibfnamefont {L.}~\bibnamefont {Bruno}},\ }\bibfield  {title}
  {\bibinfo {title} {Characterization of microtubule buckling in living
  cells},\ }\href@noop {} {\bibfield  {journal} {\bibinfo  {journal} {Eur.
  Biophys. J.}\ }\textbf {\bibinfo {volume} {46}},\ \bibinfo {pages} {581}
  (\bibinfo {year} {2017})}\BibitemShut {NoStop}%
\bibitem [{\citenamefont {Kratky}\ and\ \citenamefont
  {Porod}(1949)}]{kratky1949rontgenuntersuchung}%
  \BibitemOpen
  \bibfield  {author} {\bibinfo {author} {\bibfnamefont {O.}~\bibnamefont
  {Kratky}}\ and\ \bibinfo {author} {\bibfnamefont {G.}~\bibnamefont {Porod}},\
  }\bibfield  {title} {\bibinfo {title} {R{\"o}ntgenuntersuchung gel{\"o}ster
  fadenmolek{\"u}le},\ }\href@noop {} {\bibfield  {journal} {\bibinfo
  {journal} {Rec. Trav. Chim.}\ }\textbf {\bibinfo {volume} {68}},\ \bibinfo
  {pages} {1106} (\bibinfo {year} {1949})}\BibitemShut {NoStop}%
\bibitem [{\citenamefont {Wiggins}\ and\ \citenamefont
  {Goldstein}(1998)}]{wiggins1998flexive}%
  \BibitemOpen
  \bibfield  {author} {\bibinfo {author} {\bibfnamefont {C.~H.}\ \bibnamefont
  {Wiggins}}\ and\ \bibinfo {author} {\bibfnamefont {R.~E.}\ \bibnamefont
  {Goldstein}},\ }\bibfield  {title} {\bibinfo {title} {Flexive and propulsive
  dynamics of elastica at low reynolds number},\ }\href@noop {} {\bibfield
  {journal} {\bibinfo  {journal} {Physical Review Letters}\ }\textbf {\bibinfo
  {volume} {80}},\ \bibinfo {pages} {3879} (\bibinfo {year}
  {1998})}\BibitemShut {NoStop}%
\bibitem [{\citenamefont {Hohenberg}\ and\ \citenamefont
  {Halperin}(1977)}]{hohenberg1977theory}%
  \BibitemOpen
  \bibfield  {author} {\bibinfo {author} {\bibfnamefont {P.~C.}\ \bibnamefont
  {Hohenberg}}\ and\ \bibinfo {author} {\bibfnamefont {B.~I.}\ \bibnamefont
  {Halperin}},\ }\bibfield  {title} {\bibinfo {title} {Theory of dynamic
  critical phenomena},\ }\href@noop {} {\bibfield  {journal} {\bibinfo
  {journal} {Rev. Mod. Phys.}\ }\textbf {\bibinfo {volume} {49}},\ \bibinfo
  {pages} {435} (\bibinfo {year} {1977})}\BibitemShut {NoStop}%
\bibitem [{\citenamefont {Chaikin}\ and\ \citenamefont
  {Lubensky}(1995)}]{chaikin1995principles}%
  \BibitemOpen
  \bibfield  {author} {\bibinfo {author} {\bibfnamefont {P.~M.}\ \bibnamefont
  {Chaikin}}\ and\ \bibinfo {author} {\bibfnamefont {T.~C.}\ \bibnamefont
  {Lubensky}},\ }\href@noop {} {\emph {\bibinfo {title} {Principles of
  condensed matter physics}}}\ (\bibinfo  {publisher} {Cambridge university
  press Cambridge},\ \bibinfo {year} {1995})\BibitemShut {NoStop}%
\bibitem [{\citenamefont {Lamb}(1932)}]{lamb1932hydrodynamics}%
  \BibitemOpen
  \bibfield  {author} {\bibinfo {author} {\bibfnamefont {H.}~\bibnamefont
  {Lamb}},\ }\href@noop {} {\emph {\bibinfo {title} {Hydrodynamics}}}\
  (\bibinfo  {publisher} {Cambridge University Press},\ \bibinfo {year}
  {1932})\BibitemShut {NoStop}%
\bibitem [{\citenamefont {Jayaweera}\ and\ \citenamefont
  {Mason}(1965)}]{jayaweera1965behaviour}%
  \BibitemOpen
  \bibfield  {author} {\bibinfo {author} {\bibfnamefont {K.~O. L.~F.}\
  \bibnamefont {Jayaweera}}\ and\ \bibinfo {author} {\bibfnamefont {B.~J.}\
  \bibnamefont {Mason}},\ }\bibfield  {title} {\bibinfo {title} {The behaviour
  of freely falling cylinders and cones in a viscous fluid},\ }\href@noop {}
  {\bibfield  {journal} {\bibinfo  {journal} {J. Fluid Mech.}\ }\textbf
  {\bibinfo {volume} {22}},\ \bibinfo {pages} {709} (\bibinfo {year}
  {1965})}\BibitemShut {NoStop}%
\bibitem [{\citenamefont {Broedersz}\ and\ \citenamefont
  {MacKintosh}(2014)}]{broedersz2014modeling}%
  \BibitemOpen
  \bibfield  {author} {\bibinfo {author} {\bibfnamefont {C.~P.}\ \bibnamefont
  {Broedersz}}\ and\ \bibinfo {author} {\bibfnamefont {F.~C.}\ \bibnamefont
  {MacKintosh}},\ }\bibfield  {title} {\bibinfo {title} {Modeling semiflexible
  polymer networks},\ }\href@noop {} {\bibfield  {journal} {\bibinfo  {journal}
  {Rev. Mod. Phys.}\ }\textbf {\bibinfo {volume} {86}},\ \bibinfo {pages} {995}
  (\bibinfo {year} {2014})}\BibitemShut {NoStop}%
\bibitem [{\citenamefont {Hallatschek}\ \emph {et~al.}(2005)\citenamefont
  {Hallatschek}, \citenamefont {Frey},\ and\ \citenamefont
  {Kroy}}]{hallatschek2005propagation}%
  \BibitemOpen
  \bibfield  {author} {\bibinfo {author} {\bibfnamefont {O.}~\bibnamefont
  {Hallatschek}}, \bibinfo {author} {\bibfnamefont {E.}~\bibnamefont {Frey}},\
  and\ \bibinfo {author} {\bibfnamefont {K.}~\bibnamefont {Kroy}},\ }\bibfield
  {title} {\bibinfo {title} {Propagation and relaxation of tension in stiff
  polymers},\ }\href@noop {} {\bibfield  {journal} {\bibinfo  {journal} {Phys.
  Rev. Lett.}\ }\textbf {\bibinfo {volume} {94}},\ \bibinfo {pages} {077804}
  (\bibinfo {year} {2005})}\BibitemShut {NoStop}%
\bibitem [{\citenamefont {Hallatschek}\ \emph {et~al.}(2007)\citenamefont
  {Hallatschek}, \citenamefont {Frey},\ and\ \citenamefont
  {Kroy}}]{hallatschek2007tension}%
  \BibitemOpen
  \bibfield  {author} {\bibinfo {author} {\bibfnamefont {O.}~\bibnamefont
  {Hallatschek}}, \bibinfo {author} {\bibfnamefont {E.}~\bibnamefont {Frey}},\
  and\ \bibinfo {author} {\bibfnamefont {K.}~\bibnamefont {Kroy}},\ }\bibfield
  {title} {\bibinfo {title} {Tension dynamics in semiflexible polymers. i.
  coarse-grained equations of motion},\ }\href@noop {} {\bibfield  {journal}
  {\bibinfo  {journal} {Phys. Rev. E}\ }\textbf {\bibinfo {volume} {75}},\
  \bibinfo {pages} {031905} (\bibinfo {year} {2007})}\BibitemShut {NoStop}%
\bibitem [{\citenamefont {Winkler}\ and\ \citenamefont
  {Gompper}(2020)}]{winkler2020physics}%
  \BibitemOpen
  \bibfield  {author} {\bibinfo {author} {\bibfnamefont {R.~G.}\ \bibnamefont
  {Winkler}}\ and\ \bibinfo {author} {\bibfnamefont {G.}~\bibnamefont
  {Gompper}},\ }\bibfield  {title} {\bibinfo {title} {The physics of active
  polymers and filaments},\ }\href@noop {} {\bibfield  {journal} {\bibinfo
  {journal} {J. Chem. Phys.}\ }\textbf {\bibinfo {volume} {153}} (\bibinfo
  {year} {2020})}\BibitemShut {NoStop}%
\bibitem [{\citenamefont {Janmey}\ \emph {et~al.}(1991)\citenamefont {Janmey},
  \citenamefont {Euteneuer}, \citenamefont {Traub},\ and\ \citenamefont
  {Schliwa}}]{janmey1991viscoelastic}%
  \BibitemOpen
  \bibfield  {author} {\bibinfo {author} {\bibfnamefont {P.~A.}\ \bibnamefont
  {Janmey}}, \bibinfo {author} {\bibfnamefont {U.}~\bibnamefont {Euteneuer}},
  \bibinfo {author} {\bibfnamefont {P.}~\bibnamefont {Traub}},\ and\ \bibinfo
  {author} {\bibfnamefont {M.}~\bibnamefont {Schliwa}},\ }\bibfield  {title}
  {\bibinfo {title} {Viscoelastic properties of vimentin compared with other
  filamentous biopolymer networks.},\ }\href@noop {} {\bibfield  {journal}
  {\bibinfo  {journal} {J. Cell Biol.}\ }\textbf {\bibinfo {volume} {113}},\
  \bibinfo {pages} {155} (\bibinfo {year} {1991})}\BibitemShut {NoStop}%
\bibitem [{\citenamefont {Boal}(2012)}]{boal2012mechanics}%
  \BibitemOpen
  \bibfield  {author} {\bibinfo {author} {\bibfnamefont {D.~H.}\ \bibnamefont
  {Boal}},\ }\href@noop {} {\emph {\bibinfo {title} {Mechanics of the Cell}}}\
  (\bibinfo  {publisher} {Cambridge University Press},\ \bibinfo {year}
  {2012})\BibitemShut {NoStop}%
\bibitem [{\citenamefont {Mofrad}(2009)}]{mofrad2009rheology}%
  \BibitemOpen
  \bibfield  {author} {\bibinfo {author} {\bibfnamefont {M.~R.~K.}\
  \bibnamefont {Mofrad}},\ }\bibfield  {title} {\bibinfo {title} {Rheology of
  the cytoskeleton},\ }\href@noop {} {\bibfield  {journal} {\bibinfo  {journal}
  {Annu. Rev. Fluid Mech.}\ }\textbf {\bibinfo {volume} {41}},\ \bibinfo
  {pages} {433} (\bibinfo {year} {2009})}\BibitemShut {NoStop}%
\bibitem [{\citenamefont {Gittes}\ \emph {et~al.}(1993)\citenamefont {Gittes},
  \citenamefont {Mickey}, \citenamefont {Nettleton},\ and\ \citenamefont
  {Howard}}]{gittes1993flexural}%
  \BibitemOpen
  \bibfield  {author} {\bibinfo {author} {\bibfnamefont {F.}~\bibnamefont
  {Gittes}}, \bibinfo {author} {\bibfnamefont {B.}~\bibnamefont {Mickey}},
  \bibinfo {author} {\bibfnamefont {J.}~\bibnamefont {Nettleton}},\ and\
  \bibinfo {author} {\bibfnamefont {J.}~\bibnamefont {Howard}},\ }\bibfield
  {title} {\bibinfo {title} {Flexural rigidity of microtubules and actin
  filaments measured from thermal fluctuations in shape.},\ }\href@noop {}
  {\bibfield  {journal} {\bibinfo  {journal} {J. Cell Bio.}\ }\textbf {\bibinfo
  {volume} {120}},\ \bibinfo {pages} {923} (\bibinfo {year}
  {1993})}\BibitemShut {NoStop}%
\bibitem [{\citenamefont {Collet}\ \emph {et~al.}(2005)\citenamefont {Collet},
  \citenamefont {Shuman}, \citenamefont {Ledger}, \citenamefont {Lee},\ and\
  \citenamefont {Weisel}}]{collet2005elasticity}%
  \BibitemOpen
  \bibfield  {author} {\bibinfo {author} {\bibfnamefont {J.-P.}\ \bibnamefont
  {Collet}}, \bibinfo {author} {\bibfnamefont {H.}~\bibnamefont {Shuman}},
  \bibinfo {author} {\bibfnamefont {R.~E.}\ \bibnamefont {Ledger}}, \bibinfo
  {author} {\bibfnamefont {S.}~\bibnamefont {Lee}},\ and\ \bibinfo {author}
  {\bibfnamefont {J.~W.}\ \bibnamefont {Weisel}},\ }\bibfield  {title}
  {\bibinfo {title} {The elasticity of an individual fibrin fiber in a clot},\
  }\href@noop {} {\bibfield  {journal} {\bibinfo  {journal} {Proc. Natl. Acad.
  Sci. USA}\ }\textbf {\bibinfo {volume} {102}},\ \bibinfo {pages} {9133}
  (\bibinfo {year} {2005})}\BibitemShut {NoStop}%
\bibitem [{\citenamefont {Schopferer}\ \emph {et~al.}(2009)\citenamefont
  {Schopferer}, \citenamefont {B{\"a}r}, \citenamefont {Hochstein},
  \citenamefont {Sharma}, \citenamefont {M{\"u}cke}, \citenamefont {Herrmann},\
  and\ \citenamefont {Willenbacher}}]{schopferer2009desmin}%
  \BibitemOpen
  \bibfield  {author} {\bibinfo {author} {\bibfnamefont {M.}~\bibnamefont
  {Schopferer}}, \bibinfo {author} {\bibfnamefont {H.}~\bibnamefont {B{\"a}r}},
  \bibinfo {author} {\bibfnamefont {B.}~\bibnamefont {Hochstein}}, \bibinfo
  {author} {\bibfnamefont {S.}~\bibnamefont {Sharma}}, \bibinfo {author}
  {\bibfnamefont {N.}~\bibnamefont {M{\"u}cke}}, \bibinfo {author}
  {\bibfnamefont {H.}~\bibnamefont {Herrmann}},\ and\ \bibinfo {author}
  {\bibfnamefont {N.}~\bibnamefont {Willenbacher}},\ }\bibfield  {title}
  {\bibinfo {title} {Desmin and vimentin intermediate filament networks: their
  viscoelastic properties investigated by mechanical rheometry},\ }\href@noop
  {} {\bibfield  {journal} {\bibinfo  {journal} {J. Mol. Biol.}\ }\textbf
  {\bibinfo {volume} {388}},\ \bibinfo {pages} {133} (\bibinfo {year}
  {2009})}\BibitemShut {NoStop}%
\bibitem [{\citenamefont {Walter}\ \emph {et~al.}(2011)\citenamefont {Walter},
  \citenamefont {Busch}, \citenamefont {Seufferlein},\ and\ \citenamefont
  {Spatz}}]{walter2011elastic}%
  \BibitemOpen
  \bibfield  {author} {\bibinfo {author} {\bibfnamefont {N.}~\bibnamefont
  {Walter}}, \bibinfo {author} {\bibfnamefont {T.}~\bibnamefont {Busch}},
  \bibinfo {author} {\bibfnamefont {T.}~\bibnamefont {Seufferlein}},\ and\
  \bibinfo {author} {\bibfnamefont {J.~P.}\ \bibnamefont {Spatz}},\ }\bibfield
  {title} {\bibinfo {title} {Elastic moduli of living epithelial pancreatic
  cancer cells and their skeletonized keratin intermediate filament network},\
  }\href@noop {} {\bibfield  {journal} {\bibinfo  {journal} {Biointerphases}\
  }\textbf {\bibinfo {volume} {6}},\ \bibinfo {pages} {79} (\bibinfo {year}
  {2011})}\BibitemShut {NoStop}%
\bibitem [{\citenamefont {Lichtenstern}\ \emph {et~al.}(2012)\citenamefont
  {Lichtenstern}, \citenamefont {M{\"u}cke}, \citenamefont {Aebi},
  \citenamefont {Mauermann},\ and\ \citenamefont
  {Herrmann}}]{lichtenstern2012complex}%
  \BibitemOpen
  \bibfield  {author} {\bibinfo {author} {\bibfnamefont {T.}~\bibnamefont
  {Lichtenstern}}, \bibinfo {author} {\bibfnamefont {N.}~\bibnamefont
  {M{\"u}cke}}, \bibinfo {author} {\bibfnamefont {U.}~\bibnamefont {Aebi}},
  \bibinfo {author} {\bibfnamefont {M.}~\bibnamefont {Mauermann}},\ and\
  \bibinfo {author} {\bibfnamefont {H.}~\bibnamefont {Herrmann}},\ }\bibfield
  {title} {\bibinfo {title} {Complex formation and kinetics of filament
  assembly exhibited by the simple epithelial keratins {K}8 and {K}18},\
  }\href@noop {} {\bibfield  {journal} {\bibinfo  {journal} {J. Struct. Biol.}\
  }\textbf {\bibinfo {volume} {177}},\ \bibinfo {pages} {54} (\bibinfo {year}
  {2012})}\BibitemShut {NoStop}%
\bibitem [{\citenamefont {Fudge}\ \emph {et~al.}(2003)\citenamefont {Fudge},
  \citenamefont {Gardner}, \citenamefont {Forsyth}, \citenamefont {Riekel},\
  and\ \citenamefont {Gosline}}]{fudge2003mechanical}%
  \BibitemOpen
  \bibfield  {author} {\bibinfo {author} {\bibfnamefont {D.~S.}\ \bibnamefont
  {Fudge}}, \bibinfo {author} {\bibfnamefont {K.~H.}\ \bibnamefont {Gardner}},
  \bibinfo {author} {\bibfnamefont {V.~T.}\ \bibnamefont {Forsyth}}, \bibinfo
  {author} {\bibfnamefont {C.}~\bibnamefont {Riekel}},\ and\ \bibinfo {author}
  {\bibfnamefont {J.~M.}\ \bibnamefont {Gosline}},\ }\bibfield  {title}
  {\bibinfo {title} {The mechanical properties of hydrated intermediate
  filaments: insights from hagfish slime threads},\ }\href@noop {} {\bibfield
  {journal} {\bibinfo  {journal} {Biophys. J.}\ }\textbf {\bibinfo {volume}
  {85}},\ \bibinfo {pages} {2015} (\bibinfo {year} {2003})}\BibitemShut
  {NoStop}%
\bibitem [{\citenamefont {McKittrick}\ \emph {et~al.}(2012)\citenamefont
  {McKittrick}, \citenamefont {Chen}, \citenamefont {Bodde}, \citenamefont
  {Yang}, \citenamefont {Novitskaya},\ and\ \citenamefont
  {Meyers}}]{mckittrick2012structure}%
  \BibitemOpen
  \bibfield  {author} {\bibinfo {author} {\bibfnamefont {J.}~\bibnamefont
  {McKittrick}}, \bibinfo {author} {\bibfnamefont {P.-Y.}\ \bibnamefont
  {Chen}}, \bibinfo {author} {\bibfnamefont {S.~G.}\ \bibnamefont {Bodde}},
  \bibinfo {author} {\bibfnamefont {W.}~\bibnamefont {Yang}}, \bibinfo {author}
  {\bibfnamefont {E.~E.}\ \bibnamefont {Novitskaya}},\ and\ \bibinfo {author}
  {\bibfnamefont {M.~A.}\ \bibnamefont {Meyers}},\ }\bibfield  {title}
  {\bibinfo {title} {The structure, functions, and mechanical properties of
  keratin},\ }\href@noop {} {\bibfield  {journal} {\bibinfo  {journal} {JOM}\
  }\textbf {\bibinfo {volume} {64}},\ \bibinfo {pages} {449} (\bibinfo {year}
  {2012})}\BibitemShut {NoStop}%
\bibitem [{\citenamefont {Sharma}\ and\ \citenamefont
  {Vershinin}(2020)}]{sharma2020length}%
  \BibitemOpen
  \bibfield  {author} {\bibinfo {author} {\bibfnamefont {A.}~\bibnamefont
  {Sharma}}\ and\ \bibinfo {author} {\bibfnamefont {M.}~\bibnamefont
  {Vershinin}},\ }\bibfield  {title} {\bibinfo {title} {Length dependence of
  the rigidity of microtubules in small networks},\ }\href@noop {} {\bibfield
  {journal} {\bibinfo  {journal} {Biochem. Biophys. Res. Commun.}\ }\textbf
  {\bibinfo {volume} {529}},\ \bibinfo {pages} {303} (\bibinfo {year}
  {2020})}\BibitemShut {NoStop}%
\bibitem [{\citenamefont {Nishino}\ \emph {et~al.}(1995)\citenamefont
  {Nishino}, \citenamefont {Takano},\ and\ \citenamefont
  {Nakamae}}]{nishino1995elastic}%
  \BibitemOpen
  \bibfield  {author} {\bibinfo {author} {\bibfnamefont {T.}~\bibnamefont
  {Nishino}}, \bibinfo {author} {\bibfnamefont {K.}~\bibnamefont {Takano}},\
  and\ \bibinfo {author} {\bibfnamefont {K.}~\bibnamefont {Nakamae}},\
  }\bibfield  {title} {\bibinfo {title} {Elastic modulus of the crystalline
  regions of cellulose polymorphs},\ }\href@noop {} {\bibfield  {journal}
  {\bibinfo  {journal} {J. Polym. Sci. B: Polym. Phys.}\ }\textbf {\bibinfo
  {volume} {33}},\ \bibinfo {pages} {1647} (\bibinfo {year}
  {1995})}\BibitemShut {NoStop}%
\bibitem [{\citenamefont {Trachtenberg}\ and\ \citenamefont
  {Hammel}(1992)}]{trachtenberg1992rigidity}%
  \BibitemOpen
  \bibfield  {author} {\bibinfo {author} {\bibfnamefont {S.}~\bibnamefont
  {Trachtenberg}}\ and\ \bibinfo {author} {\bibfnamefont {I.}~\bibnamefont
  {Hammel}},\ }\bibfield  {title} {\bibinfo {title} {The rigidity of bacterial
  flagellar filaments and its relation to filament polymorphism},\ }\href@noop
  {} {\bibfield  {journal} {\bibinfo  {journal} {J. Struct. Biol.}\ }\textbf
  {\bibinfo {volume} {109}},\ \bibinfo {pages} {18} (\bibinfo {year}
  {1992})}\BibitemShut {NoStop}%
\bibitem [{\citenamefont {Shen}\ \emph {et~al.}(2022)\citenamefont {Shen},
  \citenamefont {Tran}, \citenamefont {Tay} \emph {et~al.}}]{shen2022bending}%
  \BibitemOpen
  \bibfield  {author} {\bibinfo {author} {\bibfnamefont {X.}~\bibnamefont
  {Shen}}, \bibinfo {author} {\bibfnamefont {P.~N.}\ \bibnamefont {Tran}},
  \bibinfo {author} {\bibfnamefont {B.~Z.}\ \bibnamefont {Tay}}, \emph
  {et~al.},\ }\bibfield  {title} {\bibinfo {title} {Bending stiffness
  characterization of bacillus subtilis’ flagellar filament},\ }\href@noop {}
  {\bibfield  {journal} {\bibinfo  {journal} {Biophysical Journal}\ }\textbf
  {\bibinfo {volume} {121}},\ \bibinfo {pages} {1975} (\bibinfo {year}
  {2022})}\BibitemShut {NoStop}%
\bibitem [{\citenamefont {Sen}\ \emph {et~al.}(2004)\citenamefont {Sen},
  \citenamefont {Nandy},\ and\ \citenamefont {Ghosh}}]{sen2004elasticity}%
  \BibitemOpen
  \bibfield  {author} {\bibinfo {author} {\bibfnamefont {A.}~\bibnamefont
  {Sen}}, \bibinfo {author} {\bibfnamefont {R.~K.}\ \bibnamefont {Nandy}},\
  and\ \bibinfo {author} {\bibfnamefont {A.~N.}\ \bibnamefont {Ghosh}},\
  }\bibfield  {title} {\bibinfo {title} {Elasticity of flagellar hooks},\
  }\href@noop {} {\bibfield  {journal} {\bibinfo  {journal} {Microscopy}\
  }\textbf {\bibinfo {volume} {53}},\ \bibinfo {pages} {305} (\bibinfo {year}
  {2004})}\BibitemShut {NoStop}%
\bibitem [{\citenamefont {Nord}\ \emph {et~al.}(2022)\citenamefont {Nord},
  \citenamefont {Biquet-Bisquert}, \citenamefont {Abkarian}, \citenamefont
  {Pigaglio}, \citenamefont {Seduk}, \citenamefont {Magalon},\ and\
  \citenamefont {Pedaci}}]{nord2022dynamic}%
  \BibitemOpen
  \bibfield  {author} {\bibinfo {author} {\bibfnamefont {A.~L.}\ \bibnamefont
  {Nord}}, \bibinfo {author} {\bibfnamefont {A.}~\bibnamefont
  {Biquet-Bisquert}}, \bibinfo {author} {\bibfnamefont {M.}~\bibnamefont
  {Abkarian}}, \bibinfo {author} {\bibfnamefont {T.}~\bibnamefont {Pigaglio}},
  \bibinfo {author} {\bibfnamefont {F.}~\bibnamefont {Seduk}}, \bibinfo
  {author} {\bibfnamefont {A.}~\bibnamefont {Magalon}},\ and\ \bibinfo {author}
  {\bibfnamefont {F.}~\bibnamefont {Pedaci}},\ }\bibfield  {title} {\bibinfo
  {title} {Dynamic stiffening of the flagellar hook},\ }\href@noop {}
  {\bibfield  {journal} {\bibinfo  {journal} {Nature Communications}\ }\textbf
  {\bibinfo {volume} {13}},\ \bibinfo {pages} {2925} (\bibinfo {year}
  {2022})}\BibitemShut {NoStop}%
\bibitem [{\citenamefont {Okuno}\ and\ \citenamefont
  {Hiramoto}(1979)}]{okuno1979direct}%
  \BibitemOpen
  \bibfield  {author} {\bibinfo {author} {\bibfnamefont {M.}~\bibnamefont
  {Okuno}}\ and\ \bibinfo {author} {\bibfnamefont {Y.}~\bibnamefont
  {Hiramoto}},\ }\bibfield  {title} {\bibinfo {title} {Direct measurements of
  the stiffness of echinoderm sperm flagella},\ }\href@noop {} {\bibfield
  {journal} {\bibinfo  {journal} {J. Exp. Biol.}\ }\textbf {\bibinfo {volume}
  {79}},\ \bibinfo {pages} {235} (\bibinfo {year} {1979})}\BibitemShut
  {NoStop}%
\bibitem [{\citenamefont {Ishijima}\ and\ \citenamefont
  {Hiramoto}(1994)}]{ishijima1994flexural}%
  \BibitemOpen
  \bibfield  {author} {\bibinfo {author} {\bibfnamefont {S.}~\bibnamefont
  {Ishijima}}\ and\ \bibinfo {author} {\bibfnamefont {Y.}~\bibnamefont
  {Hiramoto}},\ }\bibfield  {title} {\bibinfo {title} {Flexural rigidity of
  echinoderm sperm flagella},\ }\href@noop {} {\bibfield  {journal} {\bibinfo
  {journal} {Cell Structure and Function}\ }\textbf {\bibinfo {volume} {19}},\
  \bibinfo {pages} {349} (\bibinfo {year} {1994})}\BibitemShut {NoStop}%
\bibitem [{\citenamefont {Faluweki}\ and\ \citenamefont
  {Goehring}(2022)}]{faluweki2022structural}%
  \BibitemOpen
  \bibfield  {author} {\bibinfo {author} {\bibfnamefont {M.~K.}\ \bibnamefont
  {Faluweki}}\ and\ \bibinfo {author} {\bibfnamefont {L.}~\bibnamefont
  {Goehring}},\ }\bibfield  {title} {\bibinfo {title} {Structural mechanics of
  filamentous cyanobacteria},\ }\href@noop {} {\bibfield  {journal} {\bibinfo
  {journal} {Journal of the Royal Society Interface}\ }\textbf {\bibinfo
  {volume} {19}},\ \bibinfo {pages} {20220268} (\bibinfo {year}
  {2022})}\BibitemShut {NoStop}%
\bibitem [{\citenamefont {Boal}\ and\ \citenamefont
  {Ng}(2010)}]{boal2010shape}%
  \BibitemOpen
  \bibfield  {author} {\bibinfo {author} {\bibfnamefont {D.}~\bibnamefont
  {Boal}}\ and\ \bibinfo {author} {\bibfnamefont {R.}~\bibnamefont {Ng}},\
  }\bibfield  {title} {\bibinfo {title} {Shape analysis of filamentous
  precambrian microfossils and modern cyanobacteria},\ }\href@noop {}
  {\bibfield  {journal} {\bibinfo  {journal} {Paleobiology}\ }\textbf {\bibinfo
  {volume} {36}},\ \bibinfo {pages} {555} (\bibinfo {year} {2010})}\BibitemShut
  {NoStop}%
\bibitem [{\citenamefont {Kurjahn}\ \emph
  {et~al.}(2024{\natexlab{a}})\citenamefont {Kurjahn}, \citenamefont {Deka},
  \citenamefont {Girot}, \citenamefont {Abbaspour}, \citenamefont {Klumpp},
  \citenamefont {Lorenz}, \citenamefont {B{\"a}umchen},\ and\ \citenamefont
  {Karpitschka}}]{kurjahn2024quantifying}%
  \BibitemOpen
  \bibfield  {author} {\bibinfo {author} {\bibfnamefont {M.}~\bibnamefont
  {Kurjahn}}, \bibinfo {author} {\bibfnamefont {A.}~\bibnamefont {Deka}},
  \bibinfo {author} {\bibfnamefont {A.}~\bibnamefont {Girot}}, \bibinfo
  {author} {\bibfnamefont {L.}~\bibnamefont {Abbaspour}}, \bibinfo {author}
  {\bibfnamefont {S.}~\bibnamefont {Klumpp}}, \bibinfo {author} {\bibfnamefont
  {M.}~\bibnamefont {Lorenz}}, \bibinfo {author} {\bibfnamefont
  {O.}~\bibnamefont {B{\"a}umchen}},\ and\ \bibinfo {author} {\bibfnamefont
  {S.}~\bibnamefont {Karpitschka}},\ }\bibfield  {title} {\bibinfo {title}
  {Quantifying gliding forces of filamentous cyanobacteria by self-buckling},\
  }\href@noop {} {\bibfield  {journal} {\bibinfo  {journal} {eLife}\ }\textbf
  {\bibinfo {volume} {12}},\ \bibinfo {pages} {RP87450} (\bibinfo {year}
  {2024}{\natexlab{a}})}\BibitemShut {NoStop}%
\bibitem [{\citenamefont {Backholm}\ \emph {et~al.}(2013)\citenamefont
  {Backholm}, \citenamefont {Ryu},\ and\ \citenamefont
  {Dalnoki-Veress}}]{backholm2013viscoelastic}%
  \BibitemOpen
  \bibfield  {author} {\bibinfo {author} {\bibfnamefont {M.}~\bibnamefont
  {Backholm}}, \bibinfo {author} {\bibfnamefont {W.~S.}\ \bibnamefont {Ryu}},\
  and\ \bibinfo {author} {\bibfnamefont {K.}~\bibnamefont {Dalnoki-Veress}},\
  }\bibfield  {title} {\bibinfo {title} {Viscoelastic properties of the
  nematode caenorhabditis elegans, a self-similar, shear-thinning worm},\
  }\href@noop {} {\bibfield  {journal} {\bibinfo  {journal} {Proc. Natl. Acad.
  Sci. USA}\ }\textbf {\bibinfo {volume} {110}},\ \bibinfo {pages} {4528}
  (\bibinfo {year} {2013})}\BibitemShut {NoStop}%
\bibitem [{\citenamefont {Guo}\ and\ \citenamefont
  {Mahadevan}(2008)}]{guo2008limbless}%
  \BibitemOpen
  \bibfield  {author} {\bibinfo {author} {\bibfnamefont {Z.~V.}\ \bibnamefont
  {Guo}}\ and\ \bibinfo {author} {\bibfnamefont {L.}~\bibnamefont
  {Mahadevan}},\ }\bibfield  {title} {\bibinfo {title} {Limbless undulatory
  propulsion on land},\ }\href@noop {} {\bibfield  {journal} {\bibinfo
  {journal} {Proc. Natl. Acad. Sci. USA}\ }\textbf {\bibinfo {volume} {105}},\
  \bibinfo {pages} {3179} (\bibinfo {year} {2008})}\BibitemShut {NoStop}%
\bibitem [{\citenamefont {Zhang}\ \emph {et~al.}(2021)\citenamefont {Zhang},
  \citenamefont {Naughton}, \citenamefont {Parthasarathy},\ and\ \citenamefont
  {Gazzola}}]{zhang2021friction}%
  \BibitemOpen
  \bibfield  {author} {\bibinfo {author} {\bibfnamefont {X.}~\bibnamefont
  {Zhang}}, \bibinfo {author} {\bibfnamefont {N.}~\bibnamefont {Naughton}},
  \bibinfo {author} {\bibfnamefont {T.}~\bibnamefont {Parthasarathy}},\ and\
  \bibinfo {author} {\bibfnamefont {M.}~\bibnamefont {Gazzola}},\ }\bibfield
  {title} {\bibinfo {title} {Friction modulation in limbless, three-dimensional
  gaits and heterogeneous terrains},\ }\href@noop {} {\bibfield  {journal}
  {\bibinfo  {journal} {Nat. Commun.}\ }\textbf {\bibinfo {volume} {12}},\
  \bibinfo {pages} {6076} (\bibinfo {year} {2021})}\BibitemShut {NoStop}%
\bibitem [{\citenamefont {Long~Jr}(1998)}]{long1998muscles}%
  \BibitemOpen
  \bibfield  {author} {\bibinfo {author} {\bibfnamefont {J.~H.}\ \bibnamefont
  {Long~Jr}},\ }\bibfield  {title} {\bibinfo {title} {Muscles, elastic energy,
  and the dynamics of body stiffness in swimming eels},\ }\href@noop {}
  {\bibfield  {journal} {\bibinfo  {journal} {Am. Zool.}\ }\textbf {\bibinfo
  {volume} {38}},\ \bibinfo {pages} {771} (\bibinfo {year} {1998})}\BibitemShut
  {NoStop}%
\bibitem [{\citenamefont {Zia}\ and\ \citenamefont
  {Schmittmann}(2007)}]{zia2007probability}%
  \BibitemOpen
  \bibfield  {author} {\bibinfo {author} {\bibfnamefont {R.~K.~P.}\
  \bibnamefont {Zia}}\ and\ \bibinfo {author} {\bibfnamefont {B.}~\bibnamefont
  {Schmittmann}},\ }\bibfield  {title} {\bibinfo {title} {Probability currents
  as principal characteristics in the statistical mechanics of non-equilibrium
  steady states},\ }\href@noop {} {\bibfield  {journal} {\bibinfo  {journal}
  {J. Stat. Mech.}\ }\textbf {\bibinfo {volume} {2007}},\ \bibinfo {pages}
  {P07012} (\bibinfo {year} {2007})}\BibitemShut {NoStop}%
\bibitem [{\citenamefont {Da~Costa}\ and\ \citenamefont
  {Pavliotis}(2023)}]{da2023entropy}%
  \BibitemOpen
  \bibfield  {author} {\bibinfo {author} {\bibfnamefont {L.}~\bibnamefont
  {Da~Costa}}\ and\ \bibinfo {author} {\bibfnamefont {G.~A.}\ \bibnamefont
  {Pavliotis}},\ }\bibfield  {title} {\bibinfo {title} {The entropy production
  of stationary diffusions},\ }\href@noop {} {\bibfield  {journal} {\bibinfo
  {journal} {J. Phys. A}\ }\textbf {\bibinfo {volume} {56}},\ \bibinfo {pages}
  {365001} (\bibinfo {year} {2023})}\BibitemShut {NoStop}%
\bibitem [{\citenamefont {Battle}\ \emph {et~al.}(2016)\citenamefont {Battle},
  \citenamefont {Broedersz}, \citenamefont {Fakhri}, \citenamefont {Geyer},
  \citenamefont {Howard}, \citenamefont {Schmidt},\ and\ \citenamefont
  {MacKintosh}}]{battle2016broken}%
  \BibitemOpen
  \bibfield  {author} {\bibinfo {author} {\bibfnamefont {C.}~\bibnamefont
  {Battle}}, \bibinfo {author} {\bibfnamefont {C.~P.}\ \bibnamefont
  {Broedersz}}, \bibinfo {author} {\bibfnamefont {N.}~\bibnamefont {Fakhri}},
  \bibinfo {author} {\bibfnamefont {V.~F.}\ \bibnamefont {Geyer}}, \bibinfo
  {author} {\bibfnamefont {J.}~\bibnamefont {Howard}}, \bibinfo {author}
  {\bibfnamefont {C.~F.}\ \bibnamefont {Schmidt}},\ and\ \bibinfo {author}
  {\bibfnamefont {F.~C.}\ \bibnamefont {MacKintosh}},\ }\bibfield  {title}
  {\bibinfo {title} {Broken detailed balance at mesoscopic scales in active
  biological systems},\ }\href@noop {} {\bibfield  {journal} {\bibinfo
  {journal} {Science}\ }\textbf {\bibinfo {volume} {352}},\ \bibinfo {pages}
  {604} (\bibinfo {year} {2016})}\BibitemShut {NoStop}%
\bibitem [{\citenamefont {Herminghaus}\ and\ \citenamefont
  {Mazza}(2017)}]{herminghaus2017phase}%
  \BibitemOpen
  \bibfield  {author} {\bibinfo {author} {\bibfnamefont {S.}~\bibnamefont
  {Herminghaus}}\ and\ \bibinfo {author} {\bibfnamefont {M.~G.}\ \bibnamefont
  {Mazza}},\ }\bibfield  {title} {\bibinfo {title} {Phase separation in driven
  granular gases: exploring the elusive character of nonequilibrium steady
  states},\ }\href@noop {} {\bibfield  {journal} {\bibinfo  {journal} {Soft
  Matter}\ }\textbf {\bibinfo {volume} {13}},\ \bibinfo {pages} {898} (\bibinfo
  {year} {2017})}\BibitemShut {NoStop}%
\bibitem [{\citenamefont {Cammann}\ \emph {et~al.}(2021)\citenamefont
  {Cammann}, \citenamefont {Schwarzendahl}, \citenamefont {Ostapenko},
  \citenamefont {Lavrentovich}, \citenamefont {B{\"a}umchen},\ and\
  \citenamefont {Mazza}}]{cammann2021emergent}%
  \BibitemOpen
  \bibfield  {author} {\bibinfo {author} {\bibfnamefont {J.}~\bibnamefont
  {Cammann}}, \bibinfo {author} {\bibfnamefont {F.~J.}\ \bibnamefont
  {Schwarzendahl}}, \bibinfo {author} {\bibfnamefont {T.}~\bibnamefont
  {Ostapenko}}, \bibinfo {author} {\bibfnamefont {D.}~\bibnamefont
  {Lavrentovich}}, \bibinfo {author} {\bibfnamefont {O.}~\bibnamefont
  {B{\"a}umchen}},\ and\ \bibinfo {author} {\bibfnamefont {M.~G.}\ \bibnamefont
  {Mazza}},\ }\bibfield  {title} {\bibinfo {title} {Emergent probability fluxes
  in confined microbial navigation},\ }\href@noop {} {\bibfield  {journal}
  {\bibinfo  {journal} {Proc. Natl. Acad. Sci. USA}\ }\textbf {\bibinfo
  {volume} {118}},\ \bibinfo {pages} {e2024752118} (\bibinfo {year}
  {2021})}\BibitemShut {NoStop}%
\bibitem [{\citenamefont {Brouhard}\ and\ \citenamefont
  {Rice}(2018)}]{brouhard2018microtubule}%
  \BibitemOpen
  \bibfield  {author} {\bibinfo {author} {\bibfnamefont {G.~J.}\ \bibnamefont
  {Brouhard}}\ and\ \bibinfo {author} {\bibfnamefont {L.~M.}\ \bibnamefont
  {Rice}},\ }\bibfield  {title} {\bibinfo {title} {Microtubule dynamics: an
  interplay of biochemistry and mechanics},\ }\href@noop {} {\bibfield
  {journal} {\bibinfo  {journal} {Nat. Rev. Mol. Cell Biol.}\ }\textbf
  {\bibinfo {volume} {19}},\ \bibinfo {pages} {451} (\bibinfo {year}
  {2018})}\BibitemShut {NoStop}%
\bibitem [{\citenamefont {Zhou}\ \emph {et~al.}(2023)\citenamefont {Zhou},
  \citenamefont {Wang}, \citenamefont {Song}, \citenamefont {Liu},
  \citenamefont {Wang}, \citenamefont {Li}, \citenamefont {Liang},
  \citenamefont {Li}, \citenamefont {Chu},\ and\ \citenamefont
  {Wang}}]{zhou2023structural}%
  \BibitemOpen
  \bibfield  {author} {\bibinfo {author} {\bibfnamefont {J.}~\bibnamefont
  {Zhou}}, \bibinfo {author} {\bibfnamefont {A.}~\bibnamefont {Wang}}, \bibinfo
  {author} {\bibfnamefont {Y.}~\bibnamefont {Song}}, \bibinfo {author}
  {\bibfnamefont {N.}~\bibnamefont {Liu}}, \bibinfo {author} {\bibfnamefont
  {J.}~\bibnamefont {Wang}}, \bibinfo {author} {\bibfnamefont {Y.}~\bibnamefont
  {Li}}, \bibinfo {author} {\bibfnamefont {X.}~\bibnamefont {Liang}}, \bibinfo
  {author} {\bibfnamefont {G.}~\bibnamefont {Li}}, \bibinfo {author}
  {\bibfnamefont {H.}~\bibnamefont {Chu}},\ and\ \bibinfo {author}
  {\bibfnamefont {H.-W.}\ \bibnamefont {Wang}},\ }\bibfield  {title} {\bibinfo
  {title} {Structural insights into the mechanism of gtp initiation of
  microtubule assembly},\ }\href@noop {} {\bibfield  {journal} {\bibinfo
  {journal} {Nat. Commun.}\ }\textbf {\bibinfo {volume} {14}},\ \bibinfo
  {pages} {5980} (\bibinfo {year} {2023})}\BibitemShut {NoStop}%
\bibitem [{\citenamefont {Dustin}(2012)}]{dustin2012microtubules}%
  \BibitemOpen
  \bibfield  {author} {\bibinfo {author} {\bibfnamefont {P.}~\bibnamefont
  {Dustin}},\ }\href@noop {} {\emph {\bibinfo {title} {Microtubules}}}\
  (\bibinfo  {publisher} {Springer Science \& Business Media},\ \bibinfo {year}
  {2012})\BibitemShut {NoStop}%
\bibitem [{\citenamefont {Azimzadeh}\ and\ \citenamefont
  {Marshall}(2010)}]{azimzadeh2010building}%
  \BibitemOpen
  \bibfield  {author} {\bibinfo {author} {\bibfnamefont {J.}~\bibnamefont
  {Azimzadeh}}\ and\ \bibinfo {author} {\bibfnamefont {W.~F.}\ \bibnamefont
  {Marshall}},\ }\bibfield  {title} {\bibinfo {title} {Building the
  centriole},\ }\href@noop {} {\bibfield  {journal} {\bibinfo  {journal} {Curr.
  Biol.}\ }\textbf {\bibinfo {volume} {20}},\ \bibinfo {pages} {R816} (\bibinfo
  {year} {2010})}\BibitemShut {NoStop}%
\bibitem [{\citenamefont {Grimstone}\ and\ \citenamefont
  {Klug}(1966)}]{grimstone1966observations}%
  \BibitemOpen
  \bibfield  {author} {\bibinfo {author} {\bibfnamefont {A.~V.}\ \bibnamefont
  {Grimstone}}\ and\ \bibinfo {author} {\bibfnamefont {A.}~\bibnamefont
  {Klug}},\ }\bibfield  {title} {\bibinfo {title} {Observations on the
  substructure of flagellar fibres},\ }\href@noop {} {\bibfield  {journal}
  {\bibinfo  {journal} {J. Cell Sci.}\ }\textbf {\bibinfo {volume} {1}},\
  \bibinfo {pages} {351} (\bibinfo {year} {1966})}\BibitemShut {NoStop}%
\bibitem [{\citenamefont {Amos}\ and\ \citenamefont
  {Klug}(1974)}]{amos1974arrangement}%
  \BibitemOpen
  \bibfield  {author} {\bibinfo {author} {\bibfnamefont {L.~A.}\ \bibnamefont
  {Amos}}\ and\ \bibinfo {author} {\bibfnamefont {A.}~\bibnamefont {Klug}},\
  }\bibfield  {title} {\bibinfo {title} {Arrangement of subunits in flagellar
  microtubules},\ }\href@noop {} {\bibfield  {journal} {\bibinfo  {journal} {J.
  Cell Sci.}\ }\textbf {\bibinfo {volume} {14}},\ \bibinfo {pages} {523}
  (\bibinfo {year} {1974})}\BibitemShut {NoStop}%
\bibitem [{\citenamefont {Pampaloni}\ \emph {et~al.}(2006)\citenamefont
  {Pampaloni}, \citenamefont {Lattanzi}, \citenamefont {Jon{\'a}{\v{s}}},
  \citenamefont {Surrey}, \citenamefont {Frey},\ and\ \citenamefont
  {Florin}}]{pampaloni2006thermal}%
  \BibitemOpen
  \bibfield  {author} {\bibinfo {author} {\bibfnamefont {F.}~\bibnamefont
  {Pampaloni}}, \bibinfo {author} {\bibfnamefont {G.}~\bibnamefont {Lattanzi}},
  \bibinfo {author} {\bibfnamefont {A.}~\bibnamefont {Jon{\'a}{\v{s}}}},
  \bibinfo {author} {\bibfnamefont {T.}~\bibnamefont {Surrey}}, \bibinfo
  {author} {\bibfnamefont {E.}~\bibnamefont {Frey}},\ and\ \bibinfo {author}
  {\bibfnamefont {E.-L.}\ \bibnamefont {Florin}},\ }\bibfield  {title}
  {\bibinfo {title} {Thermal fluctuations of grafted microtubules provide
  evidence of a length-dependent persistence length},\ }\href@noop {}
  {\bibfield  {journal} {\bibinfo  {journal} {Proc. Natl. Acad. Sci. USA}\
  }\textbf {\bibinfo {volume} {103}},\ \bibinfo {pages} {10248} (\bibinfo
  {year} {2006})}\BibitemShut {NoStop}%
\bibitem [{\citenamefont {Bray}(2000)}]{bray2000cell}%
  \BibitemOpen
  \bibfield  {author} {\bibinfo {author} {\bibfnamefont {D.}~\bibnamefont
  {Bray}},\ }\href@noop {} {\emph {\bibinfo {title} {Cell movements: from
  molecules to motility}}}\ (\bibinfo  {publisher} {Garland Science},\ \bibinfo
  {year} {2000})\BibitemShut {NoStop}%
\bibitem [{\citenamefont {Welte}(2004)}]{welte2004bidirectional}%
  \BibitemOpen
  \bibfield  {author} {\bibinfo {author} {\bibfnamefont {M.~A.}\ \bibnamefont
  {Welte}},\ }\bibfield  {title} {\bibinfo {title} {Bidirectional transport
  along microtubules},\ }\href@noop {} {\bibfield  {journal} {\bibinfo
  {journal} {Curr. Biol.}\ }\textbf {\bibinfo {volume} {14}},\ \bibinfo {pages}
  {R525} (\bibinfo {year} {2004})}\BibitemShut {NoStop}%
\bibitem [{\citenamefont {Barlan}\ and\ \citenamefont
  {Gelfand}(2017)}]{barlan2017microtubule}%
  \BibitemOpen
  \bibfield  {author} {\bibinfo {author} {\bibfnamefont {K.}~\bibnamefont
  {Barlan}}\ and\ \bibinfo {author} {\bibfnamefont {V.~I.}\ \bibnamefont
  {Gelfand}},\ }\bibfield  {title} {\bibinfo {title} {Microtubule-based
  transport and the distribution, tethering, and organization of organelles},\
  }\href@noop {} {\bibfield  {journal} {\bibinfo  {journal} {Cold Spring Harb.
  Perspect. Biol.}\ }\textbf {\bibinfo {volume} {9}},\ \bibinfo {pages}
  {a025817} (\bibinfo {year} {2017})}\BibitemShut {NoStop}%
\bibitem [{\citenamefont {Singh}\ \emph {et~al.}(2018)\citenamefont {Singh},
  \citenamefont {Saha}, \citenamefont {Begemann}, \citenamefont {Ricker},
  \citenamefont {N{\"u}sse}, \citenamefont {Thorn-Seshold}, \citenamefont
  {Klingauf}, \citenamefont {Galic},\ and\ \citenamefont
  {Matis}}]{singh2018polarized}%
  \BibitemOpen
  \bibfield  {author} {\bibinfo {author} {\bibfnamefont {A.}~\bibnamefont
  {Singh}}, \bibinfo {author} {\bibfnamefont {T.}~\bibnamefont {Saha}},
  \bibinfo {author} {\bibfnamefont {I.}~\bibnamefont {Begemann}}, \bibinfo
  {author} {\bibfnamefont {A.}~\bibnamefont {Ricker}}, \bibinfo {author}
  {\bibfnamefont {H.}~\bibnamefont {N{\"u}sse}}, \bibinfo {author}
  {\bibfnamefont {O.}~\bibnamefont {Thorn-Seshold}}, \bibinfo {author}
  {\bibfnamefont {J.}~\bibnamefont {Klingauf}}, \bibinfo {author}
  {\bibfnamefont {M.}~\bibnamefont {Galic}},\ and\ \bibinfo {author}
  {\bibfnamefont {M.}~\bibnamefont {Matis}},\ }\bibfield  {title} {\bibinfo
  {title} {Polarized microtubule dynamics directs cell mechanics and
  coordinates forces during epithelial morphogenesis},\ }\href@noop {}
  {\bibfield  {journal} {\bibinfo  {journal} {Nat. Cell Biol.}\ }\textbf
  {\bibinfo {volume} {20}},\ \bibinfo {pages} {1126} (\bibinfo {year}
  {2018})}\BibitemShut {NoStop}%
\bibitem [{\citenamefont {R{\"o}per}(2020)}]{roper2020microtubules}%
  \BibitemOpen
  \bibfield  {author} {\bibinfo {author} {\bibfnamefont {K.}~\bibnamefont
  {R{\"o}per}},\ }\bibfield  {title} {\bibinfo {title} {Microtubules enter
  centre stage for morphogenesis},\ }\href@noop {} {\bibfield  {journal}
  {\bibinfo  {journal} {Philos. Trans. R. Soc. B}\ }\textbf {\bibinfo {volume}
  {375}},\ \bibinfo {pages} {20190557} (\bibinfo {year} {2020})}\BibitemShut
  {NoStop}%
\bibitem [{\citenamefont {Dogterom}\ and\ \citenamefont
  {Leibler}(1993)}]{dogterom1993physical}%
  \BibitemOpen
  \bibfield  {author} {\bibinfo {author} {\bibfnamefont {M.}~\bibnamefont
  {Dogterom}}\ and\ \bibinfo {author} {\bibfnamefont {S.}~\bibnamefont
  {Leibler}},\ }\bibfield  {title} {\bibinfo {title} {Physical aspects of the
  growth and regulation of microtubule structures},\ }\href@noop {} {\bibfield
  {journal} {\bibinfo  {journal} {Phys. Rev. Lett.}\ }\textbf {\bibinfo
  {volume} {70}},\ \bibinfo {pages} {1347} (\bibinfo {year}
  {1993})}\BibitemShut {NoStop}%
\bibitem [{\citenamefont {Fygenson}\ \emph {et~al.}(1994)\citenamefont
  {Fygenson}, \citenamefont {Braun},\ and\ \citenamefont
  {Libchaber}}]{fygenson1994phase}%
  \BibitemOpen
  \bibfield  {author} {\bibinfo {author} {\bibfnamefont {D.~K.}\ \bibnamefont
  {Fygenson}}, \bibinfo {author} {\bibfnamefont {E.}~\bibnamefont {Braun}},\
  and\ \bibinfo {author} {\bibfnamefont {A.}~\bibnamefont {Libchaber}},\
  }\bibfield  {title} {\bibinfo {title} {Phase diagram of microtubules},\
  }\href@noop {} {\bibfield  {journal} {\bibinfo  {journal} {Phys. Rev. E}\
  }\textbf {\bibinfo {volume} {50}},\ \bibinfo {pages} {1579} (\bibinfo {year}
  {1994})}\BibitemShut {NoStop}%
\bibitem [{\citenamefont {Mitchison}\ and\ \citenamefont
  {Kirschner}(1984)}]{mitchison1984dynamic}%
  \BibitemOpen
  \bibfield  {author} {\bibinfo {author} {\bibfnamefont {T.}~\bibnamefont
  {Mitchison}}\ and\ \bibinfo {author} {\bibfnamefont {M.}~\bibnamefont
  {Kirschner}},\ }\bibfield  {title} {\bibinfo {title} {Dynamic instability of
  microtubule growth},\ }\href@noop {} {\bibfield  {journal} {\bibinfo
  {journal} {Nature}\ }\textbf {\bibinfo {volume} {312}},\ \bibinfo {pages}
  {237} (\bibinfo {year} {1984})}\BibitemShut {NoStop}%
\bibitem [{\citenamefont {Weingarten}\ \emph {et~al.}(1975)\citenamefont
  {Weingarten}, \citenamefont {Lockwood}, \citenamefont {Hwo},\ and\
  \citenamefont {Kirschner}}]{weingarten1975protein}%
  \BibitemOpen
  \bibfield  {author} {\bibinfo {author} {\bibfnamefont {M.~D.}\ \bibnamefont
  {Weingarten}}, \bibinfo {author} {\bibfnamefont {A.~H.}\ \bibnamefont
  {Lockwood}}, \bibinfo {author} {\bibfnamefont {S.-Y.}\ \bibnamefont {Hwo}},\
  and\ \bibinfo {author} {\bibfnamefont {M.~W.}\ \bibnamefont {Kirschner}},\
  }\bibfield  {title} {\bibinfo {title} {A protein factor essential for
  microtubule assembly.},\ }\href@noop {} {\bibfield  {journal} {\bibinfo
  {journal} {Proc. Natl. Acad. Sci. USA}\ }\textbf {\bibinfo {volume} {72}},\
  \bibinfo {pages} {1858} (\bibinfo {year} {1975})}\BibitemShut {NoStop}%
\bibitem [{\citenamefont {Palmqvist}\ \emph {et~al.}(2020)\citenamefont
  {Palmqvist}, \citenamefont {Janelidze}, \citenamefont {Quiroz}, \citenamefont
  {Zetterberg}, \citenamefont {Lopera}, \citenamefont {Stomrud}, \citenamefont
  {Su}, \citenamefont {Chen}, \citenamefont {Serrano}, \citenamefont {Leuzy}
  \emph {et~al.}}]{palmqvist2020discriminative}%
  \BibitemOpen
  \bibfield  {author} {\bibinfo {author} {\bibfnamefont {S.}~\bibnamefont
  {Palmqvist}}, \bibinfo {author} {\bibfnamefont {S.}~\bibnamefont
  {Janelidze}}, \bibinfo {author} {\bibfnamefont {Y.~T.}\ \bibnamefont
  {Quiroz}}, \bibinfo {author} {\bibfnamefont {H.}~\bibnamefont {Zetterberg}},
  \bibinfo {author} {\bibfnamefont {F.}~\bibnamefont {Lopera}}, \bibinfo
  {author} {\bibfnamefont {E.}~\bibnamefont {Stomrud}}, \bibinfo {author}
  {\bibfnamefont {Y.}~\bibnamefont {Su}}, \bibinfo {author} {\bibfnamefont
  {Y.}~\bibnamefont {Chen}}, \bibinfo {author} {\bibfnamefont {G.~E.}\
  \bibnamefont {Serrano}}, \bibinfo {author} {\bibfnamefont {A.}~\bibnamefont
  {Leuzy}}, \emph {et~al.},\ }\bibfield  {title} {\bibinfo {title}
  {Discriminative accuracy of plasma phospho-tau217 for {A}lzheimer disease vs
  other neurodegenerative disorders},\ }\href@noop {} {\bibfield  {journal}
  {\bibinfo  {journal} {JAMA}\ }\textbf {\bibinfo {volume} {324}},\ \bibinfo
  {pages} {772} (\bibinfo {year} {2020})}\BibitemShut {NoStop}%
\bibitem [{\citenamefont {Chen}\ \emph {et~al.}(2024)\citenamefont {Chen},
  \citenamefont {Tseng},\ and\ \citenamefont {Ward}}]{chen2024mathematical}%
  \BibitemOpen
  \bibfield  {author} {\bibinfo {author} {\bibfnamefont {C.~Y.}\ \bibnamefont
  {Chen}}, \bibinfo {author} {\bibfnamefont {Y.~H.}\ \bibnamefont {Tseng}},\
  and\ \bibinfo {author} {\bibfnamefont {J.~P.}\ \bibnamefont {Ward}},\
  }\bibfield  {title} {\bibinfo {title} {A mathematical model on the
  propagation of tau pathology in neurodegenerative diseases},\ }\href@noop {}
  {\bibfield  {journal} {\bibinfo  {journal} {J. Math. Biol.}\ }\textbf
  {\bibinfo {volume} {89}},\ \bibinfo {pages} {4} (\bibinfo {year}
  {2024})}\BibitemShut {NoStop}%
\bibitem [{\citenamefont {Desai}\ and\ \citenamefont
  {Mitchison}(1997)}]{desai1997microtubule}%
  \BibitemOpen
  \bibfield  {author} {\bibinfo {author} {\bibfnamefont {A.}~\bibnamefont
  {Desai}}\ and\ \bibinfo {author} {\bibfnamefont {T.~J.}\ \bibnamefont
  {Mitchison}},\ }\bibfield  {title} {\bibinfo {title} {Microtubule
  polymerization dynamics},\ }\href@noop {} {\bibfield  {journal} {\bibinfo
  {journal} {Annu. Rev. Cell Dev. Biol.}\ }\textbf {\bibinfo {volume} {13}},\
  \bibinfo {pages} {83} (\bibinfo {year} {1997})}\BibitemShut {NoStop}%
\bibitem [{\citenamefont {Lehninger}\ \emph {et~al.}(1993)\citenamefont
  {Lehninger}, \citenamefont {Nelson},\ and\ \citenamefont
  {Cox}}]{nelson2008lehninger}%
  \BibitemOpen
  \bibfield  {author} {\bibinfo {author} {\bibfnamefont {A.~L.}\ \bibnamefont
  {Lehninger}}, \bibinfo {author} {\bibfnamefont {D.~L.}\ \bibnamefont
  {Nelson}},\ and\ \bibinfo {author} {\bibfnamefont {M.~M.}\ \bibnamefont
  {Cox}},\ }\href@noop {} {\emph {\bibinfo {title} {Principles of
  Biochemistry.}}}\ (\bibinfo  {publisher} {New York: Worth},\ \bibinfo {year}
  {1993})\ \bibinfo {note} {2nd ed.}\BibitemShut {Stop}%
\bibitem [{\citenamefont {Coue}\ \emph {et~al.}(1991)\citenamefont {Coue},
  \citenamefont {Lombillo},\ and\ \citenamefont
  {McIntosh}}]{coue1991microtubule}%
  \BibitemOpen
  \bibfield  {author} {\bibinfo {author} {\bibfnamefont {M.}~\bibnamefont
  {Coue}}, \bibinfo {author} {\bibfnamefont {V.~A.}\ \bibnamefont {Lombillo}},\
  and\ \bibinfo {author} {\bibfnamefont {J.~R.}\ \bibnamefont {McIntosh}},\
  }\bibfield  {title} {\bibinfo {title} {Microtubule depolymerization promotes
  particle and chromosome movement in vitro},\ }\href@noop {} {\bibfield
  {journal} {\bibinfo  {journal} {J. Cell. Biol.}\ }\textbf {\bibinfo {volume}
  {112}},\ \bibinfo {pages} {1165} (\bibinfo {year} {1991})}\BibitemShut
  {NoStop}%
\bibitem [{\citenamefont {Koshland}\ \emph {et~al.}(1988)\citenamefont
  {Koshland}, \citenamefont {Mitchison},\ and\ \citenamefont
  {Kirschner}}]{koshland1988polewards}%
  \BibitemOpen
  \bibfield  {author} {\bibinfo {author} {\bibfnamefont {D.~E.}\ \bibnamefont
  {Koshland}}, \bibinfo {author} {\bibfnamefont {T.~J.}\ \bibnamefont
  {Mitchison}},\ and\ \bibinfo {author} {\bibfnamefont {M.~W.}\ \bibnamefont
  {Kirschner}},\ }\bibfield  {title} {\bibinfo {title} {Polewards chromosome
  movement driven by microtubule depolymerization in vitro},\ }\href@noop {}
  {\bibfield  {journal} {\bibinfo  {journal} {Nature}\ }\textbf {\bibinfo
  {volume} {331}},\ \bibinfo {pages} {499} (\bibinfo {year}
  {1988})}\BibitemShut {NoStop}%
\bibitem [{\citenamefont {Wittmann}\ \emph {et~al.}(2001)\citenamefont
  {Wittmann}, \citenamefont {Hyman},\ and\ \citenamefont
  {Desai}}]{wittmann2001spindle}%
  \BibitemOpen
  \bibfield  {author} {\bibinfo {author} {\bibfnamefont {T.}~\bibnamefont
  {Wittmann}}, \bibinfo {author} {\bibfnamefont {A.}~\bibnamefont {Hyman}},\
  and\ \bibinfo {author} {\bibfnamefont {A.}~\bibnamefont {Desai}},\ }\bibfield
   {title} {\bibinfo {title} {The spindle: a dynamic assembly of microtubules
  and motors},\ }\href@noop {} {\bibfield  {journal} {\bibinfo  {journal} {Nat.
  Cell Biol.}\ }\textbf {\bibinfo {volume} {3}},\ \bibinfo {pages} {E28}
  (\bibinfo {year} {2001})}\BibitemShut {NoStop}%
\bibitem [{\citenamefont {Gadde}\ and\ \citenamefont
  {Heald}(2004)}]{gadde2004mechanisms}%
  \BibitemOpen
  \bibfield  {author} {\bibinfo {author} {\bibfnamefont {S.}~\bibnamefont
  {Gadde}}\ and\ \bibinfo {author} {\bibfnamefont {R.}~\bibnamefont {Heald}},\
  }\bibfield  {title} {\bibinfo {title} {Mechanisms and molecules of the
  mitotic spindle},\ }\href@noop {} {\bibfield  {journal} {\bibinfo  {journal}
  {Curr. Biol.}\ }\textbf {\bibinfo {volume} {14}},\ \bibinfo {pages} {R797}
  (\bibinfo {year} {2004})}\BibitemShut {NoStop}%
\bibitem [{\citenamefont {Kirschner}\ and\ \citenamefont
  {Mitchison}(1986)}]{kirschner1986beyond}%
  \BibitemOpen
  \bibfield  {author} {\bibinfo {author} {\bibfnamefont {M.}~\bibnamefont
  {Kirschner}}\ and\ \bibinfo {author} {\bibfnamefont {T.}~\bibnamefont
  {Mitchison}},\ }\bibfield  {title} {\bibinfo {title} {Beyond self-assembly:
  from microtubules to morphogenesis},\ }\href@noop {} {\bibfield  {journal}
  {\bibinfo  {journal} {Cell}\ }\textbf {\bibinfo {volume} {45}},\ \bibinfo
  {pages} {329} (\bibinfo {year} {1986})}\BibitemShut {NoStop}%
\bibitem [{\citenamefont {Dogterom}\ and\ \citenamefont
  {Koenderink}(2019)}]{dogterom2019actin}%
  \BibitemOpen
  \bibfield  {author} {\bibinfo {author} {\bibfnamefont {M.}~\bibnamefont
  {Dogterom}}\ and\ \bibinfo {author} {\bibfnamefont {G.~H.}\ \bibnamefont
  {Koenderink}},\ }\bibfield  {title} {\bibinfo {title} {Actin--microtubule
  crosstalk in cell biology},\ }\href@noop {} {\bibfield  {journal} {\bibinfo
  {journal} {Nat. Rev. Mol. Cell Biol.}\ }\textbf {\bibinfo {volume} {20}},\
  \bibinfo {pages} {38} (\bibinfo {year} {2019})}\BibitemShut {NoStop}%
\bibitem [{\citenamefont {Xie}\ \emph {et~al.}(2022)\citenamefont {Xie},
  \citenamefont {Najafi}, \citenamefont {Le~Borgne}, \citenamefont {Verbavatz},
  \citenamefont {Durieu}, \citenamefont {Sall{\'e}},\ and\ \citenamefont
  {Minc}}]{xie2022contribution}%
  \BibitemOpen
  \bibfield  {author} {\bibinfo {author} {\bibfnamefont {J.}~\bibnamefont
  {Xie}}, \bibinfo {author} {\bibfnamefont {J.}~\bibnamefont {Najafi}},
  \bibinfo {author} {\bibfnamefont {R.}~\bibnamefont {Le~Borgne}}, \bibinfo
  {author} {\bibfnamefont {J.-M.}\ \bibnamefont {Verbavatz}}, \bibinfo {author}
  {\bibfnamefont {C.}~\bibnamefont {Durieu}}, \bibinfo {author} {\bibfnamefont
  {J.}~\bibnamefont {Sall{\'e}}},\ and\ \bibinfo {author} {\bibfnamefont
  {N.}~\bibnamefont {Minc}},\ }\bibfield  {title} {\bibinfo {title}
  {Contribution of cytoplasm viscoelastic properties to mitotic spindle
  positioning},\ }\href@noop {} {\bibfield  {journal} {\bibinfo  {journal}
  {Proc. Natl. Acad. Sci. USA}\ }\textbf {\bibinfo {volume} {119}},\ \bibinfo
  {pages} {e2115593119} (\bibinfo {year} {2022})}\BibitemShut {NoStop}%
\bibitem [{\citenamefont {Schaeffer}\ \emph {et~al.}(2025)\citenamefont
  {Schaeffer}, \citenamefont {Burraco}, \citenamefont {Gazzola}, \citenamefont
  {Gelin}, \citenamefont {Vianay}, \citenamefont {{d}e Pascalis}, \citenamefont
  {Blanchoin},\ and\ \citenamefont {Th{\'e}ry}}]{schaeffer2025Microtubule}%
  \BibitemOpen
  \bibfield  {author} {\bibinfo {author} {\bibfnamefont {A.}~\bibnamefont
  {Schaeffer}}, \bibinfo {author} {\bibfnamefont {S.}~\bibnamefont {Burraco}},
  \bibinfo {author} {\bibfnamefont {M.}~\bibnamefont {Gazzola}}, \bibinfo
  {author} {\bibfnamefont {M.}~\bibnamefont {Gelin}}, \bibinfo {author}
  {\bibfnamefont {B.}~\bibnamefont {Vianay}}, \bibinfo {author} {\bibfnamefont
  {C.}~\bibnamefont {{d}e Pascalis}}, \bibinfo {author} {\bibfnamefont
  {L.}~\bibnamefont {Blanchoin}},\ and\ \bibinfo {author} {\bibfnamefont
  {M.}~\bibnamefont {Th{\'e}ry}},\ }\bibfield  {title} {\bibinfo {title}
  {Microtubule-driven cell shape changes and actomyosin flow synergize to
  position the centrosome},\ }\href {https://doi.org/doi:
  https://doi.org/10.1101/2024.06.11.598451} {\bibfield  {journal} {\bibinfo
  {journal} {J. Cell Biol.}\ }\textbf {\bibinfo {volume} {224}},\ \bibinfo
  {pages} {e202405126} (\bibinfo {year} {2025})}\BibitemShut {NoStop}%
\bibitem [{\citenamefont {Lepore}\ \emph {et~al.}(2016)\citenamefont {Lepore},
  \citenamefont {Isaia}, \citenamefont {Mammola},\ and\ \citenamefont
  {Pugno}}]{lepore2016effect}%
  \BibitemOpen
  \bibfield  {author} {\bibinfo {author} {\bibfnamefont {E.}~\bibnamefont
  {Lepore}}, \bibinfo {author} {\bibfnamefont {M.}~\bibnamefont {Isaia}},
  \bibinfo {author} {\bibfnamefont {S.}~\bibnamefont {Mammola}},\ and\ \bibinfo
  {author} {\bibfnamefont {N.}~\bibnamefont {Pugno}},\ }\bibfield  {title}
  {\bibinfo {title} {The effect of ageing on the mechanical properties of the
  silk of the bridge spider \emph{{L}arinioides cornutus} ({C}lerck, 1757)},\
  }\href@noop {} {\bibfield  {journal} {\bibinfo  {journal} {Sci. Rep.}\
  }\textbf {\bibinfo {volume} {6}},\ \bibinfo {pages} {24699} (\bibinfo {year}
  {2016})}\BibitemShut {NoStop}%
\bibitem [{\citenamefont {Greco}\ \emph {et~al.}(2022)\citenamefont {Greco},
  \citenamefont {Mirbaha}, \citenamefont {Schmuck}, \citenamefont {Rising},\
  and\ \citenamefont {Pugno}}]{greco2022artificial}%
  \BibitemOpen
  \bibfield  {author} {\bibinfo {author} {\bibfnamefont {G.}~\bibnamefont
  {Greco}}, \bibinfo {author} {\bibfnamefont {H.}~\bibnamefont {Mirbaha}},
  \bibinfo {author} {\bibfnamefont {B.}~\bibnamefont {Schmuck}}, \bibinfo
  {author} {\bibfnamefont {A.}~\bibnamefont {Rising}},\ and\ \bibinfo {author}
  {\bibfnamefont {N.~M.}\ \bibnamefont {Pugno}},\ }\bibfield  {title} {\bibinfo
  {title} {Artificial and natural silk materials have high mechanical property
  variability regardless of sample size},\ }\href@noop {} {\bibfield  {journal}
  {\bibinfo  {journal} {Sci. Rep.}\ }\textbf {\bibinfo {volume} {12}},\
  \bibinfo {pages} {3507} (\bibinfo {year} {2022})}\BibitemShut {NoStop}%
\bibitem [{\citenamefont {Wainwright}(1982)}]{wainwright1982mechanical}%
  \BibitemOpen
  \bibfield  {author} {\bibinfo {author} {\bibfnamefont {S.~A.}\ \bibnamefont
  {Wainwright}},\ }\href@noop {} {\emph {\bibinfo {title} {Mechanical design in
  organisms}}}\ (\bibinfo  {publisher} {Princeton University Press},\ \bibinfo
  {year} {1982})\BibitemShut {NoStop}%
\bibitem [{\citenamefont {Bonser}\ and\ \citenamefont
  {Purslow}(1995)}]{bonser1995young}%
  \BibitemOpen
  \bibfield  {author} {\bibinfo {author} {\bibfnamefont {R.~H.~C.}\
  \bibnamefont {Bonser}}\ and\ \bibinfo {author} {\bibfnamefont {P.~P.}\
  \bibnamefont {Purslow}},\ }\bibfield  {title} {\bibinfo {title} {The
  {Y}oung’s modulus of feather keratin},\ }\href@noop {} {\bibfield
  {journal} {\bibinfo  {journal} {J. Exp. Biol.}\ }\textbf {\bibinfo {volume}
  {198}},\ \bibinfo {pages} {1029} (\bibinfo {year} {1995})}\BibitemShut
  {NoStop}%
\bibitem [{\citenamefont {Burlacu}\ \emph {et~al.}(1992)\citenamefont
  {Burlacu}, \citenamefont {Janmey},\ and\ \citenamefont
  {Borejdo}}]{burlacu1992distribution}%
  \BibitemOpen
  \bibfield  {author} {\bibinfo {author} {\bibfnamefont {S.}~\bibnamefont
  {Burlacu}}, \bibinfo {author} {\bibfnamefont {P.~A.}\ \bibnamefont
  {Janmey}},\ and\ \bibinfo {author} {\bibfnamefont {J.}~\bibnamefont
  {Borejdo}},\ }\bibfield  {title} {\bibinfo {title} {Distribution of actin
  filament lengths measured by fluorescence microscopy},\ }\href@noop {}
  {\bibfield  {journal} {\bibinfo  {journal} {Am. J. Physiol. Cell Physiol.}\
  }\textbf {\bibinfo {volume} {262}},\ \bibinfo {pages} {C569} (\bibinfo {year}
  {1992})}\BibitemShut {NoStop}%
\bibitem [{\citenamefont {Karimi}\ \emph {et~al.}(2021)\citenamefont {Karimi},
  \citenamefont {Alam},\ and\ \citenamefont {Mofrad}}]{karimi2021hydrodynamic}%
  \BibitemOpen
  \bibfield  {author} {\bibinfo {author} {\bibfnamefont {R.}~\bibnamefont
  {Karimi}}, \bibinfo {author} {\bibfnamefont {M.~R.}\ \bibnamefont {Alam}},\
  and\ \bibinfo {author} {\bibfnamefont {M.~R.~K.}\ \bibnamefont {Mofrad}},\
  }\bibfield  {title} {\bibinfo {title} {Hydrodynamic interactions
  significantly alter the dynamics of actin networks and result in a length
  scale dependent loss modulus},\ }\href@noop {} {\bibfield  {journal}
  {\bibinfo  {journal} {J. Biomech.}\ }\textbf {\bibinfo {volume} {120}},\
  \bibinfo {pages} {110352} (\bibinfo {year} {2021})}\BibitemShut {NoStop}%
\bibitem [{\citenamefont {Hartman}\ and\ \citenamefont
  {Spudich}(2012)}]{hartman2012myosin}%
  \BibitemOpen
  \bibfield  {author} {\bibinfo {author} {\bibfnamefont {M.~A.}\ \bibnamefont
  {Hartman}}\ and\ \bibinfo {author} {\bibfnamefont {J.~A.}\ \bibnamefont
  {Spudich}},\ }\bibfield  {title} {\bibinfo {title} {The myosin superfamily at
  a glance},\ }\href@noop {} {\bibfield  {journal} {\bibinfo  {journal} {J.
  Cell Sci.}\ }\textbf {\bibinfo {volume} {125}},\ \bibinfo {pages} {1627}
  (\bibinfo {year} {2012})}\BibitemShut {NoStop}%
\bibitem [{\citenamefont {Odronitz}\ and\ \citenamefont
  {Kollmar}(2007)}]{odronitz2007drawing}%
  \BibitemOpen
  \bibfield  {author} {\bibinfo {author} {\bibfnamefont {F.}~\bibnamefont
  {Odronitz}}\ and\ \bibinfo {author} {\bibfnamefont {M.}~\bibnamefont
  {Kollmar}},\ }\bibfield  {title} {\bibinfo {title} {Drawing the tree of
  eukaryotic life based on the analysis of 2,269 manually annotated myosins
  from 328 species},\ }\href@noop {} {\bibfield  {journal} {\bibinfo  {journal}
  {Genome Biol.}\ }\textbf {\bibinfo {volume} {8}},\ \bibinfo {pages} {1}
  (\bibinfo {year} {2007})}\BibitemShut {NoStop}%
\bibitem [{\citenamefont {Sweeney}\ and\ \citenamefont
  {Houdusse}(2010)}]{sweeney2010structural}%
  \BibitemOpen
  \bibfield  {author} {\bibinfo {author} {\bibfnamefont {H.~L.}\ \bibnamefont
  {Sweeney}}\ and\ \bibinfo {author} {\bibfnamefont {A.}~\bibnamefont
  {Houdusse}},\ }\bibfield  {title} {\bibinfo {title} {Structural and
  functional insights into the myosin motor mechanism},\ }\href@noop {}
  {\bibfield  {journal} {\bibinfo  {journal} {Annu. Rev. Biophys.}\ }\textbf
  {\bibinfo {volume} {39}},\ \bibinfo {pages} {539} (\bibinfo {year}
  {2010})}\BibitemShut {NoStop}%
\bibitem [{\citenamefont {Vicente-Manzanares}\ \emph
  {et~al.}(2009)\citenamefont {Vicente-Manzanares}, \citenamefont {Ma},
  \citenamefont {Adelstein},\ and\ \citenamefont {Horwitz}}]{vicente2009non}%
  \BibitemOpen
  \bibfield  {author} {\bibinfo {author} {\bibfnamefont {M.}~\bibnamefont
  {Vicente-Manzanares}}, \bibinfo {author} {\bibfnamefont {X.}~\bibnamefont
  {Ma}}, \bibinfo {author} {\bibfnamefont {R.~S.}\ \bibnamefont {Adelstein}},\
  and\ \bibinfo {author} {\bibfnamefont {A.~R.}\ \bibnamefont {Horwitz}},\
  }\bibfield  {title} {\bibinfo {title} {Non-muscle myosin ii takes centre
  stage in cell adhesion and migration},\ }\href@noop {} {\bibfield  {journal}
  {\bibinfo  {journal} {Nat. Rev. Mol. Cell Biol.}\ }\textbf {\bibinfo {volume}
  {10}},\ \bibinfo {pages} {778} (\bibinfo {year} {2009})}\BibitemShut
  {NoStop}%
\bibitem [{\citenamefont {{Tina Carvalho}}(2011)}]{CIL219}%
  \BibitemOpen
  \bibfield  {author} {\bibinfo {author} {\bibnamefont {{Tina Carvalho}}},\
  }\href@noop {} {\bibinfo {title} {{CIL}:219}},\ \bibinfo {howpublished}
  {{CIL} Dataset, \url{https://doi.org/doi:10.7295/W9CIL219}} (\bibinfo {year}
  {2011})\BibitemShut {NoStop}%
\bibitem [{\citenamefont {{David Becker}}(2011)}]{CIL39004}%
  \BibitemOpen
  \bibfield  {author} {\bibinfo {author} {\bibnamefont {{David Becker}}},\
  }\href@noop {} {\bibinfo {title} {{CIL}:39004}},\ \bibinfo {howpublished}
  {{CIL} Dataset, \url{https://doi.org/doi:10.7295/W9CIL39004}} (\bibinfo
  {year} {2011})\BibitemShut {NoStop}%
\bibitem [{\citenamefont {{\'E}tienne}\ \emph {et~al.}(2015)\citenamefont
  {{\'E}tienne}, \citenamefont {Fouchard}, \citenamefont {Mitrossilis},
  \citenamefont {Bufi}, \citenamefont {Durand-Smet},\ and\ \citenamefont
  {Asnacios}}]{etienne2015cells}%
  \BibitemOpen
  \bibfield  {author} {\bibinfo {author} {\bibfnamefont {J.}~\bibnamefont
  {{\'E}tienne}}, \bibinfo {author} {\bibfnamefont {J.}~\bibnamefont
  {Fouchard}}, \bibinfo {author} {\bibfnamefont {D.}~\bibnamefont
  {Mitrossilis}}, \bibinfo {author} {\bibfnamefont {N.}~\bibnamefont {Bufi}},
  \bibinfo {author} {\bibfnamefont {P.}~\bibnamefont {Durand-Smet}},\ and\
  \bibinfo {author} {\bibfnamefont {A.}~\bibnamefont {Asnacios}},\ }\bibfield
  {title} {\bibinfo {title} {Cells as liquid motors: Mechanosensitivity emerges
  from collective dynamics of actomyosin cortex},\ }\href@noop {} {\bibfield
  {journal} {\bibinfo  {journal} {Proc. Natl. Acad. Sci. USA}\ }\textbf
  {\bibinfo {volume} {112}},\ \bibinfo {pages} {2740} (\bibinfo {year}
  {2015})}\BibitemShut {NoStop}%
\bibitem [{\citenamefont {Saha}\ \emph {et~al.}(2016)\citenamefont {Saha},
  \citenamefont {Nishikawa}, \citenamefont {Behrndt}, \citenamefont
  {Heisenberg}, \citenamefont {J{\"u}licher},\ and\ \citenamefont
  {Grill}}]{saha2016determining}%
  \BibitemOpen
  \bibfield  {author} {\bibinfo {author} {\bibfnamefont {A.}~\bibnamefont
  {Saha}}, \bibinfo {author} {\bibfnamefont {M.}~\bibnamefont {Nishikawa}},
  \bibinfo {author} {\bibfnamefont {M.}~\bibnamefont {Behrndt}}, \bibinfo
  {author} {\bibfnamefont {C.-P.}\ \bibnamefont {Heisenberg}}, \bibinfo
  {author} {\bibfnamefont {F.}~\bibnamefont {J{\"u}licher}},\ and\ \bibinfo
  {author} {\bibfnamefont {S.~W.}\ \bibnamefont {Grill}},\ }\bibfield  {title}
  {\bibinfo {title} {Determining physical properties of the cell cortex},\
  }\href@noop {} {\bibfield  {journal} {\bibinfo  {journal} {Biophys. J.}\
  }\textbf {\bibinfo {volume} {110}},\ \bibinfo {pages} {1421} (\bibinfo {year}
  {2016})}\BibitemShut {NoStop}%
\bibitem [{\citenamefont {Kumar}(2021)}]{kumar2021actomyosin}%
  \BibitemOpen
  \bibfield  {author} {\bibinfo {author} {\bibfnamefont {K.~V.}\ \bibnamefont
  {Kumar}},\ }\bibfield  {title} {\bibinfo {title} {The actomyosin cortex of
  cells: a thin film of active matter},\ }\href@noop {} {\bibfield  {journal}
  {\bibinfo  {journal} {J. Indian Inst. Sci.}\ }\textbf {\bibinfo {volume}
  {101}},\ \bibinfo {pages} {97} (\bibinfo {year} {2021})}\BibitemShut
  {NoStop}%
\bibitem [{\citenamefont {Cardamone}\ \emph {et~al.}(2011)\citenamefont
  {Cardamone}, \citenamefont {Laio}, \citenamefont {Torre}, \citenamefont
  {Shahapure},\ and\ \citenamefont {DeSimone}}]{cardamone2011cytoskeletal}%
  \BibitemOpen
  \bibfield  {author} {\bibinfo {author} {\bibfnamefont {L.}~\bibnamefont
  {Cardamone}}, \bibinfo {author} {\bibfnamefont {A.}~\bibnamefont {Laio}},
  \bibinfo {author} {\bibfnamefont {V.}~\bibnamefont {Torre}}, \bibinfo
  {author} {\bibfnamefont {R.}~\bibnamefont {Shahapure}},\ and\ \bibinfo
  {author} {\bibfnamefont {A.}~\bibnamefont {DeSimone}},\ }\bibfield  {title}
  {\bibinfo {title} {Cytoskeletal actin networks in motile cells are critically
  self-organized systems synchronized by mechanical interactions},\ }\href@noop
  {} {\bibfield  {journal} {\bibinfo  {journal} {Proc. Natl. Acad. Sci. USA}\
  }\textbf {\bibinfo {volume} {108}},\ \bibinfo {pages} {13978} (\bibinfo
  {year} {2011})}\BibitemShut {NoStop}%
\bibitem [{\citenamefont {Schwarz}\ and\ \citenamefont
  {Safran}(2013)}]{Schwarz2013}%
  \BibitemOpen
  \bibfield  {author} {\bibinfo {author} {\bibfnamefont {U.~S.}\ \bibnamefont
  {Schwarz}}\ and\ \bibinfo {author} {\bibfnamefont {S.~A.}\ \bibnamefont
  {Safran}},\ }\bibfield  {title} {\bibinfo {title} {Physics of adherent
  cells},\ }\href {https://doi.org/10.1103/RevModPhys.85.1327} {\bibfield
  {journal} {\bibinfo  {journal} {Rev. Mod. Phys.}\ }\textbf {\bibinfo {volume}
  {85}},\ \bibinfo {pages} {1327} (\bibinfo {year} {2013})}\BibitemShut
  {NoStop}%
\bibitem [{\citenamefont {Garcia}\ \emph {et~al.}(2015)\citenamefont {Garcia},
  \citenamefont {Hannezo}, \citenamefont {Elgeti}, \citenamefont {Joanny},
  \citenamefont {Silberzan},\ and\ \citenamefont {Gov}}]{garcia2015physics}%
  \BibitemOpen
  \bibfield  {author} {\bibinfo {author} {\bibfnamefont {S.}~\bibnamefont
  {Garcia}}, \bibinfo {author} {\bibfnamefont {E.}~\bibnamefont {Hannezo}},
  \bibinfo {author} {\bibfnamefont {J.}~\bibnamefont {Elgeti}}, \bibinfo
  {author} {\bibfnamefont {J.-F.}\ \bibnamefont {Joanny}}, \bibinfo {author}
  {\bibfnamefont {P.}~\bibnamefont {Silberzan}},\ and\ \bibinfo {author}
  {\bibfnamefont {N.~S.}\ \bibnamefont {Gov}},\ }\bibfield  {title} {\bibinfo
  {title} {Physics of active jamming during collective cellular motion in a
  monolayer},\ }\href@noop {} {\bibfield  {journal} {\bibinfo  {journal} {Proc.
  Natl. Acad. Sci. USA}\ }\textbf {\bibinfo {volume} {112}},\ \bibinfo {pages}
  {15314} (\bibinfo {year} {2015})}\BibitemShut {NoStop}%
\bibitem [{\citenamefont {Knowles}\ \emph {et~al.}(2007)\citenamefont
  {Knowles}, \citenamefont {Fitzpatrick}, \citenamefont {Meehan}, \citenamefont
  {Mott}, \citenamefont {Vendruscolo}, \citenamefont {Dobson},\ and\
  \citenamefont {Welland}}]{knowles2007role}%
  \BibitemOpen
  \bibfield  {author} {\bibinfo {author} {\bibfnamefont {T.~P.}\ \bibnamefont
  {Knowles}}, \bibinfo {author} {\bibfnamefont {A.~W.}\ \bibnamefont
  {Fitzpatrick}}, \bibinfo {author} {\bibfnamefont {S.}~\bibnamefont {Meehan}},
  \bibinfo {author} {\bibfnamefont {H.~R.}\ \bibnamefont {Mott}}, \bibinfo
  {author} {\bibfnamefont {M.}~\bibnamefont {Vendruscolo}}, \bibinfo {author}
  {\bibfnamefont {C.~M.}\ \bibnamefont {Dobson}},\ and\ \bibinfo {author}
  {\bibfnamefont {M.~E.}\ \bibnamefont {Welland}},\ }\bibfield  {title}
  {\bibinfo {title} {Role of intermolecular forces in defining material
  properties of protein nanofibrils},\ }\href@noop {} {\bibfield  {journal}
  {\bibinfo  {journal} {Science}\ }\textbf {\bibinfo {volume} {318}},\ \bibinfo
  {pages} {1900} (\bibinfo {year} {2007})}\BibitemShut {NoStop}%
\bibitem [{\citenamefont {Heisenberg}\ and\ \citenamefont
  {Bella{\"\i}che}(2013)}]{heisenberg2013forces}%
  \BibitemOpen
  \bibfield  {author} {\bibinfo {author} {\bibfnamefont {C.-P.}\ \bibnamefont
  {Heisenberg}}\ and\ \bibinfo {author} {\bibfnamefont {Y.}~\bibnamefont
  {Bella{\"\i}che}},\ }\bibfield  {title} {\bibinfo {title} {Forces in tissue
  morphogenesis and patterning},\ }\href@noop {} {\bibfield  {journal}
  {\bibinfo  {journal} {Cell}\ }\textbf {\bibinfo {volume} {153}},\ \bibinfo
  {pages} {948} (\bibinfo {year} {2013})}\BibitemShut {NoStop}%
\bibitem [{\citenamefont {Aigouy}\ \emph {et~al.}(2010)\citenamefont {Aigouy},
  \citenamefont {Farhadifar}, \citenamefont {Staple}, \citenamefont {Sagner},
  \citenamefont {R{\"o}per}, \citenamefont {J{\"u}licher},\ and\ \citenamefont
  {Eaton}}]{aigouy2010cell}%
  \BibitemOpen
  \bibfield  {author} {\bibinfo {author} {\bibfnamefont {B.}~\bibnamefont
  {Aigouy}}, \bibinfo {author} {\bibfnamefont {R.}~\bibnamefont {Farhadifar}},
  \bibinfo {author} {\bibfnamefont {D.~B.}\ \bibnamefont {Staple}}, \bibinfo
  {author} {\bibfnamefont {A.}~\bibnamefont {Sagner}}, \bibinfo {author}
  {\bibfnamefont {J.-C.}\ \bibnamefont {R{\"o}per}}, \bibinfo {author}
  {\bibfnamefont {F.}~\bibnamefont {J{\"u}licher}},\ and\ \bibinfo {author}
  {\bibfnamefont {S.}~\bibnamefont {Eaton}},\ }\bibfield  {title} {\bibinfo
  {title} {Cell flow reorients the axis of planar polarity in the wing
  epithelium of drosophila},\ }\href@noop {} {\bibfield  {journal} {\bibinfo
  {journal} {Cell}\ }\textbf {\bibinfo {volume} {142}},\ \bibinfo {pages} {773}
  (\bibinfo {year} {2010})}\BibitemShut {NoStop}%
\bibitem [{\citenamefont {Bray}\ and\ \citenamefont
  {White}(1988)}]{bray1988cortical}%
  \BibitemOpen
  \bibfield  {author} {\bibinfo {author} {\bibfnamefont {D.}~\bibnamefont
  {Bray}}\ and\ \bibinfo {author} {\bibfnamefont {J.~G.}\ \bibnamefont
  {White}},\ }\bibfield  {title} {\bibinfo {title} {Cortical flow in animal
  cells},\ }\href@noop {} {\bibfield  {journal} {\bibinfo  {journal} {Science}\
  }\textbf {\bibinfo {volume} {239}},\ \bibinfo {pages} {883} (\bibinfo {year}
  {1988})}\BibitemShut {NoStop}%
\bibitem [{\citenamefont {K{\"o}ster}\ \emph {et~al.}(2015)\citenamefont
  {K{\"o}ster}, \citenamefont {Weitz}, \citenamefont {Goldman}, \citenamefont
  {Aebi},\ and\ \citenamefont {Herrmann}}]{koster2015intermediate}%
  \BibitemOpen
  \bibfield  {author} {\bibinfo {author} {\bibfnamefont {S.}~\bibnamefont
  {K{\"o}ster}}, \bibinfo {author} {\bibfnamefont {D.~A.}\ \bibnamefont
  {Weitz}}, \bibinfo {author} {\bibfnamefont {R.~D.}\ \bibnamefont {Goldman}},
  \bibinfo {author} {\bibfnamefont {U.}~\bibnamefont {Aebi}},\ and\ \bibinfo
  {author} {\bibfnamefont {H.}~\bibnamefont {Herrmann}},\ }\bibfield  {title}
  {\bibinfo {title} {Intermediate filament mechanics in vitro and in the cell:
  from coiled coils to filaments, fibers and networks},\ }\href@noop {}
  {\bibfield  {journal} {\bibinfo  {journal} {Curr. Opin. Struct. Biol.}\
  }\textbf {\bibinfo {volume} {32}},\ \bibinfo {pages} {82} (\bibinfo {year}
  {2015})}\BibitemShut {NoStop}%
\bibitem [{\citenamefont {N\"oding}\ and\ \citenamefont
  {K\"oster}(2012)}]{Noeding2012}%
  \BibitemOpen
  \bibfield  {author} {\bibinfo {author} {\bibfnamefont {B.}~\bibnamefont
  {N\"oding}}\ and\ \bibinfo {author} {\bibfnamefont {S.}~\bibnamefont
  {K\"oster}},\ }\bibfield  {title} {\bibinfo {title} {Intermediate filaments
  in small configuration spaces},\ }\href
  {https://doi.org/10.1103/PhysRevLett.108.088101} {\bibfield  {journal}
  {\bibinfo  {journal} {Phys. Rev. Lett.}\ }\textbf {\bibinfo {volume} {108}},\
  \bibinfo {pages} {088101} (\bibinfo {year} {2012})}\BibitemShut {NoStop}%
\bibitem [{\citenamefont {Lowery}\ \emph {et~al.}(2015)\citenamefont {Lowery},
  \citenamefont {Kuczmarski}, \citenamefont {Herrmann},\ and\ \citenamefont
  {Goldman}}]{lowery2015intermediate}%
  \BibitemOpen
  \bibfield  {author} {\bibinfo {author} {\bibfnamefont {J.}~\bibnamefont
  {Lowery}}, \bibinfo {author} {\bibfnamefont {E.~R.}\ \bibnamefont
  {Kuczmarski}}, \bibinfo {author} {\bibfnamefont {H.}~\bibnamefont
  {Herrmann}},\ and\ \bibinfo {author} {\bibfnamefont {R.~D.}\ \bibnamefont
  {Goldman}},\ }\bibfield  {title} {\bibinfo {title} {Intermediate filaments
  play a pivotal role in regulating cell architecture and function},\
  }\href@noop {} {\bibfield  {journal} {\bibinfo  {journal} {J. Biol. Chem.}\
  }\textbf {\bibinfo {volume} {290}},\ \bibinfo {pages} {17145} (\bibinfo
  {year} {2015})}\BibitemShut {NoStop}%
\bibitem [{\citenamefont {Parry}\ and\ \citenamefont
  {Steinert}(1992)}]{parry1992intermediate}%
  \BibitemOpen
  \bibfield  {author} {\bibinfo {author} {\bibfnamefont {D.~A.~D.}\
  \bibnamefont {Parry}}\ and\ \bibinfo {author} {\bibfnamefont {P.~M.}\
  \bibnamefont {Steinert}},\ }\bibfield  {title} {\bibinfo {title}
  {Intermediate filament structure},\ }\href@noop {} {\bibfield  {journal}
  {\bibinfo  {journal} {Curr. Opin. Cell Biol.}\ }\textbf {\bibinfo {volume}
  {4}},\ \bibinfo {pages} {94} (\bibinfo {year} {1992})}\BibitemShut {NoStop}%
\bibitem [{\citenamefont {Chernyatina}\ \emph {et~al.}(2015)\citenamefont
  {Chernyatina}, \citenamefont {Guzenko},\ and\ \citenamefont
  {Strelkov}}]{chernyatina2015intermediate}%
  \BibitemOpen
  \bibfield  {author} {\bibinfo {author} {\bibfnamefont {A.~A.}\ \bibnamefont
  {Chernyatina}}, \bibinfo {author} {\bibfnamefont {D.}~\bibnamefont
  {Guzenko}},\ and\ \bibinfo {author} {\bibfnamefont {S.~V.}\ \bibnamefont
  {Strelkov}},\ }\bibfield  {title} {\bibinfo {title} {Intermediate filament
  structure: the bottom-up approach},\ }\href@noop {} {\bibfield  {journal}
  {\bibinfo  {journal} {Curr. Opin. Cell Biol.}\ }\textbf {\bibinfo {volume}
  {32}},\ \bibinfo {pages} {65} (\bibinfo {year} {2015})}\BibitemShut {NoStop}%
\bibitem [{\citenamefont {Herrmann}\ and\ \citenamefont
  {Aebi}(2016)}]{herrmann2016intermediate}%
  \BibitemOpen
  \bibfield  {author} {\bibinfo {author} {\bibfnamefont {H.}~\bibnamefont
  {Herrmann}}\ and\ \bibinfo {author} {\bibfnamefont {U.}~\bibnamefont
  {Aebi}},\ }\bibfield  {title} {\bibinfo {title} {Intermediate filaments:
  structure and assembly},\ }\href@noop {} {\bibfield  {journal} {\bibinfo
  {journal} {Cold Spring Harb. Perspect. Biol.}\ }\textbf {\bibinfo {volume}
  {8}},\ \bibinfo {pages} {a018242} (\bibinfo {year} {2016})}\BibitemShut
  {NoStop}%
\bibitem [{\citenamefont {Wen}\ and\ \citenamefont
  {Janmey}(2011)}]{wen2011polymer}%
  \BibitemOpen
  \bibfield  {author} {\bibinfo {author} {\bibfnamefont {Q.}~\bibnamefont
  {Wen}}\ and\ \bibinfo {author} {\bibfnamefont {P.~A.}\ \bibnamefont
  {Janmey}},\ }\bibfield  {title} {\bibinfo {title} {Polymer physics of the
  cytoskeleton},\ }\href@noop {} {\bibfield  {journal} {\bibinfo  {journal}
  {Curr. Opin. Solid State Mater. Sci.}\ }\textbf {\bibinfo {volume} {15}},\
  \bibinfo {pages} {177} (\bibinfo {year} {2011})}\BibitemShut {NoStop}%
\bibitem [{\citenamefont {Pritchard}\ \emph {et~al.}(2014)\citenamefont
  {Pritchard}, \citenamefont {Huang},\ and\ \citenamefont
  {Terentjev}}]{pritchard2014mechanics}%
  \BibitemOpen
  \bibfield  {author} {\bibinfo {author} {\bibfnamefont {R.~H.}\ \bibnamefont
  {Pritchard}}, \bibinfo {author} {\bibfnamefont {Y.~Y.~S.}\ \bibnamefont
  {Huang}},\ and\ \bibinfo {author} {\bibfnamefont {E.~M.}\ \bibnamefont
  {Terentjev}},\ }\bibfield  {title} {\bibinfo {title} {Mechanics of biological
  networks: from the cell cytoskeleton to connective tissue},\ }\href@noop {}
  {\bibfield  {journal} {\bibinfo  {journal} {Soft Matter}\ }\textbf {\bibinfo
  {volume} {10}},\ \bibinfo {pages} {1864} (\bibinfo {year}
  {2014})}\BibitemShut {NoStop}%
\bibitem [{\citenamefont {Kerst}\ \emph {et~al.}(1990)\citenamefont {Kerst},
  \citenamefont {Chmielewski}, \citenamefont {Livesay}, \citenamefont
  {Buxbaum},\ and\ \citenamefont {Heidemann}}]{kerst1990liquid}%
  \BibitemOpen
  \bibfield  {author} {\bibinfo {author} {\bibfnamefont {A.}~\bibnamefont
  {Kerst}}, \bibinfo {author} {\bibfnamefont {C.}~\bibnamefont {Chmielewski}},
  \bibinfo {author} {\bibfnamefont {C.}~\bibnamefont {Livesay}}, \bibinfo
  {author} {\bibfnamefont {R.~E.}\ \bibnamefont {Buxbaum}},\ and\ \bibinfo
  {author} {\bibfnamefont {S.~R.}\ \bibnamefont {Heidemann}},\ }\bibfield
  {title} {\bibinfo {title} {Liquid crystal domains and thixotropy of
  filamentous actin suspensions.},\ }\href@noop {} {\bibfield  {journal}
  {\bibinfo  {journal} {Proc. Natl. Acad. Sci. USA}\ }\textbf {\bibinfo
  {volume} {87}},\ \bibinfo {pages} {4241} (\bibinfo {year}
  {1990})}\BibitemShut {NoStop}%
\bibitem [{\citenamefont {Suzuki}\ \emph {et~al.}(1989)\citenamefont {Suzuki},
  \citenamefont {Yamazaki},\ and\ \citenamefont {Ito}}]{suzuki1989osmoelastic}%
  \BibitemOpen
  \bibfield  {author} {\bibinfo {author} {\bibfnamefont {A.}~\bibnamefont
  {Suzuki}}, \bibinfo {author} {\bibfnamefont {M.}~\bibnamefont {Yamazaki}},\
  and\ \bibinfo {author} {\bibfnamefont {T.}~\bibnamefont {Ito}},\ }\bibfield
  {title} {\bibinfo {title} {Osmoelastic coupling in biological structures:
  formation of parallel bundles of actin filaments in a crystalline-like
  structure caused by osmotic stress},\ }\href@noop {} {\bibfield  {journal}
  {\bibinfo  {journal} {Biochemistry}\ }\textbf {\bibinfo {volume} {28}},\
  \bibinfo {pages} {6513} (\bibinfo {year} {1989})}\BibitemShut {NoStop}%
\bibitem [{\citenamefont {Hitt}\ \emph {et~al.}(1990)\citenamefont {Hitt},
  \citenamefont {Cross},\ and\ \citenamefont
  {Williams~{Jr}}}]{hitt1990microtubule}%
  \BibitemOpen
  \bibfield  {author} {\bibinfo {author} {\bibfnamefont {A.~L.}\ \bibnamefont
  {Hitt}}, \bibinfo {author} {\bibfnamefont {A.~R.}\ \bibnamefont {Cross}},\
  and\ \bibinfo {author} {\bibfnamefont {R.~C.}\ \bibnamefont
  {Williams~{Jr}}},\ }\bibfield  {title} {\bibinfo {title} {Microtubule
  solutions display nematic liquid crystalline structure.},\ }\href@noop {}
  {\bibfield  {journal} {\bibinfo  {journal} {J. Biol. Chem.}\ }\textbf
  {\bibinfo {volume} {265}},\ \bibinfo {pages} {1639} (\bibinfo {year}
  {1990})}\BibitemShut {NoStop}%
\bibitem [{\citenamefont {Viamontes}\ \emph {et~al.}(2006)\citenamefont
  {Viamontes}, \citenamefont {Oakes},\ and\ \citenamefont
  {Tang}}]{Viamontes2006}%
  \BibitemOpen
  \bibfield  {author} {\bibinfo {author} {\bibfnamefont {J.}~\bibnamefont
  {Viamontes}}, \bibinfo {author} {\bibfnamefont {P.~W.}\ \bibnamefont
  {Oakes}},\ and\ \bibinfo {author} {\bibfnamefont {J.~X.}\ \bibnamefont
  {Tang}},\ }\bibfield  {title} {\bibinfo {title} {Isotropic to nematic liquid
  crystalline phase transition of $f$-actin varies from continuous to first
  order},\ }\href {https://doi.org/10.1103/PhysRevLett.97.118103} {\bibfield
  {journal} {\bibinfo  {journal} {Phys. Rev. Lett.}\ }\textbf {\bibinfo
  {volume} {97}},\ \bibinfo {pages} {118103} (\bibinfo {year}
  {2006})}\BibitemShut {NoStop}%
\bibitem [{\citenamefont {Hess}\ and\ \citenamefont
  {Ross}(2017)}]{hess2017non}%
  \BibitemOpen
  \bibfield  {author} {\bibinfo {author} {\bibfnamefont {H.}~\bibnamefont
  {Hess}}\ and\ \bibinfo {author} {\bibfnamefont {J.~L.}\ \bibnamefont
  {Ross}},\ }\bibfield  {title} {\bibinfo {title} {Non-equilibrium assembly of
  microtubules: from molecules to autonomous chemical robots},\ }\href@noop {}
  {\bibfield  {journal} {\bibinfo  {journal} {Chem. Soc. Rev.}\ }\textbf
  {\bibinfo {volume} {46}},\ \bibinfo {pages} {5570} (\bibinfo {year}
  {2017})}\BibitemShut {NoStop}%
\bibitem [{\citenamefont {Onsager}(1949)}]{onsager1949effects}%
  \BibitemOpen
  \bibfield  {author} {\bibinfo {author} {\bibfnamefont {L.}~\bibnamefont
  {Onsager}},\ }\bibfield  {title} {\bibinfo {title} {The effects of shape on
  the interaction of colloidal particles},\ }\href@noop {} {\bibfield
  {journal} {\bibinfo  {journal} {Ann. N. Y. Acad. Sci.}\ }\textbf {\bibinfo
  {volume} {51}},\ \bibinfo {pages} {627} (\bibinfo {year} {1949})}\BibitemShut
  {NoStop}%
\bibitem [{\citenamefont {Flory}(1956)}]{flory1956phase}%
  \BibitemOpen
  \bibfield  {author} {\bibinfo {author} {\bibfnamefont {P.~J.}\ \bibnamefont
  {Flory}},\ }\bibfield  {title} {\bibinfo {title} {Phase equilibria in
  solutions of rod-like particles},\ }\href@noop {} {\bibfield  {journal}
  {\bibinfo  {journal} {Proc. R. Soc. Lond. A}\ }\textbf {\bibinfo {volume}
  {234}},\ \bibinfo {pages} {73} (\bibinfo {year} {1956})}\BibitemShut
  {NoStop}%
\bibitem [{\citenamefont {Wagner}\ \emph {et~al.}(2006)\citenamefont {Wagner},
  \citenamefont {Tharmann}, \citenamefont {Haase}, \citenamefont {Fischer},\
  and\ \citenamefont {Bausch}}]{wagner2006cytoskeletal}%
  \BibitemOpen
  \bibfield  {author} {\bibinfo {author} {\bibfnamefont {B.}~\bibnamefont
  {Wagner}}, \bibinfo {author} {\bibfnamefont {R.}~\bibnamefont {Tharmann}},
  \bibinfo {author} {\bibfnamefont {I.}~\bibnamefont {Haase}}, \bibinfo
  {author} {\bibfnamefont {M.}~\bibnamefont {Fischer}},\ and\ \bibinfo {author}
  {\bibfnamefont {A.~R.}\ \bibnamefont {Bausch}},\ }\bibfield  {title}
  {\bibinfo {title} {Cytoskeletal polymer networks: the molecular structure of
  cross-linkers determines macroscopic properties},\ }\href@noop {} {\bibfield
  {journal} {\bibinfo  {journal} {Proc. Natl. Acad. Sci. USA}\ }\textbf
  {\bibinfo {volume} {103}},\ \bibinfo {pages} {13974} (\bibinfo {year}
  {2006})}\BibitemShut {NoStop}%
\bibitem [{\citenamefont {Gardel}\ \emph {et~al.}(2008)\citenamefont {Gardel},
  \citenamefont {Kasza}, \citenamefont {Brangwynne}, \citenamefont {Liu},\ and\
  \citenamefont {Weitz}}]{gardel2008mechanical}%
  \BibitemOpen
  \bibfield  {author} {\bibinfo {author} {\bibfnamefont {M.~L.}\ \bibnamefont
  {Gardel}}, \bibinfo {author} {\bibfnamefont {K.~E.}\ \bibnamefont {Kasza}},
  \bibinfo {author} {\bibfnamefont {C.~P.}\ \bibnamefont {Brangwynne}},
  \bibinfo {author} {\bibfnamefont {J.}~\bibnamefont {Liu}},\ and\ \bibinfo
  {author} {\bibfnamefont {D.~A.}\ \bibnamefont {Weitz}},\ }\bibfield  {title}
  {\bibinfo {title} {Mechanical response of cytoskeletal networks},\
  }\href@noop {} {\bibfield  {journal} {\bibinfo  {journal} {Methods Cell
  Biol.}\ }\textbf {\bibinfo {volume} {89}},\ \bibinfo {pages} {487} (\bibinfo
  {year} {2008})}\BibitemShut {NoStop}%
\bibitem [{\citenamefont {Pollard}\ and\ \citenamefont
  {Borisy}(2003)}]{pollard2003cellular}%
  \BibitemOpen
  \bibfield  {author} {\bibinfo {author} {\bibfnamefont {T.~D.}\ \bibnamefont
  {Pollard}}\ and\ \bibinfo {author} {\bibfnamefont {G.~G.}\ \bibnamefont
  {Borisy}},\ }\bibfield  {title} {\bibinfo {title} {Cellular motility driven
  by assembly and disassembly of actin filaments},\ }\href@noop {} {\bibfield
  {journal} {\bibinfo  {journal} {Cell}\ }\textbf {\bibinfo {volume} {112}},\
  \bibinfo {pages} {453} (\bibinfo {year} {2003})}\BibitemShut {NoStop}%
\bibitem [{\citenamefont {Marko}\ and\ \citenamefont
  {Siggia}(1995)}]{marko1995stretching}%
  \BibitemOpen
  \bibfield  {author} {\bibinfo {author} {\bibfnamefont {J.~F.}\ \bibnamefont
  {Marko}}\ and\ \bibinfo {author} {\bibfnamefont {E.~D.}\ \bibnamefont
  {Siggia}},\ }\bibfield  {title} {\bibinfo {title} {Stretching {DNA}},\
  }\href@noop {} {\bibfield  {journal} {\bibinfo  {journal} {Macromolecules}\
  }\textbf {\bibinfo {volume} {28}},\ \bibinfo {pages} {8759} (\bibinfo {year}
  {1995})}\BibitemShut {NoStop}%
\bibitem [{\citenamefont {Bustamante}\ \emph {et~al.}(1994)\citenamefont
  {Bustamante}, \citenamefont {Marko}, \citenamefont {Siggia},\ and\
  \citenamefont {Smith}}]{bustamante1994entropic}%
  \BibitemOpen
  \bibfield  {author} {\bibinfo {author} {\bibfnamefont {C.}~\bibnamefont
  {Bustamante}}, \bibinfo {author} {\bibfnamefont {J.~F.}\ \bibnamefont
  {Marko}}, \bibinfo {author} {\bibfnamefont {E.~D.}\ \bibnamefont {Siggia}},\
  and\ \bibinfo {author} {\bibfnamefont {S.}~\bibnamefont {Smith}},\ }\bibfield
   {title} {\bibinfo {title} {Entropic elasticity of $\lambda$-phage {DNA}},\
  }\href@noop {} {\bibfield  {journal} {\bibinfo  {journal} {Science}\ }\textbf
  {\bibinfo {volume} {265}},\ \bibinfo {pages} {1599} (\bibinfo {year}
  {1994})}\BibitemShut {NoStop}%
\bibitem [{\citenamefont {Fixman}\ and\ \citenamefont
  {Kovac}(1973)}]{fixman1973polymer}%
  \BibitemOpen
  \bibfield  {author} {\bibinfo {author} {\bibfnamefont {M.}~\bibnamefont
  {Fixman}}\ and\ \bibinfo {author} {\bibfnamefont {J.}~\bibnamefont {Kovac}},\
  }\bibfield  {title} {\bibinfo {title} {Polymer conformational statistics.
  {III}. {Modified Gaussian} models of stiff chains},\ }\href@noop {}
  {\bibfield  {journal} {\bibinfo  {journal} {J. Chem. Phys.}\ }\textbf
  {\bibinfo {volume} {58}},\ \bibinfo {pages} {1564} (\bibinfo {year}
  {1973})}\BibitemShut {NoStop}%
\bibitem [{\citenamefont {Lin}\ \emph {et~al.}(2010)\citenamefont {Lin},
  \citenamefont {Yao}, \citenamefont {Broedersz}, \citenamefont {Herrmann},
  \citenamefont {MacKintosh},\ and\ \citenamefont {Weitz}}]{lin2010origins}%
  \BibitemOpen
  \bibfield  {author} {\bibinfo {author} {\bibfnamefont {Y.-C.}\ \bibnamefont
  {Lin}}, \bibinfo {author} {\bibfnamefont {N.~Y.}\ \bibnamefont {Yao}},
  \bibinfo {author} {\bibfnamefont {C.~P.}\ \bibnamefont {Broedersz}}, \bibinfo
  {author} {\bibfnamefont {H.}~\bibnamefont {Herrmann}}, \bibinfo {author}
  {\bibfnamefont {F.~C.}\ \bibnamefont {MacKintosh}},\ and\ \bibinfo {author}
  {\bibfnamefont {D.~A.}\ \bibnamefont {Weitz}},\ }\bibfield  {title} {\bibinfo
  {title} {Origins of elasticity in intermediate filament networks},\
  }\href@noop {} {\bibfield  {journal} {\bibinfo  {journal} {Phys. Rev. Lett.}\
  }\textbf {\bibinfo {volume} {104}},\ \bibinfo {pages} {058101} (\bibinfo
  {year} {2010})}\BibitemShut {NoStop}%
\bibitem [{\citenamefont {Burla}\ \emph {et~al.}(2019)\citenamefont {Burla},
  \citenamefont {Mulla}, \citenamefont {Vos}, \citenamefont
  {Aufderhorst-Roberts},\ and\ \citenamefont
  {Koenderink}}]{burla2019mechanical}%
  \BibitemOpen
  \bibfield  {author} {\bibinfo {author} {\bibfnamefont {F.}~\bibnamefont
  {Burla}}, \bibinfo {author} {\bibfnamefont {Y.}~\bibnamefont {Mulla}},
  \bibinfo {author} {\bibfnamefont {B.~E.}\ \bibnamefont {Vos}}, \bibinfo
  {author} {\bibfnamefont {A.}~\bibnamefont {Aufderhorst-Roberts}},\ and\
  \bibinfo {author} {\bibfnamefont {G.~H.}\ \bibnamefont {Koenderink}},\
  }\bibfield  {title} {\bibinfo {title} {From mechanical resilience to active
  material properties in biopolymer networks},\ }\href@noop {} {\bibfield
  {journal} {\bibinfo  {journal} {Nat. Rev. Phys.}\ }\textbf {\bibinfo {volume}
  {1}},\ \bibinfo {pages} {249} (\bibinfo {year} {2019})}\BibitemShut {NoStop}%
\bibitem [{\citenamefont {Storm}\ \emph {et~al.}(2005)\citenamefont {Storm},
  \citenamefont {Pastore}, \citenamefont {MacKintosh}, \citenamefont
  {Lubensky},\ and\ \citenamefont {Janmey}}]{storm2005nonlinear}%
  \BibitemOpen
  \bibfield  {author} {\bibinfo {author} {\bibfnamefont {C.}~\bibnamefont
  {Storm}}, \bibinfo {author} {\bibfnamefont {J.~J.}\ \bibnamefont {Pastore}},
  \bibinfo {author} {\bibfnamefont {F.~C.}\ \bibnamefont {MacKintosh}},
  \bibinfo {author} {\bibfnamefont {T.~C.}\ \bibnamefont {Lubensky}},\ and\
  \bibinfo {author} {\bibfnamefont {P.~A.}\ \bibnamefont {Janmey}},\ }\bibfield
   {title} {\bibinfo {title} {Nonlinear elasticity in biological gels},\
  }\href@noop {} {\bibfield  {journal} {\bibinfo  {journal} {Nature}\ }\textbf
  {\bibinfo {volume} {435}},\ \bibinfo {pages} {191} (\bibinfo {year}
  {2005})}\BibitemShut {NoStop}%
\bibitem [{\citenamefont {Chaubet}\ \emph {et~al.}(2020)\citenamefont
  {Chaubet}, \citenamefont {Chaudhary}, \citenamefont {Heris}, \citenamefont
  {Ehrlicher},\ and\ \citenamefont {Hendricks}}]{chaubet2020dynamic}%
  \BibitemOpen
  \bibfield  {author} {\bibinfo {author} {\bibfnamefont {L.}~\bibnamefont
  {Chaubet}}, \bibinfo {author} {\bibfnamefont {A.~R.}\ \bibnamefont
  {Chaudhary}}, \bibinfo {author} {\bibfnamefont {H.~K.}\ \bibnamefont
  {Heris}}, \bibinfo {author} {\bibfnamefont {A.~J.}\ \bibnamefont
  {Ehrlicher}},\ and\ \bibinfo {author} {\bibfnamefont {A.~G.}\ \bibnamefont
  {Hendricks}},\ }\bibfield  {title} {\bibinfo {title} {Dynamic actin
  cross-linking governs the cytoplasm’s transition to fluid-like behavior},\
  }\href@noop {} {\bibfield  {journal} {\bibinfo  {journal} {Mol. Biol. Cell}\
  }\textbf {\bibinfo {volume} {31}},\ \bibinfo {pages} {1744} (\bibinfo {year}
  {2020})}\BibitemShut {NoStop}%
\bibitem [{\citenamefont {Lorenz}\ and\ \citenamefont
  {K{\"o}ster}(2022)}]{lorenz2022multiscale}%
  \BibitemOpen
  \bibfield  {author} {\bibinfo {author} {\bibfnamefont {C.}~\bibnamefont
  {Lorenz}}\ and\ \bibinfo {author} {\bibfnamefont {S.}~\bibnamefont
  {K{\"o}ster}},\ }\bibfield  {title} {\bibinfo {title} {Multiscale
  architecture: Mechanics of composite cytoskeletal networks},\ }\href@noop {}
  {\bibfield  {journal} {\bibinfo  {journal} {Biophys. Rev.}\ }\textbf
  {\bibinfo {volume} {3}} (\bibinfo {year} {2022})}\BibitemShut {NoStop}%
\bibitem [{\citenamefont {Chaudhuri}\ \emph {et~al.}(2007)\citenamefont
  {Chaudhuri}, \citenamefont {Parekh},\ and\ \citenamefont
  {Fletcher}}]{chaudhuri2007reversible}%
  \BibitemOpen
  \bibfield  {author} {\bibinfo {author} {\bibfnamefont {O.}~\bibnamefont
  {Chaudhuri}}, \bibinfo {author} {\bibfnamefont {S.~H.}\ \bibnamefont
  {Parekh}},\ and\ \bibinfo {author} {\bibfnamefont {D.~A.}\ \bibnamefont
  {Fletcher}},\ }\bibfield  {title} {\bibinfo {title} {Reversible stress
  softening of actin networks},\ }\href@noop {} {\bibfield  {journal} {\bibinfo
   {journal} {Nature}\ }\textbf {\bibinfo {volume} {445}},\ \bibinfo {pages}
  {295} (\bibinfo {year} {2007})}\BibitemShut {NoStop}%
\bibitem [{\citenamefont {Janmey}\ \emph {et~al.}(2007)\citenamefont {Janmey},
  \citenamefont {McCormick}, \citenamefont {Rammensee}, \citenamefont {Leight},
  \citenamefont {Georges},\ and\ \citenamefont
  {MacKintosh}}]{janmey2007negative}%
  \BibitemOpen
  \bibfield  {author} {\bibinfo {author} {\bibfnamefont {P.~A.}\ \bibnamefont
  {Janmey}}, \bibinfo {author} {\bibfnamefont {M.~E.}\ \bibnamefont
  {McCormick}}, \bibinfo {author} {\bibfnamefont {S.}~\bibnamefont
  {Rammensee}}, \bibinfo {author} {\bibfnamefont {J.~L.}\ \bibnamefont
  {Leight}}, \bibinfo {author} {\bibfnamefont {P.~C.}\ \bibnamefont
  {Georges}},\ and\ \bibinfo {author} {\bibfnamefont {F.~C.}\ \bibnamefont
  {MacKintosh}},\ }\bibfield  {title} {\bibinfo {title} {Negative normal stress
  in semiflexible biopolymer gels},\ }\href@noop {} {\bibfield  {journal}
  {\bibinfo  {journal} {Nature Mater.}\ }\textbf {\bibinfo {volume} {6}},\
  \bibinfo {pages} {48} (\bibinfo {year} {2007})}\BibitemShut {NoStop}%
\bibitem [{\citenamefont {MacKintosh}\ \emph {et~al.}(1995)\citenamefont
  {MacKintosh}, \citenamefont {K{\"a}s},\ and\ \citenamefont
  {Janmey}}]{mackintosh1995elasticity}%
  \BibitemOpen
  \bibfield  {author} {\bibinfo {author} {\bibfnamefont {F.~C.}\ \bibnamefont
  {MacKintosh}}, \bibinfo {author} {\bibfnamefont {J.}~\bibnamefont
  {K{\"a}s}},\ and\ \bibinfo {author} {\bibfnamefont {P.~A.}\ \bibnamefont
  {Janmey}},\ }\bibfield  {title} {\bibinfo {title} {Elasticity of semiflexible
  biopolymer networks},\ }\href@noop {} {\bibfield  {journal} {\bibinfo
  {journal} {Phys. Rev. Lett.}\ }\textbf {\bibinfo {volume} {75}},\ \bibinfo
  {pages} {4425} (\bibinfo {year} {1995})}\BibitemShut {NoStop}%
\bibitem [{\citenamefont {Sun}(2021)}]{sun2021mechanics}%
  \BibitemOpen
  \bibfield  {author} {\bibinfo {author} {\bibfnamefont {B.}~\bibnamefont
  {Sun}},\ }\bibfield  {title} {\bibinfo {title} {The mechanics of fibrillar
  collagen extracellular matrix},\ }\href@noop {} {\bibfield  {journal}
  {\bibinfo  {journal} {Cell Rep. Phys. Sci.}\ }\textbf {\bibinfo {volume} {2}}
  (\bibinfo {year} {2021})}\BibitemShut {NoStop}%
\bibitem [{\citenamefont {Sheu}\ \emph {et~al.}(2001)\citenamefont {Sheu},
  \citenamefont {Huang}, \citenamefont {Yeh},\ and\ \citenamefont
  {Ho}}]{sheu2001characterization}%
  \BibitemOpen
  \bibfield  {author} {\bibinfo {author} {\bibfnamefont {M.-T.}\ \bibnamefont
  {Sheu}}, \bibinfo {author} {\bibfnamefont {J.-C.}\ \bibnamefont {Huang}},
  \bibinfo {author} {\bibfnamefont {G.-C.}\ \bibnamefont {Yeh}},\ and\ \bibinfo
  {author} {\bibfnamefont {H.-O.}\ \bibnamefont {Ho}},\ }\bibfield  {title}
  {\bibinfo {title} {Characterization of collagen gel solutions and collagen
  matrices for cell culture},\ }\href@noop {} {\bibfield  {journal} {\bibinfo
  {journal} {Biomater.}\ }\textbf {\bibinfo {volume} {22}},\ \bibinfo {pages}
  {1713} (\bibinfo {year} {2001})}\BibitemShut {NoStop}%
\bibitem [{\citenamefont {Dutov}\ \emph {et~al.}(2016)\citenamefont {Dutov},
  \citenamefont {Antipova}, \citenamefont {Varma}, \citenamefont {Orgel},\ and\
  \citenamefont {Schieber}}]{dutov2016measurement}%
  \BibitemOpen
  \bibfield  {author} {\bibinfo {author} {\bibfnamefont {P.}~\bibnamefont
  {Dutov}}, \bibinfo {author} {\bibfnamefont {O.}~\bibnamefont {Antipova}},
  \bibinfo {author} {\bibfnamefont {S.}~\bibnamefont {Varma}}, \bibinfo
  {author} {\bibfnamefont {J.~P. R.~O.}\ \bibnamefont {Orgel}},\ and\ \bibinfo
  {author} {\bibfnamefont {J.~D.}\ \bibnamefont {Schieber}},\ }\bibfield
  {title} {\bibinfo {title} {Measurement of elastic modulus of collagen type i
  single fiber},\ }\href@noop {} {\bibfield  {journal} {\bibinfo  {journal}
  {PloS One}\ }\textbf {\bibinfo {volume} {11}},\ \bibinfo {pages} {e0145711}
  (\bibinfo {year} {2016})}\BibitemShut {NoStop}%
\bibitem [{\citenamefont {Eichhorn}\ and\ \citenamefont
  {Young}(2001)}]{eichhorn2001young}%
  \BibitemOpen
  \bibfield  {author} {\bibinfo {author} {\bibfnamefont {S.~J.}\ \bibnamefont
  {Eichhorn}}\ and\ \bibinfo {author} {\bibfnamefont {R.~J.}\ \bibnamefont
  {Young}},\ }\bibfield  {title} {\bibinfo {title} {The young's modulus of a
  microcrystalline cellulose},\ }\href@noop {} {\bibfield  {journal} {\bibinfo
  {journal} {Cellulose}\ }\textbf {\bibinfo {volume} {8}},\ \bibinfo {pages}
  {197} (\bibinfo {year} {2001})}\BibitemShut {NoStop}%
\bibitem [{\citenamefont {Hsieh}\ \emph {et~al.}(2008)\citenamefont {Hsieh},
  \citenamefont {Yano}, \citenamefont {Nogi},\ and\ \citenamefont
  {Eichhorn}}]{hsieh2008estimation}%
  \BibitemOpen
  \bibfield  {author} {\bibinfo {author} {\bibfnamefont {Y.-C.}\ \bibnamefont
  {Hsieh}}, \bibinfo {author} {\bibfnamefont {H.}~\bibnamefont {Yano}},
  \bibinfo {author} {\bibfnamefont {M.}~\bibnamefont {Nogi}},\ and\ \bibinfo
  {author} {\bibfnamefont {S.~J.}\ \bibnamefont {Eichhorn}},\ }\bibfield
  {title} {\bibinfo {title} {An estimation of the young’s modulus of
  bacterial cellulose filaments},\ }\href@noop {} {\bibfield  {journal}
  {\bibinfo  {journal} {Cellulose}\ }\textbf {\bibinfo {volume} {15}},\
  \bibinfo {pages} {507} (\bibinfo {year} {2008})}\BibitemShut {NoStop}%
\bibitem [{\citenamefont {Strzelecka}\ \emph {et~al.}(1988)\citenamefont
  {Strzelecka}, \citenamefont {Davidson},\ and\ \citenamefont
  {Rill}}]{strzelecka1988multiple}%
  \BibitemOpen
  \bibfield  {author} {\bibinfo {author} {\bibfnamefont {T.~E.}\ \bibnamefont
  {Strzelecka}}, \bibinfo {author} {\bibfnamefont {M.~W.}\ \bibnamefont
  {Davidson}},\ and\ \bibinfo {author} {\bibfnamefont {R.~L.}\ \bibnamefont
  {Rill}},\ }\bibfield  {title} {\bibinfo {title} {Multiple liquid crystal
  phases of {DNA} at high concentrations},\ }\href@noop {} {\bibfield
  {journal} {\bibinfo  {journal} {Nature}\ }\textbf {\bibinfo {volume} {331}},\
  \bibinfo {pages} {457} (\bibinfo {year} {1988})}\BibitemShut {NoStop}%
\bibitem [{\citenamefont {Ramsay}(1998)}]{ramsay1998dna}%
  \BibitemOpen
  \bibfield  {author} {\bibinfo {author} {\bibfnamefont {G.}~\bibnamefont
  {Ramsay}},\ }\bibfield  {title} {\bibinfo {title} {{DNA} chips: state-of-the
  art},\ }\href@noop {} {\bibfield  {journal} {\bibinfo  {journal} {Nat.
  Biotechnol.}\ }\textbf {\bibinfo {volume} {16}},\ \bibinfo {pages} {40}
  (\bibinfo {year} {1998})}\BibitemShut {NoStop}%
\bibitem [{\citenamefont {Seeman}(2010)}]{seeman2010nanomaterials}%
  \BibitemOpen
  \bibfield  {author} {\bibinfo {author} {\bibfnamefont {N.~C.}\ \bibnamefont
  {Seeman}},\ }\bibfield  {title} {\bibinfo {title} {Nanomaterials based on
  {DNA}},\ }\href@noop {} {\bibfield  {journal} {\bibinfo  {journal} {Annu.
  Rev. Biochem.}\ }\textbf {\bibinfo {volume} {79}},\ \bibinfo {pages} {65}
  (\bibinfo {year} {2010})}\BibitemShut {NoStop}%
\bibitem [{\citenamefont {Sacc{\`a}}\ and\ \citenamefont
  {Niemeyer}(2012)}]{sacca2012dna}%
  \BibitemOpen
  \bibfield  {author} {\bibinfo {author} {\bibfnamefont {B.}~\bibnamefont
  {Sacc{\`a}}}\ and\ \bibinfo {author} {\bibfnamefont {C.~M.}\ \bibnamefont
  {Niemeyer}},\ }\bibfield  {title} {\bibinfo {title} {{DNA} origami: the art
  of folding {DNA}},\ }\href@noop {} {\bibfield  {journal} {\bibinfo  {journal}
  {Angew. Chem. Int. Ed.}\ }\textbf {\bibinfo {volume} {51}},\ \bibinfo {pages}
  {58} (\bibinfo {year} {2012})}\BibitemShut {NoStop}%
\bibitem [{\citenamefont {Seeman}\ and\ \citenamefont
  {Sleiman}(2017)}]{seeman2017dna}%
  \BibitemOpen
  \bibfield  {author} {\bibinfo {author} {\bibfnamefont {N.~C.}\ \bibnamefont
  {Seeman}}\ and\ \bibinfo {author} {\bibfnamefont {H.~F.}\ \bibnamefont
  {Sleiman}},\ }\bibfield  {title} {\bibinfo {title} {{DNA} nanotechnology},\
  }\href@noop {} {\bibfield  {journal} {\bibinfo  {journal} {Nat. Rev. Mater.}\
  }\textbf {\bibinfo {volume} {3}},\ \bibinfo {pages} {1} (\bibinfo {year}
  {2017})}\BibitemShut {NoStop}%
\bibitem [{\citenamefont {Praetorius}\ \emph {et~al.}(2017)\citenamefont
  {Praetorius}, \citenamefont {Kick}, \citenamefont {Behler}, \citenamefont
  {Honemann}, \citenamefont {Weuster-Botz},\ and\ \citenamefont
  {Dietz}}]{praetorius2017biotechnological}%
  \BibitemOpen
  \bibfield  {author} {\bibinfo {author} {\bibfnamefont {F.}~\bibnamefont
  {Praetorius}}, \bibinfo {author} {\bibfnamefont {B.}~\bibnamefont {Kick}},
  \bibinfo {author} {\bibfnamefont {K.~L.}\ \bibnamefont {Behler}}, \bibinfo
  {author} {\bibfnamefont {M.~N.}\ \bibnamefont {Honemann}}, \bibinfo {author}
  {\bibfnamefont {D.}~\bibnamefont {Weuster-Botz}},\ and\ \bibinfo {author}
  {\bibfnamefont {H.}~\bibnamefont {Dietz}},\ }\bibfield  {title} {\bibinfo
  {title} {Biotechnological mass production of {DNA} origami},\ }\href@noop {}
  {\bibfield  {journal} {\bibinfo  {journal} {Nature}\ }\textbf {\bibinfo
  {volume} {552}},\ \bibinfo {pages} {84} (\bibinfo {year} {2017})}\BibitemShut
  {NoStop}%
\bibitem [{\citenamefont {Engelen}\ and\ \citenamefont
  {Dietz}(2021)}]{engelen2021advancing}%
  \BibitemOpen
  \bibfield  {author} {\bibinfo {author} {\bibfnamefont {W.}~\bibnamefont
  {Engelen}}\ and\ \bibinfo {author} {\bibfnamefont {H.}~\bibnamefont
  {Dietz}},\ }\bibfield  {title} {\bibinfo {title} {Advancing biophysics using
  {DNA} origami},\ }\href@noop {} {\bibfield  {journal} {\bibinfo  {journal}
  {Annu. Rev. Biophys.}\ }\textbf {\bibinfo {volume} {50}},\ \bibinfo {pages}
  {469} (\bibinfo {year} {2021})}\BibitemShut {NoStop}%
\bibitem [{\citenamefont {Kornyshev}\ \emph {et~al.}(2007)\citenamefont
  {Kornyshev}, \citenamefont {Lee}, \citenamefont {Leikin},\ and\ \citenamefont
  {Wynveen}}]{kornyshev2007structure}%
  \BibitemOpen
  \bibfield  {author} {\bibinfo {author} {\bibfnamefont {A.~A.}\ \bibnamefont
  {Kornyshev}}, \bibinfo {author} {\bibfnamefont {D.~J.}\ \bibnamefont {Lee}},
  \bibinfo {author} {\bibfnamefont {S.}~\bibnamefont {Leikin}},\ and\ \bibinfo
  {author} {\bibfnamefont {A.}~\bibnamefont {Wynveen}},\ }\bibfield  {title}
  {\bibinfo {title} {Structure and interactions of biological helices},\
  }\href@noop {} {\bibfield  {journal} {\bibinfo  {journal} {Rev. Mod. Phys.}\
  }\textbf {\bibinfo {volume} {79}},\ \bibinfo {pages} {943} (\bibinfo {year}
  {2007})}\BibitemShut {NoStop}%
\bibitem [{\citenamefont {Wong}\ and\ \citenamefont
  {Pollack}(2010)}]{wong2010electrostatics}%
  \BibitemOpen
  \bibfield  {author} {\bibinfo {author} {\bibfnamefont {G.~C.~L.}\
  \bibnamefont {Wong}}\ and\ \bibinfo {author} {\bibfnamefont {L.}~\bibnamefont
  {Pollack}},\ }\bibfield  {title} {\bibinfo {title} {Electrostatics of
  strongly charged biological polymers: ion-mediated interactions and
  self-organization in nucleic acids and proteins},\ }\href@noop {} {\bibfield
  {journal} {\bibinfo  {journal} {Annu. Rev. Phys. Chem.}\ }\textbf {\bibinfo
  {volume} {61}},\ \bibinfo {pages} {171} (\bibinfo {year} {2010})}\BibitemShut
  {NoStop}%
\bibitem [{\citenamefont {Teif}\ and\ \citenamefont
  {Bohinc}(2011)}]{teif2011condensed}%
  \BibitemOpen
  \bibfield  {author} {\bibinfo {author} {\bibfnamefont {V.~B.}\ \bibnamefont
  {Teif}}\ and\ \bibinfo {author} {\bibfnamefont {K.}~\bibnamefont {Bohinc}},\
  }\bibfield  {title} {\bibinfo {title} {Condensed {DNA}: condensing the
  concepts},\ }\href@noop {} {\bibfield  {journal} {\bibinfo  {journal} {Prog.
  Biophys. Mol. Biol.}\ }\textbf {\bibinfo {volume} {105}},\ \bibinfo {pages}
  {208} (\bibinfo {year} {2011})}\BibitemShut {NoStop}%
\bibitem [{\citenamefont {Travers}\ and\ \citenamefont
  {Muskhelishvili}(2015)}]{travers2015dna}%
  \BibitemOpen
  \bibfield  {author} {\bibinfo {author} {\bibfnamefont {A.}~\bibnamefont
  {Travers}}\ and\ \bibinfo {author} {\bibfnamefont {G.}~\bibnamefont
  {Muskhelishvili}},\ }\bibfield  {title} {\bibinfo {title} {{DNA} structure
  and function},\ }\href@noop {} {\bibfield  {journal} {\bibinfo  {journal}
  {FEBS J.}\ }\textbf {\bibinfo {volume} {282}},\ \bibinfo {pages} {2279}
  (\bibinfo {year} {2015})}\BibitemShut {NoStop}%
\bibitem [{\citenamefont {Michieletto}\ and\ \citenamefont
  {Gilbert}(2019)}]{michieletto2019role}%
  \BibitemOpen
  \bibfield  {author} {\bibinfo {author} {\bibfnamefont {D.}~\bibnamefont
  {Michieletto}}\ and\ \bibinfo {author} {\bibfnamefont {N.}~\bibnamefont
  {Gilbert}},\ }\bibfield  {title} {\bibinfo {title} {Role of nuclear {RNA} in
  regulating chromatin structure and transcription},\ }\href@noop {} {\bibfield
   {journal} {\bibinfo  {journal} {Curr. Opin. Cell Biol.}\ }\textbf {\bibinfo
  {volume} {58}},\ \bibinfo {pages} {120} (\bibinfo {year} {2019})}\BibitemShut
  {NoStop}%
\bibitem [{\citenamefont {Chevizovich}\ \emph {et~al.}(2020)\citenamefont
  {Chevizovich}, \citenamefont {Michieletto}, \citenamefont {Mvogo},
  \citenamefont {Zakiryanov},\ and\ \citenamefont
  {Zdravkovi{\'c}}}]{chevizovich2020review}%
  \BibitemOpen
  \bibfield  {author} {\bibinfo {author} {\bibfnamefont {D.}~\bibnamefont
  {Chevizovich}}, \bibinfo {author} {\bibfnamefont {D.}~\bibnamefont
  {Michieletto}}, \bibinfo {author} {\bibfnamefont {A.}~\bibnamefont {Mvogo}},
  \bibinfo {author} {\bibfnamefont {F.}~\bibnamefont {Zakiryanov}},\ and\
  \bibinfo {author} {\bibfnamefont {S.}~\bibnamefont {Zdravkovi{\'c}}},\
  }\bibfield  {title} {\bibinfo {title} {A review on nonlinear {DNA} physics},\
  }\href@noop {} {\bibfield  {journal} {\bibinfo  {journal} {R. Soc. Open
  Sci.}\ }\textbf {\bibinfo {volume} {7}},\ \bibinfo {pages} {200774} (\bibinfo
  {year} {2020})}\BibitemShut {NoStop}%
\bibitem [{\citenamefont {Michieletto}(2024)}]{michieletto2024kinetoplast}%
  \BibitemOpen
  \bibfield  {author} {\bibinfo {author} {\bibfnamefont {D.}~\bibnamefont
  {Michieletto}},\ }\bibfield  {title} {\bibinfo {title} {Kinetoplast {DNA}: a
  polymer physicist’s topological {O}lympic dream},\ }\href@noop {}
  {\bibfield  {journal} {\bibinfo  {journal} {Nucleic Acids Res.}\ ,\ \bibinfo
  {pages} {gkae1206}} (\bibinfo {year} {2024})}\BibitemShut {NoStop}%
\bibitem [{\citenamefont {Gao}\ \emph {et~al.}(2015)\citenamefont {Gao},
  \citenamefont {Blackwell}, \citenamefont {Glaser}, \citenamefont
  {Betterton},\ and\ \citenamefont {Shelley}}]{gao2015multiscale}%
  \BibitemOpen
  \bibfield  {author} {\bibinfo {author} {\bibfnamefont {T.}~\bibnamefont
  {Gao}}, \bibinfo {author} {\bibfnamefont {R.}~\bibnamefont {Blackwell}},
  \bibinfo {author} {\bibfnamefont {M.~A.}\ \bibnamefont {Glaser}}, \bibinfo
  {author} {\bibfnamefont {M.~D.}\ \bibnamefont {Betterton}},\ and\ \bibinfo
  {author} {\bibfnamefont {M.~J.}\ \bibnamefont {Shelley}},\ }\bibfield
  {title} {\bibinfo {title} {Multiscale polar theory of microtubule and
  motor-protein assemblies},\ }\href@noop {} {\bibfield  {journal} {\bibinfo
  {journal} {Phys. Rev. Lett.}\ }\textbf {\bibinfo {volume} {114}},\ \bibinfo
  {pages} {048101} (\bibinfo {year} {2015})}\BibitemShut {NoStop}%
\bibitem [{\citenamefont {Rincon}\ \emph {et~al.}(2017)\citenamefont {Rincon},
  \citenamefont {Lamson}, \citenamefont {Blackwell}, \citenamefont
  {Syrovatkina}, \citenamefont {Fraisier}, \citenamefont {Paoletti},
  \citenamefont {Betterton},\ and\ \citenamefont {Tran}}]{rincon2017kinesin}%
  \BibitemOpen
  \bibfield  {author} {\bibinfo {author} {\bibfnamefont {S.~A.}\ \bibnamefont
  {Rincon}}, \bibinfo {author} {\bibfnamefont {A.}~\bibnamefont {Lamson}},
  \bibinfo {author} {\bibfnamefont {R.}~\bibnamefont {Blackwell}}, \bibinfo
  {author} {\bibfnamefont {V.}~\bibnamefont {Syrovatkina}}, \bibinfo {author}
  {\bibfnamefont {V.}~\bibnamefont {Fraisier}}, \bibinfo {author}
  {\bibfnamefont {A.}~\bibnamefont {Paoletti}}, \bibinfo {author}
  {\bibfnamefont {M.~D.}\ \bibnamefont {Betterton}},\ and\ \bibinfo {author}
  {\bibfnamefont {P.~T.}\ \bibnamefont {Tran}},\ }\bibfield  {title} {\bibinfo
  {title} {Kinesin-5-independent mitotic spindle assembly requires the
  antiparallel microtubule crosslinker ase1 in fission yeast},\ }\href@noop {}
  {\bibfield  {journal} {\bibinfo  {journal} {Nat. Commun.}\ }\textbf {\bibinfo
  {volume} {8}},\ \bibinfo {pages} {15286} (\bibinfo {year}
  {2017})}\BibitemShut {NoStop}%
\bibitem [{\citenamefont {Bun}\ \emph {et~al.}(2018)\citenamefont {Bun},
  \citenamefont {Dmitrieff}, \citenamefont {Belmonte}, \citenamefont
  {N{\'e}d{\'e}lec},\ and\ \citenamefont {Lenart}}]{bun2018disassembly}%
  \BibitemOpen
  \bibfield  {author} {\bibinfo {author} {\bibfnamefont {P.}~\bibnamefont
  {Bun}}, \bibinfo {author} {\bibfnamefont {S.}~\bibnamefont {Dmitrieff}},
  \bibinfo {author} {\bibfnamefont {J.~M.}\ \bibnamefont {Belmonte}}, \bibinfo
  {author} {\bibfnamefont {F.~J.}\ \bibnamefont {N{\'e}d{\'e}lec}},\ and\
  \bibinfo {author} {\bibfnamefont {P.}~\bibnamefont {Lenart}},\ }\bibfield
  {title} {\bibinfo {title} {A disassembly-driven mechanism explains
  f-actin-mediated chromosome transport in starfish oocytes},\ }\href@noop {}
  {\bibfield  {journal} {\bibinfo  {journal} {{eL}ife}\ }\textbf {\bibinfo
  {volume} {7}},\ \bibinfo {pages} {e31469} (\bibinfo {year}
  {2018})}\BibitemShut {NoStop}%
\bibitem [{\citenamefont {Nedelec}\ and\ \citenamefont
  {Foethke}(2007)}]{nedelec2007collective}%
  \BibitemOpen
  \bibfield  {author} {\bibinfo {author} {\bibfnamefont {F.}~\bibnamefont
  {Nedelec}}\ and\ \bibinfo {author} {\bibfnamefont {D.}~\bibnamefont
  {Foethke}},\ }\bibfield  {title} {\bibinfo {title} {Collective langevin
  dynamics of flexible cytoskeletal fibers},\ }\href@noop {} {\bibfield
  {journal} {\bibinfo  {journal} {New J. Phys.}\ }\textbf {\bibinfo {volume}
  {9}},\ \bibinfo {pages} {427} (\bibinfo {year} {2007})}\BibitemShut {NoStop}%
\bibitem [{\citenamefont {Popov}\ \emph {et~al.}(2016)\citenamefont {Popov},
  \citenamefont {Komianos},\ and\ \citenamefont {Papoian}}]{popov2016medyan}%
  \BibitemOpen
  \bibfield  {author} {\bibinfo {author} {\bibfnamefont {K.}~\bibnamefont
  {Popov}}, \bibinfo {author} {\bibfnamefont {J.}~\bibnamefont {Komianos}},\
  and\ \bibinfo {author} {\bibfnamefont {G.~A.}\ \bibnamefont {Papoian}},\
  }\bibfield  {title} {\bibinfo {title} {Medyan: mechanochemical simulations of
  contraction and polarity alignment in actomyosin networks},\ }\href@noop {}
  {\bibfield  {journal} {\bibinfo  {journal} {{PLoS} Comput. Biol.}\ }\textbf
  {\bibinfo {volume} {12}},\ \bibinfo {pages} {e1004877} (\bibinfo {year}
  {2016})}\BibitemShut {NoStop}%
\bibitem [{\citenamefont {Freedman}\ \emph {et~al.}(2017)\citenamefont
  {Freedman}, \citenamefont {Banerjee}, \citenamefont {Hocky},\ and\
  \citenamefont {Dinner}}]{freedman2017versatile}%
  \BibitemOpen
  \bibfield  {author} {\bibinfo {author} {\bibfnamefont {S.~L.}\ \bibnamefont
  {Freedman}}, \bibinfo {author} {\bibfnamefont {S.}~\bibnamefont {Banerjee}},
  \bibinfo {author} {\bibfnamefont {G.~M.}\ \bibnamefont {Hocky}},\ and\
  \bibinfo {author} {\bibfnamefont {A.~R.}\ \bibnamefont {Dinner}},\ }\bibfield
   {title} {\bibinfo {title} {A versatile framework for simulating the dynamic
  mechanical structure of cytoskeletal networks},\ }\href@noop {} {\bibfield
  {journal} {\bibinfo  {journal} {Biophys. J.}\ }\textbf {\bibinfo {volume}
  {113}},\ \bibinfo {pages} {448} (\bibinfo {year} {2017})}\BibitemShut
  {NoStop}%
\bibitem [{\citenamefont {Fiorenza}\ \emph {et~al.}(2021)\citenamefont
  {Fiorenza}, \citenamefont {Steckhahn},\ and\ \citenamefont
  {Betterton}}]{fiorenza2021modeling}%
  \BibitemOpen
  \bibfield  {author} {\bibinfo {author} {\bibfnamefont {S.~A.}\ \bibnamefont
  {Fiorenza}}, \bibinfo {author} {\bibfnamefont {D.~G.}\ \bibnamefont
  {Steckhahn}},\ and\ \bibinfo {author} {\bibfnamefont {M.~D.}\ \bibnamefont
  {Betterton}},\ }\bibfield  {title} {\bibinfo {title} {Modeling
  spatiotemporally varying protein--protein interactions in cylaks, the
  cytoskeleton lattice-based kinetic simulator},\ }\href@noop {} {\bibfield
  {journal} {\bibinfo  {journal} {Eur. Phys. J. E}\ }\textbf {\bibinfo {volume}
  {44}},\ \bibinfo {pages} {105} (\bibinfo {year} {2021})}\BibitemShut
  {NoStop}%
\bibitem [{\citenamefont {Yan}\ \emph {et~al.}(2022)\citenamefont {Yan},
  \citenamefont {Ansari}, \citenamefont {Lamson}, \citenamefont {Glaser},
  \citenamefont {Blackwell}, \citenamefont {Betterton},\ and\ \citenamefont
  {Shelley}}]{yan2022toward}%
  \BibitemOpen
  \bibfield  {author} {\bibinfo {author} {\bibfnamefont {W.}~\bibnamefont
  {Yan}}, \bibinfo {author} {\bibfnamefont {S.}~\bibnamefont {Ansari}},
  \bibinfo {author} {\bibfnamefont {A.}~\bibnamefont {Lamson}}, \bibinfo
  {author} {\bibfnamefont {M.~A.}\ \bibnamefont {Glaser}}, \bibinfo {author}
  {\bibfnamefont {R.}~\bibnamefont {Blackwell}}, \bibinfo {author}
  {\bibfnamefont {M.~D.}\ \bibnamefont {Betterton}},\ and\ \bibinfo {author}
  {\bibfnamefont {M.}~\bibnamefont {Shelley}},\ }\bibfield  {title} {\bibinfo
  {title} {Toward the cellular-scale simulation of motor-driven cytoskeletal
  assemblies},\ }\href@noop {} {\bibfield  {journal} {\bibinfo  {journal}
  {{eL}ife}\ }\textbf {\bibinfo {volume} {11}},\ \bibinfo {pages} {e74160}
  (\bibinfo {year} {2022})}\BibitemShut {NoStop}%
\bibitem [{\citenamefont {Isele-Holder}\ \emph {et~al.}(2015)\citenamefont
  {Isele-Holder}, \citenamefont {Elgeti},\ and\ \citenamefont
  {Gompper}}]{isele2015self}%
  \BibitemOpen
  \bibfield  {author} {\bibinfo {author} {\bibfnamefont {R.~E.}\ \bibnamefont
  {Isele-Holder}}, \bibinfo {author} {\bibfnamefont {J.}~\bibnamefont
  {Elgeti}},\ and\ \bibinfo {author} {\bibfnamefont {G.}~\bibnamefont
  {Gompper}},\ }\bibfield  {title} {\bibinfo {title} {Self-propelled worm-like
  filaments: spontaneous spiral formation, structure, and dynamics},\
  }\href@noop {} {\bibfield  {journal} {\bibinfo  {journal} {Soft matter}\
  }\textbf {\bibinfo {volume} {11}},\ \bibinfo {pages} {7181} (\bibinfo {year}
  {2015})}\BibitemShut {NoStop}%
\bibitem [{\citenamefont {Winkler}\ \emph {et~al.}(2017)\citenamefont
  {Winkler}, \citenamefont {Elgeti},\ and\ \citenamefont
  {Gompper}}]{winkler2017active}%
  \BibitemOpen
  \bibfield  {author} {\bibinfo {author} {\bibfnamefont {R.~G.}\ \bibnamefont
  {Winkler}}, \bibinfo {author} {\bibfnamefont {J.}~\bibnamefont {Elgeti}},\
  and\ \bibinfo {author} {\bibfnamefont {G.}~\bibnamefont {Gompper}},\
  }\bibfield  {title} {\bibinfo {title} {Active polymers—emergent
  conformational and dynamical properties: A brief review},\ }\href@noop {}
  {\bibfield  {journal} {\bibinfo  {journal} {J. Phys. Soc. Jpn.}\ }\textbf
  {\bibinfo {volume} {86}},\ \bibinfo {pages} {101014} (\bibinfo {year}
  {2017})}\BibitemShut {NoStop}%
\bibitem [{\citenamefont {Philipps}\ \emph {et~al.}(2022)\citenamefont
  {Philipps}, \citenamefont {Gompper},\ and\ \citenamefont
  {Winkler}}]{philipps2022tangentially}%
  \BibitemOpen
  \bibfield  {author} {\bibinfo {author} {\bibfnamefont {C.~A.}\ \bibnamefont
  {Philipps}}, \bibinfo {author} {\bibfnamefont {G.}~\bibnamefont {Gompper}},\
  and\ \bibinfo {author} {\bibfnamefont {R.~G.}\ \bibnamefont {Winkler}},\
  }\bibfield  {title} {\bibinfo {title} {Tangentially driven active polar
  linear polymers—an analytical study},\ }\href@noop {} {\bibfield  {journal}
  {\bibinfo  {journal} {The Journal of Chemical Physics}\ }\textbf {\bibinfo
  {volume} {157}} (\bibinfo {year} {2022})}\BibitemShut {NoStop}%
\bibitem [{\citenamefont {Sinaasappel}\ \emph {et~al.}(2025)\citenamefont
  {Sinaasappel}, \citenamefont {Fazelzadeh}, \citenamefont {Hooijschuur},
  \citenamefont {Di}, \citenamefont {Jabbari-Farouji},\ and\ \citenamefont
  {Deblais}}]{sinaasappel2024locomotion}%
  \BibitemOpen
  \bibfield  {author} {\bibinfo {author} {\bibfnamefont {R.}~\bibnamefont
  {Sinaasappel}}, \bibinfo {author} {\bibfnamefont {M.}~\bibnamefont
  {Fazelzadeh}}, \bibinfo {author} {\bibfnamefont {T.}~\bibnamefont
  {Hooijschuur}}, \bibinfo {author} {\bibfnamefont {Q.}~\bibnamefont {Di}},
  \bibinfo {author} {\bibfnamefont {S.}~\bibnamefont {Jabbari-Farouji}},\ and\
  \bibinfo {author} {\bibfnamefont {A.}~\bibnamefont {Deblais}},\ }\bibfield
  {title} {\bibinfo {title} {Locomotion of active polymerlike worms in porous
  media},\ }\href {https://doi.org/10.1103/PhysRevLett.134.128303} {\bibfield
  {journal} {\bibinfo  {journal} {Phys. Rev. Lett.}\ }\textbf {\bibinfo
  {volume} {134}},\ \bibinfo {pages} {128303} (\bibinfo {year}
  {2025})}\BibitemShut {NoStop}%
\bibitem [{\citenamefont {Zhu}\ \emph {et~al.}(2024)\citenamefont {Zhu},
  \citenamefont {Gao}, \citenamefont {Sun}, \citenamefont {Wei},\ and\
  \citenamefont {Yan}}]{zhu2024non}%
  \BibitemOpen
  \bibfield  {author} {\bibinfo {author} {\bibfnamefont {G.}~\bibnamefont
  {Zhu}}, \bibinfo {author} {\bibfnamefont {L.}~\bibnamefont {Gao}}, \bibinfo
  {author} {\bibfnamefont {Y.}~\bibnamefont {Sun}}, \bibinfo {author}
  {\bibfnamefont {W.}~\bibnamefont {Wei}},\ and\ \bibinfo {author}
  {\bibfnamefont {L.-T.}\ \bibnamefont {Yan}},\ }\bibfield  {title} {\bibinfo
  {title} {Non-equilibrium structural and dynamic behaviors of active polymers
  in complex and crowded environments},\ }\href@noop {} {\bibfield  {journal}
  {\bibinfo  {journal} {Rep. Prog. Phys.}\ }\textbf {\bibinfo {volume} {87}},\
  \bibinfo {pages} {054601} (\bibinfo {year} {2024})}\BibitemShut {NoStop}%
\bibitem [{\citenamefont {Mokhtari}\ and\ \citenamefont
  {Zippelius}(2019)}]{mokhtari2019dynamics}%
  \BibitemOpen
  \bibfield  {author} {\bibinfo {author} {\bibfnamefont {Z.}~\bibnamefont
  {Mokhtari}}\ and\ \bibinfo {author} {\bibfnamefont {A.}~\bibnamefont
  {Zippelius}},\ }\bibfield  {title} {\bibinfo {title} {Dynamics of active
  filaments in porous media},\ }\href@noop {} {\bibfield  {journal} {\bibinfo
  {journal} {Phys. Rev. Lett.}\ }\textbf {\bibinfo {volume} {123}},\ \bibinfo
  {pages} {028001} (\bibinfo {year} {2019})}\BibitemShut {NoStop}%
\bibitem [{\citenamefont {Theeyancheri}\ \emph {et~al.}(2023)\citenamefont
  {Theeyancheri}, \citenamefont {Chaki}, \citenamefont {Bhattacharjee},\ and\
  \citenamefont {Chakrabarti}}]{theeyancheri2023active}%
  \BibitemOpen
  \bibfield  {author} {\bibinfo {author} {\bibfnamefont {L.}~\bibnamefont
  {Theeyancheri}}, \bibinfo {author} {\bibfnamefont {S.}~\bibnamefont {Chaki}},
  \bibinfo {author} {\bibfnamefont {T.}~\bibnamefont {Bhattacharjee}},\ and\
  \bibinfo {author} {\bibfnamefont {R.}~\bibnamefont {Chakrabarti}},\
  }\bibfield  {title} {\bibinfo {title} {Active dynamics of linear chains and
  rings in porous media},\ }\href@noop {} {\bibfield  {journal} {\bibinfo
  {journal} {J. Chem. Phys.}\ }\textbf {\bibinfo {volume} {159}} (\bibinfo
  {year} {2023})}\BibitemShut {NoStop}%
\bibitem [{\citenamefont {Weeks}\ \emph {et~al.}(1971)\citenamefont {Weeks},
  \citenamefont {Chandler},\ and\ \citenamefont {Andersen}}]{weeks1971role}%
  \BibitemOpen
  \bibfield  {author} {\bibinfo {author} {\bibfnamefont {J.~D.}\ \bibnamefont
  {Weeks}}, \bibinfo {author} {\bibfnamefont {D.}~\bibnamefont {Chandler}},\
  and\ \bibinfo {author} {\bibfnamefont {H.~C.}\ \bibnamefont {Andersen}},\
  }\bibfield  {title} {\bibinfo {title} {Role of repulsive forces in
  determining the equilibrium structure of simple liquids},\ }\href@noop {}
  {\bibfield  {journal} {\bibinfo  {journal} {J. Chem. Phys.}\ }\textbf
  {\bibinfo {volume} {54}},\ \bibinfo {pages} {5237} (\bibinfo {year}
  {1971})}\BibitemShut {NoStop}%
\bibitem [{\citenamefont {Chelakkot}\ \emph {et~al.}(2014)\citenamefont
  {Chelakkot}, \citenamefont {Gopinath}, \citenamefont {Mahadevan},\ and\
  \citenamefont {Hagan}}]{chelakkot2014flagellar}%
  \BibitemOpen
  \bibfield  {author} {\bibinfo {author} {\bibfnamefont {R.}~\bibnamefont
  {Chelakkot}}, \bibinfo {author} {\bibfnamefont {A.}~\bibnamefont {Gopinath}},
  \bibinfo {author} {\bibfnamefont {L.}~\bibnamefont {Mahadevan}},\ and\
  \bibinfo {author} {\bibfnamefont {M.~F.}\ \bibnamefont {Hagan}},\ }\bibfield
  {title} {\bibinfo {title} {Flagellar dynamics of a connected chain of active,
  polar, brownian particles},\ }\href@noop {} {\bibfield  {journal} {\bibinfo
  {journal} {J. R. Soc. Interface}\ }\textbf {\bibinfo {volume} {11}},\
  \bibinfo {pages} {20130884} (\bibinfo {year} {2014})}\BibitemShut {NoStop}%
\bibitem [{\citenamefont {Kurjahn}\ \emph
  {et~al.}(2024{\natexlab{b}})\citenamefont {Kurjahn}, \citenamefont
  {Abbaspour}, \citenamefont {Papenfu{\ss}}, \citenamefont {Bittihn},
  \citenamefont {Golestanian}, \citenamefont {Mahault},\ and\ \citenamefont
  {Karpitschka}}]{kurjahn2024collective}%
  \BibitemOpen
  \bibfield  {author} {\bibinfo {author} {\bibfnamefont {M.}~\bibnamefont
  {Kurjahn}}, \bibinfo {author} {\bibfnamefont {L.}~\bibnamefont {Abbaspour}},
  \bibinfo {author} {\bibfnamefont {F.}~\bibnamefont {Papenfu{\ss}}}, \bibinfo
  {author} {\bibfnamefont {P.}~\bibnamefont {Bittihn}}, \bibinfo {author}
  {\bibfnamefont {R.}~\bibnamefont {Golestanian}}, \bibinfo {author}
  {\bibfnamefont {B.}~\bibnamefont {Mahault}},\ and\ \bibinfo {author}
  {\bibfnamefont {S.}~\bibnamefont {Karpitschka}},\ }\bibfield  {title}
  {\bibinfo {title} {Collective self-caging of active filaments in virtual
  confinement},\ }\href@noop {} {\bibfield  {journal} {\bibinfo  {journal}
  {Nat. Commun.}\ }\textbf {\bibinfo {volume} {15}},\ \bibinfo {pages} {9122}
  (\bibinfo {year} {2024}{\natexlab{b}})}\BibitemShut {NoStop}%
\bibitem [{\citenamefont {Nguyen}\ \emph {et~al.}(2021)\citenamefont {Nguyen},
  \citenamefont {Ozkan-Aydin}, \citenamefont {Tuazon}, \citenamefont {Goldman},
  \citenamefont {Bhamla},\ and\ \citenamefont {Peleg}}]{nguyen2021emergent}%
  \BibitemOpen
  \bibfield  {author} {\bibinfo {author} {\bibfnamefont {C.}~\bibnamefont
  {Nguyen}}, \bibinfo {author} {\bibfnamefont {Y.}~\bibnamefont {Ozkan-Aydin}},
  \bibinfo {author} {\bibfnamefont {H.}~\bibnamefont {Tuazon}}, \bibinfo
  {author} {\bibfnamefont {D.~I.}\ \bibnamefont {Goldman}}, \bibinfo {author}
  {\bibfnamefont {M.~S.}\ \bibnamefont {Bhamla}},\ and\ \bibinfo {author}
  {\bibfnamefont {O.}~\bibnamefont {Peleg}},\ }\bibfield  {title} {\bibinfo
  {title} {Emergent collective locomotion in an active polymer model of
  entangled worm blobs},\ }\href@noop {} {\bibfield  {journal} {\bibinfo
  {journal} {Front. Phys.}\ }\textbf {\bibinfo {volume} {9}},\ \bibinfo {pages}
  {734499} (\bibinfo {year} {2021})}\BibitemShut {NoStop}%
\bibitem [{\citenamefont {N\'ed\'elec}\ \emph {et~al.}(1997)\citenamefont
  {N\'ed\'elec}, \citenamefont {Surrey}, \citenamefont {Maggs},\ and\
  \citenamefont {Leibler}}]{ndlec1997self}%
  \BibitemOpen
  \bibfield  {author} {\bibinfo {author} {\bibfnamefont {F.~J.}\ \bibnamefont
  {N\'ed\'elec}}, \bibinfo {author} {\bibfnamefont {T.}~\bibnamefont {Surrey}},
  \bibinfo {author} {\bibfnamefont {A.~C.}\ \bibnamefont {Maggs}},\ and\
  \bibinfo {author} {\bibfnamefont {S.}~\bibnamefont {Leibler}},\ }\bibfield
  {title} {\bibinfo {title} {Self-organization of microtubules and motors},\
  }\href@noop {} {\bibfield  {journal} {\bibinfo  {journal} {Nature}\ }\textbf
  {\bibinfo {volume} {389}},\ \bibinfo {pages} {305} (\bibinfo {year}
  {1997})}\BibitemShut {NoStop}%
\bibitem [{\citenamefont {Sanchez}\ \emph {et~al.}(2012)\citenamefont
  {Sanchez}, \citenamefont {Chen}, \citenamefont {DeCamp}, \citenamefont
  {Heymann},\ and\ \citenamefont {Dogic}}]{sanchez2012spontaneous}%
  \BibitemOpen
  \bibfield  {author} {\bibinfo {author} {\bibfnamefont {T.}~\bibnamefont
  {Sanchez}}, \bibinfo {author} {\bibfnamefont {D.~T.~N.}\ \bibnamefont
  {Chen}}, \bibinfo {author} {\bibfnamefont {S.~J.}\ \bibnamefont {DeCamp}},
  \bibinfo {author} {\bibfnamefont {M.}~\bibnamefont {Heymann}},\ and\ \bibinfo
  {author} {\bibfnamefont {Z.}~\bibnamefont {Dogic}},\ }\bibfield  {title}
  {\bibinfo {title} {Spontaneous motion in hierarchically assembled active
  matter},\ }\href@noop {} {\bibfield  {journal} {\bibinfo  {journal} {Nature}\
  }\textbf {\bibinfo {volume} {491}},\ \bibinfo {pages} {431} (\bibinfo {year}
  {2012})}\BibitemShut {NoStop}%
\bibitem [{\citenamefont {Doostmohammadi}\ \emph {et~al.}(2018)\citenamefont
  {Doostmohammadi}, \citenamefont {Ign{\'e}s-Mullol}, \citenamefont {Yeomans},\
  and\ \citenamefont {Sagu{\'e}s}}]{doostmohammadi2018active}%
  \BibitemOpen
  \bibfield  {author} {\bibinfo {author} {\bibfnamefont {A.}~\bibnamefont
  {Doostmohammadi}}, \bibinfo {author} {\bibfnamefont {J.}~\bibnamefont
  {Ign{\'e}s-Mullol}}, \bibinfo {author} {\bibfnamefont {J.~M.}\ \bibnamefont
  {Yeomans}},\ and\ \bibinfo {author} {\bibfnamefont {F.}~\bibnamefont
  {Sagu{\'e}s}},\ }\bibfield  {title} {\bibinfo {title} {Active nematics},\
  }\href@noop {} {\bibfield  {journal} {\bibinfo  {journal} {Nat. Commun.}\
  }\textbf {\bibinfo {volume} {9}},\ \bibinfo {pages} {3246} (\bibinfo {year}
  {2018})}\BibitemShut {NoStop}%
\bibitem [{\citenamefont {Marenduzzo}\ \emph {et~al.}(2007)\citenamefont
  {Marenduzzo}, \citenamefont {Orlandini},\ and\ \citenamefont
  {Yeomans}}]{marenduzzo2007hydrodynamics}%
  \BibitemOpen
  \bibfield  {author} {\bibinfo {author} {\bibfnamefont {D.}~\bibnamefont
  {Marenduzzo}}, \bibinfo {author} {\bibfnamefont {E.}~\bibnamefont
  {Orlandini}},\ and\ \bibinfo {author} {\bibfnamefont {J.~M.}\ \bibnamefont
  {Yeomans}},\ }\bibfield  {title} {\bibinfo {title} {Hydrodynamics and
  rheology of active liquid crystals: a numerical investigation},\ }\href@noop
  {} {\bibfield  {journal} {\bibinfo  {journal} {Phys. Rev. Lett.}\ }\textbf
  {\bibinfo {volume} {98}},\ \bibinfo {pages} {118102} (\bibinfo {year}
  {2007})}\BibitemShut {NoStop}%
\bibitem [{\citenamefont {Giomi}\ \emph {et~al.}(2013)\citenamefont {Giomi},
  \citenamefont {Bowick}, \citenamefont {Ma},\ and\ \citenamefont
  {Marchetti}}]{giomi2013defect}%
  \BibitemOpen
  \bibfield  {author} {\bibinfo {author} {\bibfnamefont {L.}~\bibnamefont
  {Giomi}}, \bibinfo {author} {\bibfnamefont {M.~J.}\ \bibnamefont {Bowick}},
  \bibinfo {author} {\bibfnamefont {X.}~\bibnamefont {Ma}},\ and\ \bibinfo
  {author} {\bibfnamefont {M.~C.}\ \bibnamefont {Marchetti}},\ }\bibfield
  {title} {\bibinfo {title} {Defect annihilation and proliferation in active
  nematics},\ }\href@noop {} {\bibfield  {journal} {\bibinfo  {journal} {Phys.
  Rev. Lett.}\ }\textbf {\bibinfo {volume} {110}},\ \bibinfo {pages} {228101}
  (\bibinfo {year} {2013})}\BibitemShut {NoStop}%
\bibitem [{\citenamefont {Thampi}\ \emph {et~al.}(2013)\citenamefont {Thampi},
  \citenamefont {Golestanian},\ and\ \citenamefont
  {Yeomans}}]{thampi2013velocity}%
  \BibitemOpen
  \bibfield  {author} {\bibinfo {author} {\bibfnamefont {S.~P.}\ \bibnamefont
  {Thampi}}, \bibinfo {author} {\bibfnamefont {R.}~\bibnamefont
  {Golestanian}},\ and\ \bibinfo {author} {\bibfnamefont {J.~M.}\ \bibnamefont
  {Yeomans}},\ }\bibfield  {title} {\bibinfo {title} {Velocity correlations in
  an active nematic},\ }\href@noop {} {\bibfield  {journal} {\bibinfo
  {journal} {Phys. Rev. Lett.}\ }\textbf {\bibinfo {volume} {111}},\ \bibinfo
  {pages} {118101} (\bibinfo {year} {2013})}\BibitemShut {NoStop}%
\bibitem [{\citenamefont {Prost}\ \emph {et~al.}(2015)\citenamefont {Prost},
  \citenamefont {J{\"u}licher},\ and\ \citenamefont
  {Joanny}}]{prost2015active}%
  \BibitemOpen
  \bibfield  {author} {\bibinfo {author} {\bibfnamefont {J.}~\bibnamefont
  {Prost}}, \bibinfo {author} {\bibfnamefont {F.}~\bibnamefont
  {J{\"u}licher}},\ and\ \bibinfo {author} {\bibfnamefont {J.-F.}\ \bibnamefont
  {Joanny}},\ }\bibfield  {title} {\bibinfo {title} {Active gel physics},\
  }\href@noop {} {\bibfield  {journal} {\bibinfo  {journal} {Nat. Phys.}\
  }\textbf {\bibinfo {volume} {11}},\ \bibinfo {pages} {111} (\bibinfo {year}
  {2015})}\BibitemShut {NoStop}%
\bibitem [{\citenamefont {DeCamp}\ \emph {et~al.}(2015)\citenamefont {DeCamp},
  \citenamefont {Redner}, \citenamefont {Baskaran}, \citenamefont {Hagan},\
  and\ \citenamefont {Dogic}}]{decamp2015orientational}%
  \BibitemOpen
  \bibfield  {author} {\bibinfo {author} {\bibfnamefont {S.~J.}\ \bibnamefont
  {DeCamp}}, \bibinfo {author} {\bibfnamefont {G.~S.}\ \bibnamefont {Redner}},
  \bibinfo {author} {\bibfnamefont {A.}~\bibnamefont {Baskaran}}, \bibinfo
  {author} {\bibfnamefont {M.~F.}\ \bibnamefont {Hagan}},\ and\ \bibinfo
  {author} {\bibfnamefont {Z.}~\bibnamefont {Dogic}},\ }\bibfield  {title}
  {\bibinfo {title} {Orientational order of motile defects in active
  nematics},\ }\href@noop {} {\bibfield  {journal} {\bibinfo  {journal} {Nat.
  Mater.}\ }\textbf {\bibinfo {volume} {14}},\ \bibinfo {pages} {1110}
  (\bibinfo {year} {2015})}\BibitemShut {NoStop}%
\bibitem [{\citenamefont {Wu}\ \emph {et~al.}(2017)\citenamefont {Wu},
  \citenamefont {Hishamunda}, \citenamefont {Chen}, \citenamefont {DeCamp},
  \citenamefont {Chang}, \citenamefont {Fern{\'a}ndez-Nieves}, \citenamefont
  {Fraden},\ and\ \citenamefont {Dogic}}]{wu2017transition}%
  \BibitemOpen
  \bibfield  {author} {\bibinfo {author} {\bibfnamefont {K.-T.}\ \bibnamefont
  {Wu}}, \bibinfo {author} {\bibfnamefont {J.~B.}\ \bibnamefont {Hishamunda}},
  \bibinfo {author} {\bibfnamefont {D.~T.}\ \bibnamefont {Chen}}, \bibinfo
  {author} {\bibfnamefont {S.~J.}\ \bibnamefont {DeCamp}}, \bibinfo {author}
  {\bibfnamefont {Y.-W.}\ \bibnamefont {Chang}}, \bibinfo {author}
  {\bibfnamefont {A.}~\bibnamefont {Fern{\'a}ndez-Nieves}}, \bibinfo {author}
  {\bibfnamefont {S.}~\bibnamefont {Fraden}},\ and\ \bibinfo {author}
  {\bibfnamefont {Z.}~\bibnamefont {Dogic}},\ }\bibfield  {title} {\bibinfo
  {title} {Transition from turbulent to coherent flows in confined
  three-dimensional active fluids},\ }\href@noop {} {\bibfield  {journal}
  {\bibinfo  {journal} {Science}\ }\textbf {\bibinfo {volume} {355}},\ \bibinfo
  {pages} {eaal1979} (\bibinfo {year} {2017})}\BibitemShut {NoStop}%
\bibitem [{\citenamefont {Duclos}\ \emph {et~al.}(2020)\citenamefont {Duclos},
  \citenamefont {Adkins}, \citenamefont {Banerjee}, \citenamefont {Peterson},
  \citenamefont {Varghese}, \citenamefont {Kolvin}, \citenamefont {Baskaran},
  \citenamefont {Pelcovits}, \citenamefont {Powers}, \citenamefont {Baskaran},
  \citenamefont {Toschi}, \citenamefont {Hagan}, \citenamefont {Streichan},
  \citenamefont {Vitelli}, \citenamefont {Beller},\ and\ \citenamefont
  {Dogic}}]{duclos2020topological}%
  \BibitemOpen
  \bibfield  {author} {\bibinfo {author} {\bibfnamefont {G.}~\bibnamefont
  {Duclos}}, \bibinfo {author} {\bibfnamefont {R.}~\bibnamefont {Adkins}},
  \bibinfo {author} {\bibfnamefont {D.}~\bibnamefont {Banerjee}}, \bibinfo
  {author} {\bibfnamefont {M.~S.~E.}\ \bibnamefont {Peterson}}, \bibinfo
  {author} {\bibfnamefont {M.}~\bibnamefont {Varghese}}, \bibinfo {author}
  {\bibfnamefont {I.}~\bibnamefont {Kolvin}}, \bibinfo {author} {\bibfnamefont
  {A.}~\bibnamefont {Baskaran}}, \bibinfo {author} {\bibfnamefont {R.~A.}\
  \bibnamefont {Pelcovits}}, \bibinfo {author} {\bibfnamefont {T.~R.}\
  \bibnamefont {Powers}}, \bibinfo {author} {\bibfnamefont {A.}~\bibnamefont
  {Baskaran}}, \bibinfo {author} {\bibfnamefont {F.}~\bibnamefont {Toschi}},
  \bibinfo {author} {\bibfnamefont {M.~F.}\ \bibnamefont {Hagan}}, \bibinfo
  {author} {\bibfnamefont {S.~J.}\ \bibnamefont {Streichan}}, \bibinfo {author}
  {\bibfnamefont {V.}~\bibnamefont {Vitelli}}, \bibinfo {author} {\bibfnamefont
  {D.~A.}\ \bibnamefont {Beller}},\ and\ \bibinfo {author} {\bibfnamefont
  {Z.}~\bibnamefont {Dogic}},\ }\bibfield  {title} {\bibinfo {title}
  {Topological structure and dynamics of three-dimensional active nematics},\
  }\href@noop {} {\bibfield  {journal} {\bibinfo  {journal} {Science}\ }\textbf
  {\bibinfo {volume} {367}},\ \bibinfo {pages} {1120} (\bibinfo {year}
  {2020})}\BibitemShut {NoStop}%
\bibitem [{\citenamefont {Serra}\ \emph {et~al.}(2023)\citenamefont {Serra},
  \citenamefont {Lemma}, \citenamefont {Giomi}, \citenamefont {Dogic},\ and\
  \citenamefont {Mahadevan}}]{serra2023defect}%
  \BibitemOpen
  \bibfield  {author} {\bibinfo {author} {\bibfnamefont {M.}~\bibnamefont
  {Serra}}, \bibinfo {author} {\bibfnamefont {L.}~\bibnamefont {Lemma}},
  \bibinfo {author} {\bibfnamefont {L.}~\bibnamefont {Giomi}}, \bibinfo
  {author} {\bibfnamefont {Z.}~\bibnamefont {Dogic}},\ and\ \bibinfo {author}
  {\bibfnamefont {L.}~\bibnamefont {Mahadevan}},\ }\bibfield  {title} {\bibinfo
  {title} {Defect-mediated dynamics of coherent structures in active
  nematics},\ }\href@noop {} {\bibfield  {journal} {\bibinfo  {journal} {Nat.
  Phys.}\ }\textbf {\bibinfo {volume} {19}},\ \bibinfo {pages} {1355} (\bibinfo
  {year} {2023})}\BibitemShut {NoStop}%
\bibitem [{\citenamefont {Hancock}(1953)}]{hancock1953self}%
  \BibitemOpen
  \bibfield  {author} {\bibinfo {author} {\bibfnamefont {G.~J.}\ \bibnamefont
  {Hancock}},\ }\bibfield  {title} {\bibinfo {title} {The self-propulsion of
  microscopic organisms through liquids},\ }\href@noop {} {\bibfield  {journal}
  {\bibinfo  {journal} {Proceedings of the Royal Society of London. Series A.
  Mathematical and Physical Sciences}\ }\textbf {\bibinfo {volume} {217}},\
  \bibinfo {pages} {96} (\bibinfo {year} {1953})}\BibitemShut {NoStop}%
\bibitem [{\citenamefont {Dobell}(1958)}]{dobell1958antony}%
  \BibitemOpen
  \bibfield  {author} {\bibinfo {author} {\bibfnamefont {C.}~\bibnamefont
  {Dobell}},\ }\href@noop {} {\emph {\bibinfo {title} {Antony van Leeuwenhoek
  and his ``{L}ittle {A}nimals.''}}}\ (\bibinfo  {publisher} {Russell \&
  Russell New York},\ \bibinfo {year} {1958})\BibitemShut {NoStop}%
\bibitem [{\citenamefont {Haimo}\ and\ \citenamefont
  {Rosenbaum}(1981)}]{haimo1981cilia}%
  \BibitemOpen
  \bibfield  {author} {\bibinfo {author} {\bibfnamefont {L.~T.}\ \bibnamefont
  {Haimo}}\ and\ \bibinfo {author} {\bibfnamefont {J.~L.}\ \bibnamefont
  {Rosenbaum}},\ }\bibfield  {title} {\bibinfo {title} {Cilia, flagella, and
  microtubules.},\ }\href@noop {} {\bibfield  {journal} {\bibinfo  {journal}
  {J. Cell Biol.}\ }\textbf {\bibinfo {volume} {91}},\ \bibinfo {pages} {125s}
  (\bibinfo {year} {1981})}\BibitemShut {NoStop}%
\bibitem [{\citenamefont {Beeby}\ \emph {et~al.}(2020)\citenamefont {Beeby},
  \citenamefont {Ferreira}, \citenamefont {Tripp}, \citenamefont {Albers},\
  and\ \citenamefont {Mitchell}}]{beeby2020propulsive}%
  \BibitemOpen
  \bibfield  {author} {\bibinfo {author} {\bibfnamefont {M.}~\bibnamefont
  {Beeby}}, \bibinfo {author} {\bibfnamefont {J.~L.}\ \bibnamefont {Ferreira}},
  \bibinfo {author} {\bibfnamefont {P.}~\bibnamefont {Tripp}}, \bibinfo
  {author} {\bibfnamefont {S.-V.}\ \bibnamefont {Albers}},\ and\ \bibinfo
  {author} {\bibfnamefont {D.~R.}\ \bibnamefont {Mitchell}},\ }\bibfield
  {title} {\bibinfo {title} {Propulsive nanomachines: the convergent evolution
  of archaella, flagella and cilia},\ }\href
  {https://doi.org/10.1093/femsre/fuaa006} {\bibfield  {journal} {\bibinfo
  {journal} {FEMS Microbiol. Rev.}\ }\textbf {\bibinfo {volume} {44}},\
  \bibinfo {pages} {253} (\bibinfo {year} {2020})}\BibitemShut {NoStop}%
\bibitem [{\citenamefont {Bondoc-Naumovitz}\ \emph {et~al.}(2023)\citenamefont
  {Bondoc-Naumovitz}, \citenamefont {Laeverenz-Schlogelhofer}, \citenamefont
  {Poon}, \citenamefont {Boggon}, \citenamefont {Bentley}, \citenamefont
  {Cortese},\ and\ \citenamefont {Wan}}]{bondoc2023methods}%
  \BibitemOpen
  \bibfield  {author} {\bibinfo {author} {\bibfnamefont {K.~G.}\ \bibnamefont
  {Bondoc-Naumovitz}}, \bibinfo {author} {\bibfnamefont {H.}~\bibnamefont
  {Laeverenz-Schlogelhofer}}, \bibinfo {author} {\bibfnamefont {R.~N.}\
  \bibnamefont {Poon}}, \bibinfo {author} {\bibfnamefont {A.~K.}\ \bibnamefont
  {Boggon}}, \bibinfo {author} {\bibfnamefont {S.~A.}\ \bibnamefont {Bentley}},
  \bibinfo {author} {\bibfnamefont {D.}~\bibnamefont {Cortese}},\ and\ \bibinfo
  {author} {\bibfnamefont {K.~Y.}\ \bibnamefont {Wan}},\ }\bibfield  {title}
  {\bibinfo {title} {Methods and measures for investigating microscale
  motility},\ }\href@noop {} {\bibfield  {journal} {\bibinfo  {journal}
  {Integrative and Comparative Biology}\ }\textbf {\bibinfo {volume} {63}},\
  \bibinfo {pages} {1485} (\bibinfo {year} {2023})}\BibitemShut {NoStop}%
\bibitem [{\citenamefont {Eichele}\ \emph {et~al.}(2020)\citenamefont
  {Eichele}, \citenamefont {Bodenschatz}, \citenamefont {Ditte}, \citenamefont
  {G{\"u}nther}, \citenamefont {Kapoor}, \citenamefont {Wang},\ and\
  \citenamefont {Westendorf}}]{eichele2020cilia}%
  \BibitemOpen
  \bibfield  {author} {\bibinfo {author} {\bibfnamefont {G.}~\bibnamefont
  {Eichele}}, \bibinfo {author} {\bibfnamefont {E.}~\bibnamefont
  {Bodenschatz}}, \bibinfo {author} {\bibfnamefont {Z.}~\bibnamefont {Ditte}},
  \bibinfo {author} {\bibfnamefont {A.-K.}\ \bibnamefont {G{\"u}nther}},
  \bibinfo {author} {\bibfnamefont {S.}~\bibnamefont {Kapoor}}, \bibinfo
  {author} {\bibfnamefont {Y.}~\bibnamefont {Wang}},\ and\ \bibinfo {author}
  {\bibfnamefont {C.}~\bibnamefont {Westendorf}},\ }\bibfield  {title}
  {\bibinfo {title} {Cilia-driven flows in the brain third ventricle},\
  }\href@noop {} {\bibfield  {journal} {\bibinfo  {journal} {Philosophical
  Transactions of the Royal Society B}\ }\textbf {\bibinfo {volume} {375}},\
  \bibinfo {pages} {20190154} (\bibinfo {year} {2020})}\BibitemShut {NoStop}%
\bibitem [{\citenamefont {Huang}\ and\ \citenamefont
  {Choma}(2015)}]{huang2015microscale}%
  \BibitemOpen
  \bibfield  {author} {\bibinfo {author} {\bibfnamefont {B.~K.}\ \bibnamefont
  {Huang}}\ and\ \bibinfo {author} {\bibfnamefont {M.~A.}\ \bibnamefont
  {Choma}},\ }\bibfield  {title} {\bibinfo {title} {Microscale imaging of
  cilia-driven fluid flow},\ }\href@noop {} {\bibfield  {journal} {\bibinfo
  {journal} {Cellular and molecular life sciences}\ }\textbf {\bibinfo {volume}
  {72}},\ \bibinfo {pages} {1095} (\bibinfo {year} {2015})}\BibitemShut
  {NoStop}%
\bibitem [{\citenamefont {Ling}\ \emph {et~al.}(2024)\citenamefont {Ling},
  \citenamefont {Essock-Burns}, \citenamefont {McFall-Ngai}, \citenamefont
  {Katija}, \citenamefont {Nawroth},\ and\ \citenamefont
  {Kanso}}]{ling2024flow}%
  \BibitemOpen
  \bibfield  {author} {\bibinfo {author} {\bibfnamefont {F.}~\bibnamefont
  {Ling}}, \bibinfo {author} {\bibfnamefont {T.}~\bibnamefont {Essock-Burns}},
  \bibinfo {author} {\bibfnamefont {M.}~\bibnamefont {McFall-Ngai}}, \bibinfo
  {author} {\bibfnamefont {K.}~\bibnamefont {Katija}}, \bibinfo {author}
  {\bibfnamefont {J.~C.}\ \bibnamefont {Nawroth}},\ and\ \bibinfo {author}
  {\bibfnamefont {E.}~\bibnamefont {Kanso}},\ }\bibfield  {title} {\bibinfo
  {title} {Flow physics guides morphology of ciliated organs},\ }\href@noop {}
  {\bibfield  {journal} {\bibinfo  {journal} {Nature Physics}\ }\textbf
  {\bibinfo {volume} {20}},\ \bibinfo {pages} {1679} (\bibinfo {year}
  {2024})}\BibitemShut {NoStop}%
\bibitem [{\citenamefont {Marshall}\ and\ \citenamefont
  {Nonaka}(2006)}]{marshall2006cilia}%
  \BibitemOpen
  \bibfield  {author} {\bibinfo {author} {\bibfnamefont {W.~F.}\ \bibnamefont
  {Marshall}}\ and\ \bibinfo {author} {\bibfnamefont {S.}~\bibnamefont
  {Nonaka}},\ }\bibfield  {title} {\bibinfo {title} {Cilia: tuning in to the
  cell's antenna},\ }\href@noop {} {\bibfield  {journal} {\bibinfo  {journal}
  {Current Biology}\ }\textbf {\bibinfo {volume} {16}},\ \bibinfo {pages}
  {R604} (\bibinfo {year} {2006})}\BibitemShut {NoStop}%
\bibitem [{\citenamefont {R.~Ferreira}\ \emph {et~al.}(2019)\citenamefont
  {R.~Ferreira}, \citenamefont {Fukui}, \citenamefont {Chow}, \citenamefont
  {Vilfan},\ and\ \citenamefont {Vermot}}]{ferreira2019cilium}%
  \BibitemOpen
  \bibfield  {author} {\bibinfo {author} {\bibfnamefont {R.}~\bibnamefont
  {R.~Ferreira}}, \bibinfo {author} {\bibfnamefont {H.}~\bibnamefont {Fukui}},
  \bibinfo {author} {\bibfnamefont {R.}~\bibnamefont {Chow}}, \bibinfo {author}
  {\bibfnamefont {A.}~\bibnamefont {Vilfan}},\ and\ \bibinfo {author}
  {\bibfnamefont {J.}~\bibnamefont {Vermot}},\ }\bibfield  {title} {\bibinfo
  {title} {The cilium as a force sensor- myth versus reality},\ }\href@noop {}
  {\bibfield  {journal} {\bibinfo  {journal} {J. Cell Sci.}\ }\textbf {\bibinfo
  {volume} {132}},\ \bibinfo {pages} {jcs213496} (\bibinfo {year}
  {2019})}\BibitemShut {NoStop}%
\bibitem [{\citenamefont {Ryu}\ and\ \citenamefont
  {Matsudaira}(2010)}]{ryu2010unsteady}%
  \BibitemOpen
  \bibfield  {author} {\bibinfo {author} {\bibfnamefont {S.}~\bibnamefont
  {Ryu}}\ and\ \bibinfo {author} {\bibfnamefont {P.}~\bibnamefont
  {Matsudaira}},\ }\bibfield  {title} {\bibinfo {title} {Unsteady motion,
  finite reynolds numbers, and wall effect on vorticella convallaria contribute
  contraction force greater than the stokes drag},\ }\href@noop {} {\bibfield
  {journal} {\bibinfo  {journal} {Biophys. J.}\ }\textbf {\bibinfo {volume}
  {98}},\ \bibinfo {pages} {2574} (\bibinfo {year} {2010})}\BibitemShut
  {NoStop}%
\bibitem [{\citenamefont {Jarrell}\ \emph {et~al.}(2013)\citenamefont
  {Jarrell}, \citenamefont {Ding}, \citenamefont {Nair},\ and\ \citenamefont
  {Siu}}]{jarrell2013surface}%
  \BibitemOpen
  \bibfield  {author} {\bibinfo {author} {\bibfnamefont {K.~F.}\ \bibnamefont
  {Jarrell}}, \bibinfo {author} {\bibfnamefont {Y.}~\bibnamefont {Ding}},
  \bibinfo {author} {\bibfnamefont {D.~B.}\ \bibnamefont {Nair}},\ and\
  \bibinfo {author} {\bibfnamefont {S.}~\bibnamefont {Siu}},\ }\bibfield
  {title} {\bibinfo {title} {Surface appendages of archaea: structure,
  function, genetics and assembly},\ }\href@noop {} {\bibfield  {journal}
  {\bibinfo  {journal} {Life}\ }\textbf {\bibinfo {volume} {3}},\ \bibinfo
  {pages} {86} (\bibinfo {year} {2013})}\BibitemShut {NoStop}%
\bibitem [{\citenamefont {Purcell}(1977)}]{purcell1997life}%
  \BibitemOpen
  \bibfield  {author} {\bibinfo {author} {\bibfnamefont {E.~M.}\ \bibnamefont
  {Purcell}},\ }\bibfield  {title} {\bibinfo {title} {{Life at low Reynolds
  number}},\ }\href {https://doi.org/10.1119/1.10903} {\bibfield  {journal}
  {\bibinfo  {journal} {Am. J. Phys.}\ }\textbf {\bibinfo {volume} {45}},\
  \bibinfo {pages} {3} (\bibinfo {year} {1977})}\BibitemShut {NoStop}%
\bibitem [{\citenamefont {Chang}\ and\ \citenamefont
  {Prakash}(2024)}]{chang2024biophysical}%
  \BibitemOpen
  \bibfield  {author} {\bibinfo {author} {\bibfnamefont {R.}~\bibnamefont
  {Chang}}\ and\ \bibinfo {author} {\bibfnamefont {M.}~\bibnamefont
  {Prakash}},\ }\bibfield  {title} {\bibinfo {title} {Biophysical limits of
  ultrafast cellular motility},\ }\href@noop {} {\bibfield  {journal} {\bibinfo
   {journal} {bioRxiv}\ ,\ \bibinfo {pages} {2024}} (\bibinfo {year}
  {2024})}\BibitemShut {NoStop}%
\bibitem [{\citenamefont {Batchelor}(1970)}]{batchelor1970slender}%
  \BibitemOpen
  \bibfield  {author} {\bibinfo {author} {\bibfnamefont {G.~K.}\ \bibnamefont
  {Batchelor}},\ }\bibfield  {title} {\bibinfo {title} {Slender-body theory for
  particles of arbitrary cross-section in {S}tokes flow},\ }\href@noop {}
  {\bibfield  {journal} {\bibinfo  {journal} {J. Fluid Mech.}\ }\textbf
  {\bibinfo {volume} {44}},\ \bibinfo {pages} {419} (\bibinfo {year}
  {1970})}\BibitemShut {NoStop}%
\bibitem [{\citenamefont {Keller}\ and\ \citenamefont
  {Rubinow}(1976)}]{keller1976slender}%
  \BibitemOpen
  \bibfield  {author} {\bibinfo {author} {\bibfnamefont {J.~B.}\ \bibnamefont
  {Keller}}\ and\ \bibinfo {author} {\bibfnamefont {S.~I.}\ \bibnamefont
  {Rubinow}},\ }\bibfield  {title} {\bibinfo {title} {Slender-body theory for
  slow viscous flow},\ }\href@noop {} {\bibfield  {journal} {\bibinfo
  {journal} {J. Fluid Mech.}\ }\textbf {\bibinfo {volume} {75}},\ \bibinfo
  {pages} {705} (\bibinfo {year} {1976})}\BibitemShut {NoStop}%
\bibitem [{\citenamefont {Rodenborn}\ \emph {et~al.}(2013)\citenamefont
  {Rodenborn}, \citenamefont {Chen}, \citenamefont {Swinney}, \citenamefont
  {Liu},\ and\ \citenamefont {Zhang}}]{rodenborn2013propulsion}%
  \BibitemOpen
  \bibfield  {author} {\bibinfo {author} {\bibfnamefont {B.}~\bibnamefont
  {Rodenborn}}, \bibinfo {author} {\bibfnamefont {C.-H.}\ \bibnamefont {Chen}},
  \bibinfo {author} {\bibfnamefont {H.~L.}\ \bibnamefont {Swinney}}, \bibinfo
  {author} {\bibfnamefont {B.}~\bibnamefont {Liu}},\ and\ \bibinfo {author}
  {\bibfnamefont {H.~P.}\ \bibnamefont {Zhang}},\ }\bibfield  {title} {\bibinfo
  {title} {Propulsion of microorganisms by a helical flagellum},\ }\href@noop
  {} {\bibfield  {journal} {\bibinfo  {journal} {Proc. Natl. Acad. Sci. USA}\
  }\textbf {\bibinfo {volume} {110}},\ \bibinfo {pages} {E338} (\bibinfo {year}
  {2013})}\BibitemShut {NoStop}%
\bibitem [{\citenamefont {Wan}\ and\ \citenamefont
  {Poon}(2024)}]{wan2024mechanisms}%
  \BibitemOpen
  \bibfield  {author} {\bibinfo {author} {\bibfnamefont {K.~Y.}\ \bibnamefont
  {Wan}}\ and\ \bibinfo {author} {\bibfnamefont {R.~N.}\ \bibnamefont {Poon}},\
  }\bibfield  {title} {\bibinfo {title} {Mechanisms and functions of
  multiciliary coordination},\ }\href@noop {} {\bibfield  {journal} {\bibinfo
  {journal} {Curr. Opin. Cell Biol.}\ }\textbf {\bibinfo {volume} {86}},\
  \bibinfo {pages} {102286} (\bibinfo {year} {2024})}\BibitemShut {NoStop}%
\bibitem [{\citenamefont {Wan}(2018)}]{wan2018coordination}%
  \BibitemOpen
  \bibfield  {author} {\bibinfo {author} {\bibfnamefont {K.~Y.}\ \bibnamefont
  {Wan}},\ }\bibfield  {title} {\bibinfo {title} {Coordination of eukaryotic
  cilia and flagella},\ }\href@noop {} {\bibfield  {journal} {\bibinfo
  {journal} {Essays Biochem.}\ }\textbf {\bibinfo {volume} {62}},\ \bibinfo
  {pages} {829} (\bibinfo {year} {2018})}\BibitemShut {NoStop}%
\bibitem [{\citenamefont {Uchida}\ \emph {et~al.}(2017)\citenamefont {Uchida},
  \citenamefont {Golestanian},\ and\ \citenamefont
  {Bennett}}]{uchida2017synchronization}%
  \BibitemOpen
  \bibfield  {author} {\bibinfo {author} {\bibfnamefont {N.}~\bibnamefont
  {Uchida}}, \bibinfo {author} {\bibfnamefont {R.}~\bibnamefont
  {Golestanian}},\ and\ \bibinfo {author} {\bibfnamefont {R.~R.}\ \bibnamefont
  {Bennett}},\ }\bibfield  {title} {\bibinfo {title} {Synchronization and
  collective dynamics of flagella and cilia as hydrodynamically coupled
  oscillators},\ }\href@noop {} {\bibfield  {journal} {\bibinfo  {journal} {J.
  Phys. Soc. Jpn.}\ }\textbf {\bibinfo {volume} {86}},\ \bibinfo {pages}
  {101007} (\bibinfo {year} {2017})}\BibitemShut {NoStop}%
\bibitem [{\citenamefont {Bruot}\ and\ \citenamefont
  {Cicuta}(2016)}]{bruot2016realizing}%
  \BibitemOpen
  \bibfield  {author} {\bibinfo {author} {\bibfnamefont {N.}~\bibnamefont
  {Bruot}}\ and\ \bibinfo {author} {\bibfnamefont {P.}~\bibnamefont {Cicuta}},\
  }\bibfield  {title} {\bibinfo {title} {Realizing the physics of motile cilia
  synchronization with driven colloids},\ }\href@noop {} {\bibfield  {journal}
  {\bibinfo  {journal} {Annu. Rev. Condens. Matter Phys.}\ }\textbf {\bibinfo
  {volume} {7}},\ \bibinfo {pages} {323} (\bibinfo {year} {2016})}\BibitemShut
  {NoStop}%
\bibitem [{\citenamefont {Poon}\ \emph {et~al.}(2023)\citenamefont {Poon},
  \citenamefont {Westwood}, \citenamefont {Laeverenz-Schlogelhofer},
  \citenamefont {Brodrick}, \citenamefont {Craggs}, \citenamefont {Keaveny},
  \citenamefont {J{\'e}kely},\ and\ \citenamefont {Wan}}]{poon2023ciliary}%
  \BibitemOpen
  \bibfield  {author} {\bibinfo {author} {\bibfnamefont {R.~N.}\ \bibnamefont
  {Poon}}, \bibinfo {author} {\bibfnamefont {T.~A.}\ \bibnamefont {Westwood}},
  \bibinfo {author} {\bibfnamefont {H.}~\bibnamefont
  {Laeverenz-Schlogelhofer}}, \bibinfo {author} {\bibfnamefont
  {E.}~\bibnamefont {Brodrick}}, \bibinfo {author} {\bibfnamefont
  {J.}~\bibnamefont {Craggs}}, \bibinfo {author} {\bibfnamefont {E.~E.}\
  \bibnamefont {Keaveny}}, \bibinfo {author} {\bibfnamefont {G.}~\bibnamefont
  {J{\'e}kely}},\ and\ \bibinfo {author} {\bibfnamefont {K.~Y.}\ \bibnamefont
  {Wan}},\ }\bibfield  {title} {\bibinfo {title} {Ciliary propulsion and
  metachronal coordination in reef coral larvae},\ }\href@noop {} {\bibfield
  {journal} {\bibinfo  {journal} {Phys. Rev. Res.}\ }\textbf {\bibinfo {volume}
  {5}},\ \bibinfo {pages} {L042037} (\bibinfo {year} {2023})}\BibitemShut
  {NoStop}%
\bibitem [{\citenamefont {Meng}\ \emph {et~al.}(2021)\citenamefont {Meng},
  \citenamefont {Bennett}, \citenamefont {Uchida},\ and\ \citenamefont
  {Golestanian}}]{meng2021conditions}%
  \BibitemOpen
  \bibfield  {author} {\bibinfo {author} {\bibfnamefont {F.}~\bibnamefont
  {Meng}}, \bibinfo {author} {\bibfnamefont {R.~R.}\ \bibnamefont {Bennett}},
  \bibinfo {author} {\bibfnamefont {N.}~\bibnamefont {Uchida}},\ and\ \bibinfo
  {author} {\bibfnamefont {R.}~\bibnamefont {Golestanian}},\ }\bibfield
  {title} {\bibinfo {title} {Conditions for metachronal coordination in arrays
  of model cilia},\ }\href@noop {} {\bibfield  {journal} {\bibinfo  {journal}
  {Proceedings of the National Academy of Sciences}\ }\textbf {\bibinfo
  {volume} {118}},\ \bibinfo {pages} {e2102828118} (\bibinfo {year}
  {2021})}\BibitemShut {NoStop}%
\bibitem [{\citenamefont {Elgeti}\ and\ \citenamefont
  {Gompper}(2013)}]{elgeti2013emergence}%
  \BibitemOpen
  \bibfield  {author} {\bibinfo {author} {\bibfnamefont {J.}~\bibnamefont
  {Elgeti}}\ and\ \bibinfo {author} {\bibfnamefont {G.}~\bibnamefont
  {Gompper}},\ }\bibfield  {title} {\bibinfo {title} {Emergence of metachronal
  waves in cilia arrays},\ }\href@noop {} {\bibfield  {journal} {\bibinfo
  {journal} {Proc. Natl. Acad. Sci. USA}\ }\textbf {\bibinfo {volume} {110}},\
  \bibinfo {pages} {4470} (\bibinfo {year} {2013})}\BibitemShut {NoStop}%
\bibitem [{\citenamefont {Grognot}\ and\ \citenamefont
  {Taute}(2021)}]{grognot2021morethanpropellors}%
  \BibitemOpen
  \bibfield  {author} {\bibinfo {author} {\bibfnamefont {M.}~\bibnamefont
  {Grognot}}\ and\ \bibinfo {author} {\bibfnamefont {K.~M.}\ \bibnamefont
  {Taute}},\ }\bibfield  {title} {\bibinfo {title} {More than propellers: how
  flagella shape bacterial motility behaviors},\ }\href@noop {} {\bibfield
  {journal} {\bibinfo  {journal} {Current Opinion in Microbiology}\ }\textbf
  {\bibinfo {volume} {61}},\ \bibinfo {pages} {73} (\bibinfo {year}
  {2021})}\BibitemShut {NoStop}%
\bibitem [{\citenamefont {Wadhwa}\ and\ \citenamefont
  {Berg}(2022)}]{wadhwa2022bacterial}%
  \BibitemOpen
  \bibfield  {author} {\bibinfo {author} {\bibfnamefont {N.}~\bibnamefont
  {Wadhwa}}\ and\ \bibinfo {author} {\bibfnamefont {H.~C.}\ \bibnamefont
  {Berg}},\ }\bibfield  {title} {\bibinfo {title} {Bacterial motility:
  machinery and mechanisms},\ }\href@noop {} {\bibfield  {journal} {\bibinfo
  {journal} {Nat. Rev. Microbiol.}\ }\textbf {\bibinfo {volume} {20}},\
  \bibinfo {pages} {161} (\bibinfo {year} {2022})}\BibitemShut {NoStop}%
\bibitem [{\citenamefont {Laeverenz-Schlogelhofer}\ and\ \citenamefont
  {Wan}(2024)}]{laeverenz2024bioelectric}%
  \BibitemOpen
  \bibfield  {author} {\bibinfo {author} {\bibfnamefont {H.}~\bibnamefont
  {Laeverenz-Schlogelhofer}}\ and\ \bibinfo {author} {\bibfnamefont {K.~Y.}\
  \bibnamefont {Wan}},\ }\bibfield  {title} {\bibinfo {title} {Bioelectric
  control of locomotor gaits in the walking ciliate euplotes},\ }\href@noop {}
  {\bibfield  {journal} {\bibinfo  {journal} {Current Biology}\ }\textbf
  {\bibinfo {volume} {34}},\ \bibinfo {pages} {697} (\bibinfo {year}
  {2024})}\BibitemShut {NoStop}%
\bibitem [{\citenamefont {Jokura}\ \emph {et~al.}(2022)\citenamefont {Jokura},
  \citenamefont {Sato}, \citenamefont {Shiba},\ and\ \citenamefont
  {Inaba}}]{jokura2022two}%
  \BibitemOpen
  \bibfield  {author} {\bibinfo {author} {\bibfnamefont {K.}~\bibnamefont
  {Jokura}}, \bibinfo {author} {\bibfnamefont {Y.}~\bibnamefont {Sato}},
  \bibinfo {author} {\bibfnamefont {K.}~\bibnamefont {Shiba}},\ and\ \bibinfo
  {author} {\bibfnamefont {K.}~\bibnamefont {Inaba}},\ }\bibfield  {title}
  {\bibinfo {title} {Two distinct compartments of a ctenophore comb plate
  provide structural and functional integrity for the motility of giant
  multicilia},\ }\href@noop {} {\bibfield  {journal} {\bibinfo  {journal}
  {Current Biology}\ }\textbf {\bibinfo {volume} {32}},\ \bibinfo {pages}
  {5144} (\bibinfo {year} {2022})}\BibitemShut {NoStop}%
\bibitem [{\citenamefont {Be’er}\ and\ \citenamefont
  {Ariel}(2019)}]{beer2019statistical}%
  \BibitemOpen
  \bibfield  {author} {\bibinfo {author} {\bibfnamefont {A.}~\bibnamefont
  {Be’er}}\ and\ \bibinfo {author} {\bibfnamefont {G.}~\bibnamefont
  {Ariel}},\ }\bibfield  {title} {\bibinfo {title} {A statistical physics view
  of swarming bacteria},\ }\href@noop {} {\bibfield  {journal} {\bibinfo
  {journal} {Movement ecology}\ }\textbf {\bibinfo {volume} {7}},\ \bibinfo
  {pages} {1} (\bibinfo {year} {2019})}\BibitemShut {NoStop}%
\bibitem [{\citenamefont {Bees}(2020)}]{bees2020advances}%
  \BibitemOpen
  \bibfield  {author} {\bibinfo {author} {\bibfnamefont {M.~A.}\ \bibnamefont
  {Bees}},\ }\bibfield  {title} {\bibinfo {title} {Advances in bioconvection},\
  }\href@noop {} {\bibfield  {journal} {\bibinfo  {journal} {Annual Review of
  Fluid Mechanics}\ }\textbf {\bibinfo {volume} {52}},\ \bibinfo {pages} {449}
  (\bibinfo {year} {2020})}\BibitemShut {NoStop}%
\bibitem [{\citenamefont {Schwarzendahl}\ and\ \citenamefont
  {Mazza}(2018)}]{schwarzendahl2018maximum}%
  \BibitemOpen
  \bibfield  {author} {\bibinfo {author} {\bibfnamefont {F.~J.}\ \bibnamefont
  {Schwarzendahl}}\ and\ \bibinfo {author} {\bibfnamefont {M.~G.}\ \bibnamefont
  {Mazza}},\ }\bibfield  {title} {\bibinfo {title} {Maximum in density
  heterogeneities of active swimmers},\ }\href@noop {} {\bibfield  {journal}
  {\bibinfo  {journal} {Soft Matter}\ }\textbf {\bibinfo {volume} {14}},\
  \bibinfo {pages} {4666} (\bibinfo {year} {2018})}\BibitemShut {NoStop}%
\bibitem [{\citenamefont {Vachier}\ and\ \citenamefont
  {Mazza}(2019)}]{vachier2019dynamics}%
  \BibitemOpen
  \bibfield  {author} {\bibinfo {author} {\bibfnamefont {J.}~\bibnamefont
  {Vachier}}\ and\ \bibinfo {author} {\bibfnamefont {M.~G.}\ \bibnamefont
  {Mazza}},\ }\bibfield  {title} {\bibinfo {title} {Dynamics of sedimenting
  active brownian particles},\ }\href@noop {} {\bibfield  {journal} {\bibinfo
  {journal} {Eur. Phys. J. E}\ }\textbf {\bibinfo {volume} {42}},\ \bibinfo
  {pages} {1} (\bibinfo {year} {2019})}\BibitemShut {NoStop}%
\bibitem [{\citenamefont {Schwarzendahl}\ and\ \citenamefont
  {Mazza}(2019)}]{schwarzendahl2019hydrodynamic}%
  \BibitemOpen
  \bibfield  {author} {\bibinfo {author} {\bibfnamefont {F.~J.}\ \bibnamefont
  {Schwarzendahl}}\ and\ \bibinfo {author} {\bibfnamefont {M.~G.}\ \bibnamefont
  {Mazza}},\ }\bibfield  {title} {\bibinfo {title} {Hydrodynamic interactions
  dominate the structure of active swimmers’ pair distribution functions},\
  }\href@noop {} {\bibfield  {journal} {\bibinfo  {journal} {J. Chem. Phys.}\
  }\textbf {\bibinfo {volume} {150}} (\bibinfo {year} {2019})}\BibitemShut
  {NoStop}%
\bibitem [{\citenamefont {Schwarzendahl}\ and\ \citenamefont
  {Mazza}(2022)}]{schwarzendahl2022active}%
  \BibitemOpen
  \bibfield  {author} {\bibinfo {author} {\bibfnamefont {F.~J.}\ \bibnamefont
  {Schwarzendahl}}\ and\ \bibinfo {author} {\bibfnamefont {M.~G.}\ \bibnamefont
  {Mazza}},\ }\bibfield  {title} {\bibinfo {title} {Active percolation in
  pusher-type microswimmers},\ }\href@noop {} {\bibfield  {journal} {\bibinfo
  {journal} {EPL}\ }\textbf {\bibinfo {volume} {140}},\ \bibinfo {pages}
  {47001} (\bibinfo {year} {2022})}\BibitemShut {NoStop}%
\bibitem [{\citenamefont {Turner}\ \emph {et~al.}(2000)\citenamefont {Turner},
  \citenamefont {Ryu},\ and\ \citenamefont {Berg}}]{turner2000real}%
  \BibitemOpen
  \bibfield  {author} {\bibinfo {author} {\bibfnamefont {L.}~\bibnamefont
  {Turner}}, \bibinfo {author} {\bibfnamefont {W.~S.}\ \bibnamefont {Ryu}},\
  and\ \bibinfo {author} {\bibfnamefont {H.~C.}\ \bibnamefont {Berg}},\
  }\bibfield  {title} {\bibinfo {title} {Real-time imaging of fluorescent
  flagellar filaments},\ }\href@noop {} {\bibfield  {journal} {\bibinfo
  {journal} {J. Bacteriol.}\ }\textbf {\bibinfo {volume} {182}},\ \bibinfo
  {pages} {2793} (\bibinfo {year} {2000})}\BibitemShut {NoStop}%
\bibitem [{\citenamefont {Neuhaus}\ \emph {et~al.}(2020)\citenamefont
  {Neuhaus}, \citenamefont {Selvaraj}, \citenamefont {Salzer}, \citenamefont
  {Langer}, \citenamefont {Kruse}, \citenamefont {Kirchner}, \citenamefont
  {Sanders}, \citenamefont {Daum}, \citenamefont {Averhoff},\ and\
  \citenamefont {Gold}}]{neuhaus2020cryo}%
  \BibitemOpen
  \bibfield  {author} {\bibinfo {author} {\bibfnamefont {A.}~\bibnamefont
  {Neuhaus}}, \bibinfo {author} {\bibfnamefont {M.}~\bibnamefont {Selvaraj}},
  \bibinfo {author} {\bibfnamefont {R.}~\bibnamefont {Salzer}}, \bibinfo
  {author} {\bibfnamefont {J.~D.}\ \bibnamefont {Langer}}, \bibinfo {author}
  {\bibfnamefont {K.}~\bibnamefont {Kruse}}, \bibinfo {author} {\bibfnamefont
  {L.}~\bibnamefont {Kirchner}}, \bibinfo {author} {\bibfnamefont
  {K.}~\bibnamefont {Sanders}}, \bibinfo {author} {\bibfnamefont
  {B.}~\bibnamefont {Daum}}, \bibinfo {author} {\bibfnamefont {B.}~\bibnamefont
  {Averhoff}},\ and\ \bibinfo {author} {\bibfnamefont {V.~A.}\ \bibnamefont
  {Gold}},\ }\bibfield  {title} {\bibinfo {title} {Cryo-electron microscopy
  reveals two distinct type iv pili assembled by the same bacterium},\
  }\href@noop {} {\bibfield  {journal} {\bibinfo  {journal} {Nature
  communications}\ }\textbf {\bibinfo {volume} {11}},\ \bibinfo {pages} {2231}
  (\bibinfo {year} {2020})}\BibitemShut {NoStop}%
\bibitem [{\citenamefont {Nomura}\ \emph {et~al.}(2019)\citenamefont {Nomura},
  \citenamefont {Atsuji}, \citenamefont {Hirose}, \citenamefont {Shiba},
  \citenamefont {Yanase}, \citenamefont {Nakayama}, \citenamefont {Ishida},\
  and\ \citenamefont {Inaba}}]{nomura2019microtubule}%
  \BibitemOpen
  \bibfield  {author} {\bibinfo {author} {\bibfnamefont {M.}~\bibnamefont
  {Nomura}}, \bibinfo {author} {\bibfnamefont {K.}~\bibnamefont {Atsuji}},
  \bibinfo {author} {\bibfnamefont {K.}~\bibnamefont {Hirose}}, \bibinfo
  {author} {\bibfnamefont {K.}~\bibnamefont {Shiba}}, \bibinfo {author}
  {\bibfnamefont {R.}~\bibnamefont {Yanase}}, \bibinfo {author} {\bibfnamefont
  {T.}~\bibnamefont {Nakayama}}, \bibinfo {author} {\bibfnamefont {K.-i.}\
  \bibnamefont {Ishida}},\ and\ \bibinfo {author} {\bibfnamefont
  {K.}~\bibnamefont {Inaba}},\ }\bibfield  {title} {\bibinfo {title}
  {Microtubule stabilizer reveals requirement of {Ca2}+-dependent
  conformational changes of microtubules for rapid coiling of haptonema in
  haptophyte algae},\ }\href {https://doi.org/10.1242/bio.036590} {\bibfield
  {journal} {\bibinfo  {journal} {Biology Open}\ }\textbf {\bibinfo {volume}
  {8}},\ \bibinfo {pages} {bio036590} (\bibinfo {year} {2019})}\BibitemShut
  {NoStop}%
\bibitem [{\citenamefont {Chwang}\ and\ \citenamefont
  {Wu}(1971)}]{chwang1971note}%
  \BibitemOpen
  \bibfield  {author} {\bibinfo {author} {\bibfnamefont {A.~T.}\ \bibnamefont
  {Chwang}}\ and\ \bibinfo {author} {\bibfnamefont {T.~Y.}\ \bibnamefont
  {Wu}},\ }\bibfield  {title} {\bibinfo {title} {A note on the helical movement
  of micro-organisms},\ }\href@noop {} {\bibfield  {journal} {\bibinfo
  {journal} {Proc. R. Soc. Lond. B.}\ }\textbf {\bibinfo {volume} {178}},\
  \bibinfo {pages} {327} (\bibinfo {year} {1971})}\BibitemShut {NoStop}%
\bibitem [{\citenamefont {Silverman}\ and\ \citenamefont
  {Simon}(1977)}]{silverman1977bacterial}%
  \BibitemOpen
  \bibfield  {author} {\bibinfo {author} {\bibfnamefont {M.}~\bibnamefont
  {Silverman}}\ and\ \bibinfo {author} {\bibfnamefont {M.~I.}\ \bibnamefont
  {Simon}},\ }\bibfield  {title} {\bibinfo {title} {Bacterial flagella},\
  }\href@noop {} {\bibfield  {journal} {\bibinfo  {journal} {Ann. Rev.
  Microbiol.}\ }\textbf {\bibinfo {volume} {31}},\ \bibinfo {pages} {397}
  (\bibinfo {year} {1977})}\BibitemShut {NoStop}%
\bibitem [{\citenamefont {Coombs}\ \emph {et~al.}(2002)\citenamefont {Coombs},
  \citenamefont {Huber}, \citenamefont {Kessler},\ and\ \citenamefont
  {Goldstein}}]{coombs2002periodic}%
  \BibitemOpen
  \bibfield  {author} {\bibinfo {author} {\bibfnamefont {D.}~\bibnamefont
  {Coombs}}, \bibinfo {author} {\bibfnamefont {G.}~\bibnamefont {Huber}},
  \bibinfo {author} {\bibfnamefont {J.~O.}\ \bibnamefont {Kessler}},\ and\
  \bibinfo {author} {\bibfnamefont {R.~E.}\ \bibnamefont {Goldstein}},\
  }\bibfield  {title} {\bibinfo {title} {Periodic chirality transformations
  propagating on bacterial flagella},\ }\href@noop {} {\bibfield  {journal}
  {\bibinfo  {journal} {Phys. Rev. Lett.}\ }\textbf {\bibinfo {volume} {89}},\
  \bibinfo {pages} {118102} (\bibinfo {year} {2002})}\BibitemShut {NoStop}%
\bibitem [{\citenamefont {Cvirkaite-Krupovic}\ \emph
  {et~al.}(2023)\citenamefont {Cvirkaite-Krupovic}, \citenamefont {Liu},
  \citenamefont {Baquero}, \citenamefont {Liu}, \citenamefont {Sonani},
  \citenamefont {Calladine}, \citenamefont {Wang}, \citenamefont {Krupovic},\
  and\ \citenamefont {Egelman}}]{cvirkaite2023evolution}%
  \BibitemOpen
  \bibfield  {author} {\bibinfo {author} {\bibfnamefont {V.}~\bibnamefont
  {Cvirkaite-Krupovic}}, \bibinfo {author} {\bibfnamefont {Y.}~\bibnamefont
  {Liu}}, \bibinfo {author} {\bibfnamefont {D.}~\bibnamefont {Baquero}},
  \bibinfo {author} {\bibfnamefont {J.}~\bibnamefont {Liu}}, \bibinfo {author}
  {\bibfnamefont {R.~R.}\ \bibnamefont {Sonani}}, \bibinfo {author}
  {\bibfnamefont {C.~R.}\ \bibnamefont {Calladine}}, \bibinfo {author}
  {\bibfnamefont {F.}~\bibnamefont {Wang}}, \bibinfo {author} {\bibfnamefont
  {M.}~\bibnamefont {Krupovic}},\ and\ \bibinfo {author} {\bibfnamefont
  {E.}~\bibnamefont {Egelman}},\ }\bibfield  {title} {\bibinfo {title} {The
  evolution of archaeal flagellar filaments},\ }\href@noop {} {\bibfield
  {journal} {\bibinfo  {journal} {Proc. Natl. Acad. Sci. USA}\ }\textbf
  {\bibinfo {volume} {120}},\ \bibinfo {pages} {e2304256120} (\bibinfo {year}
  {2023})}\BibitemShut {NoStop}%
\bibitem [{\citenamefont {Streif}\ \emph {et~al.}(2008)\citenamefont {Streif},
  \citenamefont {Staudinger}, \citenamefont {Marwan},\ and\ \citenamefont
  {Oesterhelt}}]{streif2008flagellar}%
  \BibitemOpen
  \bibfield  {author} {\bibinfo {author} {\bibfnamefont {S.}~\bibnamefont
  {Streif}}, \bibinfo {author} {\bibfnamefont {W.~F.}\ \bibnamefont
  {Staudinger}}, \bibinfo {author} {\bibfnamefont {W.}~\bibnamefont {Marwan}},\
  and\ \bibinfo {author} {\bibfnamefont {D.}~\bibnamefont {Oesterhelt}},\
  }\bibfield  {title} {\bibinfo {title} {Flagellar rotation in the archaeon
  halobacterium salinarum depends on atp},\ }\href@noop {} {\bibfield
  {journal} {\bibinfo  {journal} {J. Mol. Biol.}\ }\textbf {\bibinfo {volume}
  {384}},\ \bibinfo {pages} {1} (\bibinfo {year} {2008})}\BibitemShut {NoStop}%
\bibitem [{\citenamefont {Albers}\ and\ \citenamefont
  {Jarrell}(2015)}]{albers2015archaellum}%
  \BibitemOpen
  \bibfield  {author} {\bibinfo {author} {\bibfnamefont {S.-V.}\ \bibnamefont
  {Albers}}\ and\ \bibinfo {author} {\bibfnamefont {K.~F.}\ \bibnamefont
  {Jarrell}},\ }\bibfield  {title} {\bibinfo {title} {The archaellum: how
  archaea swim},\ }\href@noop {} {\bibfield  {journal} {\bibinfo  {journal}
  {Frontiers in Microbiology}\ }\textbf {\bibinfo {volume} {6}},\ \bibinfo
  {pages} {23} (\bibinfo {year} {2015})}\BibitemShut {NoStop}%
\bibitem [{\citenamefont {Berg}(2003)}]{berg2003rotary}%
  \BibitemOpen
  \bibfield  {author} {\bibinfo {author} {\bibfnamefont {H.~C.}\ \bibnamefont
  {Berg}},\ }\bibfield  {title} {\bibinfo {title} {The rotary motor of
  bacterial flagella},\ }\href@noop {} {\bibfield  {journal} {\bibinfo
  {journal} {Annual review of biochemistry}\ }\textbf {\bibinfo {volume}
  {72}},\ \bibinfo {pages} {19} (\bibinfo {year} {2003})}\BibitemShut {NoStop}%
\bibitem [{\citenamefont {Manson}\ \emph {et~al.}(1977)\citenamefont {Manson},
  \citenamefont {Tedesco}, \citenamefont {Berg}, \citenamefont {Harold},\ and\
  \citenamefont {Van~der Drift}}]{manson1977protonmotive}%
  \BibitemOpen
  \bibfield  {author} {\bibinfo {author} {\bibfnamefont {M.~D.}\ \bibnamefont
  {Manson}}, \bibinfo {author} {\bibfnamefont {P.}~\bibnamefont {Tedesco}},
  \bibinfo {author} {\bibfnamefont {H.~C.}\ \bibnamefont {Berg}}, \bibinfo
  {author} {\bibfnamefont {F.~M.}\ \bibnamefont {Harold}},\ and\ \bibinfo
  {author} {\bibfnamefont {C.}~\bibnamefont {Van~der Drift}},\ }\bibfield
  {title} {\bibinfo {title} {A protonmotive force drives bacterial flagella.},\
  }\href@noop {} {\bibfield  {journal} {\bibinfo  {journal} {Proceedings of the
  National Academy of Sciences}\ }\textbf {\bibinfo {volume} {74}},\ \bibinfo
  {pages} {3060} (\bibinfo {year} {1977})}\BibitemShut {NoStop}%
\bibitem [{\citenamefont {Brumley}\ \emph {et~al.}(2015)\citenamefont
  {Brumley}, \citenamefont {Rusconi}, \citenamefont {Son},\ and\ \citenamefont
  {Stocker}}]{brumley2015flagella}%
  \BibitemOpen
  \bibfield  {author} {\bibinfo {author} {\bibfnamefont {D.~R.}\ \bibnamefont
  {Brumley}}, \bibinfo {author} {\bibfnamefont {R.}~\bibnamefont {Rusconi}},
  \bibinfo {author} {\bibfnamefont {K.}~\bibnamefont {Son}},\ and\ \bibinfo
  {author} {\bibfnamefont {R.}~\bibnamefont {Stocker}},\ }\bibfield  {title}
  {\bibinfo {title} {Flagella, flexibility and flow: Physical processes in
  microbial ecology},\ }\href@noop {} {\bibfield  {journal} {\bibinfo
  {journal} {Eur. Phys. J. Spec. Top.}\ }\textbf {\bibinfo {volume} {224}},\
  \bibinfo {pages} {3119} (\bibinfo {year} {2015})}\BibitemShut {NoStop}%
\bibitem [{\citenamefont {Brown}\ \emph {et~al.}(2012)\citenamefont {Brown},
  \citenamefont {Steel}, \citenamefont {Silvestrin}, \citenamefont {Wilkinson},
  \citenamefont {Delalez}, \citenamefont {Lumb}, \citenamefont {Obara},
  \citenamefont {Armitage},\ and\ \citenamefont {Berry}}]{brown2012flagellar}%
  \BibitemOpen
  \bibfield  {author} {\bibinfo {author} {\bibfnamefont {M.~T.}\ \bibnamefont
  {Brown}}, \bibinfo {author} {\bibfnamefont {B.~C.}\ \bibnamefont {Steel}},
  \bibinfo {author} {\bibfnamefont {C.}~\bibnamefont {Silvestrin}}, \bibinfo
  {author} {\bibfnamefont {D.~A.}\ \bibnamefont {Wilkinson}}, \bibinfo {author}
  {\bibfnamefont {N.~J.}\ \bibnamefont {Delalez}}, \bibinfo {author}
  {\bibfnamefont {C.~N.}\ \bibnamefont {Lumb}}, \bibinfo {author}
  {\bibfnamefont {B.}~\bibnamefont {Obara}}, \bibinfo {author} {\bibfnamefont
  {J.~P.}\ \bibnamefont {Armitage}},\ and\ \bibinfo {author} {\bibfnamefont
  {R.~M.}\ \bibnamefont {Berry}},\ }\bibfield  {title} {\bibinfo {title}
  {Flagellar hook flexibility is essential for bundle formation in swimming
  escherichia coli cells},\ }\href@noop {} {\bibfield  {journal} {\bibinfo
  {journal} {J. Bacteriol.}\ }\textbf {\bibinfo {volume} {194}},\ \bibinfo
  {pages} {3495} (\bibinfo {year} {2012})}\BibitemShut {NoStop}%
\bibitem [{\citenamefont {Son}\ \emph {et~al.}(2013)\citenamefont {Son},
  \citenamefont {Guasto},\ and\ \citenamefont {Stocker}}]{son2013bacteria}%
  \BibitemOpen
  \bibfield  {author} {\bibinfo {author} {\bibfnamefont {K.}~\bibnamefont
  {Son}}, \bibinfo {author} {\bibfnamefont {J.~S.}\ \bibnamefont {Guasto}},\
  and\ \bibinfo {author} {\bibfnamefont {R.}~\bibnamefont {Stocker}},\
  }\bibfield  {title} {\bibinfo {title} {Bacteria can exploit a flagellar
  buckling instability to change direction},\ }\href@noop {} {\bibfield
  {journal} {\bibinfo  {journal} {Nat. Phys.}\ }\textbf {\bibinfo {volume}
  {9}},\ \bibinfo {pages} {494} (\bibinfo {year} {2013})}\BibitemShut {NoStop}%
\bibitem [{\citenamefont {Riley}\ \emph {et~al.}(2018)\citenamefont {Riley},
  \citenamefont {Das},\ and\ \citenamefont {Lauga}}]{riley2018swimming}%
  \BibitemOpen
  \bibfield  {author} {\bibinfo {author} {\bibfnamefont {E.~E.}\ \bibnamefont
  {Riley}}, \bibinfo {author} {\bibfnamefont {D.}~\bibnamefont {Das}},\ and\
  \bibinfo {author} {\bibfnamefont {E.}~\bibnamefont {Lauga}},\ }\bibfield
  {title} {\bibinfo {title} {Swimming of peritrichous bacteria is enabled by an
  elastohydrodynamic instability},\ }\href@noop {} {\bibfield  {journal}
  {\bibinfo  {journal} {Sci. Rep.}\ }\textbf {\bibinfo {volume} {8}},\ \bibinfo
  {pages} {10728} (\bibinfo {year} {2018})}\BibitemShut {NoStop}%
\bibitem [{\citenamefont {Halte}\ \emph {et~al.}(2025)\citenamefont {Halte},
  \citenamefont {Popp}, \citenamefont {Hathcock}, \citenamefont {Severn},
  \citenamefont {Fischer}, \citenamefont {Goosmann}, \citenamefont {Ducret},
  \citenamefont {Charpentier}, \citenamefont {Tu}, \citenamefont {Lauga},
  \citenamefont {Erhardt},\ and\ \citenamefont {Renault}}]{halte2025bacterial}%
  \BibitemOpen
  \bibfield  {author} {\bibinfo {author} {\bibfnamefont {M.}~\bibnamefont
  {Halte}}, \bibinfo {author} {\bibfnamefont {P.~F.}\ \bibnamefont {Popp}},
  \bibinfo {author} {\bibfnamefont {D.}~\bibnamefont {Hathcock}}, \bibinfo
  {author} {\bibfnamefont {J.}~\bibnamefont {Severn}}, \bibinfo {author}
  {\bibfnamefont {S.}~\bibnamefont {Fischer}}, \bibinfo {author} {\bibfnamefont
  {C.}~\bibnamefont {Goosmann}}, \bibinfo {author} {\bibfnamefont
  {A.}~\bibnamefont {Ducret}}, \bibinfo {author} {\bibfnamefont
  {E.}~\bibnamefont {Charpentier}}, \bibinfo {author} {\bibfnamefont
  {Y.}~\bibnamefont {Tu}}, \bibinfo {author} {\bibfnamefont {E.}~\bibnamefont
  {Lauga}}, \bibinfo {author} {\bibfnamefont {M.}~\bibnamefont {Erhardt}},\
  and\ \bibinfo {author} {\bibfnamefont {T.~T.}\ \bibnamefont {Renault}},\
  }\bibfield  {title} {\bibinfo {title} {Bacterial motility depends on a
  critical flagellum length and energy-optimized assembly},\ }\href@noop {}
  {\bibfield  {journal} {\bibinfo  {journal} {Proc. Natl. Acad. Sci. USA}\
  }\textbf {\bibinfo {volume} {122}},\ \bibinfo {pages} {e2413488122} (\bibinfo
  {year} {2025})}\BibitemShut {NoStop}%
\bibitem [{\citenamefont {Lauga}(2016)}]{lauga2016bacterial}%
  \BibitemOpen
  \bibfield  {author} {\bibinfo {author} {\bibfnamefont {E.}~\bibnamefont
  {Lauga}},\ }\bibfield  {title} {\bibinfo {title} {Bacterial hydrodynamics},\
  }\href@noop {} {\bibfield  {journal} {\bibinfo  {journal} {Annual Review of
  Fluid Mechanics}\ }\textbf {\bibinfo {volume} {48}},\ \bibinfo {pages} {105}
  (\bibinfo {year} {2016})}\BibitemShut {NoStop}%
\bibitem [{\citenamefont {K{\"u}hn}\ \emph {et~al.}(2017)\citenamefont
  {K{\"u}hn}, \citenamefont {Schmidt}, \citenamefont {Eckhardt},\ and\
  \citenamefont {Thormann}}]{kuhn2017bacteria}%
  \BibitemOpen
  \bibfield  {author} {\bibinfo {author} {\bibfnamefont {M.~J.}\ \bibnamefont
  {K{\"u}hn}}, \bibinfo {author} {\bibfnamefont {F.~K.}\ \bibnamefont
  {Schmidt}}, \bibinfo {author} {\bibfnamefont {B.}~\bibnamefont {Eckhardt}},\
  and\ \bibinfo {author} {\bibfnamefont {K.~M.}\ \bibnamefont {Thormann}},\
  }\bibfield  {title} {\bibinfo {title} {Bacteria exploit a polymorphic
  instability of the flagellar filament to escape from traps},\ }\href@noop {}
  {\bibfield  {journal} {\bibinfo  {journal} {Proceedings of the National
  Academy of Sciences}\ }\textbf {\bibinfo {volume} {114}},\ \bibinfo {pages}
  {6340} (\bibinfo {year} {2017})}\BibitemShut {NoStop}%
\bibitem [{\citenamefont {Nicastro}\ \emph {et~al.}(2006)\citenamefont
  {Nicastro}, \citenamefont {Schwartz}, \citenamefont {Pierson}, \citenamefont
  {Gaudette}, \citenamefont {Porter},\ and\ \citenamefont
  {McIntosh}}]{nicastro2006molecular}%
  \BibitemOpen
  \bibfield  {author} {\bibinfo {author} {\bibfnamefont {D.}~\bibnamefont
  {Nicastro}}, \bibinfo {author} {\bibfnamefont {C.}~\bibnamefont {Schwartz}},
  \bibinfo {author} {\bibfnamefont {J.}~\bibnamefont {Pierson}}, \bibinfo
  {author} {\bibfnamefont {R.}~\bibnamefont {Gaudette}}, \bibinfo {author}
  {\bibfnamefont {M.~E.}\ \bibnamefont {Porter}},\ and\ \bibinfo {author}
  {\bibfnamefont {J.~R.}\ \bibnamefont {McIntosh}},\ }\bibfield  {title}
  {\bibinfo {title} {The molecular architecture of axonemes revealed by
  cryoelectron tomography},\ }\href@noop {} {\bibfield  {journal} {\bibinfo
  {journal} {Science}\ }\textbf {\bibinfo {volume} {313}},\ \bibinfo {pages}
  {944} (\bibinfo {year} {2006})}\BibitemShut {NoStop}%
\bibitem [{\citenamefont {Lindemann}\ and\ \citenamefont
  {Lesich}(2010)}]{lindemann2010flagellar}%
  \BibitemOpen
  \bibfield  {author} {\bibinfo {author} {\bibfnamefont {C.~B.}\ \bibnamefont
  {Lindemann}}\ and\ \bibinfo {author} {\bibfnamefont {K.~A.}\ \bibnamefont
  {Lesich}},\ }\bibfield  {title} {\bibinfo {title} {Flagellar and ciliary
  beating: the proven and the possible},\ }\href@noop {} {\bibfield  {journal}
  {\bibinfo  {journal} {Journal of cell science}\ }\textbf {\bibinfo {volume}
  {123}},\ \bibinfo {pages} {519} (\bibinfo {year} {2010})}\BibitemShut
  {NoStop}%
\bibitem [{\citenamefont {Satir}\ \emph {et~al.}(2014)\citenamefont {Satir},
  \citenamefont {Heuser},\ and\ \citenamefont {Sale}}]{satir2014structural}%
  \BibitemOpen
  \bibfield  {author} {\bibinfo {author} {\bibfnamefont {P.}~\bibnamefont
  {Satir}}, \bibinfo {author} {\bibfnamefont {T.}~\bibnamefont {Heuser}},\ and\
  \bibinfo {author} {\bibfnamefont {W.~S.}\ \bibnamefont {Sale}},\ }\bibfield
  {title} {\bibinfo {title} {A structural basis for how motile cilia beat},\
  }\href@noop {} {\bibfield  {journal} {\bibinfo  {journal} {Bioscience}\
  }\textbf {\bibinfo {volume} {64}},\ \bibinfo {pages} {1073} (\bibinfo {year}
  {2014})}\BibitemShut {NoStop}%
\bibitem [{\citenamefont {Riedel-Kruse}\ \emph {et~al.}(2007)\citenamefont
  {Riedel-Kruse}, \citenamefont {Hilfinger}, \citenamefont {Howard},\ and\
  \citenamefont {J{\"u}licher}}]{riedel2007molecular}%
  \BibitemOpen
  \bibfield  {author} {\bibinfo {author} {\bibfnamefont {I.~H.}\ \bibnamefont
  {Riedel-Kruse}}, \bibinfo {author} {\bibfnamefont {A.}~\bibnamefont
  {Hilfinger}}, \bibinfo {author} {\bibfnamefont {J.}~\bibnamefont {Howard}},\
  and\ \bibinfo {author} {\bibfnamefont {F.}~\bibnamefont {J{\"u}licher}},\
  }\bibfield  {title} {\bibinfo {title} {How molecular motors shape the
  flagellar beat},\ }\href@noop {} {\bibfield  {journal} {\bibinfo  {journal}
  {HFSP journal}\ }\textbf {\bibinfo {volume} {1}},\ \bibinfo {pages} {192}
  (\bibinfo {year} {2007})}\BibitemShut {NoStop}%
\bibitem [{\citenamefont {Cass}\ and\ \citenamefont
  {Bloomfield-Gad{\^e}lha}(2023)}]{cass2023reaction}%
  \BibitemOpen
  \bibfield  {author} {\bibinfo {author} {\bibfnamefont {J.~F.}\ \bibnamefont
  {Cass}}\ and\ \bibinfo {author} {\bibfnamefont {H.}~\bibnamefont
  {Bloomfield-Gad{\^e}lha}},\ }\bibfield  {title} {\bibinfo {title} {The
  reaction-diffusion basis of animated patterns in eukaryotic flagella},\
  }\href@noop {} {\bibfield  {journal} {\bibinfo  {journal} {Nat. Commun.}\
  }\textbf {\bibinfo {volume} {14}},\ \bibinfo {pages} {5638} (\bibinfo {year}
  {2023})}\BibitemShut {NoStop}%
\bibitem [{\citenamefont {Woodhams}\ \emph {et~al.}(2022)\citenamefont
  {Woodhams}, \citenamefont {Shen},\ and\ \citenamefont
  {Bayly}}]{woodhams2022generation}%
  \BibitemOpen
  \bibfield  {author} {\bibinfo {author} {\bibfnamefont {L.~G.}\ \bibnamefont
  {Woodhams}}, \bibinfo {author} {\bibfnamefont {Y.}~\bibnamefont {Shen}},\
  and\ \bibinfo {author} {\bibfnamefont {P.~V.}\ \bibnamefont {Bayly}},\
  }\bibfield  {title} {\bibinfo {title} {Generation of ciliary beating by
  steady dynein activity: the effects of inter-filament coupling in
  multi-filament models},\ }\href@noop {} {\bibfield  {journal} {\bibinfo
  {journal} {J. R. Soc. Interface}\ }\textbf {\bibinfo {volume} {19}},\
  \bibinfo {pages} {20220264} (\bibinfo {year} {2022})}\BibitemShut {NoStop}%
\bibitem [{\citenamefont {Eckert}(1972)}]{eckert1972bioelectric}%
  \BibitemOpen
  \bibfield  {author} {\bibinfo {author} {\bibfnamefont {R.}~\bibnamefont
  {Eckert}},\ }\bibfield  {title} {\bibinfo {title} {Bioelectric control of
  ciliary activity: Locomotion in the ciliated protozoa is regulated by
  membrane-limited calcium fluxes.},\ }\href@noop {} {\bibfield  {journal}
  {\bibinfo  {journal} {Science}\ }\textbf {\bibinfo {volume} {176}},\ \bibinfo
  {pages} {473} (\bibinfo {year} {1972})}\BibitemShut {NoStop}%
\bibitem [{\citenamefont {Wan}\ and\ \citenamefont
  {Goldstein}(2018)}]{wan2018time}%
  \BibitemOpen
  \bibfield  {author} {\bibinfo {author} {\bibfnamefont {K.~Y.}\ \bibnamefont
  {Wan}}\ and\ \bibinfo {author} {\bibfnamefont {R.~E.}\ \bibnamefont
  {Goldstein}},\ }\bibfield  {title} {\bibinfo {title} {Time irreversibility
  and criticality in the motility of a flagellate microorganism},\ }\href@noop
  {} {\bibfield  {journal} {\bibinfo  {journal} {Phys. Rev. Lett.}\ }\textbf
  {\bibinfo {volume} {121}},\ \bibinfo {pages} {058103} (\bibinfo {year}
  {2018})}\BibitemShut {NoStop}%
\bibitem [{\citenamefont {Ginger}\ \emph {et~al.}(2008)\citenamefont {Ginger},
  \citenamefont {Portman},\ and\ \citenamefont {McKean}}]{ginger2008swimming}%
  \BibitemOpen
  \bibfield  {author} {\bibinfo {author} {\bibfnamefont {M.~L.}\ \bibnamefont
  {Ginger}}, \bibinfo {author} {\bibfnamefont {N.}~\bibnamefont {Portman}},\
  and\ \bibinfo {author} {\bibfnamefont {P.~G.}\ \bibnamefont {McKean}},\
  }\bibfield  {title} {\bibinfo {title} {Swimming with protists: perception,
  motility and flagellum assembly},\ }\href@noop {} {\bibfield  {journal}
  {\bibinfo  {journal} {Nature Reviews Microbiology}\ }\textbf {\bibinfo
  {volume} {6}},\ \bibinfo {pages} {838} (\bibinfo {year} {2008})}\BibitemShut
  {NoStop}%
\bibitem [{\citenamefont {Langousis}\ and\ \citenamefont
  {Hill}(2014)}]{langousis2014motility}%
  \BibitemOpen
  \bibfield  {author} {\bibinfo {author} {\bibfnamefont {G.}~\bibnamefont
  {Langousis}}\ and\ \bibinfo {author} {\bibfnamefont {K.~L.}\ \bibnamefont
  {Hill}},\ }\bibfield  {title} {\bibinfo {title} {Motility and more: the
  flagellum of \textit{{T}rypanosoma brucei}},\ }\href@noop {} {\bibfield
  {journal} {\bibinfo  {journal} {Nat. Rev. Microbiol.}\ }\textbf {\bibinfo
  {volume} {12}},\ \bibinfo {pages} {505} (\bibinfo {year} {2014})}\BibitemShut
  {NoStop}%
\bibitem [{\citenamefont {Suzuki-Tellier}\ \emph {et~al.}(2024)\citenamefont
  {Suzuki-Tellier}, \citenamefont {Miano}, \citenamefont {Asadzadeh},
  \citenamefont {Simpson},\ and\ \citenamefont
  {Ki{\o}rboe}}]{suzuki2024foraging}%
  \BibitemOpen
  \bibfield  {author} {\bibinfo {author} {\bibfnamefont {S.}~\bibnamefont
  {Suzuki-Tellier}}, \bibinfo {author} {\bibfnamefont {F.}~\bibnamefont
  {Miano}}, \bibinfo {author} {\bibfnamefont {S.~S.}\ \bibnamefont
  {Asadzadeh}}, \bibinfo {author} {\bibfnamefont {A.~G.~B.}\ \bibnamefont
  {Simpson}},\ and\ \bibinfo {author} {\bibfnamefont {T.}~\bibnamefont
  {Ki{\o}rboe}},\ }\bibfield  {title} {\bibinfo {title} {Foraging mechanisms in
  excavate flagellates shed light on the functional ecology of early
  eukaryotes},\ }\href@noop {} {\bibfield  {journal} {\bibinfo  {journal}
  {Proc. Natl. Acad. Sci. USA}\ }\textbf {\bibinfo {volume} {121}},\ \bibinfo
  {pages} {e2317264121} (\bibinfo {year} {2024})}\BibitemShut {NoStop}%
\bibitem [{\citenamefont {Craig}\ \emph {et~al.}(2019)\citenamefont {Craig},
  \citenamefont {Forest},\ and\ \citenamefont {Maier}}]{craig2019typeIV}%
  \BibitemOpen
  \bibfield  {author} {\bibinfo {author} {\bibfnamefont {L.}~\bibnamefont
  {Craig}}, \bibinfo {author} {\bibfnamefont {K.~T.}\ \bibnamefont {Forest}},\
  and\ \bibinfo {author} {\bibfnamefont {B.}~\bibnamefont {Maier}},\ }\bibfield
   {title} {\bibinfo {title} {Type {IV} pili: dynamics, biophysics and
  functional consequences},\ }\href {https://doi.org/10.1038/s41579-019-0195-4}
  {\bibfield  {journal} {\bibinfo  {journal} {Nat. Rev. Microbiol.}\ }\textbf
  {\bibinfo {volume} {17}},\ \bibinfo {pages} {429} (\bibinfo {year} {2019})},\
  \bibinfo {note} {number: 7 Publisher: Nature Publishing Group}\BibitemShut
  {NoStop}%
\bibitem [{\citenamefont {Ellison}\ \emph {et~al.}(2022)\citenamefont
  {Ellison}, \citenamefont {Whitfield},\ and\ \citenamefont
  {Brun}}]{ellison2022type}%
  \BibitemOpen
  \bibfield  {author} {\bibinfo {author} {\bibfnamefont {C.~K.}\ \bibnamefont
  {Ellison}}, \bibinfo {author} {\bibfnamefont {G.~B.}\ \bibnamefont
  {Whitfield}},\ and\ \bibinfo {author} {\bibfnamefont {Y.~V.}\ \bibnamefont
  {Brun}},\ }\bibfield  {title} {\bibinfo {title} {Type iv pili: dynamic
  bacterial nanomachines},\ }\href@noop {} {\bibfield  {journal} {\bibinfo
  {journal} {FEMS Microbiology Reviews}\ }\textbf {\bibinfo {volume} {46}},\
  \bibinfo {pages} {fuab053} (\bibinfo {year} {2022})}\BibitemShut {NoStop}%
\bibitem [{\citenamefont {Mahadevan}\ and\ \citenamefont
  {Matsudaira}(2000)}]{mahadevan2000motility}%
  \BibitemOpen
  \bibfield  {author} {\bibinfo {author} {\bibfnamefont {L.}~\bibnamefont
  {Mahadevan}}\ and\ \bibinfo {author} {\bibfnamefont {P.}~\bibnamefont
  {Matsudaira}},\ }\bibfield  {title} {\bibinfo {title} {Motility {Powered} by
  {Supramolecular} {Springs} and {Ratchets}},\ }\href
  {https://doi.org/10.1126/science.288.5463.95} {\bibfield  {journal} {\bibinfo
   {journal} {Science}\ }\textbf {\bibinfo {volume} {288}},\ \bibinfo {pages}
  {95} (\bibinfo {year} {2000})},\ \bibinfo {note} {publisher: American
  Association for the Advancement of Science}\BibitemShut {NoStop}%
\bibitem [{\citenamefont {Ryu}\ \emph {et~al.}(2017)\citenamefont {Ryu},
  \citenamefont {Pepper}, \citenamefont {Nagai},\ and\ \citenamefont
  {France}}]{ryu2017vorticella}%
  \BibitemOpen
  \bibfield  {author} {\bibinfo {author} {\bibfnamefont {S.}~\bibnamefont
  {Ryu}}, \bibinfo {author} {\bibfnamefont {R.~E.}\ \bibnamefont {Pepper}},
  \bibinfo {author} {\bibfnamefont {M.}~\bibnamefont {Nagai}},\ and\ \bibinfo
  {author} {\bibfnamefont {D.~C.}\ \bibnamefont {France}},\ }\bibfield  {title}
  {\bibinfo {title} {Vorticella: {A} {Protozoan} for {Bio}-{Inspired}
  {Engineering}},\ }\href {https://doi.org/10.3390/mi8010004} {\bibfield
  {journal} {\bibinfo  {journal} {Micromachines}\ }\textbf {\bibinfo {volume}
  {8}},\ \bibinfo {pages} {4} (\bibinfo {year} {2017})}\BibitemShut {NoStop}%
\bibitem [{\citenamefont {Floyd}\ \emph {et~al.}(2023)\citenamefont {Floyd},
  \citenamefont {Molines}, \citenamefont {Lei}, \citenamefont {Honts},
  \citenamefont {Chang}, \citenamefont {Elting}, \citenamefont
  {Vaikuntanathan}, \citenamefont {Dinner},\ and\ \citenamefont
  {Bhamla}}]{floyd2023unified}%
  \BibitemOpen
  \bibfield  {author} {\bibinfo {author} {\bibfnamefont {C.}~\bibnamefont
  {Floyd}}, \bibinfo {author} {\bibfnamefont {A.~T.}\ \bibnamefont {Molines}},
  \bibinfo {author} {\bibfnamefont {X.}~\bibnamefont {Lei}}, \bibinfo {author}
  {\bibfnamefont {J.~E.}\ \bibnamefont {Honts}}, \bibinfo {author}
  {\bibfnamefont {F.}~\bibnamefont {Chang}}, \bibinfo {author} {\bibfnamefont
  {M.~W.}\ \bibnamefont {Elting}}, \bibinfo {author} {\bibfnamefont
  {S.}~\bibnamefont {Vaikuntanathan}}, \bibinfo {author} {\bibfnamefont
  {A.~R.}\ \bibnamefont {Dinner}},\ and\ \bibinfo {author} {\bibfnamefont
  {M.~S.}\ \bibnamefont {Bhamla}},\ }\bibfield  {title} {\bibinfo {title} {A
  unified model for the dynamics of atp-independent ultrafast contraction},\
  }\href@noop {} {\bibfield  {journal} {\bibinfo  {journal} {Proc. Natl. Acad.
  Sci. USA}\ }\textbf {\bibinfo {volume} {120}},\ \bibinfo {pages}
  {e2217737120} (\bibinfo {year} {2023})}\BibitemShut {NoStop}%
\bibitem [{\citenamefont {Kawachi}\ and\ \citenamefont
  {Inouye}(1994)}]{kawachi1994ca2mediated}%
  \BibitemOpen
  \bibfield  {author} {\bibinfo {author} {\bibfnamefont {M.}~\bibnamefont
  {Kawachi}}\ and\ \bibinfo {author} {\bibfnamefont {I.}~\bibnamefont
  {Inouye}},\ }\bibfield  {title} {\bibinfo {title} {Ca2+-mediated induction of
  the coiling of the haptonema in {Chrysochromulina} hirta ({Prymnesiophyta} =
  {Haptophyta})},\ }\href {https://doi.org/10.2216/i0031-8884-33-1-53.1}
  {\bibfield  {journal} {\bibinfo  {journal} {Phycologia}\ }\textbf {\bibinfo
  {volume} {33}},\ \bibinfo {pages} {53} (\bibinfo {year} {1994})}\BibitemShut
  {NoStop}%
\bibitem [{\citenamefont {Darnton}\ and\ \citenamefont
  {Berg}(2007)}]{darnton2007force}%
  \BibitemOpen
  \bibfield  {author} {\bibinfo {author} {\bibfnamefont {N.~C.}\ \bibnamefont
  {Darnton}}\ and\ \bibinfo {author} {\bibfnamefont {H.~C.}\ \bibnamefont
  {Berg}},\ }\bibfield  {title} {\bibinfo {title} {Force-extension measurements
  on bacterial flagella: triggering polymorphic transformations},\ }\href@noop
  {} {\bibfield  {journal} {\bibinfo  {journal} {Biophys. J.}\ }\textbf
  {\bibinfo {volume} {92}},\ \bibinfo {pages} {2230} (\bibinfo {year}
  {2007})}\BibitemShut {NoStop}%
\bibitem [{\citenamefont {Xu}\ \emph {et~al.}(2016)\citenamefont {Xu},
  \citenamefont {Wilson}, \citenamefont {Okamoto}, \citenamefont {Shao},
  \citenamefont {Dutcher},\ and\ \citenamefont {Bayly}}]{xu2016flexural}%
  \BibitemOpen
  \bibfield  {author} {\bibinfo {author} {\bibfnamefont {G.}~\bibnamefont
  {Xu}}, \bibinfo {author} {\bibfnamefont {K.~S.}\ \bibnamefont {Wilson}},
  \bibinfo {author} {\bibfnamefont {R.~J.}\ \bibnamefont {Okamoto}}, \bibinfo
  {author} {\bibfnamefont {J.-Y.}\ \bibnamefont {Shao}}, \bibinfo {author}
  {\bibfnamefont {S.~K.}\ \bibnamefont {Dutcher}},\ and\ \bibinfo {author}
  {\bibfnamefont {P.~V.}\ \bibnamefont {Bayly}},\ }\bibfield  {title} {\bibinfo
  {title} {Flexural rigidity and shear stiffness of flagella estimated from
  induced bends and counterbends},\ }\href@noop {} {\bibfield  {journal}
  {\bibinfo  {journal} {Biophys. J.}\ }\textbf {\bibinfo {volume} {110}},\
  \bibinfo {pages} {2759} (\bibinfo {year} {2016})}\BibitemShut {NoStop}%
\bibitem [{\citenamefont {Hill}\ \emph {et~al.}(2010)\citenamefont {Hill},
  \citenamefont {Swaminathan}, \citenamefont {Estes}, \citenamefont {Cribb},
  \citenamefont {O'Brien}, \citenamefont {Davis},\ and\ \citenamefont
  {Superfine}}]{hill2010force}%
  \BibitemOpen
  \bibfield  {author} {\bibinfo {author} {\bibfnamefont {D.~B.}\ \bibnamefont
  {Hill}}, \bibinfo {author} {\bibfnamefont {V.}~\bibnamefont {Swaminathan}},
  \bibinfo {author} {\bibfnamefont {A.}~\bibnamefont {Estes}}, \bibinfo
  {author} {\bibfnamefont {J.}~\bibnamefont {Cribb}}, \bibinfo {author}
  {\bibfnamefont {E.~T.}\ \bibnamefont {O'Brien}}, \bibinfo {author}
  {\bibfnamefont {C.~W.}\ \bibnamefont {Davis}},\ and\ \bibinfo {author}
  {\bibfnamefont {R.}~\bibnamefont {Superfine}},\ }\bibfield  {title} {\bibinfo
  {title} {Force generation and dynamics of individual cilia under external
  loading},\ }\href@noop {} {\bibfield  {journal} {\bibinfo  {journal}
  {Biophys. J.}\ }\textbf {\bibinfo {volume} {98}},\ \bibinfo {pages} {57}
  (\bibinfo {year} {2010})}\BibitemShut {NoStop}%
\bibitem [{\citenamefont {Holwill}(1965)}]{holwill1965motion}%
  \BibitemOpen
  \bibfield  {author} {\bibinfo {author} {\bibfnamefont {M.}~\bibnamefont
  {Holwill}},\ }\bibfield  {title} {\bibinfo {title} {The motion of strigomonas
  oncopelti},\ }\href@noop {} {\bibfield  {journal} {\bibinfo  {journal}
  {Journal of Experimental Biology}\ }\textbf {\bibinfo {volume} {42}},\
  \bibinfo {pages} {125} (\bibinfo {year} {1965})}\BibitemShut {NoStop}%
\bibitem [{\citenamefont {Baba}(1972)}]{baba1972flexural}%
  \BibitemOpen
  \bibfield  {author} {\bibinfo {author} {\bibfnamefont {S.~A.}\ \bibnamefont
  {Baba}},\ }\bibfield  {title} {\bibinfo {title} {Flexural rigidity and
  elastic constant of cilia},\ }\href@noop {} {\bibfield  {journal} {\bibinfo
  {journal} {J. Exp. Biol.}\ }\textbf {\bibinfo {volume} {56}},\ \bibinfo
  {pages} {459} (\bibinfo {year} {1972})}\BibitemShut {NoStop}%
\bibitem [{\citenamefont {Hickey}\ \emph {et~al.}(2021)\citenamefont {Hickey},
  \citenamefont {Vilfan},\ and\ \citenamefont
  {Golestanian}}]{hickey2021ciliary}%
  \BibitemOpen
  \bibfield  {author} {\bibinfo {author} {\bibfnamefont {D.}~\bibnamefont
  {Hickey}}, \bibinfo {author} {\bibfnamefont {A.}~\bibnamefont {Vilfan}},\
  and\ \bibinfo {author} {\bibfnamefont {R.}~\bibnamefont {Golestanian}},\
  }\bibfield  {title} {\bibinfo {title} {Ciliary chemosensitivity is enhanced
  by cilium geometry and motility},\ }\href@noop {} {\bibfield  {journal}
  {\bibinfo  {journal} {Elife}\ }\textbf {\bibinfo {volume} {10}},\ \bibinfo
  {pages} {e66322} (\bibinfo {year} {2021})}\BibitemShut {NoStop}%
\bibitem [{\citenamefont {Bloodgood}(2010)}]{bloodgood2010sensory}%
  \BibitemOpen
  \bibfield  {author} {\bibinfo {author} {\bibfnamefont {R.~A.}\ \bibnamefont
  {Bloodgood}},\ }\bibfield  {title} {\bibinfo {title} {Sensory reception is an
  attribute of both primary cilia and motile cilia},\ }\href@noop {} {\bibfield
   {journal} {\bibinfo  {journal} {J. Cell Sci.}\ }\textbf {\bibinfo {volume}
  {123}},\ \bibinfo {pages} {505} (\bibinfo {year} {2010})}\BibitemShut
  {NoStop}%
\bibitem [{\citenamefont {Mill}\ \emph {et~al.}(2023)\citenamefont {Mill},
  \citenamefont {Christensen},\ and\ \citenamefont
  {Pedersen}}]{mill2023primary}%
  \BibitemOpen
  \bibfield  {author} {\bibinfo {author} {\bibfnamefont {P.}~\bibnamefont
  {Mill}}, \bibinfo {author} {\bibfnamefont {S.~T.}\ \bibnamefont
  {Christensen}},\ and\ \bibinfo {author} {\bibfnamefont {L.~B.}\ \bibnamefont
  {Pedersen}},\ }\bibfield  {title} {\bibinfo {title} {Primary cilia as dynamic
  and diverse signalling hubs in development and disease},\ }\href@noop {}
  {\bibfield  {journal} {\bibinfo  {journal} {Nat. Rev. Genet.}\ }\textbf
  {\bibinfo {volume} {24}},\ \bibinfo {pages} {421} (\bibinfo {year}
  {2023})}\BibitemShut {NoStop}%
\bibitem [{\citenamefont {Tereshko}\ \emph {et~al.}(2021)\citenamefont
  {Tereshko}, \citenamefont {Gao}, \citenamefont {Cary}, \citenamefont
  {Turrigiano},\ and\ \citenamefont {Sengupta}}]{tereshko2021ciliary}%
  \BibitemOpen
  \bibfield  {author} {\bibinfo {author} {\bibfnamefont {L.}~\bibnamefont
  {Tereshko}}, \bibinfo {author} {\bibfnamefont {Y.}~\bibnamefont {Gao}},
  \bibinfo {author} {\bibfnamefont {B.~A.}\ \bibnamefont {Cary}}, \bibinfo
  {author} {\bibfnamefont {G.~G.}\ \bibnamefont {Turrigiano}},\ and\ \bibinfo
  {author} {\bibfnamefont {P.}~\bibnamefont {Sengupta}},\ }\bibfield  {title}
  {\bibinfo {title} {Ciliary neuropeptidergic signaling dynamically regulates
  excitatory synapses in postnatal neocortical pyramidal neurons},\ }\href@noop
  {} {\bibfield  {journal} {\bibinfo  {journal} {{eL}ife}\ }\textbf {\bibinfo
  {volume} {10}},\ \bibinfo {pages} {e65427} (\bibinfo {year}
  {2021})}\BibitemShut {NoStop}%
\bibitem [{\citenamefont {Ott}\ \emph {et~al.}(2024)\citenamefont {Ott},
  \citenamefont {Torres}, \citenamefont {Kuan}, \citenamefont {Kuan},
  \citenamefont {Buchanan}, \citenamefont {Elabbady}, \citenamefont
  {Seshamani}, \citenamefont {Bodor}, \citenamefont {Collman}, \citenamefont
  {Bock}, \citenamefont {Lee}, \citenamefont {da~Costa},\ and\ \citenamefont
  {Lippincott-Schwartz}}]{ott2024ultrastructural}%
  \BibitemOpen
  \bibfield  {author} {\bibinfo {author} {\bibfnamefont {C.~M.}\ \bibnamefont
  {Ott}}, \bibinfo {author} {\bibfnamefont {R.}~\bibnamefont {Torres}},
  \bibinfo {author} {\bibfnamefont {T.-S.}\ \bibnamefont {Kuan}}, \bibinfo
  {author} {\bibfnamefont {A.}~\bibnamefont {Kuan}}, \bibinfo {author}
  {\bibfnamefont {J.}~\bibnamefont {Buchanan}}, \bibinfo {author}
  {\bibfnamefont {L.}~\bibnamefont {Elabbady}}, \bibinfo {author}
  {\bibfnamefont {S.}~\bibnamefont {Seshamani}}, \bibinfo {author}
  {\bibfnamefont {A.~L.}\ \bibnamefont {Bodor}}, \bibinfo {author}
  {\bibfnamefont {F.}~\bibnamefont {Collman}}, \bibinfo {author} {\bibfnamefont
  {D.~D.}\ \bibnamefont {Bock}}, \bibinfo {author} {\bibfnamefont {W.~C.}\
  \bibnamefont {Lee}}, \bibinfo {author} {\bibfnamefont {N.~M.}\ \bibnamefont
  {da~Costa}},\ and\ \bibinfo {author} {\bibfnamefont {J.}~\bibnamefont
  {Lippincott-Schwartz}},\ }\bibfield  {title} {\bibinfo {title}
  {Ultrastructural differences impact cilia shape and external exposure across
  cell classes in the visual cortex},\ }\href@noop {} {\bibfield  {journal}
  {\bibinfo  {journal} {Curr. Biol.}\ }\textbf {\bibinfo {volume} {34}},\
  \bibinfo {pages} {2418} (\bibinfo {year} {2024})}\BibitemShut {NoStop}%
\bibitem [{\citenamefont {{Culture Collection of Algae, SAG}}()}]{SAG}%
  \BibitemOpen
  \bibfield  {author} {\bibinfo {author} {\bibnamefont {{Culture Collection of
  Algae, SAG}}},\ }\href@noop {} {\bibinfo {title} {Medium recipie vers.
  10.2008}},\ \bibinfo {howpublished}
  {\url{http://sagdb.uni-goettingen.de/culture_media/01\%20Basal\%20Medium.pdf}}\BibitemShut
  {NoStop}%
\bibitem [{\citenamefont {Gong}\ and\ \citenamefont
  {Prakash}(2023)}]{gong2023active}%
  \BibitemOpen
  \bibfield  {author} {\bibinfo {author} {\bibfnamefont {X.}~\bibnamefont
  {Gong}}\ and\ \bibinfo {author} {\bibfnamefont {M.}~\bibnamefont {Prakash}},\
  }\bibfield  {title} {\bibinfo {title} {Active dislocations and topological
  traps govern dynamics of spiraling filamentous cyanobacteria},\ }\href@noop
  {} {\bibfield  {journal} {\bibinfo  {journal} {arXiv preprint
  arXiv:2305.12572}\ } (\bibinfo {year} {2023})}\BibitemShut {NoStop}%
\bibitem [{\citenamefont {Garcia-Pichel}\ and\ \citenamefont
  {Wojciechowski}(2009)}]{garcia2009evolution}%
  \BibitemOpen
  \bibfield  {author} {\bibinfo {author} {\bibfnamefont {F.}~\bibnamefont
  {Garcia-Pichel}}\ and\ \bibinfo {author} {\bibfnamefont {M.~F.}\ \bibnamefont
  {Wojciechowski}},\ }\bibfield  {title} {\bibinfo {title} {The evolution of a
  capacity to build supra-cellular ropes enabled filamentous cyanobacteria to
  colonize highly erodible substrates},\ }\href@noop {} {\bibfield  {journal}
  {\bibinfo  {journal} {PloS one}\ }\textbf {\bibinfo {volume} {4}},\ \bibinfo
  {pages} {e7801} (\bibinfo {year} {2009})}\BibitemShut {NoStop}%
\bibitem [{\citenamefont {Faluweki}\ \emph {et~al.}(2023)\citenamefont
  {Faluweki}, \citenamefont {Cammann}, \citenamefont {Mazza},\ and\
  \citenamefont {Goehring}}]{faluweki2023active}%
  \BibitemOpen
  \bibfield  {author} {\bibinfo {author} {\bibfnamefont {M.~K.}\ \bibnamefont
  {Faluweki}}, \bibinfo {author} {\bibfnamefont {J.}~\bibnamefont {Cammann}},
  \bibinfo {author} {\bibfnamefont {M.~G.}\ \bibnamefont {Mazza}},\ and\
  \bibinfo {author} {\bibfnamefont {L.}~\bibnamefont {Goehring}},\ }\bibfield
  {title} {\bibinfo {title} {Active spaghetti: collective organization in
  cyanobacteria},\ }\href@noop {} {\bibfield  {journal} {\bibinfo  {journal}
  {Phys. Rev. Lett.}\ }\textbf {\bibinfo {volume} {131}},\ \bibinfo {pages}
  {158303} (\bibinfo {year} {2023})}\BibitemShut {NoStop}%
\bibitem [{\citenamefont {Cammann}\ \emph {et~al.}(2024)\citenamefont
  {Cammann}, \citenamefont {Faluweki}, \citenamefont {Dambacher}, \citenamefont
  {Goehring},\ and\ \citenamefont {Mazza}}]{cammann2024topological}%
  \BibitemOpen
  \bibfield  {author} {\bibinfo {author} {\bibfnamefont {J.}~\bibnamefont
  {Cammann}}, \bibinfo {author} {\bibfnamefont {M.~K.}\ \bibnamefont
  {Faluweki}}, \bibinfo {author} {\bibfnamefont {N.}~\bibnamefont {Dambacher}},
  \bibinfo {author} {\bibfnamefont {L.}~\bibnamefont {Goehring}},\ and\
  \bibinfo {author} {\bibfnamefont {M.~G.}\ \bibnamefont {Mazza}},\ }\bibfield
  {title} {\bibinfo {title} {Topological transition in filamentous
  cyanobacteria: from motion to structure},\ }\href@noop {} {\bibfield
  {journal} {\bibinfo  {journal} {Commun. Phys.}\ }\textbf {\bibinfo {volume}
  {7}},\ \bibinfo {pages} {376} (\bibinfo {year} {2024})}\BibitemShut {NoStop}%
\bibitem [{\citenamefont {Ehlers}\ \emph {et~al.}(1996)\citenamefont {Ehlers},
  \citenamefont {Samuel}, \citenamefont {Berg},\ and\ \citenamefont
  {Montgomery}}]{ehlers1996cyanobacteria}%
  \BibitemOpen
  \bibfield  {author} {\bibinfo {author} {\bibfnamefont {K.~M.}\ \bibnamefont
  {Ehlers}}, \bibinfo {author} {\bibfnamefont {A.~D.}\ \bibnamefont {Samuel}},
  \bibinfo {author} {\bibfnamefont {H.~C.}\ \bibnamefont {Berg}},\ and\
  \bibinfo {author} {\bibfnamefont {R.}~\bibnamefont {Montgomery}},\ }\bibfield
   {title} {\bibinfo {title} {Do cyanobacteria swim using traveling surface
  waves?},\ }\href@noop {} {\bibfield  {journal} {\bibinfo  {journal} {Proc.
  Natl. Acad. Sci. USA}\ }\textbf {\bibinfo {volume} {93}},\ \bibinfo {pages}
  {8340} (\bibinfo {year} {1996})}\BibitemShut {NoStop}%
\bibitem [{\citenamefont {Dogic}\ and\ \citenamefont
  {Fraden}(1997)}]{dogic1997smectic}%
  \BibitemOpen
  \bibfield  {author} {\bibinfo {author} {\bibfnamefont {Z.}~\bibnamefont
  {Dogic}}\ and\ \bibinfo {author} {\bibfnamefont {S.}~\bibnamefont {Fraden}},\
  }\bibfield  {title} {\bibinfo {title} {Smectic phase in a colloidal
  suspension of semiflexible virus particles},\ }\href@noop {} {\bibfield
  {journal} {\bibinfo  {journal} {Phys. Rev. Lett.}\ }\textbf {\bibinfo
  {volume} {78}},\ \bibinfo {pages} {2417} (\bibinfo {year}
  {1997})}\BibitemShut {NoStop}%
\bibitem [{\citenamefont {Dogic}\ and\ \citenamefont
  {Fraden}(2006)}]{dogic2006ordered}%
  \BibitemOpen
  \bibfield  {author} {\bibinfo {author} {\bibfnamefont {Z.}~\bibnamefont
  {Dogic}}\ and\ \bibinfo {author} {\bibfnamefont {S.}~\bibnamefont {Fraden}},\
  }\bibfield  {title} {\bibinfo {title} {Ordered phases of filamentous
  viruses},\ }\href@noop {} {\bibfield  {journal} {\bibinfo  {journal} {Curr.
  Opin. Colloid Interface Sci.}\ }\textbf {\bibinfo {volume} {11}},\ \bibinfo
  {pages} {47} (\bibinfo {year} {2006})}\BibitemShut {NoStop}%
\bibitem [{\citenamefont {Purdy}\ and\ \citenamefont
  {Fraden}(2004)}]{purdy2004isotropic}%
  \BibitemOpen
  \bibfield  {author} {\bibinfo {author} {\bibfnamefont {K.~R.}\ \bibnamefont
  {Purdy}}\ and\ \bibinfo {author} {\bibfnamefont {S.}~\bibnamefont {Fraden}},\
  }\bibfield  {title} {\bibinfo {title} {Isotropic-cholesteric phase transition
  of filamentous virus suspensions as a function of rod length and charge},\
  }\href@noop {} {\bibfield  {journal} {\bibinfo  {journal} {Phys. Rev. E}\
  }\textbf {\bibinfo {volume} {70}},\ \bibinfo {pages} {061703} (\bibinfo
  {year} {2004})}\BibitemShut {NoStop}%
\bibitem [{\citenamefont {Bilbao}\ \emph {et~al.}(2013)\citenamefont {Bilbao},
  \citenamefont {Wajnryb}, \citenamefont {Vanapalli},\ and\ \citenamefont
  {Blawzdziewicz}}]{bilbao2013nematode}%
  \BibitemOpen
  \bibfield  {author} {\bibinfo {author} {\bibfnamefont {A.}~\bibnamefont
  {Bilbao}}, \bibinfo {author} {\bibfnamefont {E.}~\bibnamefont {Wajnryb}},
  \bibinfo {author} {\bibfnamefont {S.~A.}\ \bibnamefont {Vanapalli}},\ and\
  \bibinfo {author} {\bibfnamefont {J.}~\bibnamefont {Blawzdziewicz}},\
  }\bibfield  {title} {\bibinfo {title} {Nematode locomotion in unconfined and
  confined fluids},\ }\href@noop {} {\bibfield  {journal} {\bibinfo  {journal}
  {Phys. Fluids}\ }\textbf {\bibinfo {volume} {25}} (\bibinfo {year}
  {2013})}\BibitemShut {NoStop}%
\bibitem [{\citenamefont {Gart}\ \emph {et~al.}(2011)\citenamefont {Gart},
  \citenamefont {Vella},\ and\ \citenamefont {Jung}}]{gart2011collective}%
  \BibitemOpen
  \bibfield  {author} {\bibinfo {author} {\bibfnamefont {S.}~\bibnamefont
  {Gart}}, \bibinfo {author} {\bibfnamefont {D.}~\bibnamefont {Vella}},\ and\
  \bibinfo {author} {\bibfnamefont {S.}~\bibnamefont {Jung}},\ }\bibfield
  {title} {\bibinfo {title} {The collective motion of nematodes in a thin
  liquid layer},\ }\href@noop {} {\bibfield  {journal} {\bibinfo  {journal}
  {Soft Matter}\ }\textbf {\bibinfo {volume} {7}},\ \bibinfo {pages} {2444}
  (\bibinfo {year} {2011})}\BibitemShut {NoStop}%
\bibitem [{\citenamefont {Niebur}\ and\ \citenamefont
  {Erd{\"o}s}(1991)}]{niebur1991theory}%
  \BibitemOpen
  \bibfield  {author} {\bibinfo {author} {\bibfnamefont {E.}~\bibnamefont
  {Niebur}}\ and\ \bibinfo {author} {\bibfnamefont {P.}~\bibnamefont
  {Erd{\"o}s}},\ }\bibfield  {title} {\bibinfo {title} {Theory of the
  locomotion of nematodes: dynamics of undulatory progression on a surface},\
  }\href@noop {} {\bibfield  {journal} {\bibinfo  {journal} {Biophysical
  journal}\ }\textbf {\bibinfo {volume} {60}},\ \bibinfo {pages} {1132}
  (\bibinfo {year} {1991})}\BibitemShut {NoStop}%
\bibitem [{\citenamefont {Arguedas-Leiva}\ \emph {et~al.}(2022)\citenamefont
  {Arguedas-Leiva}, \citenamefont {S{\l}omka}, \citenamefont {Lalescu},
  \citenamefont {Stocker},\ and\ \citenamefont
  {Wilczek}}]{arguedas2022elongation}%
  \BibitemOpen
  \bibfield  {author} {\bibinfo {author} {\bibfnamefont {J.-A.}\ \bibnamefont
  {Arguedas-Leiva}}, \bibinfo {author} {\bibfnamefont {J.}~\bibnamefont
  {S{\l}omka}}, \bibinfo {author} {\bibfnamefont {C.~C.}\ \bibnamefont
  {Lalescu}}, \bibinfo {author} {\bibfnamefont {R.}~\bibnamefont {Stocker}},\
  and\ \bibinfo {author} {\bibfnamefont {M.}~\bibnamefont {Wilczek}},\
  }\bibfield  {title} {\bibinfo {title} {Elongation enhances encounter rates
  between phytoplankton in turbulence},\ }\href@noop {} {\bibfield  {journal}
  {\bibinfo  {journal} {Proc. Natl. Acad. Sci. USA}\ }\textbf {\bibinfo
  {volume} {119}},\ \bibinfo {pages} {e2203191119} (\bibinfo {year}
  {2022})}\BibitemShut {NoStop}%
\bibitem [{\citenamefont {Pernthaler}(2005)}]{pernthaler2005predation}%
  \BibitemOpen
  \bibfield  {author} {\bibinfo {author} {\bibfnamefont {J.}~\bibnamefont
  {Pernthaler}},\ }\bibfield  {title} {\bibinfo {title} {Predation on
  prokaryotes in the water column and its ecological implications},\
  }\href@noop {} {\bibfield  {journal} {\bibinfo  {journal} {Nat. Rev.
  Microbiol.}\ }\textbf {\bibinfo {volume} {3}},\ \bibinfo {pages} {537}
  (\bibinfo {year} {2005})}\BibitemShut {NoStop}%
\bibitem [{\citenamefont {Young}(2006)}]{young2006selective}%
  \BibitemOpen
  \bibfield  {author} {\bibinfo {author} {\bibfnamefont {K.~D.}\ \bibnamefont
  {Young}},\ }\bibfield  {title} {\bibinfo {title} {The selective value of
  bacterial shape},\ }\href@noop {} {\bibfield  {journal} {\bibinfo  {journal}
  {Microbiol. Mol. Biol. Rev.}\ }\textbf {\bibinfo {volume} {70}},\ \bibinfo
  {pages} {660} (\bibinfo {year} {2006})}\BibitemShut {NoStop}%
\bibitem [{\citenamefont {Weiss}\ \emph {et~al.}(1995)\citenamefont {Weiss},
  \citenamefont {Mills}, \citenamefont {Hornberger},\ and\ \citenamefont
  {Herman}}]{weiss1995effect}%
  \BibitemOpen
  \bibfield  {author} {\bibinfo {author} {\bibfnamefont {T.~H.}\ \bibnamefont
  {Weiss}}, \bibinfo {author} {\bibfnamefont {A.~L.}\ \bibnamefont {Mills}},
  \bibinfo {author} {\bibfnamefont {G.~M.}\ \bibnamefont {Hornberger}},\ and\
  \bibinfo {author} {\bibfnamefont {J.~S.}\ \bibnamefont {Herman}},\ }\bibfield
   {title} {\bibinfo {title} {Effect of bacterial cell shape on transport of
  bacteria in porous media},\ }\href@noop {} {\bibfield  {journal} {\bibinfo
  {journal} {Environ. Sci. Technol.}\ }\textbf {\bibinfo {volume} {29}},\
  \bibinfo {pages} {1737} (\bibinfo {year} {1995})}\BibitemShut {NoStop}%
\bibitem [{\citenamefont {Hoiczyk}(2000)}]{hoiczyk2000gliding}%
  \BibitemOpen
  \bibfield  {author} {\bibinfo {author} {\bibfnamefont {E.}~\bibnamefont
  {Hoiczyk}},\ }\bibfield  {title} {\bibinfo {title} {Gliding motility in
  cyanobacteria: observations and possible explanations},\ }\href@noop {}
  {\bibfield  {journal} {\bibinfo  {journal} {Archives of microbiology}\
  }\textbf {\bibinfo {volume} {174}},\ \bibinfo {pages} {11} (\bibinfo {year}
  {2000})}\BibitemShut {NoStop}%
\bibitem [{\citenamefont {Wilde}\ and\ \citenamefont
  {Mullineaux}(2015)}]{wilde2015motility}%
  \BibitemOpen
  \bibfield  {author} {\bibinfo {author} {\bibfnamefont {A.}~\bibnamefont
  {Wilde}}\ and\ \bibinfo {author} {\bibfnamefont {C.~W.}\ \bibnamefont
  {Mullineaux}},\ }\bibfield  {title} {\bibinfo {title} {Motility in
  cyanobacteria: polysaccharide tracks and {T}ype {IV} pilus motors},\
  }\href@noop {} {\bibfield  {journal} {\bibinfo  {journal} {Mol. Microbiol.}\
  }\textbf {\bibinfo {volume} {98}},\ \bibinfo {pages} {998} (\bibinfo {year}
  {2015})}\BibitemShut {NoStop}%
\bibitem [{\citenamefont {Tchoufag}\ \emph {et~al.}(2019)\citenamefont
  {Tchoufag}, \citenamefont {Ghosh}, \citenamefont {Pogue}, \citenamefont
  {Nan},\ and\ \citenamefont {Mandadapu}}]{tchoufag2019mechanisms}%
  \BibitemOpen
  \bibfield  {author} {\bibinfo {author} {\bibfnamefont {J.}~\bibnamefont
  {Tchoufag}}, \bibinfo {author} {\bibfnamefont {P.}~\bibnamefont {Ghosh}},
  \bibinfo {author} {\bibfnamefont {C.~B.}\ \bibnamefont {Pogue}}, \bibinfo
  {author} {\bibfnamefont {B.}~\bibnamefont {Nan}},\ and\ \bibinfo {author}
  {\bibfnamefont {K.~K.}\ \bibnamefont {Mandadapu}},\ }\bibfield  {title}
  {\bibinfo {title} {Mechanisms for bacterial gliding motility on soft
  substrates},\ }\href@noop {} {\bibfield  {journal} {\bibinfo  {journal}
  {Proc. Natl. Acad. Sci. USA}\ }\textbf {\bibinfo {volume} {116}},\ \bibinfo
  {pages} {25087} (\bibinfo {year} {2019})}\BibitemShut {NoStop}%
\bibitem [{\citenamefont {McBride}(2001)}]{mcbride2001bacterial}%
  \BibitemOpen
  \bibfield  {author} {\bibinfo {author} {\bibfnamefont {M.~J.}\ \bibnamefont
  {McBride}},\ }\bibfield  {title} {\bibinfo {title} {Bacterial gliding
  motility: multiple mechanisms for cell movement over surfaces},\ }\href@noop
  {} {\bibfield  {journal} {\bibinfo  {journal} {Annu. Rev. Microbiol.}\
  }\textbf {\bibinfo {volume} {55}},\ \bibinfo {pages} {49} (\bibinfo {year}
  {2001})}\BibitemShut {NoStop}%
\bibitem [{\citenamefont {Read}\ \emph {et~al.}(2007)\citenamefont {Read},
  \citenamefont {Connell},\ and\ \citenamefont {Adams}}]{read2007nanoscale}%
  \BibitemOpen
  \bibfield  {author} {\bibinfo {author} {\bibfnamefont {N.}~\bibnamefont
  {Read}}, \bibinfo {author} {\bibfnamefont {S.}~\bibnamefont {Connell}},\ and\
  \bibinfo {author} {\bibfnamefont {D.~G.}\ \bibnamefont {Adams}},\ }\bibfield
  {title} {\bibinfo {title} {Nanoscale visualization of a fibrillar array in
  the cell wall of filamentous cyanobacteria and its implications for gliding
  motility},\ }\href@noop {} {\bibfield  {journal} {\bibinfo  {journal} {J.
  Bacteriol.}\ }\textbf {\bibinfo {volume} {189}},\ \bibinfo {pages} {7361}
  (\bibinfo {year} {2007})}\BibitemShut {NoStop}%
\bibitem [{\citenamefont {Dhahri}\ \emph {et~al.}(2013)\citenamefont {Dhahri},
  \citenamefont {Ramonda},\ and\ \citenamefont {Marliere}}]{dhahri2013situ}%
  \BibitemOpen
  \bibfield  {author} {\bibinfo {author} {\bibfnamefont {S.}~\bibnamefont
  {Dhahri}}, \bibinfo {author} {\bibfnamefont {M.}~\bibnamefont {Ramonda}},\
  and\ \bibinfo {author} {\bibfnamefont {C.}~\bibnamefont {Marliere}},\
  }\bibfield  {title} {\bibinfo {title} {In-situ determination of the
  mechanical properties of gliding or non-motile bacteria by atomic force
  microscopy under physiological conditions without immobilization},\
  }\href@noop {} {\bibfield  {journal} {\bibinfo  {journal} {PLoS One}\
  }\textbf {\bibinfo {volume} {8}},\ \bibinfo {pages} {e61663} (\bibinfo {year}
  {2013})}\BibitemShut {NoStop}%
\bibitem [{\citenamefont {Koiller}\ \emph {et~al.}(2010)\citenamefont
  {Koiller}, \citenamefont {Ehlers},\ and\ \citenamefont
  {Chalub}}]{koiller2010acoustic}%
  \BibitemOpen
  \bibfield  {author} {\bibinfo {author} {\bibfnamefont {J.}~\bibnamefont
  {Koiller}}, \bibinfo {author} {\bibfnamefont {K.~M.}\ \bibnamefont
  {Ehlers}},\ and\ \bibinfo {author} {\bibfnamefont {F.}~\bibnamefont
  {Chalub}},\ }\bibfield  {title} {\bibinfo {title} {Acoustic streaming, the
  ``small invention'' of cyanobacteria?},\ }\href@noop {} {\bibfield  {journal}
  {\bibinfo  {journal} {Arbor}\ }\textbf {\bibinfo {volume} {186}},\ \bibinfo
  {pages} {1089} (\bibinfo {year} {2010})}\BibitemShut {NoStop}%
\bibitem [{\citenamefont {Halfen}\ and\ \citenamefont
  {Castenholz}(1970)}]{halfen1970gliding}%
  \BibitemOpen
  \bibfield  {author} {\bibinfo {author} {\bibfnamefont {L.~N.}\ \bibnamefont
  {Halfen}}\ and\ \bibinfo {author} {\bibfnamefont {R.~W.}\ \bibnamefont
  {Castenholz}},\ }\bibfield  {title} {\bibinfo {title} {Gliding in a
  blue--green alga: a possible mechanism},\ }\href@noop {} {\bibfield
  {journal} {\bibinfo  {journal} {Nature}\ }\textbf {\bibinfo {volume} {225}},\
  \bibinfo {pages} {1163} (\bibinfo {year} {1970})}\BibitemShut {NoStop}%
\bibitem [{\citenamefont {Khayatan}\ \emph {et~al.}(2015)\citenamefont
  {Khayatan}, \citenamefont {Meeks},\ and\ \citenamefont
  {Risser}}]{khayatan2015evidence}%
  \BibitemOpen
  \bibfield  {author} {\bibinfo {author} {\bibfnamefont {B.}~\bibnamefont
  {Khayatan}}, \bibinfo {author} {\bibfnamefont {J.~C.}\ \bibnamefont
  {Meeks}},\ and\ \bibinfo {author} {\bibfnamefont {D.~D.}\ \bibnamefont
  {Risser}},\ }\bibfield  {title} {\bibinfo {title} {Evidence that a modified
  type {IV} pilus-like system powers gliding motility and polysaccharide
  secretion in filamentous cyanobacteria},\ }\href@noop {} {\bibfield
  {journal} {\bibinfo  {journal} {Mol. Microbiol.}\ }\textbf {\bibinfo {volume}
  {98}},\ \bibinfo {pages} {1021} (\bibinfo {year} {2015})}\BibitemShut
  {NoStop}%
\bibitem [{\citenamefont {Hoiczyk}\ and\ \citenamefont
  {Baumeister}(1995)}]{hoiczyk1995envelope}%
  \BibitemOpen
  \bibfield  {author} {\bibinfo {author} {\bibfnamefont {E.}~\bibnamefont
  {Hoiczyk}}\ and\ \bibinfo {author} {\bibfnamefont {W.}~\bibnamefont
  {Baumeister}},\ }\bibfield  {title} {\bibinfo {title} {Envelope structure of
  four gliding filamentous cyanobacteria},\ }\href@noop {} {\bibfield
  {journal} {\bibinfo  {journal} {J. Bacteriol.}\ }\textbf {\bibinfo {volume}
  {177}},\ \bibinfo {pages} {2387} (\bibinfo {year} {1995})}\BibitemShut
  {NoStop}%
\bibitem [{\citenamefont {Hoiczyk}\ and\ \citenamefont
  {Baumeister}(1998)}]{hoiczyk1998junctional}%
  \BibitemOpen
  \bibfield  {author} {\bibinfo {author} {\bibfnamefont {E.}~\bibnamefont
  {Hoiczyk}}\ and\ \bibinfo {author} {\bibfnamefont {W.}~\bibnamefont
  {Baumeister}},\ }\bibfield  {title} {\bibinfo {title} {The junctional pore
  complex, a prokaryotic secretion organelle, is the molecular motor underlying
  gliding motility in cyanobacteria},\ }\href@noop {} {\bibfield  {journal}
  {\bibinfo  {journal} {Curr. Biol.}\ }\textbf {\bibinfo {volume} {8}},\
  \bibinfo {pages} {1161} (\bibinfo {year} {1998})}\BibitemShut {NoStop}%
\bibitem [{\citenamefont {Wolgemuth}\ and\ \citenamefont
  {Oster}(2004)}]{wolgemuth2004junctional}%
  \BibitemOpen
  \bibfield  {author} {\bibinfo {author} {\bibfnamefont {C.~W.}\ \bibnamefont
  {Wolgemuth}}\ and\ \bibinfo {author} {\bibfnamefont {G.}~\bibnamefont
  {Oster}},\ }\bibfield  {title} {\bibinfo {title} {The junctional pore complex
  and the propulsion of bacterial cells},\ }\href@noop {} {\bibfield  {journal}
  {\bibinfo  {journal} {J. Mol. Microbiol. Biotechnol.}\ }\textbf {\bibinfo
  {volume} {7}},\ \bibinfo {pages} {72} (\bibinfo {year} {2004})}\BibitemShut
  {NoStop}%
\bibitem [{\citenamefont {Ehlers}\ and\ \citenamefont
  {Koiller}(2011)}]{ehlers2011could}%
  \BibitemOpen
  \bibfield  {author} {\bibinfo {author} {\bibfnamefont {K.~M.}\ \bibnamefont
  {Ehlers}}\ and\ \bibinfo {author} {\bibfnamefont {J.}~\bibnamefont
  {Koiller}},\ }\bibfield  {title} {\bibinfo {title} {Could cell membranes
  produce acoustic streaming? making the case for synechococcus
  self-propulsion},\ }\href@noop {} {\bibfield  {journal} {\bibinfo  {journal}
  {Math. Comput. Model.}\ }\textbf {\bibinfo {volume} {53}},\ \bibinfo {pages}
  {1489} (\bibinfo {year} {2011})}\BibitemShut {NoStop}%
\bibitem [{\citenamefont {Du}\ \emph {et~al.}(2022)\citenamefont {Du},
  \citenamefont {Kumari}, \citenamefont {Ye},\ and\ \citenamefont
  {Podgornik}}]{du2022model}%
  \BibitemOpen
  \bibfield  {author} {\bibinfo {author} {\bibfnamefont {G.}~\bibnamefont
  {Du}}, \bibinfo {author} {\bibfnamefont {S.}~\bibnamefont {Kumari}}, \bibinfo
  {author} {\bibfnamefont {F.}~\bibnamefont {Ye}},\ and\ \bibinfo {author}
  {\bibfnamefont {R.}~\bibnamefont {Podgornik}},\ }\bibfield  {title} {\bibinfo
  {title} {Model of metameric locomotion in smooth active directional filaments
  with curvature fluctuations},\ }\href@noop {} {\bibfield  {journal} {\bibinfo
   {journal} {EPL}\ }\textbf {\bibinfo {volume} {136}},\ \bibinfo {pages}
  {58003} (\bibinfo {year} {2022})}\BibitemShut {NoStop}%
\bibitem [{\citenamefont {Rosko}\ \emph {et~al.}(2024)\citenamefont {Rosko},
  \citenamefont {Cremin}, \citenamefont {Locatelli}, \citenamefont {Coates},
  \citenamefont {Duxbury}, \citenamefont {Randall}, \citenamefont {Croft},
  \citenamefont {Valeriani}, \citenamefont {Polin}, \citenamefont {Soyer} \emph
  {et~al.}}]{rosko2024cellular}%
  \BibitemOpen
  \bibfield  {author} {\bibinfo {author} {\bibfnamefont {J.}~\bibnamefont
  {Rosko}}, \bibinfo {author} {\bibfnamefont {K.}~\bibnamefont {Cremin}},
  \bibinfo {author} {\bibfnamefont {E.}~\bibnamefont {Locatelli}}, \bibinfo
  {author} {\bibfnamefont {M.}~\bibnamefont {Coates}}, \bibinfo {author}
  {\bibfnamefont {S.~J.}\ \bibnamefont {Duxbury}}, \bibinfo {author}
  {\bibfnamefont {K.}~\bibnamefont {Randall}}, \bibinfo {author} {\bibfnamefont
  {K.}~\bibnamefont {Croft}}, \bibinfo {author} {\bibfnamefont
  {C.}~\bibnamefont {Valeriani}}, \bibinfo {author} {\bibfnamefont
  {M.}~\bibnamefont {Polin}}, \bibinfo {author} {\bibfnamefont {O.~S.}\
  \bibnamefont {Soyer}}, \emph {et~al.},\ }\bibfield  {title} {\bibinfo {title}
  {Cellular (de) coordination in gliding motility and plectoneme formation},\
  }\href@noop {} {\bibfield  {journal} {\bibinfo  {journal} {bioRxiv}\ }
  (\bibinfo {year} {2024})}\BibitemShut {NoStop}%
\bibitem [{\citenamefont {Abbaspour}\ \emph {et~al.}(2023)\citenamefont
  {Abbaspour}, \citenamefont {Malek}, \citenamefont {Karpitschka},\ and\
  \citenamefont {Klumpp}}]{abbaspour2023effects}%
  \BibitemOpen
  \bibfield  {author} {\bibinfo {author} {\bibfnamefont {L.}~\bibnamefont
  {Abbaspour}}, \bibinfo {author} {\bibfnamefont {A.}~\bibnamefont {Malek}},
  \bibinfo {author} {\bibfnamefont {S.}~\bibnamefont {Karpitschka}},\ and\
  \bibinfo {author} {\bibfnamefont {S.}~\bibnamefont {Klumpp}},\ }\bibfield
  {title} {\bibinfo {title} {Effects of direction reversals on patterns of
  active filaments},\ }\href@noop {} {\bibfield  {journal} {\bibinfo  {journal}
  {Physical Review Research}\ }\textbf {\bibinfo {volume} {5}},\ \bibinfo
  {pages} {013171} (\bibinfo {year} {2023})}\BibitemShut {NoStop}%
\bibitem [{\citenamefont {Pfreundt}\ \emph {et~al.}(2023)\citenamefont
  {Pfreundt}, \citenamefont {S{\l}omka}, \citenamefont {Schneider},
  \citenamefont {Sengupta}, \citenamefont {Carrara}, \citenamefont {Fernandez},
  \citenamefont {Ackermann},\ and\ \citenamefont
  {Stocker}}]{pfreundt2023controlled}%
  \BibitemOpen
  \bibfield  {author} {\bibinfo {author} {\bibfnamefont {U.}~\bibnamefont
  {Pfreundt}}, \bibinfo {author} {\bibfnamefont {J.}~\bibnamefont {S{\l}omka}},
  \bibinfo {author} {\bibfnamefont {G.}~\bibnamefont {Schneider}}, \bibinfo
  {author} {\bibfnamefont {A.}~\bibnamefont {Sengupta}}, \bibinfo {author}
  {\bibfnamefont {F.}~\bibnamefont {Carrara}}, \bibinfo {author} {\bibfnamefont
  {V.}~\bibnamefont {Fernandez}}, \bibinfo {author} {\bibfnamefont
  {M.}~\bibnamefont {Ackermann}},\ and\ \bibinfo {author} {\bibfnamefont
  {R.}~\bibnamefont {Stocker}},\ }\bibfield  {title} {\bibinfo {title}
  {Controlled motility in the cyanobacterium trichodesmium regulates aggregate
  architecture},\ }\href@noop {} {\bibfield  {journal} {\bibinfo  {journal}
  {Science}\ }\textbf {\bibinfo {volume} {380}},\ \bibinfo {pages} {830}
  (\bibinfo {year} {2023})}\BibitemShut {NoStop}%
\bibitem [{\citenamefont {Duman}\ \emph {et~al.}(2018)\citenamefont {Duman},
  \citenamefont {Isele-Holder}, \citenamefont {Elgeti},\ and\ \citenamefont
  {Gompper}}]{duman2018collective}%
  \BibitemOpen
  \bibfield  {author} {\bibinfo {author} {\bibfnamefont {{\"O}.}~\bibnamefont
  {Duman}}, \bibinfo {author} {\bibfnamefont {R.~E.}\ \bibnamefont
  {Isele-Holder}}, \bibinfo {author} {\bibfnamefont {J.}~\bibnamefont
  {Elgeti}},\ and\ \bibinfo {author} {\bibfnamefont {G.}~\bibnamefont
  {Gompper}},\ }\bibfield  {title} {\bibinfo {title} {Collective dynamics of
  self-propelled semiflexible filaments},\ }\href@noop {} {\bibfield  {journal}
  {\bibinfo  {journal} {Soft matter}\ }\textbf {\bibinfo {volume} {14}},\
  \bibinfo {pages} {4483} (\bibinfo {year} {2018})}\BibitemShut {NoStop}%
\bibitem [{\citenamefont {Prathyusha}\ \emph {et~al.}(2018)\citenamefont
  {Prathyusha}, \citenamefont {Henkes},\ and\ \citenamefont
  {Sknepnek}}]{prathyusha2018dynamically}%
  \BibitemOpen
  \bibfield  {author} {\bibinfo {author} {\bibfnamefont {K.~R.}\ \bibnamefont
  {Prathyusha}}, \bibinfo {author} {\bibfnamefont {S.}~\bibnamefont {Henkes}},\
  and\ \bibinfo {author} {\bibfnamefont {R.}~\bibnamefont {Sknepnek}},\
  }\bibfield  {title} {\bibinfo {title} {Dynamically generated patterns in
  dense suspensions of active filaments},\ }\href@noop {} {\bibfield  {journal}
  {\bibinfo  {journal} {Phys. Rev. E}\ }\textbf {\bibinfo {volume} {97}},\
  \bibinfo {pages} {022606} (\bibinfo {year} {2018})}\BibitemShut {NoStop}%
\bibitem [{\citenamefont {Lin}\ \emph {et~al.}(2014)\citenamefont {Lin},
  \citenamefont {Lo},\ and\ \citenamefont {Lo}}]{lin2014dynamics}%
  \BibitemOpen
  \bibfield  {author} {\bibinfo {author} {\bibfnamefont {S.-N.}\ \bibnamefont
  {Lin}}, \bibinfo {author} {\bibfnamefont {W.-C.}\ \bibnamefont {Lo}},\ and\
  \bibinfo {author} {\bibfnamefont {C.-J.}\ \bibnamefont {Lo}},\ }\bibfield
  {title} {\bibinfo {title} {Dynamics of self-organized rotating spiral-coils
  in bacterial swarms},\ }\href@noop {} {\bibfield  {journal} {\bibinfo
  {journal} {Soft Matter}\ }\textbf {\bibinfo {volume} {10}},\ \bibinfo {pages}
  {760} (\bibinfo {year} {2014})}\BibitemShut {NoStop}%
\bibitem [{\citenamefont {Kurzthaler}\ \emph {et~al.}(2021)\citenamefont
  {Kurzthaler}, \citenamefont {Mandal}, \citenamefont {Bhattacharjee},
  \citenamefont {L{\"o}wen}, \citenamefont {Datta},\ and\ \citenamefont
  {Stone}}]{kurzthaler2021geometric}%
  \BibitemOpen
  \bibfield  {author} {\bibinfo {author} {\bibfnamefont {C.}~\bibnamefont
  {Kurzthaler}}, \bibinfo {author} {\bibfnamefont {S.}~\bibnamefont {Mandal}},
  \bibinfo {author} {\bibfnamefont {T.}~\bibnamefont {Bhattacharjee}}, \bibinfo
  {author} {\bibfnamefont {H.}~\bibnamefont {L{\"o}wen}}, \bibinfo {author}
  {\bibfnamefont {S.~S.}\ \bibnamefont {Datta}},\ and\ \bibinfo {author}
  {\bibfnamefont {H.~A.}\ \bibnamefont {Stone}},\ }\bibfield  {title} {\bibinfo
  {title} {A geometric criterion for the optimal spreading of active polymers
  in porous media},\ }\href@noop {} {\bibfield  {journal} {\bibinfo  {journal}
  {Nature communications}\ }\textbf {\bibinfo {volume} {12}},\ \bibinfo {pages}
  {7088} (\bibinfo {year} {2021})}\BibitemShut {NoStop}%
\bibitem [{\citenamefont {Nultsch}\ and\ \citenamefont
  {Wenderoth}(1983)}]{nultsch1983partial}%
  \BibitemOpen
  \bibfield  {author} {\bibinfo {author} {\bibfnamefont {W.}~\bibnamefont
  {Nultsch}}\ and\ \bibinfo {author} {\bibfnamefont {K.}~\bibnamefont
  {Wenderoth}},\ }\bibfield  {title} {\bibinfo {title} {Partial irradiation
  experiments with anabaena variabilis (k{\"u}tz).},\ }\href@noop {} {\bibfield
   {journal} {\bibinfo  {journal} {Zeitschrift f{\"u}r Pflanzenphysiologie}\
  }\textbf {\bibinfo {volume} {111}},\ \bibinfo {pages} {1} (\bibinfo {year}
  {1983})}\BibitemShut {NoStop}%
\bibitem [{\citenamefont {Wilde}\ and\ \citenamefont
  {Mullineaux}(2017)}]{wilde2017light}%
  \BibitemOpen
  \bibfield  {author} {\bibinfo {author} {\bibfnamefont {A.}~\bibnamefont
  {Wilde}}\ and\ \bibinfo {author} {\bibfnamefont {C.~W.}\ \bibnamefont
  {Mullineaux}},\ }\bibfield  {title} {\bibinfo {title} {Light-controlled
  motility in prokaryotes and the problem of directional light perception},\
  }\href@noop {} {\bibfield  {journal} {\bibinfo  {journal} {FEMS Microbiol.
  Rev.}\ }\textbf {\bibinfo {volume} {41}},\ \bibinfo {pages} {900} (\bibinfo
  {year} {2017})}\BibitemShut {NoStop}%
\bibitem [{\citenamefont {Pfeffer}\ \emph {et~al.}(2012)\citenamefont
  {Pfeffer}, \citenamefont {Larsen}, \citenamefont {Song}, \citenamefont
  {Dong}, \citenamefont {Besenbacher}, \citenamefont {Meyer}, \citenamefont
  {Kjeldsen}, \citenamefont {Schreiber}, \citenamefont {Gorby}, \citenamefont
  {El-Naggar} \emph {et~al.}}]{pfeffer2012filamentous}%
  \BibitemOpen
  \bibfield  {author} {\bibinfo {author} {\bibfnamefont {C.}~\bibnamefont
  {Pfeffer}}, \bibinfo {author} {\bibfnamefont {S.}~\bibnamefont {Larsen}},
  \bibinfo {author} {\bibfnamefont {J.}~\bibnamefont {Song}}, \bibinfo {author}
  {\bibfnamefont {M.}~\bibnamefont {Dong}}, \bibinfo {author} {\bibfnamefont
  {F.}~\bibnamefont {Besenbacher}}, \bibinfo {author} {\bibfnamefont {R.~L.}\
  \bibnamefont {Meyer}}, \bibinfo {author} {\bibfnamefont {K.~U.}\ \bibnamefont
  {Kjeldsen}}, \bibinfo {author} {\bibfnamefont {L.}~\bibnamefont {Schreiber}},
  \bibinfo {author} {\bibfnamefont {Y.~A.}\ \bibnamefont {Gorby}}, \bibinfo
  {author} {\bibfnamefont {M.~Y.}\ \bibnamefont {El-Naggar}}, \emph {et~al.},\
  }\bibfield  {title} {\bibinfo {title} {Filamentous bacteria transport
  electrons over centimetre distances},\ }\href@noop {} {\bibfield  {journal}
  {\bibinfo  {journal} {Nature}\ }\textbf {\bibinfo {volume} {491}},\ \bibinfo
  {pages} {218} (\bibinfo {year} {2012})}\BibitemShut {NoStop}%
\bibitem [{\citenamefont {Marzocchi}\ \emph {et~al.}(2014)\citenamefont
  {Marzocchi}, \citenamefont {Trojan}, \citenamefont {Larsen}, \citenamefont
  {Louise~Meyer}, \citenamefont {Peter~Revsbech}, \citenamefont {Schramm},
  \citenamefont {Peter~Nielsen},\ and\ \citenamefont
  {Risgaard-Petersen}}]{marzocchi2014electric}%
  \BibitemOpen
  \bibfield  {author} {\bibinfo {author} {\bibfnamefont {U.}~\bibnamefont
  {Marzocchi}}, \bibinfo {author} {\bibfnamefont {D.}~\bibnamefont {Trojan}},
  \bibinfo {author} {\bibfnamefont {S.}~\bibnamefont {Larsen}}, \bibinfo
  {author} {\bibfnamefont {R.}~\bibnamefont {Louise~Meyer}}, \bibinfo {author}
  {\bibfnamefont {N.}~\bibnamefont {Peter~Revsbech}}, \bibinfo {author}
  {\bibfnamefont {A.}~\bibnamefont {Schramm}}, \bibinfo {author} {\bibfnamefont
  {L.}~\bibnamefont {Peter~Nielsen}},\ and\ \bibinfo {author} {\bibfnamefont
  {N.}~\bibnamefont {Risgaard-Petersen}},\ }\bibfield  {title} {\bibinfo
  {title} {Electric coupling between distant nitrate reduction and sulfide
  oxidation in marine sediment},\ }\href@noop {} {\bibfield  {journal}
  {\bibinfo  {journal} {The ISME journal}\ }\textbf {\bibinfo {volume} {8}},\
  \bibinfo {pages} {1682} (\bibinfo {year} {2014})}\BibitemShut {NoStop}%
\bibitem [{\citenamefont {Nielsen}\ \emph {et~al.}(2010)\citenamefont
  {Nielsen}, \citenamefont {Risgaard-Petersen}, \citenamefont {Fossing},
  \citenamefont {Christensen},\ and\ \citenamefont
  {Sayama}}]{nielsen2010electric}%
  \BibitemOpen
  \bibfield  {author} {\bibinfo {author} {\bibfnamefont {L.~P.}\ \bibnamefont
  {Nielsen}}, \bibinfo {author} {\bibfnamefont {N.}~\bibnamefont
  {Risgaard-Petersen}}, \bibinfo {author} {\bibfnamefont {H.}~\bibnamefont
  {Fossing}}, \bibinfo {author} {\bibfnamefont {P.~B.}\ \bibnamefont
  {Christensen}},\ and\ \bibinfo {author} {\bibfnamefont {M.}~\bibnamefont
  {Sayama}},\ }\bibfield  {title} {\bibinfo {title} {Electric currents couple
  spatially separated biogeochemical processes in marine sediment},\
  }\href@noop {} {\bibfield  {journal} {\bibinfo  {journal} {Nature}\ }\textbf
  {\bibinfo {volume} {463}},\ \bibinfo {pages} {1071} (\bibinfo {year}
  {2010})}\BibitemShut {NoStop}%
\bibitem [{\citenamefont {Bjerg}\ \emph {et~al.}(2023)\citenamefont {Bjerg},
  \citenamefont {Lustermans}, \citenamefont {Marshall}, \citenamefont
  {Mueller}, \citenamefont {Brokj{\ae}r}, \citenamefont {Thorup}, \citenamefont
  {Tataru}, \citenamefont {Schmid}, \citenamefont {Wagner}, \citenamefont
  {Nielsen},\ and\ \citenamefont {Schramm}}]{bjerg2023cable}%
  \BibitemOpen
  \bibfield  {author} {\bibinfo {author} {\bibfnamefont {J.~J.}\ \bibnamefont
  {Bjerg}}, \bibinfo {author} {\bibfnamefont {J.~J.~M.}\ \bibnamefont
  {Lustermans}}, \bibinfo {author} {\bibfnamefont {I.~P.~G.}\ \bibnamefont
  {Marshall}}, \bibinfo {author} {\bibfnamefont {A.~J.}\ \bibnamefont
  {Mueller}}, \bibinfo {author} {\bibfnamefont {S.}~\bibnamefont
  {Brokj{\ae}r}}, \bibinfo {author} {\bibfnamefont {C.~A.}\ \bibnamefont
  {Thorup}}, \bibinfo {author} {\bibfnamefont {P.}~\bibnamefont {Tataru}},
  \bibinfo {author} {\bibfnamefont {M.}~\bibnamefont {Schmid}}, \bibinfo
  {author} {\bibfnamefont {M.}~\bibnamefont {Wagner}}, \bibinfo {author}
  {\bibfnamefont {L.~P.}\ \bibnamefont {Nielsen}},\ and\ \bibinfo {author}
  {\bibfnamefont {A.}~\bibnamefont {Schramm}},\ }\bibfield  {title} {\bibinfo
  {title} {Cable bacteria with electric connection to oxygen attract flocks of
  diverse bacteria},\ }\href@noop {} {\bibfield  {journal} {\bibinfo  {journal}
  {Nat. Commun.}\ }\textbf {\bibinfo {volume} {14}},\ \bibinfo {pages} {1614}
  (\bibinfo {year} {2023})}\BibitemShut {NoStop}%
\bibitem [{\citenamefont {Risgaard-Petersen}\ \emph {et~al.}(2015)\citenamefont
  {Risgaard-Petersen}, \citenamefont {Kristiansen}, \citenamefont
  {Frederiksen}, \citenamefont {Dittmer}, \citenamefont {Bjerg}, \citenamefont
  {Trojan}, \citenamefont {Schreiber}, \citenamefont {Damgaard}, \citenamefont
  {Schramm},\ and\ \citenamefont {Nielsen}}]{risgaard2015cable}%
  \BibitemOpen
  \bibfield  {author} {\bibinfo {author} {\bibfnamefont {N.}~\bibnamefont
  {Risgaard-Petersen}}, \bibinfo {author} {\bibfnamefont {M.}~\bibnamefont
  {Kristiansen}}, \bibinfo {author} {\bibfnamefont {R.~B.}\ \bibnamefont
  {Frederiksen}}, \bibinfo {author} {\bibfnamefont {A.~L.}\ \bibnamefont
  {Dittmer}}, \bibinfo {author} {\bibfnamefont {J.~T.}\ \bibnamefont {Bjerg}},
  \bibinfo {author} {\bibfnamefont {D.}~\bibnamefont {Trojan}}, \bibinfo
  {author} {\bibfnamefont {L.}~\bibnamefont {Schreiber}}, \bibinfo {author}
  {\bibfnamefont {L.~R.}\ \bibnamefont {Damgaard}}, \bibinfo {author}
  {\bibfnamefont {A.}~\bibnamefont {Schramm}},\ and\ \bibinfo {author}
  {\bibfnamefont {L.~P.}\ \bibnamefont {Nielsen}},\ }\bibfield  {title}
  {\bibinfo {title} {Cable bacteria in freshwater sediments},\ }\href@noop {}
  {\bibfield  {journal} {\bibinfo  {journal} {Appl. Environ. Microbiol.}\
  }\textbf {\bibinfo {volume} {81}},\ \bibinfo {pages} {6003} (\bibinfo {year}
  {2015})}\BibitemShut {NoStop}%
\bibitem [{\citenamefont {Reguera}\ \emph {et~al.}(2005)\citenamefont
  {Reguera}, \citenamefont {McCarthy}, \citenamefont {Mehta}, \citenamefont
  {Nicoll}, \citenamefont {Tuominen},\ and\ \citenamefont
  {Lovley}}]{reguera2005extracellular}%
  \BibitemOpen
  \bibfield  {author} {\bibinfo {author} {\bibfnamefont {G.}~\bibnamefont
  {Reguera}}, \bibinfo {author} {\bibfnamefont {K.~D.}\ \bibnamefont
  {McCarthy}}, \bibinfo {author} {\bibfnamefont {T.}~\bibnamefont {Mehta}},
  \bibinfo {author} {\bibfnamefont {J.~S.}\ \bibnamefont {Nicoll}}, \bibinfo
  {author} {\bibfnamefont {M.~T.}\ \bibnamefont {Tuominen}},\ and\ \bibinfo
  {author} {\bibfnamefont {D.~R.}\ \bibnamefont {Lovley}},\ }\bibfield  {title}
  {\bibinfo {title} {Extracellular electron transfer via microbial nanowires},\
  }\href@noop {} {\bibfield  {journal} {\bibinfo  {journal} {Nature}\ }\textbf
  {\bibinfo {volume} {435}},\ \bibinfo {pages} {1098} (\bibinfo {year}
  {2005})}\BibitemShut {NoStop}%
\bibitem [{\citenamefont {El-Naggar}\ \emph {et~al.}(2010)\citenamefont
  {El-Naggar}, \citenamefont {Wanger}, \citenamefont {Leung}, \citenamefont
  {Yuzvinsky}, \citenamefont {Southam}, \citenamefont {Yang}, \citenamefont
  {Lau}, \citenamefont {Nealson},\ and\ \citenamefont
  {Gorby}}]{el2010electrical}%
  \BibitemOpen
  \bibfield  {author} {\bibinfo {author} {\bibfnamefont {M.~Y.}\ \bibnamefont
  {El-Naggar}}, \bibinfo {author} {\bibfnamefont {G.}~\bibnamefont {Wanger}},
  \bibinfo {author} {\bibfnamefont {K.~M.}\ \bibnamefont {Leung}}, \bibinfo
  {author} {\bibfnamefont {T.~D.}\ \bibnamefont {Yuzvinsky}}, \bibinfo {author}
  {\bibfnamefont {G.}~\bibnamefont {Southam}}, \bibinfo {author} {\bibfnamefont
  {J.}~\bibnamefont {Yang}}, \bibinfo {author} {\bibfnamefont {W.~M.}\
  \bibnamefont {Lau}}, \bibinfo {author} {\bibfnamefont {K.~H.}\ \bibnamefont
  {Nealson}},\ and\ \bibinfo {author} {\bibfnamefont {Y.~A.}\ \bibnamefont
  {Gorby}},\ }\bibfield  {title} {\bibinfo {title} {Electrical transport along
  bacterial nanowires from {S}hewanella oneidensis {MR}-1},\ }\href@noop {}
  {\bibfield  {journal} {\bibinfo  {journal} {Proc. Natl. Acad. Sci. USA}\
  }\textbf {\bibinfo {volume} {107}},\ \bibinfo {pages} {18127} (\bibinfo
  {year} {2010})}\BibitemShut {NoStop}%
\bibitem [{\citenamefont {Pirbadian}\ \emph {et~al.}(2014)\citenamefont
  {Pirbadian}, \citenamefont {Barchinger}, \citenamefont {Leung}, \citenamefont
  {Byun}, \citenamefont {Jangir}, \citenamefont {Bouhenni}, \citenamefont
  {Reed}, \citenamefont {Romine}, \citenamefont {Saffarini}, \citenamefont
  {Shi} \emph {et~al.}}]{pirbadian2014shewanella}%
  \BibitemOpen
  \bibfield  {author} {\bibinfo {author} {\bibfnamefont {S.}~\bibnamefont
  {Pirbadian}}, \bibinfo {author} {\bibfnamefont {S.~E.}\ \bibnamefont
  {Barchinger}}, \bibinfo {author} {\bibfnamefont {K.~M.}\ \bibnamefont
  {Leung}}, \bibinfo {author} {\bibfnamefont {H.~S.}\ \bibnamefont {Byun}},
  \bibinfo {author} {\bibfnamefont {Y.}~\bibnamefont {Jangir}}, \bibinfo
  {author} {\bibfnamefont {R.~A.}\ \bibnamefont {Bouhenni}}, \bibinfo {author}
  {\bibfnamefont {S.~B.}\ \bibnamefont {Reed}}, \bibinfo {author}
  {\bibfnamefont {M.~F.}\ \bibnamefont {Romine}}, \bibinfo {author}
  {\bibfnamefont {D.~A.}\ \bibnamefont {Saffarini}}, \bibinfo {author}
  {\bibfnamefont {L.}~\bibnamefont {Shi}}, \emph {et~al.},\ }\bibfield  {title}
  {\bibinfo {title} {Shewanella oneidensis {MR}-1 nanowires are outer membrane
  and periplasmic extensions of the extracellular electron transport
  components},\ }\href@noop {} {\bibfield  {journal} {\bibinfo  {journal}
  {Proc. Natl. Acad. Sci. USA}\ }\textbf {\bibinfo {volume} {111}},\ \bibinfo
  {pages} {12883} (\bibinfo {year} {2014})}\BibitemShut {NoStop}%
\bibitem [{\citenamefont {Schkolnik}\ \emph {et~al.}(2015)\citenamefont
  {Schkolnik}, \citenamefont {Schmidt}, \citenamefont {Mazza}, \citenamefont
  {Harnisch},\ and\ \citenamefont {Musat}}]{schkolnik2015situ}%
  \BibitemOpen
  \bibfield  {author} {\bibinfo {author} {\bibfnamefont {G.}~\bibnamefont
  {Schkolnik}}, \bibinfo {author} {\bibfnamefont {M.}~\bibnamefont {Schmidt}},
  \bibinfo {author} {\bibfnamefont {M.~G.}\ \bibnamefont {Mazza}}, \bibinfo
  {author} {\bibfnamefont {F.}~\bibnamefont {Harnisch}},\ and\ \bibinfo
  {author} {\bibfnamefont {N.}~\bibnamefont {Musat}},\ }\bibfield  {title}
  {\bibinfo {title} {In situ analysis of a silver nanoparticle-precipitating
  shewanella biofilm by surface enhanced confocal {R}aman microscopy},\
  }\href@noop {} {\bibfield  {journal} {\bibinfo  {journal} {PLoS One}\
  }\textbf {\bibinfo {volume} {10}},\ \bibinfo {pages} {e0145871} (\bibinfo
  {year} {2015})}\BibitemShut {NoStop}%
\bibitem [{\citenamefont {Logan}\ \emph {et~al.}(2019)\citenamefont {Logan},
  \citenamefont {Rossi}, \citenamefont {Ragab},\ and\ \citenamefont
  {Saikaly}}]{logan2019electroactive}%
  \BibitemOpen
  \bibfield  {author} {\bibinfo {author} {\bibfnamefont {B.~E.}\ \bibnamefont
  {Logan}}, \bibinfo {author} {\bibfnamefont {R.}~\bibnamefont {Rossi}},
  \bibinfo {author} {\bibfnamefont {A.}~\bibnamefont {Ragab}},\ and\ \bibinfo
  {author} {\bibfnamefont {P.~E.}\ \bibnamefont {Saikaly}},\ }\bibfield
  {title} {\bibinfo {title} {Electroactive microorganisms in bioelectrochemical
  systems},\ }\href@noop {} {\bibfield  {journal} {\bibinfo  {journal} {Nat.
  Rev. Microbiol.}\ }\textbf {\bibinfo {volume} {17}},\ \bibinfo {pages} {307}
  (\bibinfo {year} {2019})}\BibitemShut {NoStop}%
\bibitem [{\citenamefont {Fazelzadeh}\ \emph {et~al.}(2023)\citenamefont
  {Fazelzadeh}, \citenamefont {Di}, \citenamefont {Irani}, \citenamefont
  {Mokhtari},\ and\ \citenamefont {Jabbari-Farouji}}]{fazelzadeh2023active}%
  \BibitemOpen
  \bibfield  {author} {\bibinfo {author} {\bibfnamefont {M.}~\bibnamefont
  {Fazelzadeh}}, \bibinfo {author} {\bibfnamefont {Q.}~\bibnamefont {Di}},
  \bibinfo {author} {\bibfnamefont {E.}~\bibnamefont {Irani}}, \bibinfo
  {author} {\bibfnamefont {Z.}~\bibnamefont {Mokhtari}},\ and\ \bibinfo
  {author} {\bibfnamefont {S.}~\bibnamefont {Jabbari-Farouji}},\ }\bibfield
  {title} {\bibinfo {title} {Active motion of tangentially driven polymers in
  periodic array of obstacles},\ }\href@noop {} {\bibfield  {journal} {\bibinfo
   {journal} {J. Chem. Phys.}\ }\textbf {\bibinfo {volume} {159}} (\bibinfo
  {year} {2023})}\BibitemShut {NoStop}%
\bibitem [{\citenamefont {Kurtz~Jr}\ and\ \citenamefont
  {Netoff}(2001)}]{kurtz2001stabilization}%
  \BibitemOpen
  \bibfield  {author} {\bibinfo {author} {\bibfnamefont {H.~D.}\ \bibnamefont
  {Kurtz~Jr}}\ and\ \bibinfo {author} {\bibfnamefont {D.~I.}\ \bibnamefont
  {Netoff}},\ }\bibfield  {title} {\bibinfo {title} {Stabilization of friable
  sandstone surfaces in a desiccating, wind-abraded environment of
  south-central utah by rock surface microorganisms},\ }\href@noop {}
  {\bibfield  {journal} {\bibinfo  {journal} {Journal of Arid Environments}\
  }\textbf {\bibinfo {volume} {48}},\ \bibinfo {pages} {89} (\bibinfo {year}
  {2001})}\BibitemShut {NoStop}%
\bibitem [{\citenamefont {Pluis}(1994)}]{pluis1994algal}%
  \BibitemOpen
  \bibfield  {author} {\bibinfo {author} {\bibfnamefont {J.~L.~A.}\
  \bibnamefont {Pluis}},\ }\bibfield  {title} {\bibinfo {title} {Algal crust
  formation in the inland dune area, {L}aarder {W}asmeer, {T}he
  {N}etherlands},\ }\href@noop {} {\bibfield  {journal} {\bibinfo  {journal}
  {Vegetatio}\ }\textbf {\bibinfo {volume} {113}},\ \bibinfo {pages} {41}
  (\bibinfo {year} {1994})}\BibitemShut {NoStop}%
\bibitem [{\citenamefont {Corte}\ \emph {et~al.}(2025)\citenamefont {Corte},
  \citenamefont {Stevens}, \citenamefont {Cárcamo-Oyarce}, \citenamefont
  {Ribbeck}, \citenamefont {Wingreen},\ and\ \citenamefont
  {Datta}}]{Gonzalez2025}%
  \BibitemOpen
  \bibfield  {author} {\bibinfo {author} {\bibfnamefont {S.~G.~L.}\
  \bibnamefont {Corte}}, \bibinfo {author} {\bibfnamefont {C.~A.}\ \bibnamefont
  {Stevens}}, \bibinfo {author} {\bibfnamefont {G.}~\bibnamefont
  {Cárcamo-Oyarce}}, \bibinfo {author} {\bibfnamefont {K.}~\bibnamefont
  {Ribbeck}}, \bibinfo {author} {\bibfnamefont {N.~S.}\ \bibnamefont
  {Wingreen}},\ and\ \bibinfo {author} {\bibfnamefont {S.~S.}\ \bibnamefont
  {Datta}},\ }\bibfield  {title} {\bibinfo {title} {Morphogenesis of bacterial
  cables in polymeric environments},\ }\href
  {https://doi.org/10.1126/sciadv.adq7797} {\bibfield  {journal} {\bibinfo
  {journal} {Science Advances}\ }\textbf {\bibinfo {volume} {11}},\ \bibinfo
  {pages} {eadq7797} (\bibinfo {year} {2025})}\BibitemShut {NoStop}%
\bibitem [{\citenamefont {Shepard}\ and\ \citenamefont
  {Sumner}(2010)}]{shepard2010undirected}%
  \BibitemOpen
  \bibfield  {author} {\bibinfo {author} {\bibfnamefont {R.}~\bibnamefont
  {Shepard}}\ and\ \bibinfo {author} {\bibfnamefont {D.}~\bibnamefont
  {Sumner}},\ }\bibfield  {title} {\bibinfo {title} {Undirected motility of
  filamentous cyanobacteria produces reticulate mats},\ }\href@noop {}
  {\bibfield  {journal} {\bibinfo  {journal} {Geobiology}\ }\textbf {\bibinfo
  {volume} {8}},\ \bibinfo {pages} {179} (\bibinfo {year} {2010})}\BibitemShut
  {NoStop}%
\bibitem [{\citenamefont {Tamulonis}\ and\ \citenamefont
  {Kaandorp}(2014)}]{tamulonis2014model}%
  \BibitemOpen
  \bibfield  {author} {\bibinfo {author} {\bibfnamefont {C.}~\bibnamefont
  {Tamulonis}}\ and\ \bibinfo {author} {\bibfnamefont {J.}~\bibnamefont
  {Kaandorp}},\ }\bibfield  {title} {\bibinfo {title} {A model of filamentous
  cyanobacteria leading to reticulate pattern formation},\ }\href@noop {}
  {\bibfield  {journal} {\bibinfo  {journal} {Life}\ }\textbf {\bibinfo
  {volume} {4}},\ \bibinfo {pages} {433} (\bibinfo {year} {2014})}\BibitemShut
  {NoStop}%
\bibitem [{\citenamefont {Cuadrado}\ and\ \citenamefont
  {Pan}(2018)}]{cuadrado2018field}%
  \BibitemOpen
  \bibfield  {author} {\bibinfo {author} {\bibfnamefont {D.~G.}\ \bibnamefont
  {Cuadrado}}\ and\ \bibinfo {author} {\bibfnamefont {J.}~\bibnamefont {Pan}},\
  }\bibfield  {title} {\bibinfo {title} {Field observations on the evolution of
  reticulate patterns in microbial mats in a modern siliciclastic coastal
  environment},\ }\href@noop {} {\bibfield  {journal} {\bibinfo  {journal}
  {Journal of Sedimentary Research}\ }\textbf {\bibinfo {volume} {88}},\
  \bibinfo {pages} {24} (\bibinfo {year} {2018})}\BibitemShut {NoStop}%
\bibitem [{\citenamefont {Mazza}(2016)}]{mazza2016physics}%
  \BibitemOpen
  \bibfield  {author} {\bibinfo {author} {\bibfnamefont {M.~G.}\ \bibnamefont
  {Mazza}},\ }\bibfield  {title} {\bibinfo {title} {The physics of biofilms--an
  introduction},\ }\href@noop {} {\bibfield  {journal} {\bibinfo  {journal} {J.
  Phys. D: Appl. Phys.}\ }\textbf {\bibinfo {volume} {49}},\ \bibinfo {pages}
  {203001} (\bibinfo {year} {2016})}\BibitemShut {NoStop}%
\bibitem [{\citenamefont {Sauer}\ \emph {et~al.}(2022)\citenamefont {Sauer},
  \citenamefont {Stoodley}, \citenamefont {Goeres}, \citenamefont
  {Hall-Stoodley}, \citenamefont {Burm{\o}lle}, \citenamefont {Stewart},\ and\
  \citenamefont {Bjarnsholt}}]{sauer2022biofilm}%
  \BibitemOpen
  \bibfield  {author} {\bibinfo {author} {\bibfnamefont {K.}~\bibnamefont
  {Sauer}}, \bibinfo {author} {\bibfnamefont {P.}~\bibnamefont {Stoodley}},
  \bibinfo {author} {\bibfnamefont {D.~M.}\ \bibnamefont {Goeres}}, \bibinfo
  {author} {\bibfnamefont {L.}~\bibnamefont {Hall-Stoodley}}, \bibinfo {author}
  {\bibfnamefont {M.}~\bibnamefont {Burm{\o}lle}}, \bibinfo {author}
  {\bibfnamefont {P.~S.}\ \bibnamefont {Stewart}},\ and\ \bibinfo {author}
  {\bibfnamefont {T.}~\bibnamefont {Bjarnsholt}},\ }\bibfield  {title}
  {\bibinfo {title} {The biofilm life cycle: expanding the conceptual model of
  biofilm formation},\ }\href@noop {} {\bibfield  {journal} {\bibinfo
  {journal} {Nat. Rev. Microbiol.}\ }\textbf {\bibinfo {volume} {20}},\
  \bibinfo {pages} {608} (\bibinfo {year} {2022})}\BibitemShut {NoStop}%
\bibitem [{\citenamefont {Costa}\ \emph {et~al.}(2024)\citenamefont {Costa},
  \citenamefont {Ahamed}, \citenamefont {Jordan},\ and\ \citenamefont
  {Stephens}}]{Costa2024}%
  \BibitemOpen
  \bibfield  {author} {\bibinfo {author} {\bibfnamefont {A.~C.}\ \bibnamefont
  {Costa}}, \bibinfo {author} {\bibfnamefont {T.}~\bibnamefont {Ahamed}},
  \bibinfo {author} {\bibfnamefont {D.}~\bibnamefont {Jordan}},\ and\ \bibinfo
  {author} {\bibfnamefont {G.~J.}\ \bibnamefont {Stephens}},\ }\bibfield
  {title} {\bibinfo {title} {A markovian dynamics for caenorhabditis elegans
  behavior across scales},\ }\href@noop {} {\bibfield  {journal} {\bibinfo
  {journal} {Proc. Natl. Acad. Sci. USA}\ }\textbf {\bibinfo {volume} {121}},\
  \bibinfo {pages} {e2318805121} (\bibinfo {year} {2024})}\BibitemShut
  {NoStop}%
\bibitem [{\citenamefont {Stephens}\ \emph {et~al.}(2008)\citenamefont
  {Stephens}, \citenamefont {Johnson-Kerner}, \citenamefont {Bialek},\ and\
  \citenamefont {Ryu}}]{Stephens2008}%
  \BibitemOpen
  \bibfield  {author} {\bibinfo {author} {\bibfnamefont {G.~J.}\ \bibnamefont
  {Stephens}}, \bibinfo {author} {\bibfnamefont {B.}~\bibnamefont
  {Johnson-Kerner}}, \bibinfo {author} {\bibfnamefont {W.}~\bibnamefont
  {Bialek}},\ and\ \bibinfo {author} {\bibfnamefont {W.~S.}\ \bibnamefont
  {Ryu}},\ }\bibfield  {title} {\bibinfo {title} {Dimensionality and dynamics
  in the behavior of c. elegans},\ }\href
  {https://doi.org/10.1371/journal.pcbi.1000028} {\bibfield  {journal}
  {\bibinfo  {journal} {PLoS Comput. Biol.}\ }\textbf {\bibinfo {volume} {4}},\
  \bibinfo {pages} {e1000028} (\bibinfo {year} {2008})}\BibitemShut {NoStop}%
\bibitem [{\citenamefont {Gillis}(1998)}]{Gillis1998}%
  \BibitemOpen
  \bibfield  {author} {\bibinfo {author} {\bibfnamefont {G.~B.}\ \bibnamefont
  {Gillis}},\ }\bibfield  {title} {\bibinfo {title} {Environmental effects on
  undulatory locomotion in the american eel anguilla rostrata: Kinematics in
  water and on land},\ }\href {https://doi.org/10.1242/jeb.201.7.949}
  {\bibfield  {journal} {\bibinfo  {journal} {J. Exp. Biol.}\ }\textbf
  {\bibinfo {volume} {201}},\ \bibinfo {pages} {949–961} (\bibinfo {year}
  {1998})}\BibitemShut {NoStop}%
\bibitem [{\citenamefont {Jayne}(2020)}]{Jayne2020}%
  \BibitemOpen
  \bibfield  {author} {\bibinfo {author} {\bibfnamefont {B.~C.}\ \bibnamefont
  {Jayne}},\ }\bibfield  {title} {\bibinfo {title} {What defines different
  modes of snake locomotion?},\ }\href {https://doi.org/10.1093/icb/icaa017}
  {\bibfield  {journal} {\bibinfo  {journal} {Integr. Comp. Biol.}\ }\textbf
  {\bibinfo {volume} {60}},\ \bibinfo {pages} {156–170} (\bibinfo {year}
  {2020})}\BibitemShut {NoStop}%
\bibitem [{\citenamefont {Grillner}\ \emph {et~al.}(1991)\citenamefont
  {Grillner}, \citenamefont {Wallen}, \citenamefont {Brodin},\ and\
  \citenamefont {Lansner}}]{Grillner1991}%
  \BibitemOpen
  \bibfield  {author} {\bibinfo {author} {\bibfnamefont {S.}~\bibnamefont
  {Grillner}}, \bibinfo {author} {\bibfnamefont {P.}~\bibnamefont {Wallen}},
  \bibinfo {author} {\bibfnamefont {L.}~\bibnamefont {Brodin}},\ and\ \bibinfo
  {author} {\bibfnamefont {A.}~\bibnamefont {Lansner}},\ }\bibfield  {title}
  {\bibinfo {title} {Neuronal network generating locomotor behavior in lamprey:
  Circuitry, transmitters, membrane properties, and simulation},\ }\href
  {https://doi.org/10.1146/annurev.ne.14.030191.001125} {\bibfield  {journal}
  {\bibinfo  {journal} {Annu. Rev. Neurosci.}\ }\textbf {\bibinfo {volume}
  {14}},\ \bibinfo {pages} {169–199} (\bibinfo {year} {1991})}\BibitemShut
  {NoStop}%
\bibitem [{\citenamefont {Katz}(2016)}]{Katz2016}%
  \BibitemOpen
  \bibfield  {author} {\bibinfo {author} {\bibfnamefont {P.~S.}\ \bibnamefont
  {Katz}},\ }\bibfield  {title} {\bibinfo {title} {Evolution of central pattern
  generators and rhythmic behaviours},\ }\href
  {https://doi.org/10.1098/rstb.2015.0057} {\bibfield  {journal} {\bibinfo
  {journal} {Philos. Trans. R. Soc. B}\ }\textbf {\bibinfo {volume} {371}},\
  \bibinfo {pages} {20150057} (\bibinfo {year} {2016})}\BibitemShut {NoStop}%
\bibitem [{\citenamefont {Ramdya}\ and\ \citenamefont
  {Ijspeert}(2023)}]{Ramdya2023}%
  \BibitemOpen
  \bibfield  {author} {\bibinfo {author} {\bibfnamefont {P.}~\bibnamefont
  {Ramdya}}\ and\ \bibinfo {author} {\bibfnamefont {A.~J.}\ \bibnamefont
  {Ijspeert}},\ }\bibfield  {title} {\bibinfo {title} {The neuromechanics of
  animal locomotion: From biology to robotics and back},\ }\href
  {http://dx.doi.org/10.1126/scirobotics.adg0279} {\bibfield  {journal}
  {\bibinfo  {journal} {Sci. Robot.}\ }\textbf {\bibinfo {volume} {8}},\
  \bibinfo {pages} {eadg0279} (\bibinfo {year} {2023})}\BibitemShut {NoStop}%
\bibitem [{\citenamefont {Hatton}\ and\ \citenamefont
  {Choset}(2011)}]{Hatton2011}%
  \BibitemOpen
  \bibfield  {author} {\bibinfo {author} {\bibfnamefont {R.~L.}\ \bibnamefont
  {Hatton}}\ and\ \bibinfo {author} {\bibfnamefont {H.}~\bibnamefont
  {Choset}},\ }\bibfield  {title} {\bibinfo {title} {Geometric motion planning:
  The local connection, stokes’ theorem, and the importance of coordinate
  choice},\ }\href {https://doi.org/10.1177/0278364910394392} {\bibfield
  {journal} {\bibinfo  {journal} {Int. J. Robot. Res.}\ }\textbf {\bibinfo
  {volume} {30}},\ \bibinfo {pages} {988–1014} (\bibinfo {year}
  {2011})}\BibitemShut {NoStop}%
\bibitem [{\citenamefont {Zhang}\ and\ \citenamefont
  {Goldman}(2014)}]{Zhang2014}%
  \BibitemOpen
  \bibfield  {author} {\bibinfo {author} {\bibfnamefont {T.}~\bibnamefont
  {Zhang}}\ and\ \bibinfo {author} {\bibfnamefont {D.~I.}\ \bibnamefont
  {Goldman}},\ }\bibfield  {title} {\bibinfo {title} {The effectiveness of
  resistive force theory in granular locomotion},\ }\href
  {http://dx.doi.org/10.1063/1.4898629} {\bibfield  {journal} {\bibinfo
  {journal} {Phys. Fluids}\ }\textbf {\bibinfo {volume} {26}},\ \bibinfo
  {pages} {101308} (\bibinfo {year} {2014})}\BibitemShut {NoStop}%
\bibitem [{\citenamefont {Padmanabhan}\ \emph {et~al.}(2012)\citenamefont
  {Padmanabhan}, \citenamefont {Khan}, \citenamefont {Solomon}, \citenamefont
  {Armstrong}, \citenamefont {Rumbaugh}, \citenamefont {Vanapalli},\ and\
  \citenamefont {Blawzdziewicz}}]{padmanabhan2012locomotion}%
  \BibitemOpen
  \bibfield  {author} {\bibinfo {author} {\bibfnamefont {V.}~\bibnamefont
  {Padmanabhan}}, \bibinfo {author} {\bibfnamefont {Z.~S.}\ \bibnamefont
  {Khan}}, \bibinfo {author} {\bibfnamefont {D.~E.}\ \bibnamefont {Solomon}},
  \bibinfo {author} {\bibfnamefont {A.}~\bibnamefont {Armstrong}}, \bibinfo
  {author} {\bibfnamefont {K.~P.}\ \bibnamefont {Rumbaugh}}, \bibinfo {author}
  {\bibfnamefont {S.~A.}\ \bibnamefont {Vanapalli}},\ and\ \bibinfo {author}
  {\bibfnamefont {J.}~\bibnamefont {Blawzdziewicz}},\ }\bibfield  {title}
  {\bibinfo {title} {Locomotion of {C}. elegans: a piecewise-harmonic curvature
  representation of nematode behavior},\ }\href@noop {} {\bibfield  {journal}
  {\bibinfo  {journal} {PloS One}\ }\textbf {\bibinfo {volume} {7}},\ \bibinfo
  {pages} {e40121} (\bibinfo {year} {2012})}\BibitemShut {NoStop}%
\bibitem [{\citenamefont {Rieser}\ \emph {et~al.}(2021)\citenamefont {Rieser},
  \citenamefont {Li}, \citenamefont {Tingle}, \citenamefont {Goldman},\ and\
  \citenamefont {Mendelson~III}}]{rieser2021functional}%
  \BibitemOpen
  \bibfield  {author} {\bibinfo {author} {\bibfnamefont {J.~M.}\ \bibnamefont
  {Rieser}}, \bibinfo {author} {\bibfnamefont {T.-D.}\ \bibnamefont {Li}},
  \bibinfo {author} {\bibfnamefont {J.~L.}\ \bibnamefont {Tingle}}, \bibinfo
  {author} {\bibfnamefont {D.~I.}\ \bibnamefont {Goldman}},\ and\ \bibinfo
  {author} {\bibfnamefont {J.~R.}\ \bibnamefont {Mendelson~III}},\ }\bibfield
  {title} {\bibinfo {title} {Functional consequences of convergently evolved
  microscopic skin features on snake locomotion},\ }\href@noop {} {\bibfield
  {journal} {\bibinfo  {journal} {Proc. Natl. Acad. Sci. USA}\ }\textbf
  {\bibinfo {volume} {118}},\ \bibinfo {pages} {e2018264118} (\bibinfo {year}
  {2021})}\BibitemShut {NoStop}%
\bibitem [{\citenamefont {Peshkov}\ \emph {et~al.}(2022)\citenamefont
  {Peshkov}, \citenamefont {McGaffigan},\ and\ \citenamefont
  {Quillen}}]{peshkov2022synchronized}%
  \BibitemOpen
  \bibfield  {author} {\bibinfo {author} {\bibfnamefont {A.}~\bibnamefont
  {Peshkov}}, \bibinfo {author} {\bibfnamefont {S.}~\bibnamefont
  {McGaffigan}},\ and\ \bibinfo {author} {\bibfnamefont {A.~C.}\ \bibnamefont
  {Quillen}},\ }\bibfield  {title} {\bibinfo {title} {Synchronized oscillations
  in swarms of nematode turbatrix aceti},\ }\href@noop {} {\bibfield  {journal}
  {\bibinfo  {journal} {Soft Matter}\ }\textbf {\bibinfo {volume} {18}},\
  \bibinfo {pages} {1174} (\bibinfo {year} {2022})}\BibitemShut {NoStop}%
\bibitem [{\citenamefont {Wu}\ and\ \citenamefont {Ma}(2010)}]{wu2010cpg}%
  \BibitemOpen
  \bibfield  {author} {\bibinfo {author} {\bibfnamefont {X.}~\bibnamefont
  {Wu}}\ and\ \bibinfo {author} {\bibfnamefont {S.}~\bibnamefont {Ma}},\
  }\bibfield  {title} {\bibinfo {title} {{CPG}-based control of serpentine
  locomotion of a snake-like robot},\ }\href@noop {} {\bibfield  {journal}
  {\bibinfo  {journal} {Mechatronics}\ }\textbf {\bibinfo {volume} {20}},\
  \bibinfo {pages} {326} (\bibinfo {year} {2010})}\BibitemShut {NoStop}%
\bibitem [{\citenamefont {Yemini}\ \emph {et~al.}(2013)\citenamefont {Yemini},
  \citenamefont {Jucikas}, \citenamefont {Grundy}, \citenamefont {Brown},\ and\
  \citenamefont {Schafer}}]{Yemini2013}%
  \BibitemOpen
  \bibfield  {author} {\bibinfo {author} {\bibfnamefont {E.}~\bibnamefont
  {Yemini}}, \bibinfo {author} {\bibfnamefont {T.}~\bibnamefont {Jucikas}},
  \bibinfo {author} {\bibfnamefont {L.~J.}\ \bibnamefont {Grundy}}, \bibinfo
  {author} {\bibfnamefont {A.~E.~X.}\ \bibnamefont {Brown}},\ and\ \bibinfo
  {author} {\bibfnamefont {W.~R.}\ \bibnamefont {Schafer}},\ }\bibfield
  {title} {\bibinfo {title} {A database of caenorhabditis elegans behavioral
  phenotypes},\ }\href {https://doi.org/10.1038/nmeth.2560} {\bibfield
  {journal} {\bibinfo  {journal} {Nat. Methods}\ }\textbf {\bibinfo {volume}
  {10}},\ \bibinfo {pages} {877–879} (\bibinfo {year} {2013})}\BibitemShut
  {NoStop}%
\bibitem [{\citenamefont {Rieser}\ \emph {et~al.}(2024)\citenamefont {Rieser},
  \citenamefont {Chong}, \citenamefont {Gong}, \citenamefont {Astley},
  \citenamefont {Schiebel}, \citenamefont {Diaz}, \citenamefont {Pierce},
  \citenamefont {Lu}, \citenamefont {Hatton}, \citenamefont {Choset},\ and\
  \citenamefont {Goldman}}]{Rieser2024}%
  \BibitemOpen
  \bibfield  {author} {\bibinfo {author} {\bibfnamefont {J.~M.}\ \bibnamefont
  {Rieser}}, \bibinfo {author} {\bibfnamefont {B.}~\bibnamefont {Chong}},
  \bibinfo {author} {\bibfnamefont {C.}~\bibnamefont {Gong}}, \bibinfo {author}
  {\bibfnamefont {H.~C.}\ \bibnamefont {Astley}}, \bibinfo {author}
  {\bibfnamefont {P.~E.}\ \bibnamefont {Schiebel}}, \bibinfo {author}
  {\bibfnamefont {K.}~\bibnamefont {Diaz}}, \bibinfo {author} {\bibfnamefont
  {C.~J.}\ \bibnamefont {Pierce}}, \bibinfo {author} {\bibfnamefont
  {H.}~\bibnamefont {Lu}}, \bibinfo {author} {\bibfnamefont {R.~L.}\
  \bibnamefont {Hatton}}, \bibinfo {author} {\bibfnamefont {H.}~\bibnamefont
  {Choset}},\ and\ \bibinfo {author} {\bibfnamefont {D.~I.}\ \bibnamefont
  {Goldman}},\ }\bibfield  {title} {\bibinfo {title} {Geometric phase predicts
  locomotion performance in undulating living systems across scales},\ }\href
  {http://dx.doi.org/10.1073/pnas.2320517121} {\bibfield  {journal} {\bibinfo
  {journal} {Proc. Natl. Acad. Sci. USA}\ }\textbf {\bibinfo {volume} {121}},\
  \bibinfo {pages} {e2320517121} (\bibinfo {year} {2024})}\BibitemShut
  {NoStop}%
\bibitem [{\citenamefont {Pierce}\ \emph {et~al.}(2024)\citenamefont {Pierce},
  \citenamefont {Irvine}, \citenamefont {Peng}, \citenamefont {Lu},
  \citenamefont {Lu},\ and\ \citenamefont {Goldman}}]{Pierce2024}%
  \BibitemOpen
  \bibfield  {author} {\bibinfo {author} {\bibfnamefont {C.~J.}\ \bibnamefont
  {Pierce}}, \bibinfo {author} {\bibfnamefont {D.}~\bibnamefont {Irvine}},
  \bibinfo {author} {\bibfnamefont {L.}~\bibnamefont {Peng}}, \bibinfo {author}
  {\bibfnamefont {X.}~\bibnamefont {Lu}}, \bibinfo {author} {\bibfnamefont
  {H.}~\bibnamefont {Lu}},\ and\ \bibinfo {author} {\bibfnamefont {D.~I.}\
  \bibnamefont {Goldman}},\ }\bibfield  {title} {\bibinfo {title} {Dispersion
  relations for active undulators in overdamped environments},\ }\href@noop {}
  {\bibfield  {journal} {\bibinfo  {journal} {arxiv.org/abs/2407.13037}\ ,\
  \bibinfo {pages} {arXiv}} (\bibinfo {year} {2024})}\BibitemShut {NoStop}%
\bibitem [{\citenamefont {Sánchez-Rodríguez}\ \emph
  {et~al.}(2023)\citenamefont {Sánchez-Rodríguez}, \citenamefont {Raufaste},\
  and\ \citenamefont {Argentina}}]{SnchezRodrguez2023}%
  \BibitemOpen
  \bibfield  {author} {\bibinfo {author} {\bibfnamefont {J.}~\bibnamefont
  {Sánchez-Rodríguez}}, \bibinfo {author} {\bibfnamefont {C.}~\bibnamefont
  {Raufaste}},\ and\ \bibinfo {author} {\bibfnamefont {M.}~\bibnamefont
  {Argentina}},\ }\bibfield  {title} {\bibinfo {title} {Scaling the tail beat
  frequency and swimming speed in underwater undulatory swimming},\ }\href
  {http://dx.doi.org/10.1038/s41467-023-41368-6} {\bibfield  {journal}
  {\bibinfo  {journal} {Nat. Commun.}\ }\textbf {\bibinfo {volume} {14}},\
  \bibinfo {pages} {5569} (\bibinfo {year} {2023})}\BibitemShut {NoStop}%
\bibitem [{\citenamefont {Ozkan-Aydin}\ \emph {et~al.}(2021)\citenamefont
  {Ozkan-Aydin}, \citenamefont {Goldman},\ and\ \citenamefont
  {Bhamla}}]{OzkanAydin2021}%
  \BibitemOpen
  \bibfield  {author} {\bibinfo {author} {\bibfnamefont {Y.}~\bibnamefont
  {Ozkan-Aydin}}, \bibinfo {author} {\bibfnamefont {D.~I.}\ \bibnamefont
  {Goldman}},\ and\ \bibinfo {author} {\bibfnamefont {M.~S.}\ \bibnamefont
  {Bhamla}},\ }\bibfield  {title} {\bibinfo {title} {Collective dynamics in
  entangled worm and robot blobs},\ }\href@noop {} {\bibfield  {journal}
  {\bibinfo  {journal} {Proc. Natl. Acad. Sci. USA}\ }\textbf {\bibinfo
  {volume} {118}},\ \bibinfo {pages} {e2010542118} (\bibinfo {year}
  {2021})}\BibitemShut {NoStop}%
\bibitem [{\citenamefont {Heeremans}\ \emph {et~al.}(2022)\citenamefont
  {Heeremans}, \citenamefont {Deblais}, \citenamefont {Bonn},\ and\
  \citenamefont {Woutersen}}]{Heeremans2022}%
  \BibitemOpen
  \bibfield  {author} {\bibinfo {author} {\bibfnamefont {T.}~\bibnamefont
  {Heeremans}}, \bibinfo {author} {\bibfnamefont {A.}~\bibnamefont {Deblais}},
  \bibinfo {author} {\bibfnamefont {D.}~\bibnamefont {Bonn}},\ and\ \bibinfo
  {author} {\bibfnamefont {S.}~\bibnamefont {Woutersen}},\ }\bibfield  {title}
  {\bibinfo {title} {Chromatographic separation of active polymer–like worm
  mixtures by contour length and activity},\ }\href
  {https://doi.org/10.1126/sciadv.abj7918} {\bibfield  {journal} {\bibinfo
  {journal} {Sci. Adv.}\ }\textbf {\bibinfo {volume} {8}},\ \bibinfo {pages}
  {eabj7918} (\bibinfo {year} {2022})}\BibitemShut {NoStop}%
\bibitem [{\citenamefont {Quillen}\ \emph {et~al.}(2021)\citenamefont
  {Quillen}, \citenamefont {Peshkov}, \citenamefont {Wright},\ and\
  \citenamefont {McGaffigan}}]{quillen2021metachronal}%
  \BibitemOpen
  \bibfield  {author} {\bibinfo {author} {\bibfnamefont {A.}~\bibnamefont
  {Quillen}}, \bibinfo {author} {\bibfnamefont {A.}~\bibnamefont {Peshkov}},
  \bibinfo {author} {\bibfnamefont {E.}~\bibnamefont {Wright}},\ and\ \bibinfo
  {author} {\bibfnamefont {S.}~\bibnamefont {McGaffigan}},\ }\bibfield  {title}
  {\bibinfo {title} {Metachronal waves in concentrations of swimming turbatrix
  aceti nematodes and an oscillator chain model for their coordinated
  motions},\ }\href@noop {} {\bibfield  {journal} {\bibinfo  {journal} {Phys.
  Rev. E}\ }\textbf {\bibinfo {volume} {104}},\ \bibinfo {pages} {014412}
  (\bibinfo {year} {2021})}\BibitemShut {NoStop}%
\bibitem [{\citenamefont {Deblais}\ \emph {et~al.}(2023)\citenamefont
  {Deblais}, \citenamefont {Prathyusha}, \citenamefont {Sinaasappel},
  \citenamefont {Tuazon}, \citenamefont {Tiwari}, \citenamefont {Patil},\ and\
  \citenamefont {Bhamla}}]{Deblais2023}%
  \BibitemOpen
  \bibfield  {author} {\bibinfo {author} {\bibfnamefont {A.}~\bibnamefont
  {Deblais}}, \bibinfo {author} {\bibfnamefont {K.~R.}\ \bibnamefont
  {Prathyusha}}, \bibinfo {author} {\bibfnamefont {R.}~\bibnamefont
  {Sinaasappel}}, \bibinfo {author} {\bibfnamefont {H.}~\bibnamefont {Tuazon}},
  \bibinfo {author} {\bibfnamefont {I.}~\bibnamefont {Tiwari}}, \bibinfo
  {author} {\bibfnamefont {V.~P.}\ \bibnamefont {Patil}},\ and\ \bibinfo
  {author} {\bibfnamefont {M.~S.}\ \bibnamefont {Bhamla}},\ }\bibfield  {title}
  {\bibinfo {title} {Worm blobs as entangled living polymers: from topological
  active matter to flexible soft robot collectives},\ }\href
  {https://doi.org/10.1039/d3sm00542a} {\bibfield  {journal} {\bibinfo
  {journal} {Soft Matter}\ }\textbf {\bibinfo {volume} {19}},\ \bibinfo {pages}
  {7057–7069} (\bibinfo {year} {2023})}\BibitemShut {NoStop}%
\bibitem [{\citenamefont {Taylor}(1952)}]{taylor1952}%
  \BibitemOpen
  \bibfield  {author} {\bibinfo {author} {\bibfnamefont {G.}~\bibnamefont
  {Taylor}},\ }\bibfield  {title} {\bibinfo {title} {Analysis of the swimming
  of long and narrow animals},\ }\href {http://www.jstor.org/stable/99081}
  {\bibfield  {journal} {\bibinfo  {journal} {Proc. R. Soc. A}\ }\textbf
  {\bibinfo {volume} {214}},\ \bibinfo {pages} {158} (\bibinfo {year}
  {1952})}\BibitemShut {NoStop}%
\bibitem [{\citenamefont {Stin}\ \emph {et~al.}(2024)\citenamefont {Stin},
  \citenamefont {Godoy‐Diana}, \citenamefont {Bonnet},\ and\ \citenamefont
  {Herrel}}]{Stin2024}%
  \BibitemOpen
  \bibfield  {author} {\bibinfo {author} {\bibfnamefont {V.}~\bibnamefont
  {Stin}}, \bibinfo {author} {\bibfnamefont {R.}~\bibnamefont {Godoy‐Diana}},
  \bibinfo {author} {\bibfnamefont {X.}~\bibnamefont {Bonnet}},\ and\ \bibinfo
  {author} {\bibfnamefont {A.}~\bibnamefont {Herrel}},\ }\bibfield  {title}
  {\bibinfo {title} {Form and function of anguilliform swimming},\ }\href
  {https://doi.org/10.1111/brv.13116} {\bibfield  {journal} {\bibinfo
  {journal} {Biol. Rev. Camb. Philos. Soc.}\ }\textbf {\bibinfo {volume}
  {99}},\ \bibinfo {pages} {2190–2210} (\bibinfo {year} {2024})}\BibitemShut
  {NoStop}%
\bibitem [{\citenamefont {Yeaton}\ \emph {et~al.}(2020)\citenamefont {Yeaton},
  \citenamefont {Ross}, \citenamefont {Baumgardner},\ and\ \citenamefont
  {Socha}}]{Yeaton2020}%
  \BibitemOpen
  \bibfield  {author} {\bibinfo {author} {\bibfnamefont {I.~J.}\ \bibnamefont
  {Yeaton}}, \bibinfo {author} {\bibfnamefont {S.~D.}\ \bibnamefont {Ross}},
  \bibinfo {author} {\bibfnamefont {G.~A.}\ \bibnamefont {Baumgardner}},\ and\
  \bibinfo {author} {\bibfnamefont {J.~J.}\ \bibnamefont {Socha}},\ }\bibfield
  {title} {\bibinfo {title} {Undulation enables gliding in flying snakes},\
  }\href@noop {} {\bibfield  {journal} {\bibinfo  {journal} {Nat. Phys.}\
  }\textbf {\bibinfo {volume} {16}},\ \bibinfo {pages} {974} (\bibinfo {year}
  {2020})}\BibitemShut {NoStop}%
\bibitem [{\citenamefont {Zheng}\ \emph {et~al.}(2023)\citenamefont {Zheng},
  \citenamefont {Brandenbourger}, \citenamefont {Robinet}, \citenamefont
  {Schall}, \citenamefont {Lerner},\ and\ \citenamefont {Coulais}}]{Zheng2023}%
  \BibitemOpen
  \bibfield  {author} {\bibinfo {author} {\bibfnamefont {E.}~\bibnamefont
  {Zheng}}, \bibinfo {author} {\bibfnamefont {M.}~\bibnamefont
  {Brandenbourger}}, \bibinfo {author} {\bibfnamefont {L.}~\bibnamefont
  {Robinet}}, \bibinfo {author} {\bibfnamefont {P.}~\bibnamefont {Schall}},
  \bibinfo {author} {\bibfnamefont {E.}~\bibnamefont {Lerner}},\ and\ \bibinfo
  {author} {\bibfnamefont {C.}~\bibnamefont {Coulais}},\ }\bibfield  {title}
  {\bibinfo {title} {Self-oscillation and synchronization transitions in
  elastoactive structures},\ }\href@noop {} {\bibfield  {journal} {\bibinfo
  {journal} {Phys. Rev. Lett.}\ }\textbf {\bibinfo {volume} {130}},\ \bibinfo
  {pages} {178202} (\bibinfo {year} {2023})}\BibitemShut {NoStop}%
\bibitem [{\citenamefont {De~Canio}\ \emph {et~al.}(2017)\citenamefont
  {De~Canio}, \citenamefont {Lauga},\ and\ \citenamefont
  {Goldstein}}]{de2017spontaneous}%
  \BibitemOpen
  \bibfield  {author} {\bibinfo {author} {\bibfnamefont {G.}~\bibnamefont
  {De~Canio}}, \bibinfo {author} {\bibfnamefont {E.}~\bibnamefont {Lauga}},\
  and\ \bibinfo {author} {\bibfnamefont {R.~E.}\ \bibnamefont {Goldstein}},\
  }\bibfield  {title} {\bibinfo {title} {Spontaneous oscillations of elastic
  filaments induced by molecular motors},\ }\href@noop {} {\bibfield  {journal}
  {\bibinfo  {journal} {J. R. Soc. Interface.}\ }\textbf {\bibinfo {volume}
  {14}},\ \bibinfo {pages} {20170491} (\bibinfo {year} {2017})}\BibitemShut
  {NoStop}%
\bibitem [{\citenamefont {Bayly}\ and\ \citenamefont
  {Dutcher}(2016)}]{bayly2016steady}%
  \BibitemOpen
  \bibfield  {author} {\bibinfo {author} {\bibfnamefont {P.~V.}\ \bibnamefont
  {Bayly}}\ and\ \bibinfo {author} {\bibfnamefont {S.~K.}\ \bibnamefont
  {Dutcher}},\ }\bibfield  {title} {\bibinfo {title} {Steady dynein forces
  induce flutter instability and propagating waves in mathematical models of
  flagella},\ }\href@noop {} {\bibfield  {journal} {\bibinfo  {journal} {J. R.
  Soc. Interface.}\ }\textbf {\bibinfo {volume} {13}},\ \bibinfo {pages}
  {20160523} (\bibinfo {year} {2016})}\BibitemShut {NoStop}%
\bibitem [{\citenamefont {Man}\ \emph {et~al.}(2020)\citenamefont {Man},
  \citenamefont {Ling},\ and\ \citenamefont {Kanso}}]{man2020cilia}%
  \BibitemOpen
  \bibfield  {author} {\bibinfo {author} {\bibfnamefont {Y.}~\bibnamefont
  {Man}}, \bibinfo {author} {\bibfnamefont {F.}~\bibnamefont {Ling}},\ and\
  \bibinfo {author} {\bibfnamefont {E.}~\bibnamefont {Kanso}},\ }\bibfield
  {title} {\bibinfo {title} {Cilia oscillations},\ }\href@noop {} {\bibfield
  {journal} {\bibinfo  {journal} {Philos. Trans. R. Soc. B.}\ }\textbf
  {\bibinfo {volume} {375}},\ \bibinfo {pages} {20190157} (\bibinfo {year}
  {2020})}\BibitemShut {NoStop}%
\bibitem [{\citenamefont {Franks}\ \emph {et~al.}(2016)\citenamefont {Franks},
  \citenamefont {Worley}, \citenamefont {Grant}, \citenamefont {Gorman},
  \citenamefont {Vizard}, \citenamefont {Plackett}, \citenamefont {Doran},
  \citenamefont {Gamble}, \citenamefont {Stumpe},\ and\ \citenamefont
  {Sendova-Franks}}]{franks2016social}%
  \BibitemOpen
  \bibfield  {author} {\bibinfo {author} {\bibfnamefont {N.~R.}\ \bibnamefont
  {Franks}}, \bibinfo {author} {\bibfnamefont {A.}~\bibnamefont {Worley}},
  \bibinfo {author} {\bibfnamefont {K.~A.}\ \bibnamefont {Grant}}, \bibinfo
  {author} {\bibfnamefont {A.~R.}\ \bibnamefont {Gorman}}, \bibinfo {author}
  {\bibfnamefont {V.}~\bibnamefont {Vizard}}, \bibinfo {author} {\bibfnamefont
  {H.}~\bibnamefont {Plackett}}, \bibinfo {author} {\bibfnamefont
  {C.}~\bibnamefont {Doran}}, \bibinfo {author} {\bibfnamefont {M.~L.}\
  \bibnamefont {Gamble}}, \bibinfo {author} {\bibfnamefont {M.~C.}\
  \bibnamefont {Stumpe}},\ and\ \bibinfo {author} {\bibfnamefont {A.~B.}\
  \bibnamefont {Sendova-Franks}},\ }\bibfield  {title} {\bibinfo {title}
  {Social behaviour and collective motion in plant-animal worms},\ }\href@noop
  {} {\bibfield  {journal} {\bibinfo  {journal} {Proc. R. Soc. B}\ }\textbf
  {\bibinfo {volume} {283}},\ \bibinfo {pages} {20152946} (\bibinfo {year}
  {2016})}\BibitemShut {NoStop}%
\bibitem [{\citenamefont {Liverpool}\ \emph {et~al.}(2001)\citenamefont
  {Liverpool}, \citenamefont {Maggs},\ and\ \citenamefont
  {Ajdari}}]{liverpool2001viscoelasticity}%
  \BibitemOpen
  \bibfield  {author} {\bibinfo {author} {\bibfnamefont {T.~B.}\ \bibnamefont
  {Liverpool}}, \bibinfo {author} {\bibfnamefont {A.~C.}\ \bibnamefont
  {Maggs}},\ and\ \bibinfo {author} {\bibfnamefont {A.}~\bibnamefont
  {Ajdari}},\ }\bibfield  {title} {\bibinfo {title} {Viscoelasticity of
  solutions of motile polymers},\ }\href@noop {} {\bibfield  {journal}
  {\bibinfo  {journal} {Physical review letters}\ }\textbf {\bibinfo {volume}
  {86}},\ \bibinfo {pages} {4171} (\bibinfo {year} {2001})}\BibitemShut
  {NoStop}%
\bibitem [{\citenamefont {Bianco}\ \emph {et~al.}(2018)\citenamefont {Bianco},
  \citenamefont {Locatelli},\ and\ \citenamefont
  {Malgaretti}}]{bianco2018globulelike}%
  \BibitemOpen
  \bibfield  {author} {\bibinfo {author} {\bibfnamefont {V.}~\bibnamefont
  {Bianco}}, \bibinfo {author} {\bibfnamefont {E.}~\bibnamefont {Locatelli}},\
  and\ \bibinfo {author} {\bibfnamefont {P.}~\bibnamefont {Malgaretti}},\
  }\bibfield  {title} {\bibinfo {title} {Globulelike conformation and enhanced
  diffusion of active polymers},\ }\href@noop {} {\bibfield  {journal}
  {\bibinfo  {journal} {Phys. Rev. Lett.}\ }\textbf {\bibinfo {volume} {121}},\
  \bibinfo {pages} {217802} (\bibinfo {year} {2018})}\BibitemShut {NoStop}%
\bibitem [{\citenamefont {Vliegenthart}\ \emph {et~al.}(2020)\citenamefont
  {Vliegenthart}, \citenamefont {Ravichandran}, \citenamefont {Ripoll},
  \citenamefont {Auth},\ and\ \citenamefont
  {Gompper}}]{vliegenthart2020filamentous}%
  \BibitemOpen
  \bibfield  {author} {\bibinfo {author} {\bibfnamefont {G.~A.}\ \bibnamefont
  {Vliegenthart}}, \bibinfo {author} {\bibfnamefont {A.}~\bibnamefont
  {Ravichandran}}, \bibinfo {author} {\bibfnamefont {M.}~\bibnamefont
  {Ripoll}}, \bibinfo {author} {\bibfnamefont {T.}~\bibnamefont {Auth}},\ and\
  \bibinfo {author} {\bibfnamefont {G.}~\bibnamefont {Gompper}},\ }\bibfield
  {title} {\bibinfo {title} {Filamentous active matter: Band formation,
  bending, buckling, and defects},\ }\href@noop {} {\bibfield  {journal}
  {\bibinfo  {journal} {Sci. Adv.}\ }\textbf {\bibinfo {volume} {6}},\ \bibinfo
  {pages} {eaaw9975} (\bibinfo {year} {2020})}\BibitemShut {NoStop}%
\bibitem [{\citenamefont {Kruse}\ \emph {et~al.}(2004)\citenamefont {Kruse},
  \citenamefont {Joanny}, \citenamefont {J{\"u}licher}, \citenamefont {Prost},\
  and\ \citenamefont {Sekimoto}}]{kruse2004asters}%
  \BibitemOpen
  \bibfield  {author} {\bibinfo {author} {\bibfnamefont {K.}~\bibnamefont
  {Kruse}}, \bibinfo {author} {\bibfnamefont {J.-F.}\ \bibnamefont {Joanny}},
  \bibinfo {author} {\bibfnamefont {F.}~\bibnamefont {J{\"u}licher}}, \bibinfo
  {author} {\bibfnamefont {J.}~\bibnamefont {Prost}},\ and\ \bibinfo {author}
  {\bibfnamefont {K.}~\bibnamefont {Sekimoto}},\ }\bibfield  {title} {\bibinfo
  {title} {Asters, vortices, and rotating spirals in active gels of polar
  filaments},\ }\href@noop {} {\bibfield  {journal} {\bibinfo  {journal} {Phys.
  Rev. Lett.}\ }\textbf {\bibinfo {volume} {92}},\ \bibinfo {pages} {078101}
  (\bibinfo {year} {2004})}\BibitemShut {NoStop}%
\bibitem [{\citenamefont {Sumino}\ \emph {et~al.}(2012)\citenamefont {Sumino},
  \citenamefont {Nagai}, \citenamefont {Shitaka}, \citenamefont {Tanaka},
  \citenamefont {Yoshikawa}, \citenamefont {Chat{\'e}},\ and\ \citenamefont
  {Oiwa}}]{sumino2012large}%
  \BibitemOpen
  \bibfield  {author} {\bibinfo {author} {\bibfnamefont {Y.}~\bibnamefont
  {Sumino}}, \bibinfo {author} {\bibfnamefont {K.~H.}\ \bibnamefont {Nagai}},
  \bibinfo {author} {\bibfnamefont {Y.}~\bibnamefont {Shitaka}}, \bibinfo
  {author} {\bibfnamefont {D.}~\bibnamefont {Tanaka}}, \bibinfo {author}
  {\bibfnamefont {K.}~\bibnamefont {Yoshikawa}}, \bibinfo {author}
  {\bibfnamefont {H.}~\bibnamefont {Chat{\'e}}},\ and\ \bibinfo {author}
  {\bibfnamefont {K.}~\bibnamefont {Oiwa}},\ }\bibfield  {title} {\bibinfo
  {title} {Large-scale vortex lattice emerging from collectively moving
  microtubules},\ }\href@noop {} {\bibfield  {journal} {\bibinfo  {journal}
  {Nature}\ }\textbf {\bibinfo {volume} {483}},\ \bibinfo {pages} {448}
  (\bibinfo {year} {2012})}\BibitemShut {NoStop}%
\bibitem [{\citenamefont {Huber}\ \emph {et~al.}(2018)\citenamefont {Huber},
  \citenamefont {Suzuki}, \citenamefont {Kr{\"u}ger}, \citenamefont {Frey},\
  and\ \citenamefont {Bausch}}]{huber2018emergence}%
  \BibitemOpen
  \bibfield  {author} {\bibinfo {author} {\bibfnamefont {L.}~\bibnamefont
  {Huber}}, \bibinfo {author} {\bibfnamefont {R.}~\bibnamefont {Suzuki}},
  \bibinfo {author} {\bibfnamefont {T.}~\bibnamefont {Kr{\"u}ger}}, \bibinfo
  {author} {\bibfnamefont {E.}~\bibnamefont {Frey}},\ and\ \bibinfo {author}
  {\bibfnamefont {A.~R.}\ \bibnamefont {Bausch}},\ }\bibfield  {title}
  {\bibinfo {title} {Emergence of coexisting ordered states in active matter
  systems},\ }\href@noop {} {\bibfield  {journal} {\bibinfo  {journal}
  {Science}\ }\textbf {\bibinfo {volume} {361}},\ \bibinfo {pages} {255}
  (\bibinfo {year} {2018})}\BibitemShut {NoStop}%
\bibitem [{\citenamefont {Sciortino}\ and\ \citenamefont
  {Bausch}(2021)}]{sciortino2021pattern}%
  \BibitemOpen
  \bibfield  {author} {\bibinfo {author} {\bibfnamefont {A.}~\bibnamefont
  {Sciortino}}\ and\ \bibinfo {author} {\bibfnamefont {A.~R.}\ \bibnamefont
  {Bausch}},\ }\bibfield  {title} {\bibinfo {title} {Pattern formation and
  polarity sorting of driven actin filaments on lipid membranes},\ }\href@noop
  {} {\bibfield  {journal} {\bibinfo  {journal} {Proc. Natl. Acad. Sci. USA}\
  }\textbf {\bibinfo {volume} {118}},\ \bibinfo {pages} {e2017047118} (\bibinfo
  {year} {2021})}\BibitemShut {NoStop}%
\bibitem [{\citenamefont {Doostmohammadi}\ \emph {et~al.}(2016)\citenamefont
  {Doostmohammadi}, \citenamefont {Thampi},\ and\ \citenamefont
  {Yeomans}}]{doostmohammadi2016defect}%
  \BibitemOpen
  \bibfield  {author} {\bibinfo {author} {\bibfnamefont {A.}~\bibnamefont
  {Doostmohammadi}}, \bibinfo {author} {\bibfnamefont {S.~P.}\ \bibnamefont
  {Thampi}},\ and\ \bibinfo {author} {\bibfnamefont {J.~M.}\ \bibnamefont
  {Yeomans}},\ }\bibfield  {title} {\bibinfo {title} {Defect-mediated
  morphologies in growing cell colonies},\ }\href@noop {} {\bibfield  {journal}
  {\bibinfo  {journal} {Phys. Rev. Lett.}\ ,\ \bibinfo {pages} {048102}}
  (\bibinfo {year} {2016})}\BibitemShut {NoStop}%
\bibitem [{\citenamefont {Repula}\ \emph {et~al.}(2024)\citenamefont {Repula},
  \citenamefont {Gates}, \citenamefont {Cameron},\ and\ \citenamefont
  {Smalyukh}}]{repula2024photosynthetically}%
  \BibitemOpen
  \bibfield  {author} {\bibinfo {author} {\bibfnamefont {A.}~\bibnamefont
  {Repula}}, \bibinfo {author} {\bibfnamefont {C.}~\bibnamefont {Gates}},
  \bibinfo {author} {\bibfnamefont {J.~C.}\ \bibnamefont {Cameron}},\ and\
  \bibinfo {author} {\bibfnamefont {I.~I.}\ \bibnamefont {Smalyukh}},\
  }\bibfield  {title} {\bibinfo {title} {Photosynthetically-powered phototactic
  active nematic liquid crystal fluids and gels},\ }\href@noop {} {\bibfield
  {journal} {\bibinfo  {journal} {Commun. Mater.}\ }\textbf {\bibinfo {volume}
  {5}},\ \bibinfo {pages} {37} (\bibinfo {year} {2024})}\BibitemShut {NoStop}%
\bibitem [{\citenamefont {Kawaguchi}\ \emph {et~al.}(2017)\citenamefont
  {Kawaguchi}, \citenamefont {Kageyama},\ and\ \citenamefont
  {Sano}}]{kawaguchi2017topological}%
  \BibitemOpen
  \bibfield  {author} {\bibinfo {author} {\bibfnamefont {K.}~\bibnamefont
  {Kawaguchi}}, \bibinfo {author} {\bibfnamefont {R.}~\bibnamefont
  {Kageyama}},\ and\ \bibinfo {author} {\bibfnamefont {M.}~\bibnamefont
  {Sano}},\ }\bibfield  {title} {\bibinfo {title} {Topological defects control
  collective dynamics in neural progenitor cell cultures},\ }\href@noop {}
  {\bibfield  {journal} {\bibinfo  {journal} {Nature}\ }\textbf {\bibinfo
  {volume} {545}},\ \bibinfo {pages} {327} (\bibinfo {year}
  {2017})}\BibitemShut {NoStop}%
\bibitem [{\citenamefont {Ishimoto}\ \emph {et~al.}(2025)\citenamefont
  {Ishimoto}, \citenamefont {Moreau},\ and\ \citenamefont
  {Herault}}]{ishimoto2025robust}%
  \BibitemOpen
  \bibfield  {author} {\bibinfo {author} {\bibfnamefont {K.}~\bibnamefont
  {Ishimoto}}, \bibinfo {author} {\bibfnamefont {C.}~\bibnamefont {Moreau}},\
  and\ \bibinfo {author} {\bibfnamefont {J.}~\bibnamefont {Herault}},\
  }\bibfield  {title} {\bibinfo {title} {Robust undulatory locomotion through
  neuromechanical adjustments in a dissipative medium},\ }\href@noop {}
  {\bibfield  {journal} {\bibinfo  {journal} {Journal of the Royal Society
  Interface}\ }\textbf {\bibinfo {volume} {22}},\ \bibinfo {pages} {20240688}
  (\bibinfo {year} {2025})}\BibitemShut {NoStop}%
\bibitem [{\citenamefont {Keaveny}\ and\ \citenamefont
  {Brown}(2017)}]{keaveny2017predicting}%
  \BibitemOpen
  \bibfield  {author} {\bibinfo {author} {\bibfnamefont {E.~E.}\ \bibnamefont
  {Keaveny}}\ and\ \bibinfo {author} {\bibfnamefont {A.~E.}\ \bibnamefont
  {Brown}},\ }\bibfield  {title} {\bibinfo {title} {Predicting path from
  undulations for c. elegans using linear and nonlinear resistive force
  theory},\ }\href@noop {} {\bibfield  {journal} {\bibinfo  {journal} {Phys.
  Biol.}\ }\textbf {\bibinfo {volume} {14}},\ \bibinfo {pages} {025001}
  (\bibinfo {year} {2017})}\BibitemShut {NoStop}%
\bibitem [{\citenamefont {Rabets}\ \emph {et~al.}(2014)\citenamefont {Rabets},
  \citenamefont {Backholm}, \citenamefont {Dalnoki-Veress},\ and\ \citenamefont
  {Ryu}}]{rabets2014direct}%
  \BibitemOpen
  \bibfield  {author} {\bibinfo {author} {\bibfnamefont {Y.}~\bibnamefont
  {Rabets}}, \bibinfo {author} {\bibfnamefont {M.}~\bibnamefont {Backholm}},
  \bibinfo {author} {\bibfnamefont {K.}~\bibnamefont {Dalnoki-Veress}},\ and\
  \bibinfo {author} {\bibfnamefont {W.~S.}\ \bibnamefont {Ryu}},\ }\bibfield
  {title} {\bibinfo {title} {Direct measurements of drag forces in c. elegans
  crawling locomotion},\ }\href@noop {} {\bibfield  {journal} {\bibinfo
  {journal} {Biophys. J.}\ }\textbf {\bibinfo {volume} {107}},\ \bibinfo
  {pages} {1980} (\bibinfo {year} {2014})}\BibitemShut {NoStop}%
\bibitem [{\citenamefont {Iosilevskii}\ and\ \citenamefont
  {Rashkovsky}(2020)}]{iosilevskii2020hydrodynamics}%
  \BibitemOpen
  \bibfield  {author} {\bibinfo {author} {\bibfnamefont {G.}~\bibnamefont
  {Iosilevskii}}\ and\ \bibinfo {author} {\bibfnamefont {A.}~\bibnamefont
  {Rashkovsky}},\ }\bibfield  {title} {\bibinfo {title} {Hydrodynamics of a
  twisting slender swimmer},\ }\href@noop {} {\bibfield  {journal} {\bibinfo
  {journal} {R. Soc. Open Sci.}\ }\textbf {\bibinfo {volume} {7}},\ \bibinfo
  {pages} {200754} (\bibinfo {year} {2020})}\BibitemShut {NoStop}%
\bibitem [{\citenamefont {Ladoux}\ and\ \citenamefont
  {M{\`e}ge}(2017)}]{ladoux2017mechanobiology}%
  \BibitemOpen
  \bibfield  {author} {\bibinfo {author} {\bibfnamefont {B.}~\bibnamefont
  {Ladoux}}\ and\ \bibinfo {author} {\bibfnamefont {R.-M.}\ \bibnamefont
  {M{\`e}ge}},\ }\bibfield  {title} {\bibinfo {title} {Mechanobiology of
  collective cell behaviours},\ }\href@noop {} {\bibfield  {journal} {\bibinfo
  {journal} {Nat. Rev. Mol. Cell Biol.}\ }\textbf {\bibinfo {volume} {18}},\
  \bibinfo {pages} {743} (\bibinfo {year} {2017})}\BibitemShut {NoStop}%
\bibitem [{\citenamefont {Martyushev}\ and\ \citenamefont
  {Seleznev}(2006)}]{martyushev2006maximum}%
  \BibitemOpen
  \bibfield  {author} {\bibinfo {author} {\bibfnamefont {L.~M.}\ \bibnamefont
  {Martyushev}}\ and\ \bibinfo {author} {\bibfnamefont {V.~D.}\ \bibnamefont
  {Seleznev}},\ }\bibfield  {title} {\bibinfo {title} {Maximum entropy
  production principle in physics, chemistry and biology},\ }\href@noop {}
  {\bibfield  {journal} {\bibinfo  {journal} {Phys. Rep.}\ }\textbf {\bibinfo
  {volume} {426}},\ \bibinfo {pages} {1} (\bibinfo {year} {2006})}\BibitemShut
  {NoStop}%
\bibitem [{\citenamefont {Davies}\ \emph {et~al.}(2013)\citenamefont {Davies},
  \citenamefont {Rieper},\ and\ \citenamefont {Tuszynski}}]{davies2013self}%
  \BibitemOpen
  \bibfield  {author} {\bibinfo {author} {\bibfnamefont {P.~C.}\ \bibnamefont
  {Davies}}, \bibinfo {author} {\bibfnamefont {E.}~\bibnamefont {Rieper}},\
  and\ \bibinfo {author} {\bibfnamefont {J.~A.}\ \bibnamefont {Tuszynski}},\
  }\bibfield  {title} {\bibinfo {title} {Self-organization and entropy
  reduction in a living cell},\ }\href@noop {} {\bibfield  {journal} {\bibinfo
  {journal} {Biosyst.}\ }\textbf {\bibinfo {volume} {111}},\ \bibinfo {pages}
  {1} (\bibinfo {year} {2013})}\BibitemShut {NoStop}%
\bibitem [{\citenamefont {Adami}\ \emph {et~al.}(2000)\citenamefont {Adami},
  \citenamefont {Ofria},\ and\ \citenamefont {Collier}}]{adami2000evolution}%
  \BibitemOpen
  \bibfield  {author} {\bibinfo {author} {\bibfnamefont {C.}~\bibnamefont
  {Adami}}, \bibinfo {author} {\bibfnamefont {C.}~\bibnamefont {Ofria}},\ and\
  \bibinfo {author} {\bibfnamefont {T.~C.}\ \bibnamefont {Collier}},\
  }\bibfield  {title} {\bibinfo {title} {Evolution of biological complexity},\
  }\href@noop {} {\bibfield  {journal} {\bibinfo  {journal} {Proc. Natl. Acad.
  Sci. USA}\ }\textbf {\bibinfo {volume} {97}},\ \bibinfo {pages} {4463}
  (\bibinfo {year} {2000})}\BibitemShut {NoStop}%
\bibitem [{\citenamefont {Adami}(2002)}]{adami2002complexity}%
  \BibitemOpen
  \bibfield  {author} {\bibinfo {author} {\bibfnamefont {C.}~\bibnamefont
  {Adami}},\ }\bibfield  {title} {\bibinfo {title} {What is complexity?},\
  }\href@noop {} {\bibfield  {journal} {\bibinfo  {journal} {Bio{E}ssays}\
  }\textbf {\bibinfo {volume} {24}},\ \bibinfo {pages} {1085} (\bibinfo {year}
  {2002})}\BibitemShut {NoStop}%
\bibitem [{\citenamefont {Bonchev}\ and\ \citenamefont
  {Buck}(2005)}]{bonchev2005quantitative}%
  \BibitemOpen
  \bibfield  {author} {\bibinfo {author} {\bibfnamefont {D.}~\bibnamefont
  {Bonchev}}\ and\ \bibinfo {author} {\bibfnamefont {G.~A.}\ \bibnamefont
  {Buck}},\ }\bibfield  {title} {\bibinfo {title} {Quantitative {M}easures of
  {N}etwork {C}omplexity},\ }in\ \href@noop {} {\emph {\bibinfo {booktitle}
  {Complexity in {C}hemistry, {B}iology, and {E}cology}}}\ (\bibinfo
  {publisher} {Springer},\ \bibinfo {year} {2005})\ pp.\ \bibinfo {pages}
  {191--235}\BibitemShut {NoStop}%
\bibitem [{\citenamefont {Wang}\ \emph {et~al.}(2023)\citenamefont {Wang},
  \citenamefont {Pierce}, \citenamefont {Kojouharov}, \citenamefont {Chong},
  \citenamefont {Diaz}, \citenamefont {Lu},\ and\ \citenamefont
  {Goldman}}]{Wang2023}%
  \BibitemOpen
  \bibfield  {author} {\bibinfo {author} {\bibfnamefont {T.}~\bibnamefont
  {Wang}}, \bibinfo {author} {\bibfnamefont {C.}~\bibnamefont {Pierce}},
  \bibinfo {author} {\bibfnamefont {V.}~\bibnamefont {Kojouharov}}, \bibinfo
  {author} {\bibfnamefont {B.}~\bibnamefont {Chong}}, \bibinfo {author}
  {\bibfnamefont {K.}~\bibnamefont {Diaz}}, \bibinfo {author} {\bibfnamefont
  {H.}~\bibnamefont {Lu}},\ and\ \bibinfo {author} {\bibfnamefont {D.~I.}\
  \bibnamefont {Goldman}},\ }\bibfield  {title} {\bibinfo {title} {Mechanical
  intelligence simplifies control in terrestrial limbless locomotion},\ }\href
  {http://dx.doi.org/10.1126/scirobotics.adi2243} {\bibfield  {journal}
  {\bibinfo  {journal} {Sci. Robot.}\ }\textbf {\bibinfo {volume} {8}},\
  \bibinfo {pages} {eadi2243} (\bibinfo {year} {2023})}\BibitemShut {NoStop}%
\bibitem [{\citenamefont {Moreau}\ \emph {et~al.}(2024)\citenamefont {Moreau},
  \citenamefont {Walker}, \citenamefont {Poon}, \citenamefont {Soto},
  \citenamefont {Goldman}, \citenamefont {Gaffney},\ and\ \citenamefont
  {Wan}}]{Moreau2024}%
  \BibitemOpen
  \bibfield  {author} {\bibinfo {author} {\bibfnamefont {C.}~\bibnamefont
  {Moreau}}, \bibinfo {author} {\bibfnamefont {B.~J.}\ \bibnamefont {Walker}},
  \bibinfo {author} {\bibfnamefont {R.~N.}\ \bibnamefont {Poon}}, \bibinfo
  {author} {\bibfnamefont {D.}~\bibnamefont {Soto}}, \bibinfo {author}
  {\bibfnamefont {D.~I.}\ \bibnamefont {Goldman}}, \bibinfo {author}
  {\bibfnamefont {E.~A.}\ \bibnamefont {Gaffney}},\ and\ \bibinfo {author}
  {\bibfnamefont {K.~Y.}\ \bibnamefont {Wan}},\ }\bibfield  {title} {\bibinfo
  {title} {Minimal design of a synthetic cilium},\ }\href@noop {} {\bibfield
  {journal} {\bibinfo  {journal} {Phys. Rev. Res.}\ }\textbf {\bibinfo {volume}
  {6}},\ \bibinfo {pages} {L042061} (\bibinfo {year} {2024})}\BibitemShut
  {NoStop}%
\end{thebibliography}

%

\end{document}